\documentclass{jpp}
\usepackage{graphicx}
\usepackage{epstopdf, epsfig}
\usepackage{hyperref}

\usepackage{amsmath}	
\usepackage{amssymb}	

\usepackage{enumitem}

\usepackage{natbib}
\usepackage[dvipsnames]{xcolor}

\newcommand{\fullplot}[1]{#1}

\shorttitle{3D reconnection with $\sigma_h= 1$}
\shortauthor{G.~R. Werner and D.~A. Uzdensky}

\title{Reconnection and particle acceleration in 3D current sheet evolution in moderately-magnetized astrophysical pair plasma}

\author{Gregory R. Werner\aff{1}
  \corresp{\email{Greg.Werner@colorado.edu}},
  \and
  Dmitri A. Uzdensky\aff{1}
 }

\affiliation{\aff{1}Center for Integrated Plasma Studies,
  Physics Department,
  390 UCB, University of Colorado, Boulder, CO  80309, USA
}

\date{\bf Draft version: \today }

\begin{document}
\maketitle

\begin{abstract}
Magnetic reconnection, a plasma process converting magnetic energy to particle kinetic energy, is often invoked to explain magnetic energy releases powering high-energy flares in astrophysical sources including pulsar wind nebulae and black hole jets.  Reconnection is usually seen as the (essentially 2D) nonlinear evolution of the tearing instability disrupting a thin current sheet. To test how this process operates in~3D, we conduct a comprehensive particle-in-cell simulation study comparing 2D and~3D evolution of long, thin current sheets in moderately-magnetized, collisionless, relativistically-hot electron-positron plasma, and find dramatic differences. We first systematically characterize this process in~2D, where classic, hierarchical plasmoid-chain reconnection determines energy release, and explore a wide range of initial configurations, guide magnetic field strengths, and system sizes. We then show that 3D simulations of similar configurations exhibit a diversity of behaviours, including some where energy release is determined by the nonlinear relativistic drift-kink instability. Thus, 3D current-sheet evolution is not always fundamentally classical reconnection with perturbing 3D effects, but, rather, a complex interplay of multiple linear and nonlinear instabilities whose relative importance depends sensitively on the ambient plasma, minor configuration details, and even stochastic events. It often yields slower but longer-lasting and ultimately greater magnetic energy release than in~2D. Intriguingly, nonthermal particle acceleration is astonishingly robust, depending on the upstream magnetization and guide field, but otherwise yielding similar particle energy spectra in~2D and~3D. Though the variety of underlying current-sheet behaviours is interesting, the similarities in overall energy release and particle spectra may be more remarkable.
\end{abstract}



\newpage
\tableofcontents

\section{Introduction}
\label{sec-intro}

Magnetic reconnection is an important plasma physical process because it can rapidly convert magnetic energy to particle kinetic energy; it does so by rearranging the magnetic field configuration, ``breaking'' and subsequently ``reconnecting'' magnetic field lines
\citep[for a review, see, e.g.,][]{Zweibel_Yamada-2009,Yamada_etal-2010}.
Magnetic reconnection is thus believed to play a fundamental role in a wide variety of rapid---and sometimes violent and spectacular---releases of magnetic energy resulting in particle energization and radiation, from solar flares to X-ray emission in coronae of accreting black holes to flaring TeV emission in blazar jets of active galactic nuclei (AGN). 
In some high-energy astrophysical sources, such as pulsar wind nebulae (PWN) and blazar jets, reconnection may operate in a somewhat extreme regime of relativistically-hot plasma containing positrons instead of, or in addition to, ions.
In such environments, reconnection in the relativistic regime has become a promising explanation for high-energy emission,
especially since particle-in-cell (PIC) simulations have convincingly demonstrated that, 
in addition to heating
plasma, reconnection can accelerate a significant fraction of particles to very high energies, yielding a nonthermal power-law energy distribution of particles
\citep[e.g.,][]{Zenitani_Hoshino-2001,Zenitani_Hoshino-2005a,Zenitani_Hoshino-2005b,Zenitani_Hoshino-2007,Zenitani_Hoshino-2008,Jaroschek_etal-2004a,Lyubarsky_Liverts-2008,Liu_etal-2011,Sironi_Spitkovsky-2011,Sironi_Spitkovsky-2014,Bessho_Bhattacharjee-2012,Cerutti_etal-2012b,Cerutti_etal-2013,Cerutti_etal-2014a,Cerutti_etal-2014b,Kagan_etal-2013,Guo_etal-2014,Guo_etal-2015,Guo_etal-2016,Guo_etal-2016b,Guo_etal-2019,Nalewajko_etal-2015,Sironi_etal-2015,Sironi_etal-2016,Dahlin_etal-2015,Dahlin_etal-2017,Werner_etal-2016,Werner_etal-2018,Werner_etal-2019,Werner_Uzdensky-2017,Petropoulou_Sironi-2018,Ball_etal-2018,Ball_etal-2019,Li_etal-2019,Schoeffler_etal-2019,Mehlhaff_etal-2020,Sironi_Beloborodov-2020,Kilian_etal-2020,Guo_etal-2020arxiv,Hakobyan_etal-2019,Hakobyan_etal-2021,Zhang_etal-2021arxiv}.
High-energy nonthermal particle acceleration (NTPA) followed by synchrotron or inverse Compton emission could explain observed nonthermal radiation spectra.

Large 2D PIC simulations of reconnection in collisionless, relativistic pair plasma have observed fast reconnection, with reconnection rates around $E_{\rm rec}\sim 0.1 B_0 v_A/c$ (where $B_0$ is the upstream, ambient, reconnecting magnetic field, and $v_A$ is the corresponding Alfv\'{e}n speed), resulting in rapid conversion of magnetic energy to plasma energy.
In these simulations, NTPA yields power-law electron energy distributions $f(\gamma)\sim \gamma^{-p}$ (where electron energy is $\gamma m_e c^2$) with a range of slopes $p$; it appears that, depending on the environment and system size, $p$ can take on values greater than~1, and $p$ may even approach~1 in highly magnetically-dominated reconnection \citep[e.g.,][]{Guo_etal-2014,Guo_etal-2015,Werner_etal-2016}.
Since $p$ can in principle be inferred from observed radiation, these simulation results may potentially elucidate plasma parameters (such as magnetization) in astrophysical sources.
Besides the power-law slope $p$, an important result of these simulations is the determination of the high-energy cutoff $\gamma_c$ of the power law, and in particular its scaling with system size $L$, since real astrophysical systems are usually much larger (with respect to kinetic scales) than we can possibly simulate.
For systems with small $L$, simulations observe ``extreme acceleration'' consistent with particles being accelerated in reconnection electric field $E_{\rm rec}\simeq 0.1(v_A/c)B_0$ (where $B_0$ is the reconnecting magnetic field) over system size $L$ \citep{Hillas-1984,Aharonian_etal-2002}, so that particles reach energies $\gamma_c m_e c^2 \simeq eE_{\rm rec} L$ \citep{Werner_etal-2016}.
However, for larger systems, this rapid acceleration seems to stall around 
$\gamma_c \sim 4\sigma$, where $\sigma\equiv B_0^2/(4\upi n_{b0} m_e c^2)$ is
the ``cold'' magnetization parameter involving the reconnecting field $B_0$ and
the ambient plasma rest-mass energy density $n_{b0} m_e c^2$ \citep{Werner_etal-2016}.
More recent PIC simulations found that acceleration can continue well beyond this limit in extremely large systems, albeit at a significantly slower rate so that cutoff energies grow with the square root of time, resulting in $\gamma_c \sim \sqrt{L}$ \citep{Petropoulou_Sironi-2018,Hakobyan_etal-2021}.
The precise mechanism of reconnection-driven particle acceleration has received much attention \citep[e.g.,][]{Zenitani_Hoshino-2001,Drake_etal-2006,Zenitani_Hoshino-2007,Jaroschek_etal-2004a,Lyubarsky_Liverts-2008,Liu_etal-2011,Bessho_Bhattacharjee-2012,Cerutti_etal-2012a,Cerutti_etal-2013,Cerutti_etal-2014b,Kagan_etal-2013,Guo_etal-2014,Guo_etal-2015,Guo_etal-2019,Bessho_etal-2015,Nalewajko_etal-2015,Dahlin_etal-2016,Sironi_etal-2016,Werner_etal-2016,Petropoulou_Sironi-2018,Ball_etal-2018,Ball_etal-2019,Li_etal-2019,Kilian_etal-2020,Guo_etal-2020arxiv,Zhang_etal-2021arxiv};
a number of mechanisms have been considered, and it remains a matter of ongoing research to determine precisely which mechanism operates most effectively in which regime.

Most studies of reconnection in relativistic plasmas have relied on 2D simulations; an important outstanding question remains whether these are applicable to 3D events in nature.
A much smaller number of 3D simulations of reconnection in relativistic pair plasmas have been done \citep{Zenitani_Hoshino-2005b,Zenitani_Hoshino-2008,Yin_etal-2008,Liu_etal-2011,Kagan_etal-2013,Cerutti_etal-2014b,Sironi_Spitkovsky-2014,Guo_etal-2014,Guo_etal-2015,Werner_Uzdensky-2017,Guo_etal-2020arxiv,Zhang_etal-2021arxiv} \citep[there have also been 3D PIC simulations of non-relativistic electron-ion reconnection, e.g.,][]{Hesse_etal-2001,Pritchett_Coroniti-2001,Pritchett_Coroniti-2004,Lapenta_etal-2006,Daughton_etal-2011,Daughton_etal-2014,Markidis_etal-2012,Lapenta_etal-2014,Lapenta_etal-2020,Nakamura_etal-2013,Wendel_etal-2013,Dahlin_etal-2015,Dahlin_etal-2017,Lapenta_etal-2017,Le_etal-2018,Le_etal-2019,Pucci_etal-2018,Li_etal-2019,Stanier_etal-2019,Lapenta_etal-2020}.
Because of computational expense, 3D simulations have been smaller in physical system size, and have covered a much narrower range of regimes.

The first 3D PIC studies of relativistic reconnection focused on the competition between the tearing instability that leads to reconnection and the relativistic drift-kink instability (RDKI), which is forbidden in 2D reconnection simulations but in 3D can grow as fast as the tearing instability \citep{Zenitani_Hoshino-2005a,Zenitani_Hoshino-2007,Zenitani_Hoshino-2008,Yin_etal-2008}.
Later, simulations with sufficiently large system sizes observed NTPA, and moreover (despite clear manifestations of RDKI) found substantial similarities between 2D and 3D relativistic reconnection in both NTPA and general reconnection dynamics, including magnetic energy release and reconnection rates \citep{Liu_etal-2011,Kagan_etal-2013,Sironi_Spitkovsky-2014,Guo_etal-2014,Guo_etal-2015,Werner_Uzdensky-2017,Guo_etal-2020arxiv,Zhang_etal-2021arxiv}.
Our previous work \citep{Werner_Uzdensky-2017} systematically compared 2D and 3D simulations in the magnetically-dominated regime (where the upstream magnetic energy dominates over the upstream plasma thermal and rest-mass energy), and found magnetic energy conversion and NTPA in 2D and 3D to be nearly indistinguishable.
This was found to be the case over a range of guide magnetic fields,
$0 \leq B_{gz}/B_0 \leq 1$ (although both energy conversion and NTPA were suppressed by stronger guide field).
In the nonrelativistic regime, where NTPA occurs but may not yield power laws, 2D and 3D reconnection (with guide field) have also been found to be similar, but with slightly enhanced NTPA in 3D \citep{Dahlin_etal-2015}.
We note in passing that all of these 3D simulations began with a thin initial Harris or force-free current sheet.
The close resemblance between 2D and 3D reconnection justifies the use of 2D simulations to model natural 3D systems, allowing us to simulate larger system sizes with lower computational expense.
However, some have supposed that 3D reconnection might behave somewhat or even drastically differently, because of 3D instabilities and turbulence \citep[e.g.,][]{Lazarian_Vishniac-1999,Zenitani_Hoshino-2008,Takamoto_etal-2015,Lazarian_etal-BookChapter2016,Beresnyak-2017,Beresnyak-2018,Munoz_Buechner-2018,Takamoto-2018,ZhouX_etal-2018,Boozer-2019,Lapenta_etal-2020,Lazarian_etal-2020} 

\vspace{\baselineskip}
{\bf In this paper,} we use PIC simulation to study both 2D and 3D magnetic reconnection in an ultrarelativistically-hot pair plasma in the moderately-magnetized regime where the upstream magnetic energy is comparable to the upstream plasma thermal energy (i.e., $\sigma_h=1$, where the ``hot'' magnetization $\sigma_h$ is defined in~\S\ref{sec:setup}).
In both 2D and 3D, we systematically explore the effects of different initial current sheet configurations, as well as the effect of guide magnetic field; in 2D, where very large simulations are feasible, we also vary the overall system size, and in 3D, we vary the system size in the third dimension (i.e., $L_z$).

The moderately-magnetized, ultrarelativistic pair plasma regime is of interest for several reasons.  
First, it is likely relevant to astrophysical sources such as AGN jets and PWN (e.g., the Crab Nebula) in which energy might be expected to be roughly equally partitioned between plasma and magnetic field.
Second, this regime lies between magnetically-dominated relativistic reconnection and non-relativistic reconnection, both of which have been much more studied (especially in 2D); this study provides an important bridge between these regimes and also connects to reconnection in semi- or trans-relativistic \emph{electron-ion} plasma, where particles may be transrelativistic or ions may be subrelativistic while electrons are relativistic \citep[e.g.,][]{Rowan_etal-2017,Rowan_etal-2019,Ball_etal-2018,Ball_etal-2019,Werner_etal-2018}.
Third, this is perhaps the most computationally tractable regime, because the kinetic (microphysical) plasma length scales all collapse to roughly the same values, allowing the greatest dynamic range between system size and the kinetic scales, with less computational expenditure.
Using pair plasma means that electron and ``ion'' (i.e., positron) scales are the same; in the ultrarelativistic limit,  the collisionless skin depth $d_e$ and Debye length $\lambda_D$ are nearly equal, $d_e=\sqrt{3}\lambda_D$; and the moderately-magnetized regime implies that the gyroradius $\rho_{b}$ in the ambient plasma is nearly the same as $d_e$.
Thus the grid spacing $\Delta x$ can be chosen just a little smaller than all these kinetic scales (which are nearly the same), maximizing the separation between the largest (hence all the) kinetic scales and the system size $L$ (for a given computational cost).
This facilitates the study of macroscopic, large-system behaviour, such as plasmoid formation and NTPA.

Reconnection in mildly relativistic, moderately magnetized pair plasma has been previously studied with 3D PIC simulations \citep{Yin_etal-2008,Liu_etal-2011,Kagan_etal-2013,Guo_etal-2020arxiv}.
However, this work features the first extensive systematic exploration of several important parameters (such as the initial current sheet configuration and guide magnetic field strength) both in 2D and 3D, running for long times with sufficiently large system sizes that allow us to investigate NTPA.
A detailed discussion of this current work in the context of those previous studies can be found in~\S\ref{sec:discussionPrevWork}.

We will show that 
the evolution of current sheets can be substantially different in 2D and 3D, in particular in the way that they evolve and convert magnetic energy to plasma energy---but despite these differences, NTPA is not much changed (if anything, it is enhanced in 3D).
First, however, we will systematically characterize 2D reconnection in the
ultrarelativistic moderately-magnetized ($\sigma_h=1$) pair-plasma regime, as a number of parameters are varied---including initial current sheet configuration, system size, and guide magnetic field.  
Then, we will characterize current sheet evolution in 3D across a similar range of parameters, comparing with 2D reconnection.
We will identify patches in 3D simulations that exhibit signatures of classical 2D reconnection, including outflows from thin current sheets, non-ideal electric fields (parallel to and/or larger than the magnetic field), and energy transfer from fields to plasma.   
However, we will see that 3D reconnection is vastly more complicated to study than 2D reconnection, displaying a greater variety of behaviours along with greater sensitivity to the initial configuration (e.g., the initial current sheet thickness or field perturbation).
The rates of magnetic energy depletion and upstream magnetic flux reduction are generally slower in~3D, but reconnection continues for much longer times.
Ultimately, more magnetic energy is released in~3D than in~2D, because
in 2D, magnetic energy in outflows is trapped forever in plasmoids, whereas in 3D the plasmoid-like structures can decay.
We also find that, in 3D, there is a new channel for magnetic energy conversion that does not necessarily involve classical reconnection at all: the kinking of the current sheet in 3D can grow to such large amplitudes that the current sheet becomes violently and chaotically deformed, resulting in rapid magnetic energy release and turbulent thickening of the current layer.
This process is not inevitable; two initially-similar simulations can end up behaving very differently over long times because one triggers such deformation and energy release, while the other does not, even though the upstream conditions may be identical in both cases.
This sensitivity to initial conditions complicates the study of 3D reconnection by increasing the parameter space that needs to be explored.

\vspace{\baselineskip}
{\bf This paper is organized as follows.}
We will define the simulation parameters and set-up in~\S\ref{sec:setup}.
Then in~\S\ref{sec:diagnostics} we describe diagnostics used to characterize current sheet evolution, starting with precise definitions of important terms (\S\ref{sec:terminology}) and followed by detailed descriptions of diagnostics, such as how we measure the amount of unreconnected flux in a simulation and how we characterize NTPA.
In~\S\ref{sec:2d} we will describe 2D reconnection in the moderately-magnetized relativistically-hot pair-plasma regime.
We will see that in 2D, this regime exhibits behaviour that is qualitatively very similar to, e.g., magnetically-dominated relativistic reconnection, and we will use this opportunity to review the basic picture of plasmoid-dominated (multiple X-line) reconnection, to be contrasted later with 3D simulations.
In addition, we quantify the effects of varying a number of parameters (such as initial current sheet thickness, system size, and guide magnetic field) on magnetic energy conversion and NTPA.
Besides expanding our knowledge of 2D reconnection in the moderately-magnetized relativistically-hot regime, this provides a baseline for our study of 3D reconnection in~\S\ref{sec:3d}.
There we find that current sheet evolution can differ distinctly from 2D reconnection, and we investigate the effects of varying the initial current sheet configuration, varying system length $L_z$ in the third dimension, and varying guide magnetic field.
We discuss differences between the moderately-magnetized regime studied in this paper and the magnetically-dominated regime (studied in 3D in previous work) in~\S\ref{sec:highSigmah}.
We list the most important findings in~\S\ref{sec:summary} before discussing some possible impacts on plasma physics and astrophysical modelling in~\S\ref{sec:discussion}, and finally conclude with~\S\ref{sec:conclusion}.
For reference, in Appendix~\ref{sec:res2d} we present a resolution convergence study justifying the cell size~$\Delta x$ used in this work.

\section{Simulation setup}
\label{sec:setup}

The simulations in this work use a standard double-periodic 
simulation box initialized with a twice-reversing magnetic field
balanced by two thin, oppositely-directed Harris current sheets 
\citep[e.g.,][]{Werner_etal-2016,Werner_Uzdensky-2017}.
The Harris sheets (which contain drifting plasma) are superimposed upon a uniform, stationary, relativistically-hot background plasma.
Once the initial state is set, the simulation code {\tt Zeltron} evolves the plasma with no external input (e.g., no driving) according to a standard explicit, relativistic electromagnetic PIC algorithm \citep{Cerutti_etal-2013}, with periodic boundary conditions; in 2D simulations, all quantities are assumed to be uniform in the third dimension ($z$) at all times---e.g., $\partial B_x/\partial z=0$, since all derivatives with respect to $z$ are zero.

In this section, we first describe the background plasma, and then the
drifting plasma that forms the Harris current sheets.  At the end,
we summarize the parameters describing the set-up, including whether they are fixed or varied throughout this study.

Ideally, we hope to associate reconnection behaviour with the uniform
background (or upstream) pair plasma, 
described by 
\begin{itemize}[leftmargin=3mm, itemindent=0mm, labelsep=1mm]
  \item $n_{b0}$, the initial (electron plus positron) particle number density, 
  \item $\theta_b \equiv T_b/m_e c^2$, the normalized temperature,
  \item $B_0$, the (reconnecting) magnetic field in the~$x$ direction, and
  \item $B_{gz}\hat{\boldsymbol{z}}$, the initially-uniform guide magnetic field.
\end{itemize} \noindent
($\boldsymbol{B}$ is initially uniform except for $B_x$ reversing around the current sheets.)
It is useful to express $n_{b0}$, $\theta_b$, $B_0$, and $B_{gz}$ in 
terms of a nominal gyroradius~$\rho_0$ and three dimensionless parameters, $\sigma$, $\sigma_h$, and $B_{gz}/B_0$.
We define the nominal relativistic gyroradius 
$\rho_0\equiv m_ec^2/eB_0$ in place of $B_0$; thus a particle with
ultrarelativistic Lorentz factor $\gamma \gg 1$ has a gyroradius of order~$\gamma \rho_0$ in
field $B_0$, hence typical background particles have gyroradii around
$3\theta_b \rho_0$ (for $\theta_b \gg 1$).
The density~$n_{b0}$ can then be expressed in terms of $\rho_0$ and
the ``cold'' magnetization
parameter $\sigma \equiv B_0^2/(4\upi n_{b0} m_e c^2)$, which is twice
the ratio of the reconnecting magnetic field's energy density, $B_0^2/8\upi$, to plasma rest-energy density, $n_{b0}m_e c^2$.
If all magnetic energy were given to particles, their Lorentz factors would increase by $\Delta \gamma = \sigma/2$---i.e., ~$\sigma$ determines the available magnetic energy per particle. Thus 
we generally normalize lengths to a characteristic scale $\rho_c \equiv \sigma \rho_0$,
which is on the order of the gyroradii of typical energized particles.
Finally, we re-express the temperature $\theta_b$ in terms of the
hot magnetization 
$\sigma_h \equiv B_0^2/(16\upi \theta_b n_{b0} m_e c^2) = \sigma/(4\theta_b)$,
the ratio of magnetic enthalpy density $B_0^2/4\upi$ to plasma
enthalpy density (including rest-mass energy), which is 
$h=4\theta_b n_{b0} m_e c^2$ in the relativistic limit $\theta_b\gg 1$ \citep{Melzani_etal-2014b}.
We note that (for $B_{gz}=0$) the Alfv\'{e}n 4-velocity is $c\sqrt{\sigma_h}$;
i.e., the Alfv\'{e}n velocity is $v_A=c\sqrt{\sigma_h/(1+\sigma_h)}$.

In this paper we focus on the specific case of reconnection with moderate magnetization, or $\sigma_h=1$ (the majority of studies of reconnection in pair plasma have explored $\sigma_h \gg 1$, and most electron-ion reconnection studies have been nonrelativistic with $\sigma_h \ll 1$).
The upstream plasma is then completely defined by specifying, in addition to $\sigma_h=1$, the values of $\rho_0$, $B_{gz}/B_0$, and $\sigma$.
The particular value of $\rho_0$ is irrelevant; the entire simulation scales trivially with $\rho_0$.  
For astrophysical relevance and theoretical simplicity, 
we choose to study the ultrarelativistically-hot regime in which $\sigma \gg 1$ and $\theta_b\gg 1$---as with $\rho_0$, the particular value of $\sigma$ does not matter (the simulation scales trivially with $\rho_c\equiv \sigma \rho_0$ as long as we stay in the ultrarelativistically-hot regime).
When $\sigma_h=1$, the upstream magnetic and plasma energy are roughly comparable; even complete conversion of magnetic energy would not drastically increase the plasma energy.
For $B_{gz}=0$, $\sigma_h=1$ implies that the plasma beta is $\beta_{\rm plasma}=1/2$ and the
expected bulk reconnection outflow velocity is $v_A=c/\sqrt{2}$.

Important plasma length scales, in terms of $\sigma\rho_0$ and $\sigma_h$, include the average background gyroradius 
$\rho_b\equiv 3\theta_b \rho_0 = (3/4)\sigma_h^{-1} \sigma \rho_0$,
the background Debye length $\lambda_D =(1/2) \sigma_h^{-1/2} \sigma\rho_0 $,
and the collisionless skin depth 
$d_e = \sqrt{3\theta_b m_e c^2/4\upi n e^2} =\sqrt{3} \lambda_D$.

Over time, the background plasma should (at least in 2D) dominate reconnection
dynamics.
However, to start in a state susceptible to reconnection, we 
need a reversing magnetic field and its associated current sheet \citep{Kirk_Skjaeraasen-2003}.
Actually we use \emph{two} oppositely-directed
Harris current sheets to allow periodic boundary conditions in all 
directions.
The simulation box, containing both current sheets, has size $L_x\times L_y \times L_z$ 
where $x$ is the direction of reconnecting
magnetic field, $y$ is perpendicular to the initial current sheet, 
and $z$ is the ``third'' dimension, parallel to the initial current.
We initialize this box with a (doubly) reversing magnetic field:
\begin{eqnarray}
  \boldsymbol{B}(t=0) &=&  \hat{\boldsymbol{z}} B_{gz} +
    \hat{\boldsymbol{x}} B_0 \times
  \left\{  \begin{array} {r@{\quad}l}
    -{\rm tanh} [(y\:-\,\phantom{3}L_y/4)/\delta] & \textrm{for } y < L_y/2 \\
    \phantom{-}{\rm tanh} [(y\:-\,3L_y/4)/\delta] & \textrm{for } y > L_y/2
  \end{array} \right.
\end{eqnarray}
where $\delta$ is the half-thickness of the initial current sheet.
All simulations in this work use $L_y=2L_x$;
although the lower and upper halves of the simulation can interact,
the separation $L_y/2=L_x$ between current sheets has been found 
to limit the interaction so that, e.g., the reconnection rate
is the same as in simulations with much larger $L_y$ (also, in 3D, the current sheets kink and stray more in the $y$ direction; it is important that $L_y$ be large enough that they never touch).
When presenting results, we often focus on just one current sheet (the lower one) for simplicity,
but display results involving total energy and particle spectra for the
entire simulation.
The system length $L_z$ in the third dimension will be varied to investigate 3D effects, but the fiducial value for 3D simulations will be $L_z=L_x$.

To balance the magnetic field reversal (according to Ampere's law), 
as well as to provide pressure balance against the upstream magnetic
field, we add a ``drifting'' plasma in addition to the background 
plasma \citep{Kirk_Skjaeraasen-2003}. 
This initially-drifting plasma has a total particle density strongly 
peaked around the magnetic field reversal, 
$n_d(y) = n_{d0} \textrm{cosh}^{-2}[(y-y_c)/\delta]$ (measured in
the simulation frame), where $y_c$ is the centre of the current sheet;
$y_c=L_y/4$ for the lower sheet, or $y_c=3L_y/4$ for the upper.
Within each sheet, the electrons and positrons drift in opposite $\pm \hat{\boldsymbol{z}}$
directions with 
speed $\beta_d c$ [hence Lorentz factor $\gamma_d=(1-\beta_d^2)^{-1/2}$], and the initial temperature in the co-moving
frame is $\theta_d m_e c^2$.
The initial current sheet configuration can be specified by the drift $\beta_d$ and the overdensity $\eta\equiv n_{d0}/n_{b0}$, with 
the remaining parameters, $\theta_d$ and $\delta$, determined by 
pressure balance and Ampere's law:
$n_{d0} \theta_d/\gamma_d = B_0^2/8\upi m_e c^2 = \sigma n_{b0}/2$
and $B_0/\delta = 4\upi e n_{d0} \beta_d$.
In terms of the characteristic length $\sigma \rho_0$, the sheet thickness is
$\delta  = (n_{b0}/n_{d0})\sigma \rho_0/\beta_d$; and in terms of the gyroradius $\rho_d \equiv 3\theta_d \rho_0$ of drifting particles,
$\delta =  2\rho_d/(3\beta_d \gamma_d)$.

In previous (especially 2D) reconnection simulations, the initial 
current sheets have often been slightly perturbed to trigger reconnection
faster (reducing computation time) and to create an initial X-point in
a predetermined location.  Although such a perturbation can have an effect in
2D \citep{Ball_etal-2019}, we find that small perturbations leave 2D 
simulations basically unchanged in many important ways 
(cf.~\S\ref{sec:pertAndEta2d}) as the simulation becomes dominated by
the background plasma; however, the effect in 3D can be more significant (cf.~\S\ref{sec:pertAndEta3d}).
While most of our simulations begin with zero perturbation, we have
explored in a few cases the effect of a small magnetic field perturbation---a tearing-type perturbation with a single X-point and a single magnetic island with a small height (in $y$) but a long width (in~$x$).
The perturbation, expressed in the vector potential, is
$\boldsymbol{A}=\hat{\boldsymbol{z}} A_z$:
\begin{eqnarray} \label{eq:pert}
  A_z(x,y) &=& \left[1 + 
         0.877 a \frac{\delta}{L_y} \cos \left( \frac{2\upi x}{L_x} \right)
         \cos^2 \left( \frac{2\upi (y-y_c)}{L_y} \right) \right]
         \nonumber \\
         & & \qquad \times
               B_0 \delta 
               \left[ \ln \textrm{cosh} \left( \frac{L_y}{4\delta} \right)
               - \ln \textrm{cosh} \left( \frac{y-y_c}{\delta} \right)
               \right]
\end{eqnarray}
where $a$ is the perturbation strength.
Although $a$ is useful for initializing $A_z(x,y)$, it is also useful to characterize the perturbation in terms of $s$,
the half-height of the initial magnetic island.
For $\delta \ll L_y$ and $s \ll L_y$, $s$ and $a$ are related approximately by
$\textrm{cosh}(s/\delta) \approx \exp (0.44 a)$.
In the above, $a$ is thus normalized so that when 
$a=1$, we have $s/\delta=1$; i.e., the separatrix extends roughly as far as the initial current sheet.

For $a \lesssim 1$, the initial separatrix lies within the initial
current sheet ($s \lesssim \delta$).
We note that the ``wavelength'' of the perturbation in $x$ is $L_x$---the
longest that can fit in the simulation.
For 3D simulations, this perturbation is uniform in $z$.
Unless specifically noted, $a=0$.

At the beginning of each simulation, particle velocities for the background (and drifting) populations are initialized
randomly, drawn from the appropriate (drifting) relativistic Maxwell-J\"{u}ttner 
distribution.  Because of this randomness, two simulations that are
identical in all macroscopic initial parameters (i.e., differing only in random initial positions and velocities) will not
yield completely identical results, even for global quantities such as the total 
magnetic energy depletion or total particle energy spectrum.
Although computational expense prohibits running large statistical ensembles of simulations for every macroscopic parameter choice, one goal of this study is the estimation of ``stochastic variation'' due to randomized particle initialization (especially in 3D; cf.~\S\ref{sec:variability3d}).
In a limited number of cases, we will therefore show the results of multiple macroscopically-identical simulations as a means of gauging stochastic variability, and distinguishing it from systematic effects correlated with macroscopic parameters.

In summary, the reconnection simulation setup is described by the following physical parameters:
\begin{itemize}[leftmargin=3mm, itemindent=0mm, labelsep=1mm]
  \item The nominal gyroradius $\rho_0=m_ec^2/eB_0$ is fixed for all simulations; its value is irrelevant---we can trivially scale simulation results to any other value of $\rho_0$.
  \item The upstream (cold) magnetization (excluding guide field) $\sigma = B_0^2/(4\upi n_{b0} m_e c^2) = 10^4 \gg 1$ is fixed in all simulations; its particular value is irrelevant, as long as $\sigma \gg 1$.
  \item The upstream hot magnetization (excluding guide field) $\sigma_h = \sigma/(4\theta_b) = 1$ is fixed
  in all simulations; its value puts this study in the moderately-magnetized regime.  This implies (for $B_{gz}=0$) plasma $\beta=1/2$ and $v_A=c/\sqrt{2}$.
  (Previous studies of reconnection in pair plasma have mostly concentrated on the $\sigma_h\gg 1$ regime, while nonrelativistic reconnection studies have $\sigma_h \ll 1$.)
  \item The initial guide magnetic field $B_{gz}$ is zero except in~\S\ref{sec:Bz2d} and in~\S\ref{sec:Bz3d}, when it is varied up to $B_{gz}=4B_0$ in 2D and up to $B_{gz}=B_0$ in 3D, respectively.
  \item The initial current sheet overdensity, $\eta=n_{d0}/n_{b0}$, is set to $\eta=5$, except when varied systematically in~\S\ref{sec:pertAndEta2d}, \S\ref{sec:pertAndEta3d}, and~\S\ref{sec:RDKIamp}.
  \item The average drift speed of particles forming the initial Harris sheet is $\beta_d c = 0.3 c$, except when varied in \S\ref{sec:pertAndEta2d}, \S\ref{sec:pertAndEta3d}, and~\S\ref{sec:RDKIamp}.
  \item The initial current sheet half-thickness $\delta=(\eta\beta_d)^{-1} \sigma\rho_0$ is $(2/3)\sigma\rho_0$ except when $\eta$ and $\beta_d$ are varied as noted above.
  \item The initial magnetic field perturbation strength $a$ (see~Eq.~\ref{eq:pert}) is zero except when varied in \S\ref{sec:pertAndEta2d} and \S\ref{sec:pertAndEta3d}.
  \item The system size $L_x$ determines the overall system size, and is desired to be as large as possible, but limited by computational resources; its value will be noted in each subsection, where we generally compare simulations with the same $L_x$.  However, the effect of varying $L_x$ is specifically investigated in~\S\ref{sec:Lx2d}.
  \item The system aspect ratio $L_y/L_x = 2$ in all simulations.
  \item The system aspect ratio $L_z/L_x$ is considered to be zero in
  all 2D simulations; in 3D simulations the value of $L_z/L_x$ is noted in each case, although we most commonly use $L_z=L_x$;
  we systematically explore the effect of varying $L_z/L_x$ in~\S\ref{sec:Lz3d}.
  \item For 2D simulations, a simulation time $T\approx 10L_x/c$ is usually sufficient; 3D simulations were run for longer times, 20--50$L_x/c$ (the longest runs are shown in~\S\ref{sec:Lz3d}).
\end{itemize}
For the various choices of above parameters, important plasma length scales will be summarized in table~\ref{tab:etaEffect} (in~\S\ref{sec:pertAndEta2d}).

Using periodic boundary conditions offers theoretical and numerical simplicity; and it is usually the simplest way to simulate a mesoscopic box---i.e., a small part of a system whose global inhomogeneity scale along the layer is much larger than the computational box.
For application to real systems, the effect of these or any boundary conditions must ultimately be studied, e.g., by examining simulation size dependence or by comparing simulations with different boundary conditions; alternatively, running simulations for less than one light-crossing time ($t\lesssim 1\,L_x/c$) can lessen effects of boundaries, but at the cost of exacerbating the effect of the initial conditions.
In this work, we run simulations well beyond~$1\,L_x/c$ for two primary reasons: 
(1) the very early evolution, $t \lesssim 1L_x/c$, may be heavily influenced by initial conditions (which are uniform in the third dimension) and may fail to capture the most interesting and important 3D phenomena, and (2) the evolution in 3D can be relatively slow.  
Over long times, boundary conditions might well be expected to affect system evolution significantly, and, this being the case, there is a trade-off between possibly more realistic but complicated boundaries \citep[like ``open'' boundaries, e.g.,][]{Daughton_etal-2006} and the simpler, better-understood periodic boundaries.  
However, the effect of periodic boundaries is arguably realistic for long simulations, e.g., $t\sim 30\,L_x/c$, if the inhomogeneity scale of the real system is larger than $\sim 30L_x$.
Determining the effect of various boundary conditions on current sheet evolution in~3D is important, but beyond the scope of this work.

Finally, we summarize the numerical parameters.
The grid resolution $\Delta x=\Delta y=\Delta z=\sigma\rho_0/3$ marginally resolves the background Debye length and skin depth
$d_e/\sqrt{3}=\lambda_D=\sigma \rho_0/2$ as well as the background
gyroradius $\rho_b = (3/4)\sigma \rho_0$ and the initial current sheet 
[usually $\delta=(2/3)\sigma \rho_0$].
This marginal resolution allows us to simulate the largest system sizes
with the computational resources available; Appendix~\ref{sec:res2d} shows
that this resolution is sufficient.
The timestep $\Delta t=\Delta x/(c\sqrt{d})$, where the dimensionality
$d$ is 2 or 3,
is determined by the Courant-Friedrichs-Lewy condition (all 2D simulations compared directly against 3D simulations were run with $d=3$).
Simulations were initialized with $20\times 4=80$ macroparticles per cell: 20 background electrons and positrons, plus 20 ``drifting'' electrons and positrons.
The initially-drifting particles were weighted (depending on their $y$-position) to represent the non-uniform current sheet density, and particles with negligibly-low weights were deleted from the simulation.
Thus our largest simulation ($L_x=L_z=512\sigma\rho_0$, with $1536\times 3072\times 1536$ cells, cf.~\S\ref{sec:overview3d}) contained about 300 billion macroparticles.
Energy is not precisely conserved by the PIC algorithm, but with these numerical parameters it is approximately conserved in all simulations to better than 1~per~cent---and in almost all but the larger 2D simulations, energy is conserved to better than 0.1~per~cent.

\section{Diagnostics}
\label{sec:diagnostics}

In this section we will define terms used to characterize reconnection (in~\S\ref{sec:terminology}) and then provide detailed descriptions of diagnostics that we will use:  
the central surface of the ``layer'' or current sheet, $y_c(x,z,t)$ (\S\ref{sec:sheetCenter}); 
global volume-integrated characteristics as functions of time---unreconnected flux (\S\ref{sec:unreconnectedFlux}) and various energy components (\S\ref{sec:diagEnergy}); 
the reconnection onset time $t_{\rm onset}$ (\S\ref{sec:onsetTime});
the local bulk velocity of the plasma (\S\ref{sec:bulkVel});
and the power-law index and high-energy cutoff of a particle energy distribution (\S\ref{sec:NTPAfit}).

\subsection{Terminology}
\label{sec:terminology}

Before describing diagnostics for reconnection, we offer explicit definitions or clarifications of some often-used terms, so that we can avoid lengthy qualifications in the text.

{\bf Transverse:} the $x$ and $y$ directions, transverse to $z$, the direction of the guide magnetic field.  

{\bf Transverse magnetic field energy $U_{Bt}$:} the energy in the $B_x$ and $B_y$ components of the magnetic field---the volume integral of $(B_x^2+B_y^2)/8\upi$ over the entire system.  
Even in simulations with substantial guide magnetic field, it is mainly the transverse magnetic energy (and not energy in $B_z$) that is depleted during reconnection.

{\bf Guide magnetic field energy $U_{Bz}$:} the energy in the $B_z$ components of the magnetic field.
Because~$B_z(x,y,z)$ is initially uniform and the flux $\int B_z dx dy$ through any transverse plane is exactly conserved in the simulation, $U_{Bz}$ can only increase from its initial value (in practice it does not increase much and the increase can often be neglected).

{\bf Plasma energy:} the total kinetic energy of individual plasma particles.

{\bf Magnetic energy conversion:} the conversion of (transverse) magnetic field energy $U_{Bt}$ to plasma energy.  Because the guide magnetic field energy~$U_{Bz}$ cannot decrease, it is essentially equivalent to refer to magnetic energy conversion or transverse magnetic energy conversion. 

{\bf Plasma energization:} the conversion of (transverse) magnetic energy to plasma energy.

{\bf Released magnetic energy:} (transverse) magnetic energy that has been
converted to plasma energy.

{\bf Magnetic energy depletion:} (transverse) magnetic energy conversion, with emphasis on the reduction in magnetic energy (as it is converted to plasma energy).
This is nearly a synonym for ``dissipation,'' except that ``dissipation'' refers to irreversible heating, whereas ``depletion'' or ''conversion'' also include conversion to bulk flow kinetic energy (in this paper, however, we will not measure the difference between depletion and dissipation).

{\bf Unreconnected (magnetic) field (line):} a magnetic field line that crosses the entire simulation domain in the $x$ direction (possibly diagonally), without reversing (in $x$).  As we discuss in~\S\ref{sec:unreconnectedFlux}, this is an imperfect, approximate, but practical definition.

{\bf ``Reconnected'' (magnetic) field (line):} any magnetic field line that is not unreconnected (according to the definition above).  In 2D reconnection, a reconnected field line is a closed loop around one or more magnetic islands or plasmoids, and our definition of a ``reconnected'' field line identifies most closed loops accurately.  In 3D, however, there may be many field lines that are ``not unreconnected'' but do not resemble closed loops or even spirals around flux ropes; we nevertheless call them ``reconnected'' and will often use quotation marks as a reminder that they do not necessarily resemble reconnected field lines in 2D reconnection.

{\bf Upstream:} the upstream region includes all points on ``unreconnected'' field lines.  By extension, upstream plasma is the plasma in this region, upstream magnetic energy is the magnetic energy in this region, etc.  

{\bf Upstream value:} When we refer to specific upstream values, such as the upstream $\sigma_h$, $B_0$, $n_{be}$, or $\theta_b$, we mean the asymptotic or far upstream values.

{\bf Current sheet (or sheets):} the region containing currents that support the reversal in magnetic field.  We often refer to the initial current sheet, which is well defined, and otherwise use the term to refer to sheet-like regions where $J_z$ is strong.

{\bf The layer (reconnection layer, current layer):} the complement of the upstream region---i.e., the region containing ``reconnected'' field lines.  This region contains the current sheet (or current sheets) as well as reconnection outflows and plasmoids.  We think of ``the layer'' as the evolved current sheet.  Each (double-periodic) simulation contains two layers---the upper and lower layers.

{\bf Separatrix:} the surface between the ``upstream'' region and ``the layer,'' i.e., that separates ``reconnected'' and ``unreconnected'' magnetic field lines.
In 2D this is a smooth, well-defined surface.  In 3D that may not be the case, but none of our analysis will be sensitive to the precise location of the separatrix.

{\bf Unreconnected flux:} the integral of $B_x$ over the $x=0$ plane intersected with the ``upstream'' region; i.e., the integral is over all (and only) unreconnected field.  See~\S\ref{sec:unreconnectedFlux} for more detail.

{\bf Reconnected flux:} the flux of magnetic field around the major (largest) plasmoid.  This concept is well defined and easily measured in 2D, where the unreconnected plus reconnected flux is conserved (cf.~\S\ref{sec:unreconnectedFlux}).  It is nontrivial, however, to define and measure this in 3D, and we will not do so.

{\bf Upstream magnetic energy $U_{Bt,\rm up}$:} the transverse magnetic energy in the upstream region.  We exclude guide field energy from this quantity, because we are primarily interested in conversion of magnetic energy to plasma energy.

{\bf Magnetic energy in the layer $U_{Bt,\rm layer}$:}, the transverse magnetic energy in the layer; $U_{Bt,\rm up} + U_{Bt,\rm layer}=U_{Bt}$.
We exclude guide field energy from this quantity.

{\bf Reconnection:} technically, ``reconnection'' should involve the conversion/rearrangement of ``unreconnected field lines'' to ``reconnected field lines'' in a way that conserves the total unreconnected plus reconnected flux.  In 2D reconnection, this conservation is easily verified.
However, in 3D it is nontrivial to decide precisely how to characterize reconnection, much less to measure true reconnection rates.  Rather than argue for any particular precise definition of reconnection, we will focus on more tangible properties of the current layer, such as magnetic energy and plasma energy.  We will use the words ``3D reconnection'' loosely to refer to any magnetic-energy-depleting evolution of thin current sheets that, in principle, could or would undergo 2D-like reconnection.
However, any discussion of reconnecting \emph{flux} or \emph{field lines} will refer specifically to the 2D-like flux-conserving process.

{\bf Flux annihilation:} the utter disappearance of upstream (unreconnected) magnetic flux by unspecified means.  In contrast, 2D reconnection does not annihilate flux (it conserves flux as described above).  For example, magnetic diffusion in a plasma with finite resistivity can directly annihilate flux; although this process is far too slow to explain observed magnetic energy releases (assuming neat laminar current sheets), it could be considerably enhanced, e.g., by effective turbulent magnetic diffusivity.
Incidentally, the flux $\int B_x dy dz$ through any $x$-plane in the simulation is exactly conserved---however, by virtue of the reversing magnetic field, this flux is zero, and it tells us nothing about how much flux is reconnected or annihilated.

{\bf (Upstream) flux depletion:} the depletion of the upstream magnetic flux, without regard to the process (e.g., reconnection or annihilation).

\subsection{Current sheet (layer) central surface}
\label{sec:sheetCenter}

It is useful to analyse some field quantities at the current sheet ``centre'' (in $y$), even as the layer kinks and deforms.
In the initial state, the current sheet (or layer) central surface $y_c(x,z,t=0)$ is the location
where $B_x(x,y_c,z,t=0)=0$, i.e., where $B_x$ crosses through zero in the $y$ direction, reversing sign.
At later times, we continue to define $y_c(x,z,t)$ in the same way, with the following complication.
The layer may become extremely distorted by the nonlinear development of the kink instability so that it folds over on itself, resulting in multiple field reversals.
Because the upstream magnetic field never changes sign, $B_x(x,y,z,t)$ will
always reverse sign at least once along $y$; however, it may reverse sign an odd number of times at each layer.
If there are multiple reversals in $y$ at some $(x,z)$ and time $t$, we take $y_c(x,z,t)$ to be the middle reversal [e.g., if there are three reversals, $y_c(x,z,t)$ will be at the second reversal].

We define the displacement of the central surface, $\Delta y_c(x,z,t) \equiv y_c(x,z,t) - y_c(x,z,0)$.

In~\S\ref{sec:Lz3d} we look at~$\tilde{y}_c(k_z,t)$, the Fourier spectrum of $\Delta y_c(x,z,t)$ in~$z$, averaged over~$x$, at a given time~$t$.
To find the ``Fourier spectrum in $z$ averaged over $x$,''
we take the power spectrum in $z$ of $\Delta y_c(x,z,t)$
at every $x$ (for a given time $t$), 
then average the power spectra over $x$, and take the
square root.  We normalize the result $\tilde{y}_c(k_z,t)$ so that
it indicates the amplitude of the sinusoidal component with $k_z$.
E.g., if $\Delta y_c(x,z,t)=A\sin(k_z z)$, then $\tilde{y}_c(k_z,t)=A$.

\subsection{Unreconnected flux and reconnection rate}
\label{sec:unreconnectedFlux}

Reconnection breaks upstream (unreconnected) field lines
and reconnects them in a different configuration in the downstream
outflows.
Defining and measuring unreconnected flux is straightforward in 2D using the vector potential $A_z(x,y)$, which is constant along field lines.
In 3D, however, it is much more difficult to define unreconnected flux precisely, and so we adopt the following practical definition.
An unreconnected field line is a field line that runs from $x=0$ to $x=L_x$ without changing direction in $x$ ($B_x$ never changes sign).
In 2D, this definition is unambiguous and almost always agrees with the method using $A_z$ (disagreement could occur in the rare case of an unreconnected field line with an $\Omega$-shaped kink; we believe such cases are negligibly rare in 2D, but we have not studied whether they are also rare in 3D). 
In 2D, a field line that runs from $x=0$ to $x=L_x$ wraps around via periodic boundary conditions exactly onto itself; therefore, this definition unambiguously determines whether the entire field line has the same sign for $B_x$.
In 3D, this is not necessarily true; nevertheless this method captures the most obviously-unreconnected upstream field lines with $|B_y|\ll |B_x|$, that 
run across the simulation box with little deflection in the $y$ direction.

Specifically, we trace a field line from each grid node $(0,y,z)$ in the $x=0$ plane, following it in the $+x$ direction; 
if the field line ever reverses direction in $x$,
then the field line is considered to be ``reconnected.'' 
If the sign of $B_x$ does not reverse, then the field line must eventually reach $x=L_x$, at which point we stop and consider the field line to be ``unreconnected.''
The unreconnected flux is then estimated as $\sum_{y,z} B_x(0,y,z) \Delta y \Delta z$ where the sum is over nodes $(y,z)$ on unreconnected field lines, 
and $\Delta y \Delta z$ is the cell face area.
We add the flux for all unreconnected field lines between the central surfaces of the two layers (cf.~\S\ref{sec:sheetCenter}) to obtain the total unreconnected flux.

During this process of tracing field lines, we note the index of every cell that is penetrated by an unreconnected field line and consider any such cell to be part of the ``upstream'' region; any cell not penetrated by any unreconnected field line is considered part of the ``layer.''
In this way we can calculate, for example, the magnetic energy in upstream and layer regions.
The boundary between regions of unreconnected and reconnected field lines is (an approximation of) the separatrix, which can be clearly seen in Fig.~\ref{fig:ndePerturb2dLx320}; (in 2D, at least) the separatrix bounds magnetic islands (O-points or plasmoids) and goes through X-points.

This analysis is rigorously based on 2D reconnection; in 3D it still paints a useful picture, even though the notions of ``unreconnected'' and ``reconnected'' are no longer precisely well-defined, and the analysis does not distinguish between reconnection and annihilation of upstream flux (i.e., between reconnected flux and annihilated flux).
In 2D, any field line that is not unreconnected is reconnected, forming a closed loop in a magnetic island (when $B_z$ is ignored).  Moreover, we find that in 2D, flux is conserved---specifically, the flux between the major O-points of the two layers (in the double-periodic simulation) is conserved.  This flux can be divided into two parts: the upstream unreconnected flux between the two layers, plus the reconnected flux between the major O-point and separatrix for each layer, and the sum of these parts remains constant, at least within our measurement precision, which is better than 1~per~cent of the initial flux.
Therefore, in 2D, flux is not annihilated or destroyed (in a more resistive plasma, flux would be annihilated; and even in a ``collisionless'' PIC simulation, numerical resistivity will eventually annihilate flux, but only over extremely long times).
In 3D, field lines that are ``not unreconnected'' do not necessarily form a closed loop around an O-point or flux rope, 
but again, for brevity we will still use the term ``reconnected'' for such field lines.
Periodically we include quotation marks around ``reconnected'' as a reminder that it really means ``not unreconnected.''
In addition (spoiler alert!), we will find that in 3D, flux is not conserved in the way it is in 2D: some flux is outright annihilated.

When we refer to a ``reconnection rate,'' we mean the rate at which the upstream magnetic flux decays---regardless of whether this flux depletion occurs due to reconnection (as always in 2D collisionless reconnection) or to annihilation (as might be happening in 3D simulations).
The reconnection rate represents the rate at which upstream flux is changing, and therefore represents an electric field along an X-line by Faraday's law, if a suitable X-line (or reasonable approximation) exists.

We use the symbol $\beta_{\rm rec}$ for the dimensionless reconnection rate normalized to $cB_0$:
$\beta_{\rm rec} \equiv -(cB_0L_z)^{-1} d\psi/dt$, where $\psi(t)$ is the 
flux upstream of one layer (or half the flux between the two layers).
Usually we obtain values of $\beta_{\rm rec}$ averaged over some 
time---for example the time over which $\psi(t)$ falls from~$0.9\psi_0$ to~$0.8\psi_0$, or alternatively, 0.8--0.7$\psi_0$ (the choice between intervals often depends on whether all simulations being compared actually reached $0.7\psi_0$).
Sometimes we also quote~$(c/v_A)\beta_{\rm rec}$, the dimensionless reconnection rate normalized to $B_0 v_A$, where $v_A$ is the Alfv\'{e}n velocity.

\subsection{Energy}
\label{sec:diagEnergy}

Energy is a powerful diagnostic because of its physical importance, because 
it is conserved (to a good approximation in these simulations), and 
because it is generally straightforward and unambiguous to measure.
We will calculate various energy components integrated over the entire simulation volume (at any given time $t$), including total particle 
energy $U_{\rm plasma}(t)$ and electric field energy $U_{E}(t)$.
The component of most frequent interest is the magnetic field energy $U_B(t)$, and more specifically the transverse magnetic field energy, $U_{Bt}(t)\equiv \int dV\, (B_x^2+B_y^2)/8\upi$;
it is mainly $U_{Bt}$ 
that gets converted to particle energy $U_{\rm plasma}$, while
$U_{Bz}\equiv U_{B}-U_{Bt}$ remains relatively constant (or at least negligibly small) over the course of a simulation.
For zero guide field, $U_{Bt}\approx U_B$ and it does not much matter which we use;
when comparing simulations with different guide fields, however, it is more illuminating to compare $U_{Bt}$.

We also calculate the upstream magnetic energy $U_{Bt,\rm up}$ in transverse magnetic
field components of unreconnected field lines, and then define
the magnetic energy in the layer 
(i.e., in ``reconnected'' field lines) to be 
$U_{Bt,\rm layer}=U_{Bt} - U_{Bt,\rm up}$ 
(see~\S\ref{sec:unreconnectedFlux}).

\subsection{Onset time $t_{\rm onset}$}
\label{sec:onsetTime}

Sometimes the initial current sheet configuration strongly affects the time it takes reconnection to start; when comparing time evolution of simulations with different initial configurations, it is sometimes helpful to compare different cases relative to the onset time rather than $t=0$.
We define the onset time as the time $t_{\rm onset}$ when the transverse magnetic energy $U_{Bt}(t)$ has declined by 1 per~cent from its initial value $U_{Bt0}$.
Although this value is somewhat arbitrary, no results will depend 
strongly on this choice, as long as it is used consistently. 
Occasionally the time axis of graphs will be shifted by $t_{\rm onset}$ to allow more revealing comparison between different simulations, which may have similar time evolution apart from different onset times.

\subsection{Bulk Velocity}
\label{sec:bulkVel}

We compute electron bulk velocities $\boldsymbol{v}_e=\boldsymbol{J}_e/\rho_e$ (in each grid cell), 
where $\boldsymbol{J}_e$ and $\rho_e$ are the current and charge density due to electrons; similarly, the positron velocity is $\boldsymbol{v}_i=
\boldsymbol{J}_i/\rho_i$.
The plasma velocity is then computed as the average of electron and ion
velocities, $\boldsymbol{v}\equiv (\boldsymbol{v}_e + \boldsymbol{v}_i)/2$.
More accurately, this average should be weighted by the mass (or $\gamma m$) of particles in each cell; however, in this paper, quasineutrality and the symmetry between electrons and positrons make the simple average a reasonable approximation.

\subsection{Particle energy spectra and power-law fitting}
\label{sec:NTPAfit}

Particle energy distributions $f(\gamma)$ are shown integrated over the entire simulation box at single time snapshots (where the Lorentz factor 
$\gamma$ is a proxy for particle energy $\gamma m_e c^2$).
The energy spectra often display power laws $f(\gamma)\sim \gamma^{-p}$ extending to high energies (well above the average energy).

We determine the power-law index~$p$ by finding the longest, straightest segment on a log-log plot \citep{Werner_etal-2018}.
We compute the local slope, $p(\gamma)=-d\ln f/d\ln \gamma$ and search exhaustively for the interval $[\gamma_1,\gamma_2]$ with the largest $\gamma_2/\gamma_1$ such that $p(\gamma)$ remains within a range of $\Delta p$ over the entire interval.
We then choose ``the'' power-law index~$p$ to be the median~$p(\gamma)$ over~$[\gamma_1,\gamma_2]$.
We do this separately for $\Delta p=$0.1, 0.2, and~0.4, as well as using spectra at different nearby times, considering variation in~$p$---whether due to time variation or choice of~$\Delta p$---as uncertainty in the measurement.
In this paper, we display ``error bars'' on~$p$ comprising the middle 68~per~cent of all the values of~$p$ measured.

To find the cutoff of the high-energy power law, we fit $f(\gamma)$ over $[\gamma_1,\gamma_2]$ to the form $A\gamma^{-p}$, where~$p$ is determined as above (thus only the normalization~$A$ must be found).  
We then consider the cutoff~$\gamma_c$ to be the energy at which~$f(\gamma_c)=e^{-1} A\gamma_c^{-p}$.

\section{2D reconnection with moderate magnetization: basic evolution and NTPA}
\label{sec:2d}

Before exploring 3D reconnection, we investigate reconnection---in
ultrarelativistically-hot pair plasma with $\sigma_h=1$---in 2D,
systematically varying a number of parameters.
We will begin with an overview of 2D reconnection (\S\ref{sec:overview2d});
then we will investigate the effects of different initial current sheet
configurations in~\S\ref{sec:pertAndEta2d}.  In~\S\ref{sec:Lx2d} we
study system-size effects, and finally in~\S\ref{sec:Bz2d} we report
on the effects of guide magnetic field.

This 2D reconnection study serves multiple purposes.
First, it is of interest in its own right to characterize 2D reconnection
in the ultrarelativistically-hot $\sigma_h=1$ regime (most previous reconnection studies have focused on $\sigma_h \ll 1$, usually in nonrelativistic electron-ion plasma, or $\sigma_h \gg 1$).
Second, much larger simulations are possible in 2D; only in 2D, therefore, can we really explore system-size dependence.
And last, to determine whether 3D reconnection is different, we need to compare with 2D simulation; this 2D study provides a baseline for the subsequent 3D study (\S\ref{sec:3d}).

\subsection{Overview of 2D reconnection with $\sigma_h=1$}
\label{sec:overview2d}

We begin by reviewing the familiar behaviour of plasmoid-dominated reconnection in 2D, for a single representative simulation with 
$L_x=1280\sigma \rho_0$; other initial parameters are: $B_{gz}=0$, $\eta=5$, $\beta_d=0.3$, $\delta=(2/3)\sigma\rho_0$, $a=0$.
This simulation exhibits the familiar behaviour of 2D plasmoid-dominated reconnection---it is qualitatively similar to plasmoid-dominated reconnection at small and large $\sigma_h$.
The simulation lingers in its initial state until the 
tearing instability triggers reconnection, resulting in a chain of plasmoids (magnetic islands or O-points) in each layer.
(In the following, we describe just one of the layers in the double-periodic system; both layers behave qualitatively similarly.)
Reconnection occurs at X-points in thin, elementary current sheets between plasmoids: at X-points, upstream magnetic field lines are broken, and reconnected in a different topology, with reconnected field lines wrapping around plasmoids in the reconnection outflows.
In this process, some of the upstream magnetic energy is converted to particle/plasma energy, while some of it ends up as magnetic energy in the layer (i.e., in ``reconnected'' field in plasmoids).
Secondary tearing breaks up the elementary current sheets when they become too long and too thin, detaching the small plasmoids from the X-point that fed them magnetic energy and reconnected flux.  Thereafter, those plasmoids no longer grow via reconnection, but instead grow by merging with other plasmoids via the coalescence instability, which conserves the flux around O-points.
A hierarchy of different-sized plasmoids thus develops, and 
eventually a single monster plasmoid dominates the simulation, continuing to grow as it consumes (i.e., merges with) smaller plasmoids.
If $L_y/L_x$ is large enough, this major plasmoid eventually grows to a size of order $L_x$, and reconnection can no longer continue; the angle of the magnetic separatrix at the major X-point opens to 90 degrees, and reconnection effectively stops (actually, systems often oscillate with low amplitudes about the final state, with magnetic energy and flux sloshing between the major plasmoid and the upstream).
If $L_y \gg L_x$ (even if $L_y \gtrsim 2L_x$), there will still be significant upstream ``unreconnected'' magnetic field that could in principle be reconnected were it not prevented from doing so by reaching a stable magnetic configuration (at least on reconnection timescales; magnetic diffusion might deplete additional magnetic energy on much, much longer timescales).

During this evolution, upstream magnetic flux is reconnected and some upstream magnetic energy is converted into plasma energy, while some ends up permanently stored in the magnetic field of the major plasmoid.
To describe this process more quantitatively, we will rely heavily on several important diagnostics.
These will be essential throughout the rest of the paper, as we explore reconnection over a very large parameter space, to help us compare and contrast reconnection with different parameters including $a$, $\eta$, $\beta_d$, $L_x$, $B_{gz}$, and (in 3D) $L_z$.
We will measure, for example, the amount of magnetic flux that is reconnected over time (cf.~\S\ref{sec:unreconnectedFlux}).
However, perhaps of even more importance is the behaviour of various energy components---for example, the plasma or magnetic energy versus time (cf.~\S\ref{sec:diagEnergy}).
In addition, we further decompose the plasma energy into the energy distribution of particles to distinguish thermal heating from NTPA (cf.~\S\ref{sec:NTPAfit}).
In the following, we describe these diagnostics for the single representative simulation. 

\begin{figure}
\centering
\includegraphics*[width=0.325\textwidth]{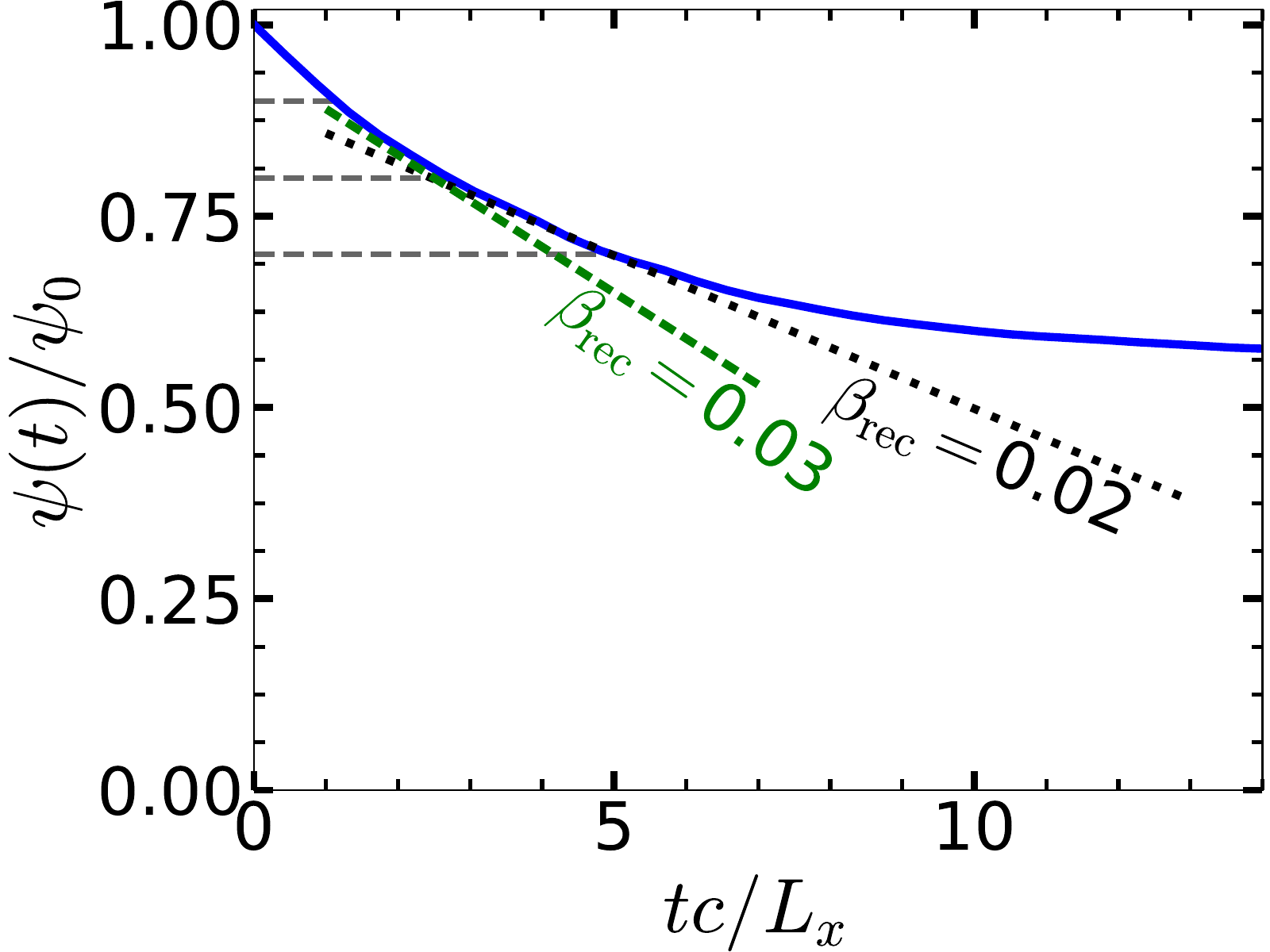}%
\hfill
\includegraphics*[width=0.325\textwidth]{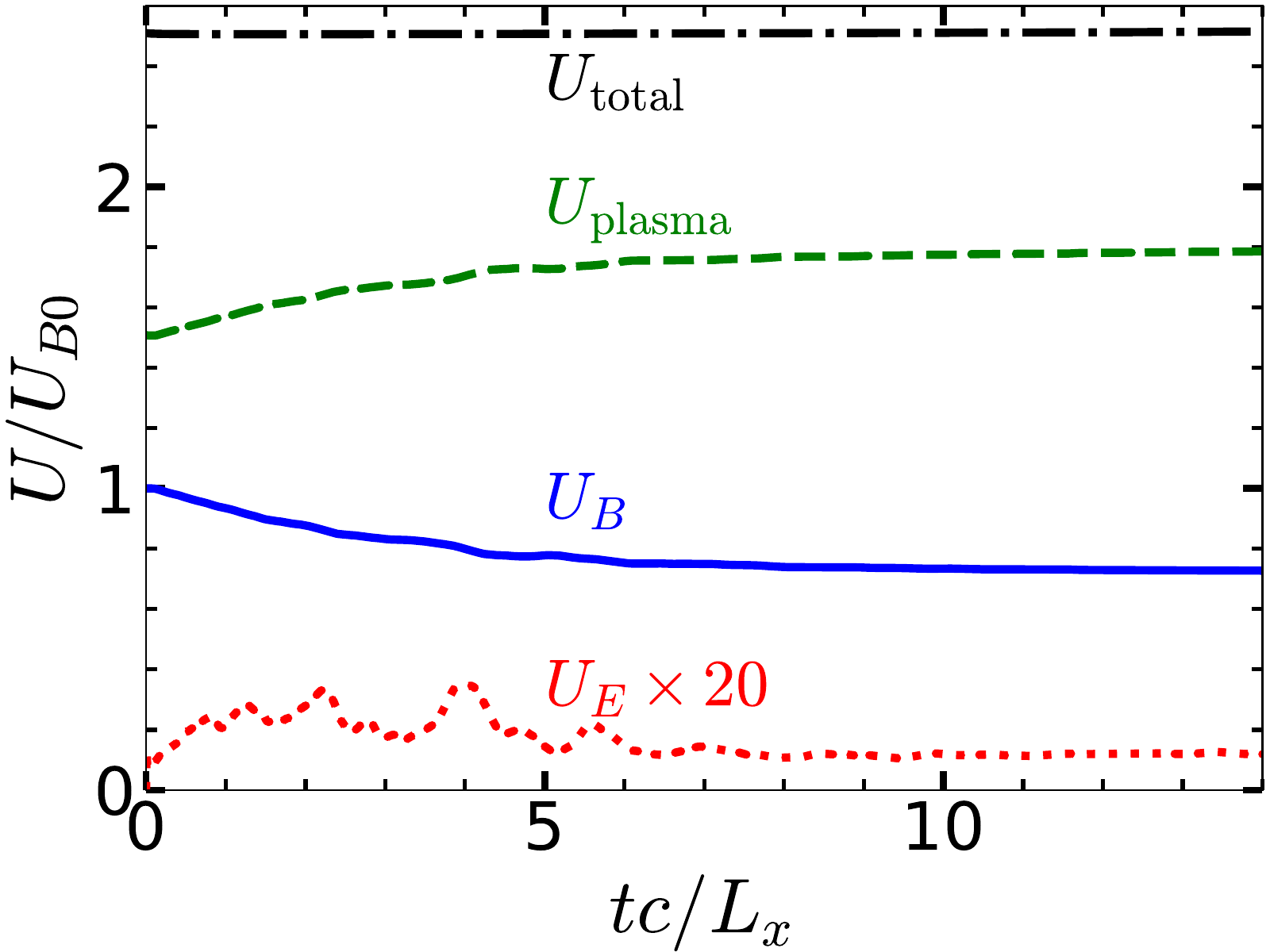}%
\hfill
\includegraphics*[width=0.325\textwidth]{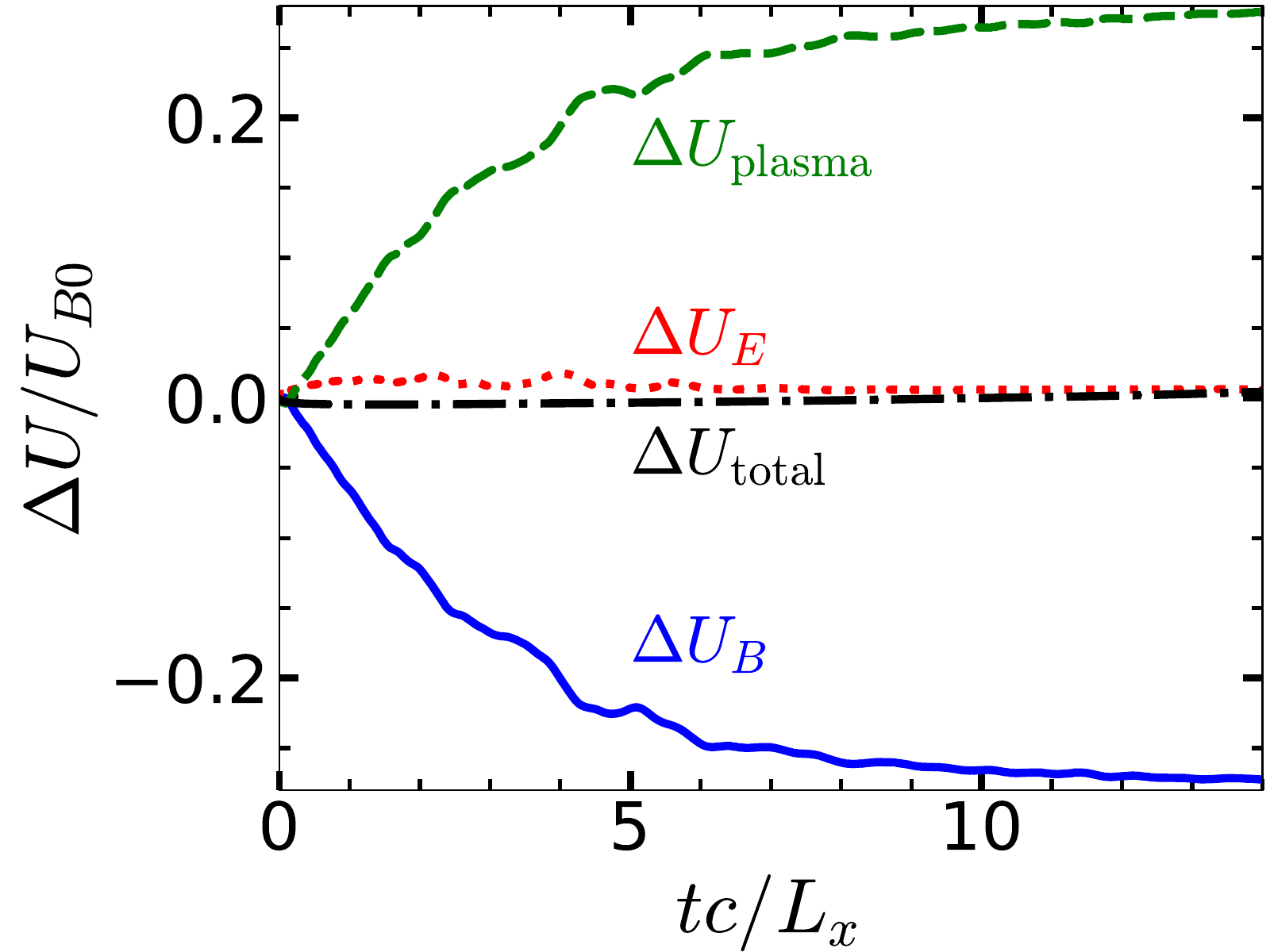}%
\caption{ \label{fig:energyFluxVsTime2D}
The left panel shows the unreconnected flux (normalized to the initial unreconnected flux~$\psi_0$) diminishing over time as reconnection occurs. Horizontal dashed lines mark the levels $\psi=0.9\psi_0$, $0.8\psi_0$, and $0.7\psi_0$; 
the dimensionless reconnection rates averaged over each of these $0.1\psi_0$ drops are $\beta_{\rm rec}=$0.03 and~0.02
(cf.~\S\ref{sec:unreconnectedFlux}),
and dashed lines with those slopes are drawn.
The middle panel shows 
the energy in various components versus time (normalized to
$\int dV \, B_0^2/8\upi$), and the right panel shows the changes in
those components from their initial values.
Energy components shown are 
total energy ($U_{\rm total}$, dash-dotted black line), 
magnetic ($U_B$, solid blue),
plasma/particle ($U_{\rm plasma}$, dashed green), and
electric ($U_E$, dotted red---in the middle panel, $U_E$ is magnified by a factor of 20).
}
\end{figure}

The behaviour of magnetic flux and energy is qualitatively similar to that observed in other 2D regimes, 
such as $\sigma_h\gg 1$ \citep[e.g.,][]{Guo_etal-2014,Werner_etal-2016,Werner_Uzdensky-2017,Werner_etal-2018}.
Figure~\ref{fig:energyFluxVsTime2D} shows unreconnected flux versus time
as well as the energy (and energy changes) versus time in various components, including the energy in magnetic fields and particles.
The unreconnected (upstream) magnetic flux $\psi(t)$
decreases over the course of reconnection, finally down to about 55~per~cent of its initial value (i.e., reconnecting 45~per~cent of upstream flux). 
(We note that these values are specific to this representative simulation; reconnection with other parameters may yield different values while being qualitatively similar.)
During this time, the magnetic energy falls to about 70~per~cent of its
initial value, with the 30~per~cent loss converted almost entirely to plasma energy.
This occurs within a time of roughly 8--12$\:L_x/c$ over which the rate of reconnection continually slows (in this closed system).
The dimensionless reconnection rates (cf.~\S\ref{sec:unreconnectedFlux}), averaged 
over the time it takes for unreconnected flux $\psi(t)$ to fall from~$0.9\psi_0$ to~$0.8\psi_0$, and from~$0.8\psi_0$ to~$0.7\psi_0$
\citep[as measured in ][]{Werner_etal-2018}, are
$\beta_{\rm rec}\equiv -(cB_0 L_z)^{-1} \dot{\psi}=0.032$ and~$0.22$, respectively (here the dimensionless rate is normalized to $B_0 c$).
Normalized to $B_0 v_A$ instead of $B_0 c$, where
$B_0 v_A=B_0 c\sqrt{\sigma_h/(1+\sigma_h)}=B_0 c/\sqrt{2}$, the
reconnection rates over these same intervals are
$(c/v_A)\beta_{\rm rec}=0.045$ and~$0.031$. 
The normalized-to-$B_0 v_A$ reconnection rate around 0.03--0.05 is on the order of, but less than the oft-cited nominal value of 0.1 for fast reconnection \citep{Cassak_etal-2017}---a value that is realized for ultrarelativistic reconnection with $\sigma_h \gg 1$.

By way of comparison, we note that, with the same aspect ratio $L_y/L_x=2$, a similar setup with $\sigma_h\gg 1$ 
depletes roughly 40~per~cent of magnetic field energy
\citep{Werner_Uzdensky-2017} and reconnects 55~per~cent of the initial flux 
\citep{Werner_etal-2018}.
Energy conversion in high-$\sigma_h$ reconnection occurs within 
about 3$\:L_x/c$, with normalized reconnection rates around 0.10--0.12.

An important aspect of 2D reconnection is that the upstream field lines are indeed broken and reconnected in the following precise sense:
the decrease in upstream flux equals the increase in flux around the O-points in magnetic islands (plasmoids).
In 2D simulations we can measure these fluxes accurately using the $z$-component of the magnetic vector potential (see~\S\ref{sec:unreconnectedFlux}), and we find that the sum of these fluxes---which is precisely equivalent to the total flux between the major O-points in each layer---is conserved to better than~1~per~cent.
This is not surprising; the only way the total flux can change is (by Faraday's law) if $E_z\neq 0$ at the major O-points
(e.g., $E_z$ could be non-zero due to resistivity, which would allow annihilation of magnetic flux via magnetic diffusion, but typically the rate of annihilation is very slow, especially in a collisionless plasma).

The reconnection rate measurements quoted in this paper are defined in terms of an upstream field of $B_0$, although in these closed systems, the actual
(far) upstream field diminishes over time (although not very much for large systems).
For simulations presented in this paper, the upstream magnetic field decays only to $\approx 0.9B_0$,
which affects the calculation less than the measurement
uncertainty.
There are also questions about whether ``the upstream field'' should be far upstream or at some upstream point closer to the current sheet \citep{Liu_etal-2017,Cassak_etal-2017}, and the difference between far upstream and near upstream might potentially depend on $\sigma_h$. 
However, we must leave a better understanding of the precise $\sigma_h$-dependence of the 2D reconnection rate to future work.

\begin{figure}
\centering
\fullplot{
\includegraphics*[width=0.49\textwidth]{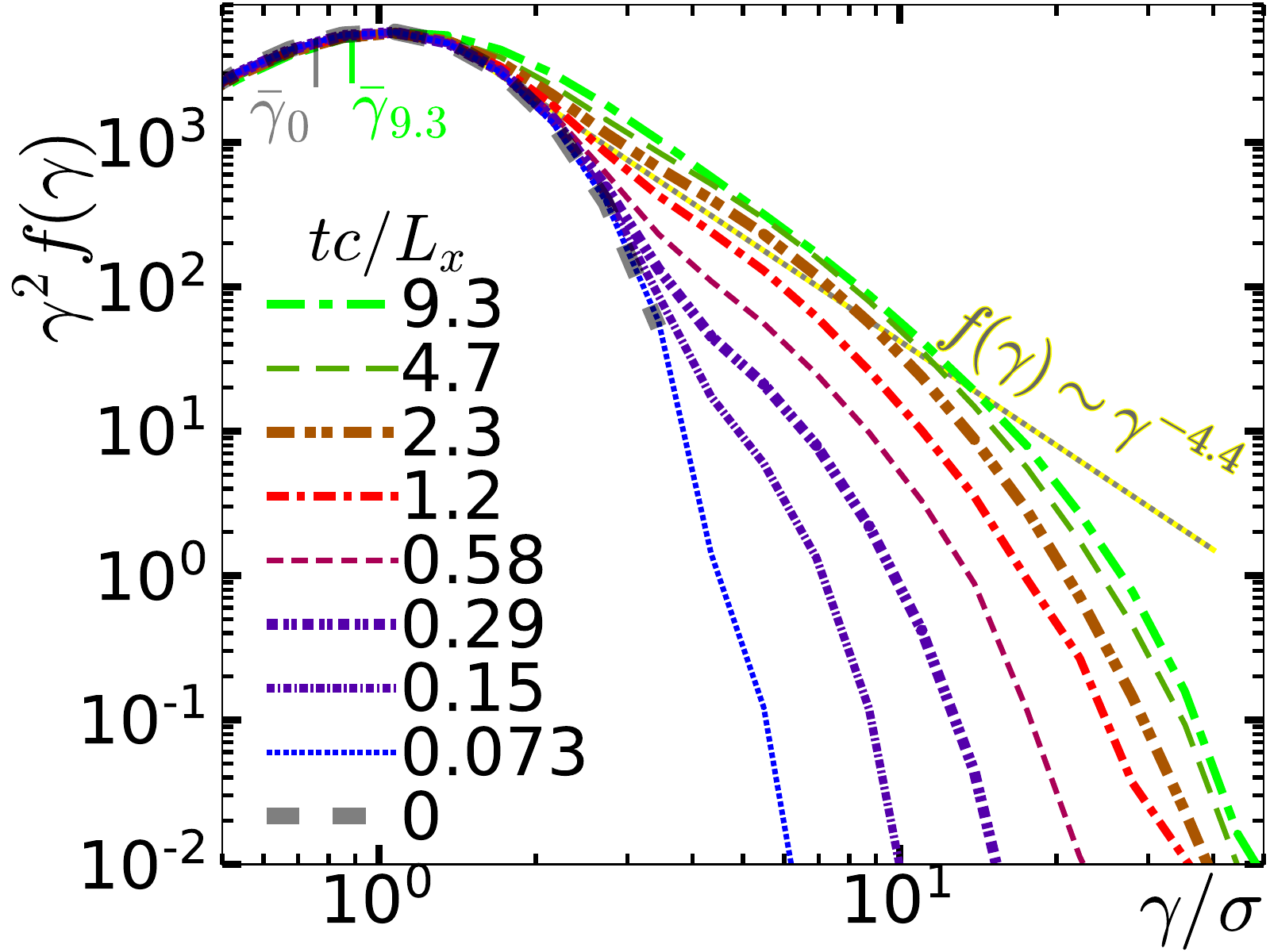}%
\hfill
\includegraphics*[width=0.49\textwidth]{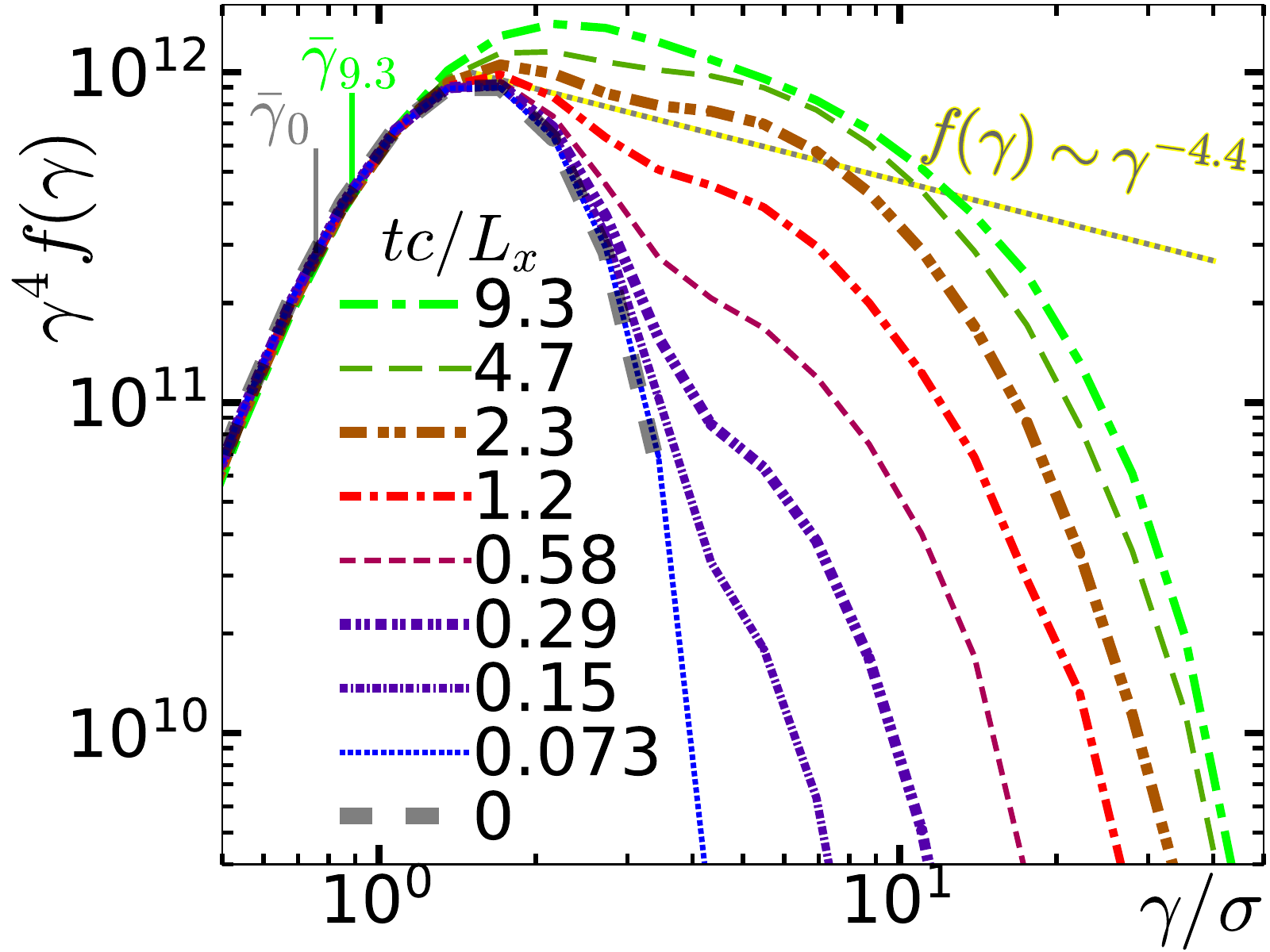}%
}
\caption{ \label{fig:spectra2D}
The particle energy spectra $f(\gamma)$ at logarithmically-spaced times, compensated by $\gamma^2$ to 
show the relative amount of energy stored in different energy ranges (left)
and by $\gamma^4$ to enhance differences (right).  Here particle energy is $\gamma m_e c^2$.
The average particle energy $\bar{\gamma}$ is indicated at $t=0$ and $tc/L_x=9.3$.  A power law $\gamma^{-4.4}$ is also shown (in the right panel, $f\sim \gamma^{-4}$ would be horizontal).
}
\end{figure}

When $\sigma_h$ is not large, particles cannot gain much energy \emph{on average} during reconnection.
E.g., for $\sigma_h=1$ and 30~per~cent of magnetic energy depleted, the average particle energy (averaged over the entire simulation) increases by a relative fraction of only $\Delta \gamma/\gamma_b = 0.30 (2/3) \sigma_h = 0.20$
(where $\gamma_b = 3\theta_b = 0.75\sigma$ is the initial average Lorentz factor).
Nevertheless, \emph{some} particles reach energies well above the average energy, $\bar{\gamma} \approx 0.7$--$0.9\sigma$,
as shown by the particle energy spectra in Fig.~\ref{fig:spectra2D}.
Reconnection in the $\sigma_h=1$ regime clearly generates NTPA;
a high-energy nonthermal power-law $f(\gamma)\sim \gamma^{-p}$
becomes apparent within a time of 1--2$L_x/c$, with
approximately the same power-law index $p\approx 4$ as at later times;
the power law extends up to a cutoff energy exceeding $4\sigma$ \citep[cf.][]{Werner_etal-2016}.
The power-law index of $p\approx 4$ is quite steep; to distinguish small changes
(when comparing other simulations), we will usually graph 
$\gamma^4 f(\gamma)$, so that if $p$ were exactly 4, the graph would be
a horizontal line.
The high-energy cutoff of the power law has been observed, for simulations with high $\sigma_h\gg 1$ and very large $L_x$, to grow rapidly to an energy $\sim 4\sigma$ \citep{Werner_etal-2016}, and then to continue to grow slowly (sublinearly) in time \citep{Petropoulou_Sironi-2018,Hakobyan_etal-2021}; our results are consistent with this.

Importantly, the high-energy cutoff measured in our simulations is far below the extreme particle acceleration limit for this simulation size and duration---i.e., below the energy a particle could gain in the nominal reconnection electric field of $E_{\rm rec}$ over the simulation duration $T$ (which is usually proportional to simulation size $L_x$).
Such an ``extremely-accelerated'' particle would gain maximum energy
$m_ec^2 \Delta \gamma_{\rm ext} = eE_{\rm rec}cT$.
We can estimate $\Delta \gamma_{\rm ext}$ using $\beta_{\rm rec}$ to relate $E_{\rm rec}$ and $B_0$, i.e.,
$E_{\rm rec}=\beta_{\rm rec}B_0$.
Since $m_e c^2/eB_0\equiv \rho_0$,
$\Delta \gamma_{\rm ext} \sim \beta_{\rm rec} cT/\rho_0
\sim 0.02 (cT/L_x) L_x/\rho_0$.
In our simulations, $\beta_{\rm rec}$ decreases over time (as reconnection slows to a stop), but above we measured $\beta_{\rm rec}\approx 0.02$ in the middle of active reconnection; and in this simulation, $L_x=1280\sigma\rho_0$, 
so $\Delta \gamma_{\rm ext} \sim 26 (cT/L_x)\sigma $.
Thus, after $T\sim 10L_x/c$, the maximum energy gain would be around
$\Delta \gamma_{\rm ext}\sim 250 \sigma$.
However, rather than estimate $\beta_{\rm rec}$ and the active reconnection time $T$, it is convenient to consider that by Faraday's law,
$\int_0^T E_{\rm rec} dt = -\Delta \psi/(cL_z)$ where $\Delta \psi$ is the change in unreconnected upstream flux; in a 2D simulation, the simulation size $L_z$ in the unsimulated third dimension is arbitrary, but the upstream flux is also proportional to $L_z$---the flux upstream of the initial current sheet is $\psi_0 \approx B_0 L_z L_y/4$.
Therefore, $\Delta \gamma_{\rm ext} \approx  (L_y/4 L_x) (\Delta \psi/\psi_0) (eB_0 L_x/m_e c^2) = 0.5(0.45)(L_x/\rho_0)$, or
$\Delta \gamma_{\rm ext}\approx 0.2(1280\sigma)\approx 250\sigma$.

In this simulation, we observe $\Delta \gamma \lesssim 40\sigma$, far below the extreme acceleration limit, $\Delta \gamma_{\rm ext}\approx 250 \sigma$. 
However, some particles undergo extreme acceleration up to at least $\gamma \sim 4\sigma$ (but not beyond $\gamma \approx 20\sigma$) in a time consistent with constant acceleration in $E_{\rm rec}$.
We can infer from Fig.~\ref{fig:spectra2D} that a tiny fraction of particles reach nearly $\gamma\approx 20\sigma$ by time $0.44L_x/c$.
At this time, $\Delta \psi/\psi_0 = 0.039$, so we estimate
$\gamma_{\rm ext} = 0.5(0.039)(1280\sigma)=25\sigma$, close to the observed $\gamma \approx 20\sigma$ (which we emphasize is not the power-law cutoff, but the maximum energy gained by any particle in the simulation).  This suggests that particles are in fact ``extremely-accelerated'' up to some energy less than but on the order of $\gamma \approx 20\sigma$.  However, no particle ever exceeds $\gamma \approx 40$ significantly, even after much longer times.
We note that: (1) we measured $\Delta \psi$ with a time cadence of $0.44 L_x/c$, so we cannot examine extreme acceleration on smaller timescales, and (2) ideally we would prefer to know about extreme acceleration in the middle of the simulation rather than at the beginning, but we present this result for the very beginning because it is the only time that we can know that a particle with $\gamma\approx 20$ accelerated to that energy within $0.44L_x/c$.
Thus our results are consistent with particles experiencing extreme acceleration up to an energy around $\gamma \sim 4\sigma$ \citep{Werner_etal-2016}, and experiencing slower acceleration thereafter \citep[as described by][]{Petropoulou_Sironi-2018,Hakobyan_etal-2021}.

Incidentally, we also note that all particle gyroradii are much smaller than the system size.  A particle with energy $\gamma$ has gyroradius $\sim \gamma\rho_0$ (in field $B_0$), and so a particle would need $\gamma=640 \sigma$ for its gyroradius to equal $L_x/2$; therefore, the system size is much larger than the gyro-orbits of even the most energetic particles.

In the rest of this paper, we will be comparing the flux reconnected versus time, as well as the conversion of magnetic to plasma energy and the resulting particle energy spectra, for reconnection simulations with different initial parameters (including 3D simulations with different lengths~$L_z$).
In the following subsections, we consider 2D simulations all with the same upstream plasma conditions as in this subsection, but with different initial current sheet configurations (\S\ref{sec:pertAndEta2d}), and different system sizes (\S\ref{sec:Lx2d}); then, we consider the addition of different guide field strengths (\S\ref{sec:Bz2d}).
We will find that for large systems, the initial current sheet configuration and system size $L_x$ have relatively weak effects on reconnection, including on energy conversion and NTPA; in contrast, (sufficiently strong) guide magnetic field will significantly slow energy conversion and inhibit NTPA.

\subsection{2D reconnection: dependence on initial current sheet configuration}
\label{sec:pertAndEta2d}

In this subsection we show that 2D reconnection
behaviour is
largely (but not completely) determined by the background (upstream)
plasma and not the configuration of the initial current sheet.
We will compare results from 2D simulations with varying initial magnetic perturbation strength $a$, and varying initial current sheet parameters (i.e., density $n_{d0}$ or, equivalently, $\eta=n_{d0}/n_{b0}$, temperature $\theta_d$, drift speed $\beta_d c$, and half-thickness $\delta$).
All simulations in this subsection have identical background plasma (i.e., $\sigma=10^4$, $\sigma_h=1$, hence upstream $\beta_{\rm plasma}=0.5$, 
and $B_{gz}=0$, as well as $L_y/L_x=2$).

The initial magnetic field perturbation $a$ [cf. Eq.~(\ref{eq:pert})], can be varied independently of all other parameters, as in the following substudy (2D-a); however, increasing $a$ shifts the system away from the Harris equilibrium.
In contrast, we always maintain the initial equilibrium when altering $\eta$, $\theta_d$, $\beta_d$, and $\delta$.
The equilibrium satisfies pressure balance, or $\eta \theta_d / \gamma_d = \sigma/2$, and Ampere's law, which requires $\delta = \sigma \rho_0 / (\eta \beta_d)$.
Thus, the initial current sheet, though fully specified through four parameters, has only two independent degrees of freedom in this study; usually we specify $\eta$ and $\beta_d$.
For reference we have listed the relevant parameters 
corresponding to different values of $\eta$ and $\beta_d$ in table~\ref{tab:etaEffect}.

\begin{table}
\centering
\begin{tabular}{l|rrrrrc|@{\hspace{0.05in}}crrrrr} 
  $\eta=n_{d0}/n_{b0}$ & 0.5 & {\bf 1} & 3.1& {\bf 5} & 10 
    && 0.33 & 0.5 & 1 & 2 & 4 \\
  $\beta_d$ & \multicolumn{5}{c}{--- 0.3 ---} 
    && 0.9 & 0.6 & 0.3 & 0.15 & 0.075 \\
  \hline
  $\theta_d/\theta_b$ & 4.2 & 2.1 &  0.68 & 0.42 & 0.21 
    && 14 & 5.0 & 2.1 & 1.01 & 0.50 \\
  $\rho_d/\sigma \rho_0$ & 3.1 & 1.6 & 0.5 &0.31 &0.16
    && 10 & 3.8 & 1.6 & 0.76 & 0.38 \\
  $d_{e,d}/\sigma \rho_0$ & 2.6 & 1.2 & 0.40 &0.24 &0.12
    && 5.5 & 2.8 & 1.2 & 0.62 & 0.31 \\
  $\delta/\sigma \rho_0$ & 6.7 & 3.3 &  1.1 &0.67 & 0.33 
    && \multicolumn{5}{c}{--- 3.3 ---} \\
  $\delta/\rho_d$ & \multicolumn{5}{c}{--- 2.1 ---} 
    && 0.32 & 0.89 & 2.1 & 4.4 & 8.9 \\
  $\delta/d_{e,d}$ & \multicolumn{5}{c}{--- 2.7 ---} 
    && 0.58 & 1.2 & 2.7 & 5.4 & 11 \\
\end{tabular}
\caption{\label{tab:etaEffect} 
Initial temperature and length scales of the drifting (current sheet) plasma resulting from various combinations of initial peak current sheet density~$n_{d0}=\eta n_{b0}$ and drift speed~$\beta_d c$ used in this study.
Listed are (normalized) values of 
the initial temperature $\theta_d m_e c^2$,
the average gyroradius $\rho_d=3\theta_d\rho_0$, and the skin depth of the drifting plasma forming the current sheet, $d_{e,d}=\sqrt{3}\lambda_{Dd}=[3\theta_d m_e c^2/(4\upi n e^2)]^{1/2}$,
as well as the initial sheet half-thickness~$\delta$ (relative to the fiducial upstream plasma length scale $\sigma \rho_0$ and drifting plasma length scales~$\rho_d$ and~$d_{e,d}$).
We note that 
$\theta_d/\theta_b=2\gamma_d \sigma_h/\eta$ 
and $\eta \delta = \sigma\rho_0/\beta_d$.
For comparison, the background plasma gyroradius is $\rho_b=0.75 \sigma\rho_0$ and the skin depth $d_e=\sqrt{3}\lambda_D=0.87 \sigma\rho_0$; the cell size is $\Delta x =0.33 \sigma\rho_0$.
Boldface entries indicate the two configurations used to study the effect of varying perturbation strength~$a$.
}
\end{table}

It is useful to compare the sheet half-thickness $\delta$ with the characteristic gyroradii of the upstream plasma and the drifting (initial current sheet) plasma.
The characteristic gyroradius of the upstream plasma is $\rho_b=3\theta_b\rho_0=(3/4)\sigma\rho_0\sim\sigma\rho_0$; the average gyroradius of the drifting plasma is $\rho_d=3\theta_d\rho_0=3\gamma_d/(2\eta)$.
The ratio of sheet thickness to upstream plasma scales is $\delta/\sigma\rho_0 = 1/(\eta\beta_d)$, and the ratio to drifting plasma scales is $\delta/\rho_d=2/(3\gamma_d\beta_d)$.
In substudy (2D-b), below, we
vary $\eta$ while keeping $\beta_d$ constant, which changes $\delta$ with respect to upstream scales but not with respect to drifting plasma scales.  On the other hand, in substudy (2D-c), we vary $\beta_d\propto \eta^{-1}$, changing $\delta/\rho_d$ but leaving $\delta/\sigma\rho_0$ constant.

\emph{(2D-a) Varying the initial perturbation.} 
For $a=0$, the magnetic field lines are initially entirely parallel to $x$.
Increasing $a$ perturbs the field so that there is a magnetic island and an X-point in the layer.
As described in~\S\ref{sec:setup}, when $a=1$, the initial island half-height $s$ (the maximum height of the separatrix above the initial midplane) equals the current sheet half-thickness, $s=\delta$; for $a\lesssim 1$, the separatrix that bounds the island is contained within the initial layer.
We investigate the effect of varying $a$ over a wide range, for 
two different initial current sheets: the first [$\beta_d=0.3$, $\eta=5$, $\delta=(2/3)\sigma\rho_0$, $\theta_d=0.10\sigma$] was chosen to be like most simulations in this paper, marginally resolving the initial sheet with $\delta/\Delta x=2$; 
the second [$\beta_d=0.3$, $\eta=1$, $\delta=(10/3)\sigma\rho_0$, $\theta_d=0.52\sigma$] was chosen to resolve the current sheet with $\delta/\Delta x=10$.
In both cases we used modest system sizes with $L_x=320\sigma\rho_0$.
To help distinguish between the effect of $a$ and stochastic variation,
we ran two simulations for each $a$, identical except for random 
initial particle velocities (the source of randomness is discussed later in detail in~\S\ref{sec:variability3d}).

\begin{figure}
\centering
\fullplot{
\raisebox{0.315\textwidth}{\large (a)}%
\includegraphics*[width=0.455\textwidth]{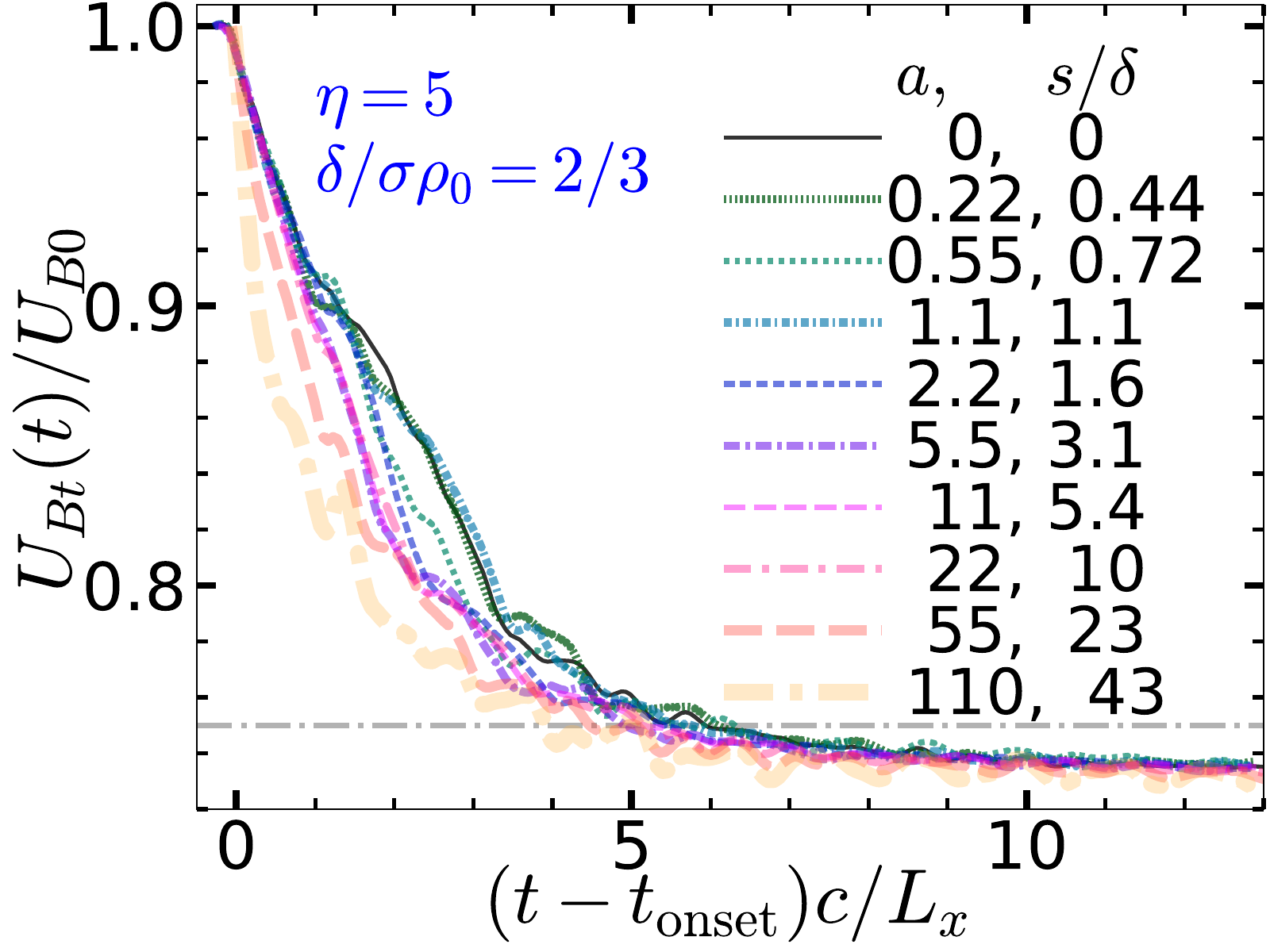}%
\hfill
\raisebox{0.315\textwidth}{\large (b)}%
\includegraphics*[width=0.455\textwidth]{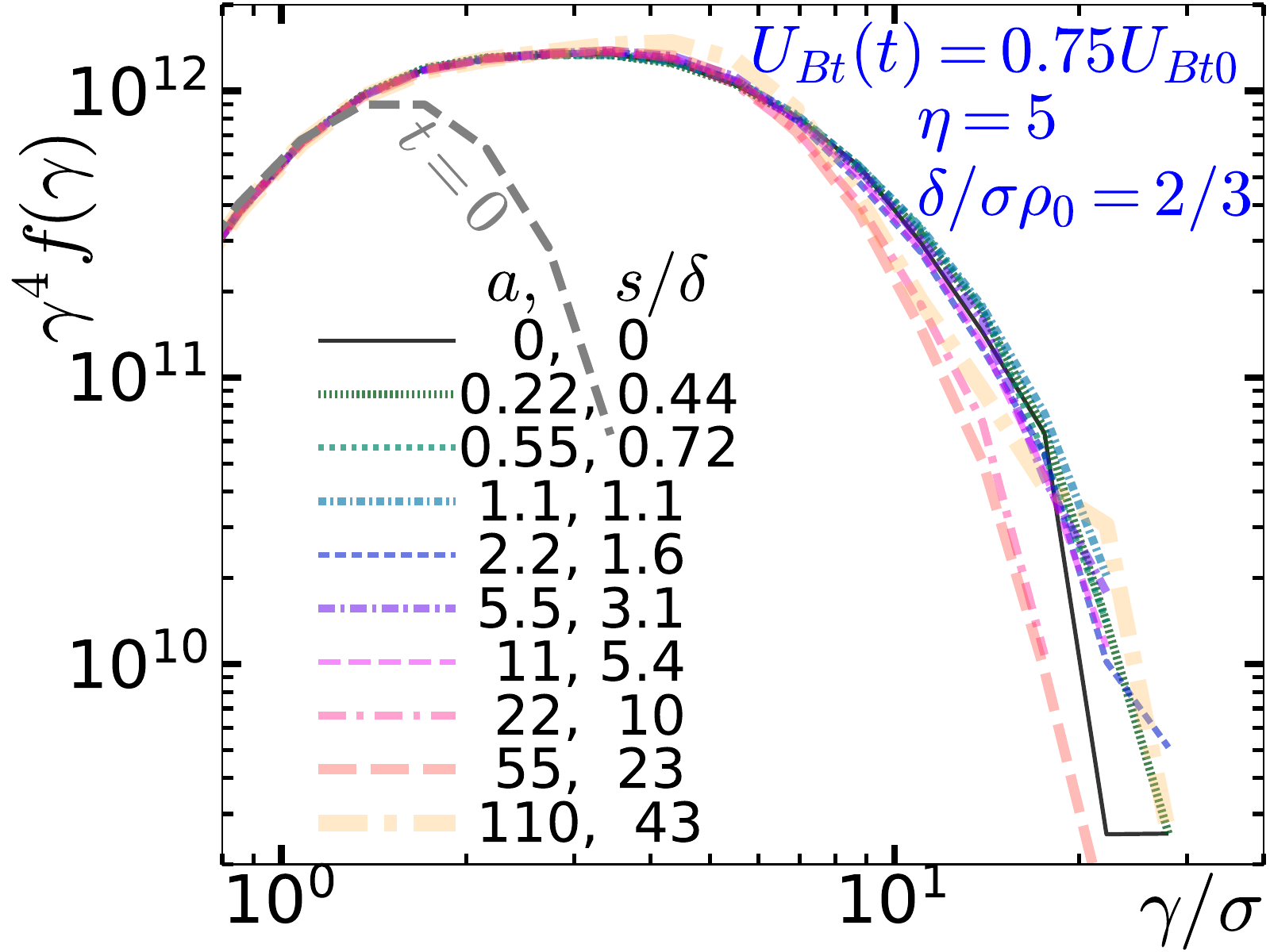}\\
\raisebox{0.315\textwidth}{\large (c)}%
\includegraphics*[width=0.455\textwidth]{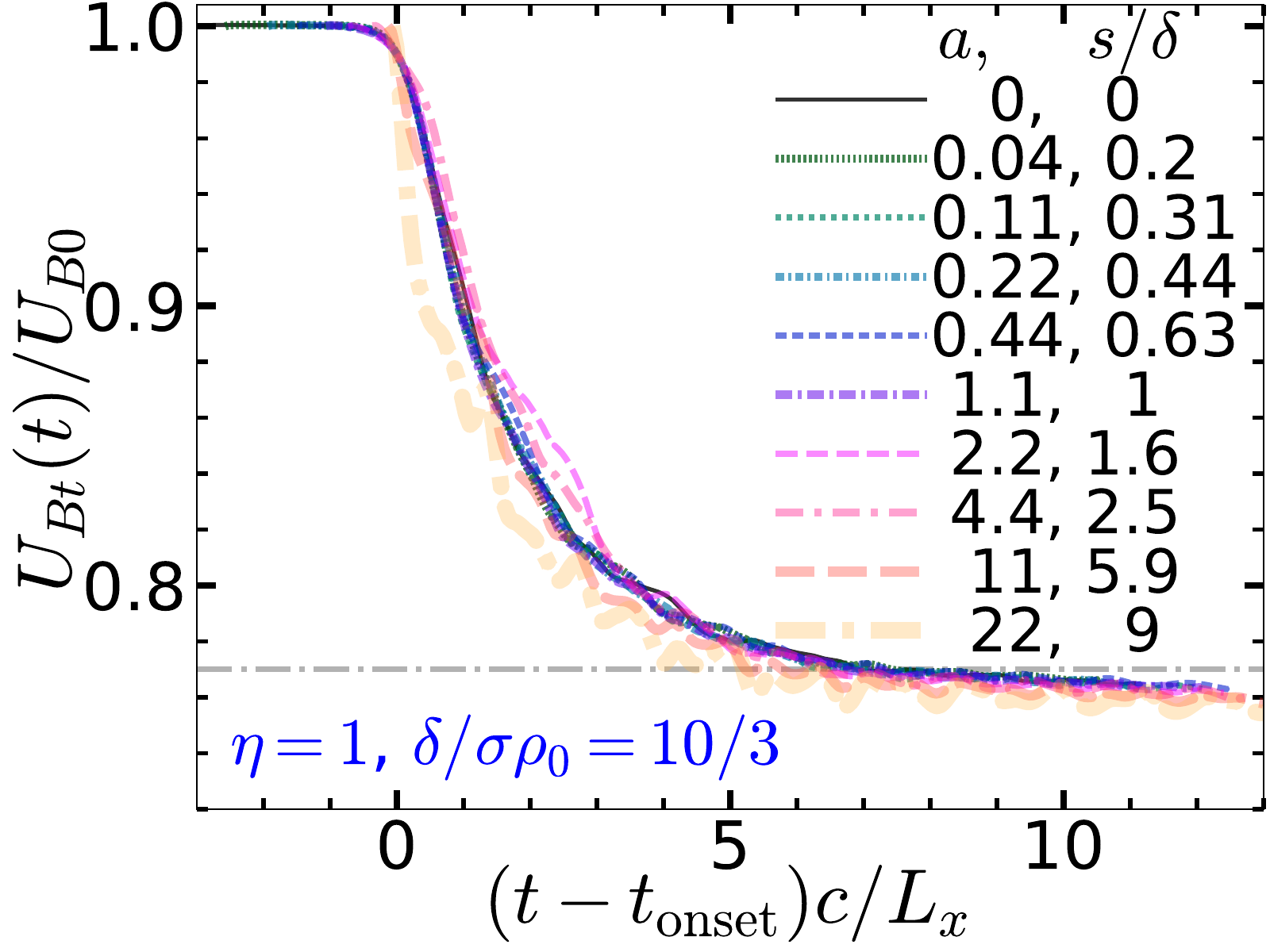}%
\hfill
\raisebox{0.315\textwidth}{\large (d)}%
\includegraphics*[width=0.455\textwidth]{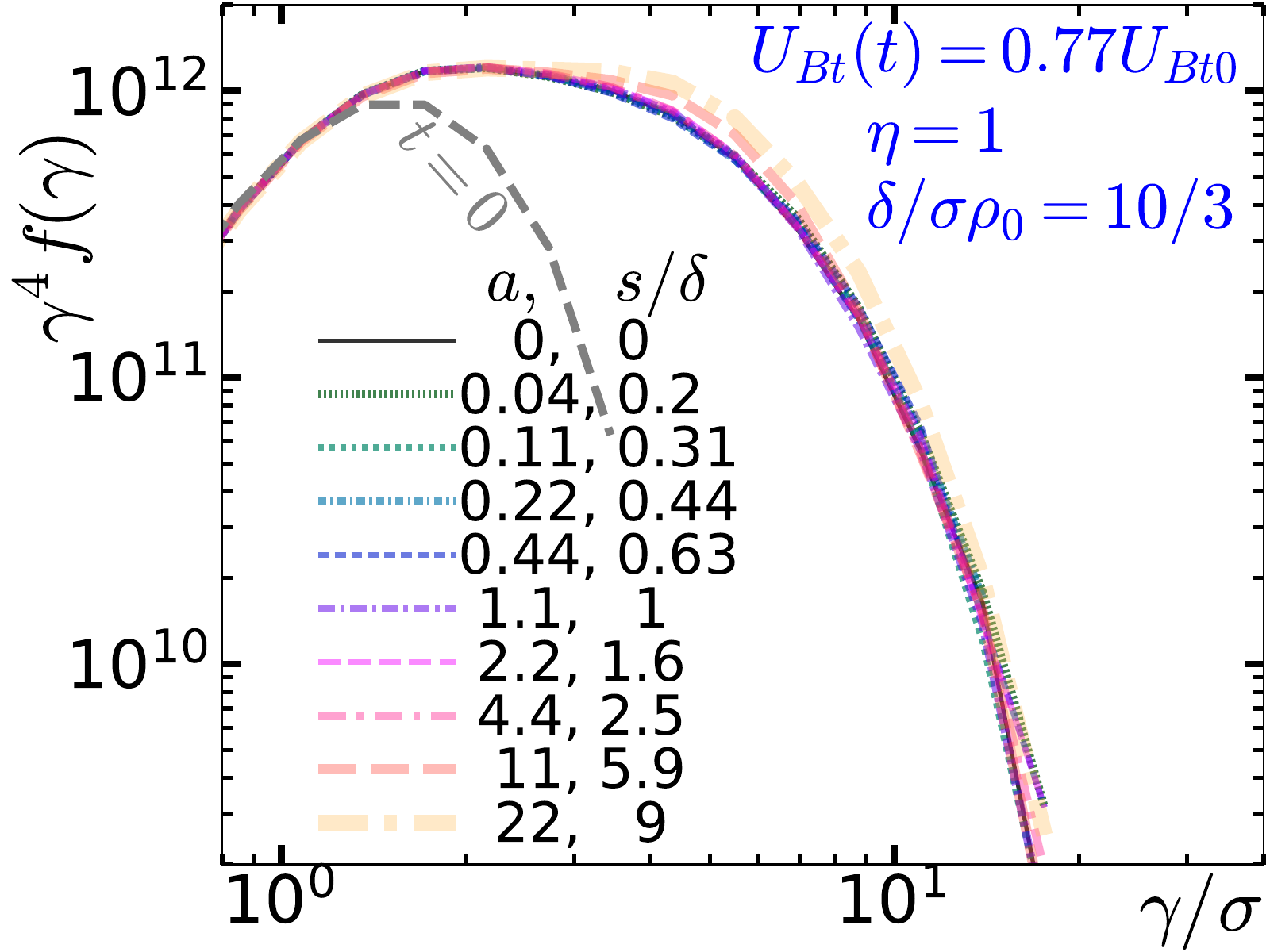}\\
}
\caption{ \label{fig:perturb2dLx320}
(a) Transverse magnetic energy versus time, and (b) electron energy spectra
$f(\gamma)$, compensated by $\gamma^4$, at times~$t$ when $U_{Bt}(t)=0.75U_{B0}$---marked by the horizontal grey line in (a)---for
different initial magnetic field perturbations in
2D simulations with an initially dense, thin current sheet  
($\eta=5$, $\delta/\sigma \rho_0=2/3$).
Panels~(c) and (d) show the same for a less dense, thicker layer
($\eta=1$, $\delta/\sigma \rho_0=10/3$), with spectra at $U_{Bt}(t)=0.77U_{B0}$.
}
\end{figure}

For the denser, thinner sheet ($\beta_d=0.3$, $\eta=5$, used in most simulations here), 
we ran simulations with  
$a \in \{$0, 0.22, 0.55, 1.1, 2.2, 5.5, 11, 22, 55, 110$\}$, corresponding to $s/\delta \in \{$0, 0.44, 0.72, 1.1, 1.6, 3.1, 5.4, 10, 23, 43$\}$, respectively.
For $a < 0.3$, we note that $s < \Delta x$, so the simulation can hardly tell the difference between
$a=0$ and $a=0.22$;
and for $a=110$, the separatrix extends out to $s=43\delta=29 \sigma\rho_0 \approx 0.1L_x$.
In Fig.~\ref{fig:perturb2dLx320}(a) we observe that the evolution of 
magnetic energy
versus time is nearly the same for the two simulations with $a < 0.3$; it is also
the same, up to stochastic variation, within the group of simulations with 
$0.55 \lesssim a \lesssim 22$.
The simulations with $a<0.3$ exhibit slightly slower magnetic energy conversion for $1 \lesssim tc/L_x \lesssim 4L_x/c$ than the simulations with $0.55 \lesssim a \lesssim 22$, but by $t\gtrsim 5L_x/c$ they have all converted the same amount of energy, so while noticeable, this is a very minor difference.
For $a\gtrsim 55$ (very large perturbations), 
we observe an increasingly quick onset of reconnection and more rapid
magnetic depletion.
The particle energy spectra in Fig.~\ref{fig:perturb2dLx320}(b) are 
all very
similar [considering that differences are enhanced by 
showing $\gamma^4 f(\gamma)$ instead of $f(\gamma)$], 
although very large perturbations result in an almost
imperceptibly harder spectrum with a slightly smaller fraction of 
particles reaching the very highest energies.
We conclude that for all but very large perturbations, the initial perturbation
does not have a very substantial effect on energy conversion or NTPA.

For the less dense, initially-thick current sheet, [$\beta_d=0.3$, $\eta=1$, $\delta=(10/3)\sigma\rho_0$, $\theta_d=0.52\sigma$],
we explored $a \in \{$0, 0.04, 0.11, 0.22, 0.44, 1.1, 2.2, 4.4, 11, 22$\}$, 
corresponding to separatrix heights $s/\delta\in \{$0, 0.2, 0.31, 0.44, 0.63, 1, 1.6, 2.5, 5.1, 9$\}$.
The results are shown in Fig.~\ref{fig:perturb2dLx320}(c,d). 
Here, $s>\Delta x$ as long as $a > 0.011$ (i.e., for all the nonzero $a$ used); for $a=22$, the separatrix extends to $s=9\delta = 30\sigma\rho_0\approx 0.1L_x$.
The less dense, thicker sheet 
leads to a slower onset of reconnection; increasing
the initial perturbation strength tends to hasten reconnection onset.
However, once reconnection is triggered, the energy evolves very similarly for zero and large perturbations, as seen in
Fig.~\ref{fig:perturb2dLx320}(c), which shows magnetic energy versus time, 
shifted relative to $t_{\rm onset}$,
the time at which the magnetic energy has fallen by 1~per~cent (cf.~\S\ref{sec:onsetTime}).
With this time shift, the magnetic energy curves are almost identical 
in all cases except for the largest perturbation, $a=22$.
Not surprisingly, the particle spectra in Fig.~\ref{fig:perturb2dLx320}(d) 
are similarly identical (within stochastic variation).
We again conclude that the strength of perturbation
(at least for $a\lesssim 20$) has negligible effect on energy conversion 
and NTPA---aside from having a strong effect on the time to reconnection 
onset.

\begin{figure}
\centering
\fullplot{
\includegraphics*[width=\textwidth]{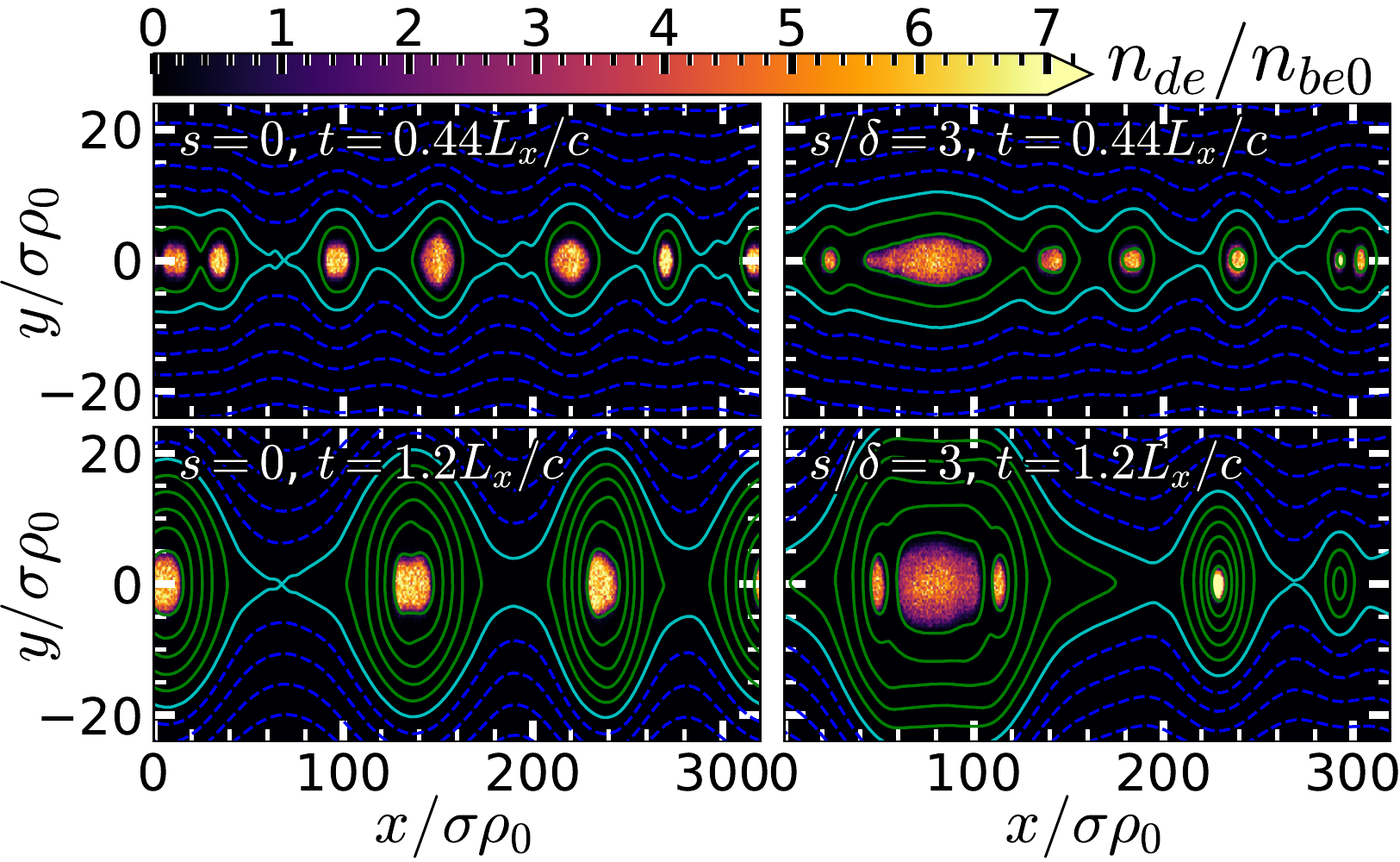}%
}
\caption{ \label{fig:ndePerturb2dLx320}
Magnetic field lines superimposed over 
the density $n_{de}$ of the initially-drifting electrons that form the initial
current sheet (i.e., excluding background electrons), for cases (left) without an initial magnetic perturbation ($a=0$), and (right) with a small
initial perturbation $a=5.5$ or $s/\delta=3$ [cf. Fig.~\ref{fig:perturb2dLx320}(a)],
at times $t=0.44L_x/c$ (top) and $t=1.2L_x/c$ (bottom).
In both cases, the drifting particles are well contained in plasmoid cores.
With zero perturbation (left), the first plasmoids formed are roughly equal in size, and they generally merge in pairs so that at any time, there is some number of equal-sized largest plasmoids, each containing a similar fraction of the drift particles.
In contrast, a small
perturbation (right) favours one plasmoid, which remains always larger than all other plasmoids, and rapidly gathers most and then all of the drifting particles.
Magnetic field lines in the layer (i.e., within the separatrix) are solid green, with the separatrix being solid cyan; upstream of the separatrix, field lines are dashed blue.
These simulations have size $L_x=320\sigma \rho_0$, with 
an initially thin current sheet, $\delta=(2/3) \sigma \rho_0$.
}
\end{figure}

For both overdensities $\eta=1$ and $\eta=5$,
the initial perturbation has significant consequences for the evolution of the plasmoid hierarchy, even though this does not seem to alter the overall energy conversion rate and NTPA.
Without perturbation (Fig.~\ref{fig:ndePerturb2dLx320}, left), 
a number of roughly equal-sized plasmoids form at the onset of 
reconnection, with sizes determined by the fastest-growing 
wavelength of the tearing instability. 
These plasmoids undergo the coalescence instability and merge in pairs, resulting in a smaller
number of larger plasmoids, 
each of which contains a similar
fraction of the initially-drifting particles that made up the initial
current sheet.
In contrast, even a very small initial perturbation will favour one plasmoid (the 
``major'' plasmoid), which is always larger than the others
(Fig.~\ref{fig:ndePerturb2dLx320}, right);
it may contain many or most of the initially-drifting particles.
In this case, subsequent plasmoid mergers tend to be between different-sized plasmoids;
plasmoids form at the major X-point, and move toward the major
plasmoid (with mergers between smaller plasmoids 
sometimes occurring before they reach the major plasmoid).
We note that the final state has just a single monster plasmoid, regardless of initial perturbation.

Summarizing (2D-a), we see that (for $L_x\gtrsim 320\sigma\rho_0$) the 
initial perturbation (if not terribly large, i.e., $a\lesssim 20$, or
$s/\delta \lesssim 9$)
does not significantly alter global energy evolution (after reconnection
onset) or NTPA.  
However, the initial perturbation can reduce the time it takes reconnection to start, and
it can also have a significant effect on the nature of plasmoid 
formation and subsequent evolution.
Similar differences in plasmoid evolution caused by artificial triggering with a locally reduced pressure have been previously studied in 2D semirelativistic electron-ion reconnection \citep{Ball_etal-2019}; there NTPA was found to be similarly insensitive to triggering, at least for weak guide field.

\emph{(2D-b) Varying current sheet density $\eta$ with fixed $\beta_d=0.3$; i.e., varying $\delta/\sigma\rho_0$ for fixed $\delta/\rho_d=2.1$}.  
We now look at the effects of different initial current sheets, keeping $\beta_d=0.3$ constant and varying $\eta \in \{$0.5, 1, 3.1, 5, 10$\}$
(considering $\eta < 1$ is unusual in reconnection research, but we include $\eta=0.5$ to show that it is not that different from $\eta=1$, although 
it does start to indicate a trend for the limit $\eta \rightarrow 0$).
Table~\ref{tab:etaEffect} shows how other quantities such as $\delta$ vary as $\eta$ is varied in this study.
Increasing $\eta$ directly increases the current sheet density $n_{d0}$ (relative to the upstream plasma density $n_{b0}$, which remains the same).
As $\eta$ increases, the temperature $\theta_d$ decreases to maintain pressure balance across the current sheet against the upstream magnetic field, $\theta_d = \gamma_d \sigma / (2\eta) \approx 0.52 \sigma/\eta$; this reduces the fundamental length scales (and hence also time scales) of the current sheet.
(We note, for context, that if background plasma with $n_{b0}$ and $\theta_b$ were 
adiabatically compressed with adiabatic index 4/3, appropriate for
relativistically-hot plasma, to a density $n_{\rm ad}$ and temperature
$\theta_{\rm ad}$ 
sufficient to balance the upstream magnetic pressure $B_0^2/8\upi$ as well as the upstream plasma pressure $n_{b0}\theta_b m_ec^2$, this would yield $\eta_{\rm ad}\equiv n_{\rm ad}/n_{b0}=2.28$ and
$\theta_{\rm ad}/\theta_b=1.3$.)

As $\eta$ varies (with fixed $\beta_d$), we have $\theta_d \propto \eta^{-1}$ and so $\delta$ as well as the fundamental length scales of the current sheet plasma (e.g., both the gyroradius $\rho_d$ and collisionless skin depth $d_{ed}$) all scale $\propto \eta^{-1}$ (as long as $\gamma_d \approx 1$). 
Therefore, $\delta/\rho_d=2.1$ and $\delta/d_{ed}=2.7$ remain constant (see table~\ref{tab:etaEffect}), although the current sheet thickness changes with respect to upstream plasma scales (e.g., $\rho_b=0.75\sigma\rho_0$) as well as $L_x$ and $\Delta x$.

For this investigation we use a larger system size, $L_x=1280\sigma\rho_0$, 
so that $L_x,L_y \gg \delta$ for all $\delta$ explored; other fixed parameters are: $\beta_d=0.3$ and $B_{gz}=0$.
We used the exact same initial magnetic field $\boldsymbol{B}(x,y)$ in all cases, with the same small 
perturbation: $a \delta /L_x=0.0052$ --- the dependence $a\sim \delta^{-1}$ ensures that the initial field is independent of $\delta$---i.e., $s/L_x$ is fixed [cf. Eq.~(\ref{eq:pert})].
Based on (2D-a) above, we expect our conclusions for this subsection would be the same for other small perturbations, including zero perturbation.

\begin{figure}
\centering
\fullplot{
\includegraphics*[width=0.49\textwidth]{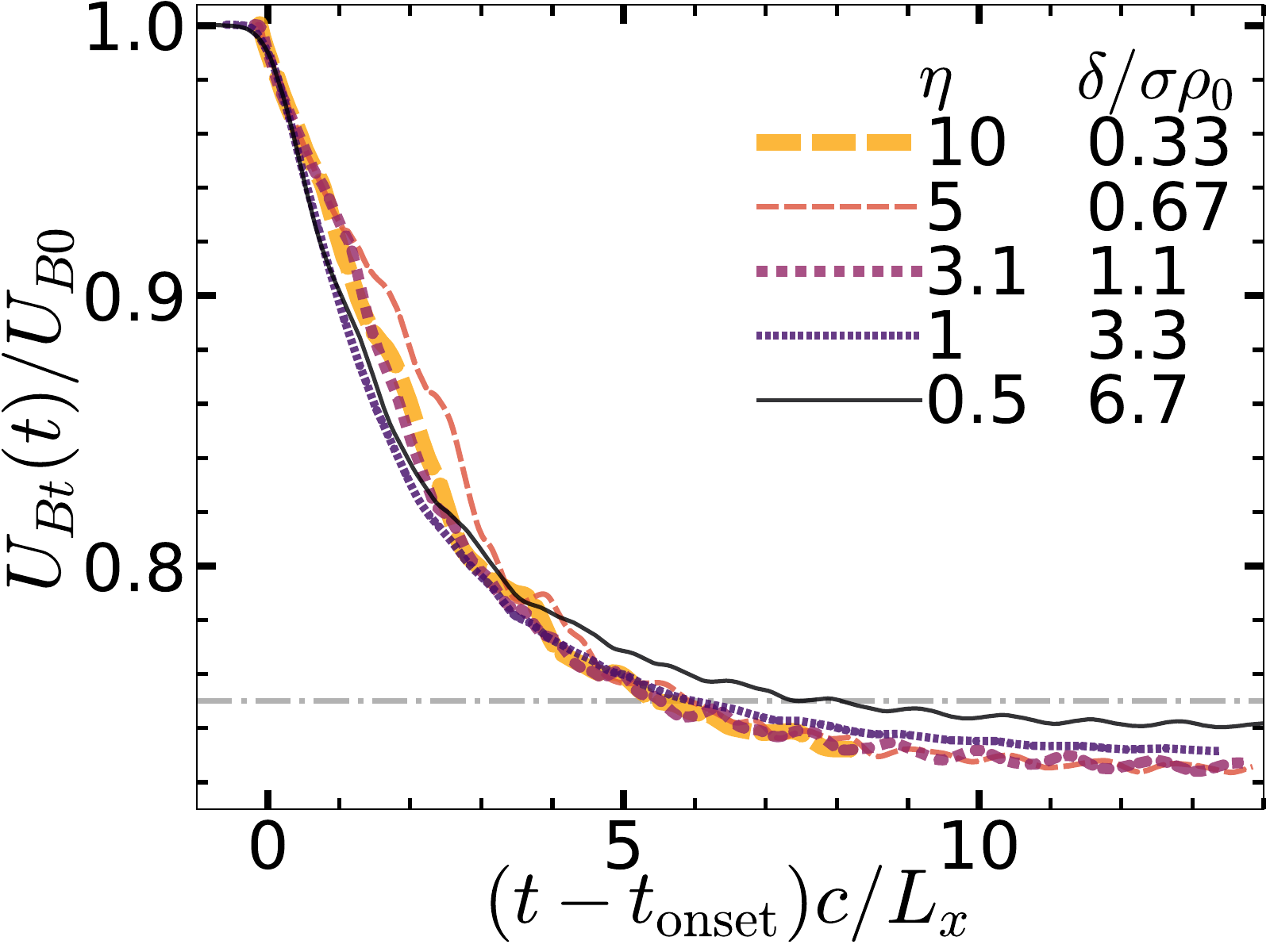}%
\hfill
\includegraphics*[width=0.49\textwidth]{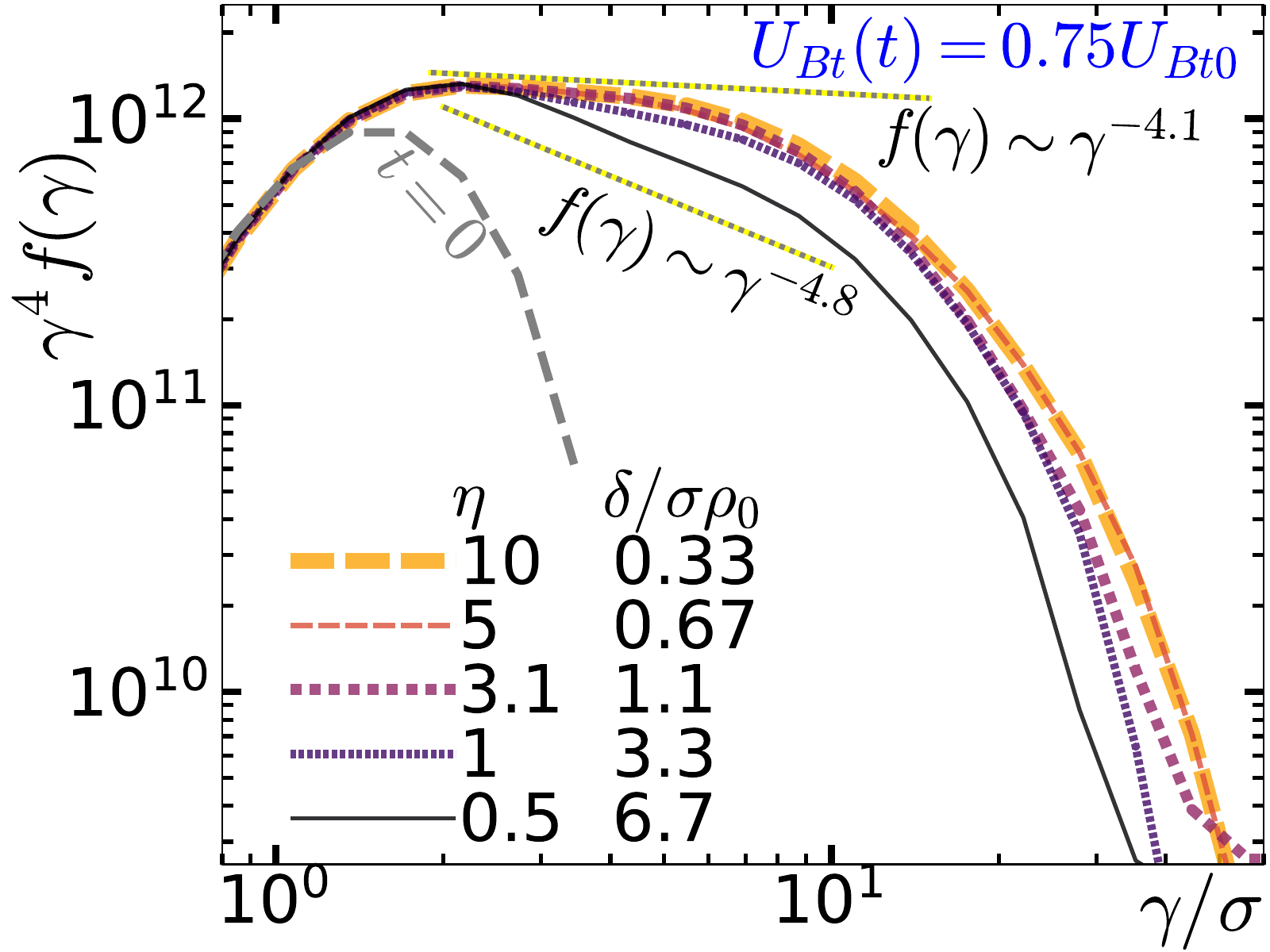}
}
\caption{ \label{fig:eta2dLx1280}
For simulations with varying $\eta$ (hence varying initial
current sheet half-thicknesses $\delta$; 
cf.~table~\ref{tab:etaEffect}):
(left) transverse magnetic energy versus time (shifted by $t_{\rm onset}$; cf.~\S\ref{sec:onsetTime}),
and (right) electron energy
spectra times $t$ when $U_{Bt}(t)=0.75 U_{B0}$ (indicated by the horizontal grey line in the left panel).  Dotted grey-on-yellow lines indicate power-law slopes of 4.1 and 4.8.
All simulations have size
$L_x=1280\sigma \rho_0$. 
}
\end{figure}

Figure~\ref{fig:eta2dLx1280} shows (left) the magnetic energy evolution and
(right) NTPA spectra for simulations 
with a range
of $\eta=0.5$--$10$, corresponding to $\delta/\sigma\rho_0=6.7$--$0.33$.
We first note, as above, that reconnection onset is delayed by less dense, thicker 
current sheets, but that once magnetic energy depletion begins,
energy evolution and NTPA are very similar and almost independent of~$\eta$.
However, closer examination shows a tentative trend of decreasing 
NTPA efficiency with decreasing $\eta$; this trend is very small for $\eta >1$, but becomes noticeable for
the $\eta =0.5$ case (with $\theta_d/\theta_b=4.2$ and 
$\delta\approx 7\sigma\rho_0$), which reconnects a little slower and generates slightly less efficient NTPA.  
Specifically, $\eta =0.5$ yields a power law $p=4.6\pm 0.1$, whereas denser layers with $\eta \geq 1$ yield $p=4.4\pm 0.1$.
Also, denser initial sheets seem to lead to slightly more overall
magnetic energy depletion.
This weak $\eta$-dependence
hints at a more severe behaviour in the limit of underdense current sheets: in a simulation (not shown here) with $\eta=0.1$ and $\delta=33\sigma\rho_0$, reconnection did not even start, despite a simulation duration longer than $10L_x/c$---hence no magnetic energy was depleted.

We thus conclude that the initial current sheet density has a negligible effect on total energy conversion and NTPA efficiency as long as $\eta \gg 0.1$; however, for small $\eta$, NTPA efficiency and total energy conversion decrease as $\eta$ decreases, and the thick, low-density current sheet may eventually become effectively stable against tearing and reconnection.

\emph{(2D-c) Varying $\beta_d\propto \eta^{-1}$; i.e., varying $\delta/\rho_d$ for fixed $\delta/\sigma\rho_0=(10/3)\sigma\rho_0$.}
The motivation for this particular parameter space exploration 
is a little complicated.
Essentially, we would like to study the influence of $\beta_d$ on reconnection.
After substudy (2D-b), which varied $\eta$ but not $\beta_d$, it would be natural to hold $\eta$ constant while varying $\beta_d$.
However, we will instead vary $\eta \propto \beta_d^{-1}$ to keep $\eta\beta_d$ constant.
Importantly, this fixes $\delta=\sigma\rho_0/(\eta\beta_d)$ with respect to upstream plasma scales.
In particular, $\delta/\Delta x=10$ and $L_x/\delta=384$ will be the same in every case, so that, as $\beta_d$ and $\eta$ vary, the initial current sheet will remain well-resolved but very thin compared with $L_y=2L_x$.
Satisfying these two criteria will be important in 3D (\S\ref{sec:pertAndEta3d}--\ref{sec:RDKIamp}) when we will focus on the evolution of the current layer at early times as it undergoes large deformations due to RDKI, requiring $\delta/\Delta x\gg 1$ and $\delta/L_y \ll 1$; in addition, it will be more straightforward to compare layer deformation and its interaction with the upstream plasma for different simulations when the current sheets begin with the same thickness.
Although neither RDKI nor extreme layer deformation will occur in 2D, this substudy provides a baseline against which to compare the 3D simulations.

Therefore, we simultaneously vary $\eta$ along with $\beta_d$ to keep the product $\eta\beta_d$ constant.
Maintaining constant $\eta\beta_d$ fixes $\delta/\sigma \rho_0$, while essentially varying the drifting plasma length scales [such as gyroradius $\rho_d=3\theta_d\rho_0$] with respect to $\delta$---i.e., varying $\delta/\rho_d = 2/(\gamma_d \beta_d)$.

We compare five configurations with the same background plasma (with zero guide field), system size $L_x=1280\sigma\rho_0=384\delta$, and zero initial magnetic perturbation ($a=0$).
We vary $\beta_d\in \{$0.075, 0.15, 0.3, 0.6, 0.9$\}$; to keep $\eta \beta_d$ constant, the overdensities $\eta$ are $\{$4, 2, 1, 0.5, 0.33$\}$, yielding the same $\delta=(10/3)\sigma \rho_0=10\Delta x$ for all five cases.
Other parameters are (see also table~\ref{tab:etaEffect}): $\theta_d/\theta_b \in \{0.5$, 1.0, 2.1, 5.0, 14$\}$ and $\delta/\rho_d \in \{8.9$, 4.4, 2.1, 0.89, 0.32$\}$, respectively.

At $\beta_d=0.075$, $\eta=4$, the gyroradius and collisionless skin depth of the drifting plasma are at their smallest values of $\rho_d=0.38\sigma\rho_0$ and $d_{ed}=0.31\sigma\rho_0$, both near the grid resolution $\Delta x$, and the half-thickness $\delta$ is 10 times larger than those scales.
At the other end of the parameter scan, $\beta_d=0.9$, $\eta=0.33$, the microphysical scales of the initial current sheet become somewhat larger than $\delta$: $\delta \approx 0.3 \rho_d \approx 0.6 d_{ed}$.

\begin{figure}
\centering
\fullplot{
\includegraphics*[width=0.49\textwidth]{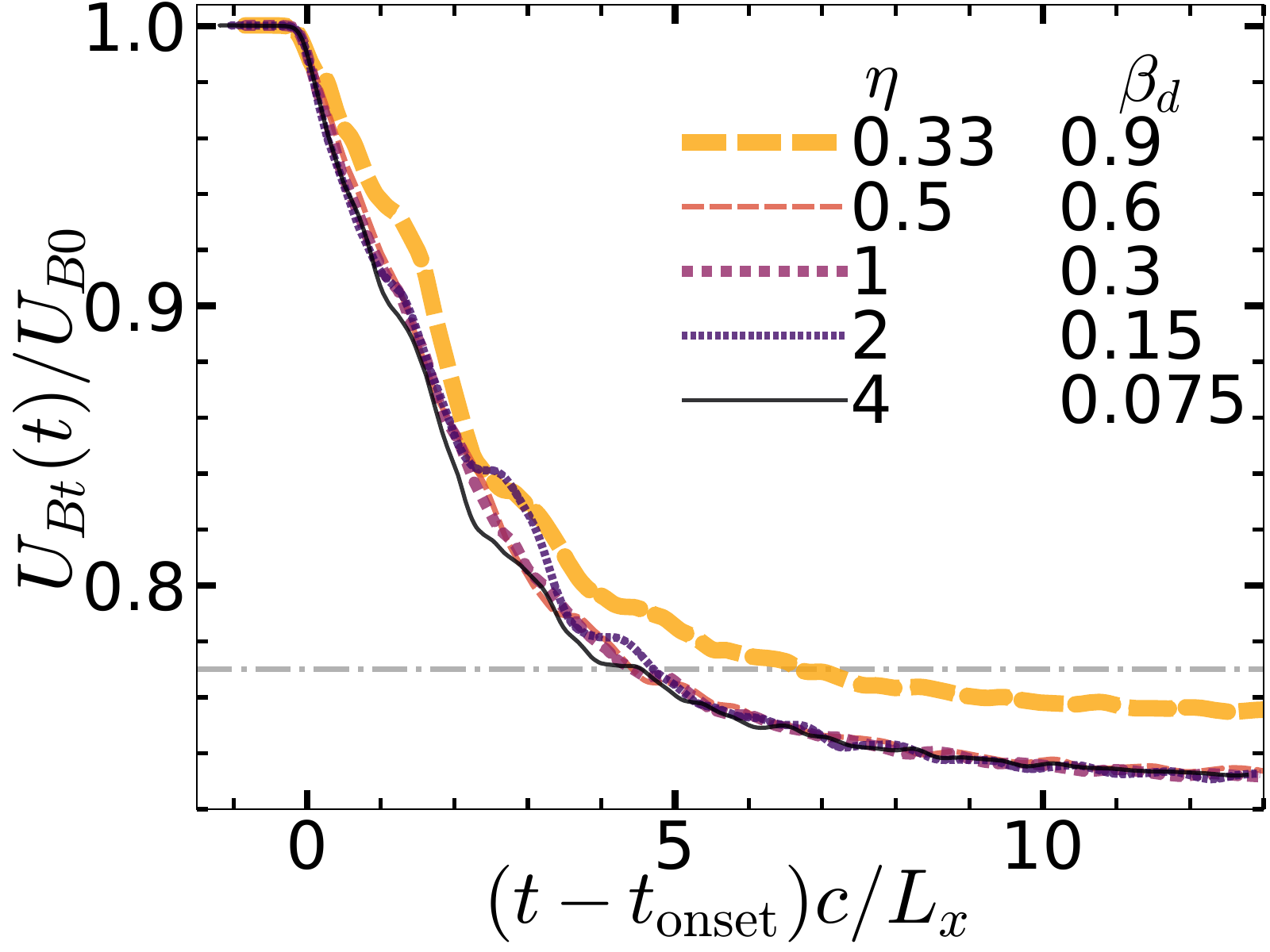}%
\hfill
\includegraphics*[width=0.49\textwidth]{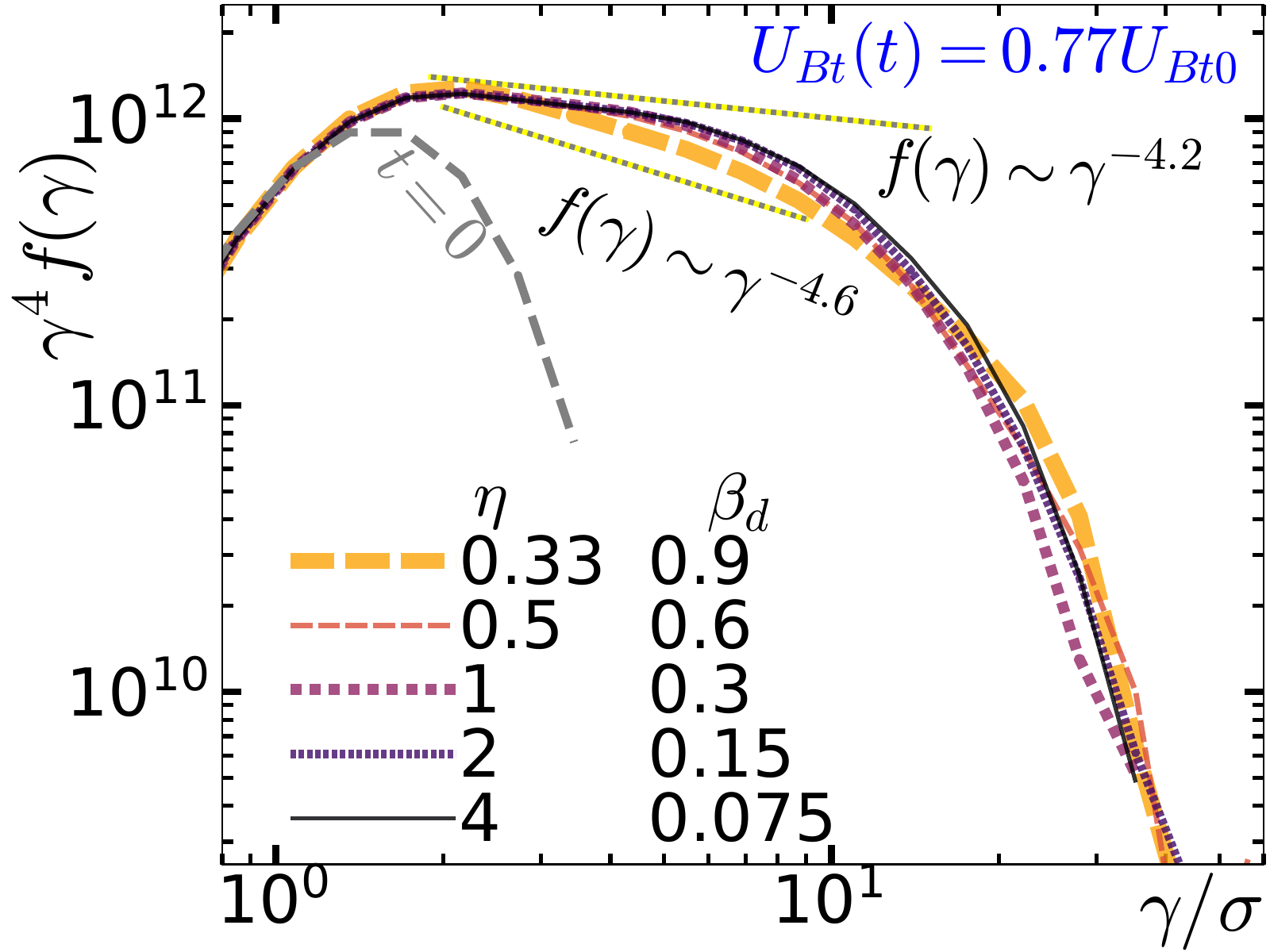}
}
\caption{ \label{fig:betad2dLx1280}
(Left) Transverse magnetic energy versus (shifted) time, 
and (right) electron energy
spectra when 23~per~cent of the initial magnetic energy has been converted (see the horizontal grey line in the left panel), for simulations with varying initial current drift $\beta_d$, but constant initial sheet thickness $\delta=(10/3)\sigma\rho_0$.  (Table~\ref{tab:etaEffect} shows how initial temperatures and gyroradii vary with $\beta_d$.)
Dotted grey-on-yellow lines indicate power-law slopes of~4.2 and~4.6.
All simulations have size
$L_x=1280\sigma \rho_0$. 
}
\end{figure}

To characterize the effects of varying $\beta_d$ (with $\eta$), we focus on the time history of magnetic and plasma energy, and on the resulting NTPA; and again, we find little variation among this set of simulations
(see Fig.~\ref{fig:betad2dLx1280}).
All cases show very similar magnetic depletion rates and NTPA (within stochastic variation), except for the case $\beta_d=0.9$ ($\eta=0.33$), which is a little different; this case ends up with a bit more energy in plasmoids, although the rate at which \emph{upstream} magnetic energy and unreconnected flux decrease is similar to the others.
The $\beta_d=0.9$ case has significantly hotter (though lower density) drift particles (cf.~table~\ref{tab:etaEffect}), which results in correspondingly hotter and less dense plasmoids cores, since the initially-drifting particles, being close to the initial midplane, are among the first to be swept into plasmoids and remain near the plasmoid cores (the original drift kinetic energy may also be converted to thermal energy, contributing to even hotter cores).

\emph{Summary of 2D-a,b,c.} 
We have seen that in 2D, the details of the initial current sheet can affect the time delay before the onset of reconnection, as well as the evolution of the plasmoid hierarchy.  Nevertheless, the initial current sheet configuration has relatively little effect on the long-term magnetic energy conversion and NTPA in 2D reconnection; after a short initial stage, reconnection energetics are dominated by the upstream plasma, regardless of the very early current sheet evolution.
This conclusion may seem to be at odds with that from \citet{Ball_etal-2019}, which found subtle but significant differences in NTPA due to initial perturbation and varying initial sheet thickness;
however, \citet{Ball_etal-2019} studied transrelativistic electron-proton reconnection, with low upstream plasma beta and non-zero guide magnetic field ($B_{gz}/B_0=$0.1 and 0.3).
They used a small pressure imbalance as an initial perturbation, and found that its influence diminished with diminishing guide field, potentially consistent with our zero-guide-field results here.
The impact of varying sheet thickness (keeping $\eta$ constant) was also seen at the larger guide field, $B_{gz}=0.3B_0$, so there are several potentially important differences between these two studies: transrelativistic electron-proton versus ultrarelativistic pairs, low upstream plasma beta versus beta of 0.5 ($\sigma_h=1$), weak-to-moderate versus zero guide field, and pressure versus magnetic field perturbation.

\subsection{2D reconnection: system-size dependence}
\label{sec:Lx2d}

In this subsection we investigate how reconnection depends on system size~$L_x$, to estimate the applicability of our kinetic simulations to astrophysical systems, which may be much larger than even the largest supercomputers can possibly simulate while including full kinetics.
We present results for system sizes ranging over 
a factor of 30, from $L_x/\sigma \rho_0=80$ to~2560.  
These simulations achieve a substantial scale separation between the system size~$L_x$ and all the microphysical scales, which are of order $\sigma\rho_0$ (cf.~\S\ref{sec:setup}).
Such large scale separations are feasible because, for $\sigma_h=1$, all the kinetic scales are comparable, hence they can all be resolved by a cell size~$\Delta x$ only marginally smaller than $\sigma\rho_0$.
All simulations presented in this subsection are 2D with zero guide field,
starting with an initial current sheet half-thickness
$\delta=(2/3)\sigma\rho_0$, $\eta=5$, $\beta_d=0.3$, and zero perturbation.

\begin{figure}
\centering
\fullplot{
\includegraphics*[width=0.49\textwidth]{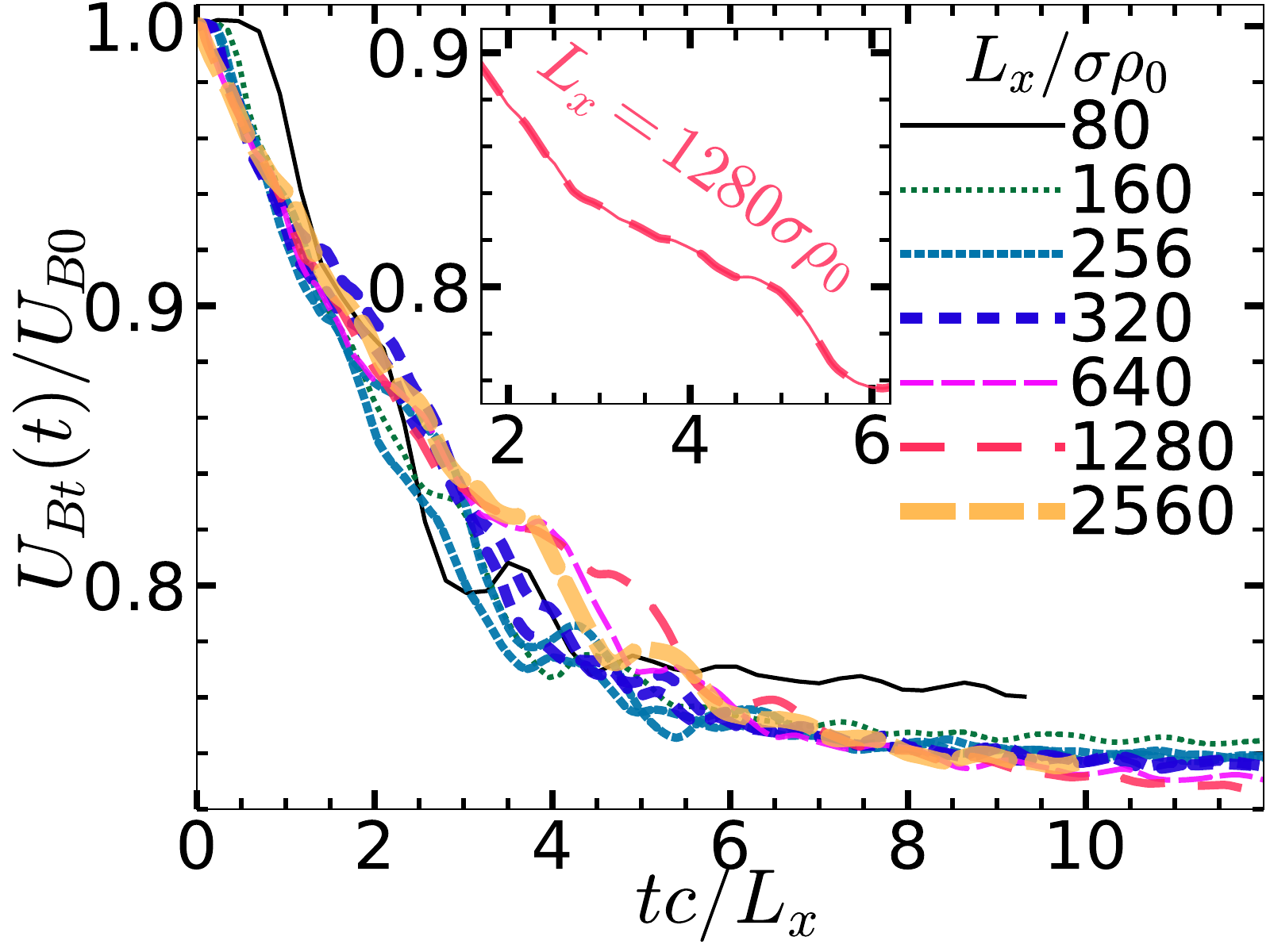}%
\hfill
\includegraphics*[width=0.49\textwidth]{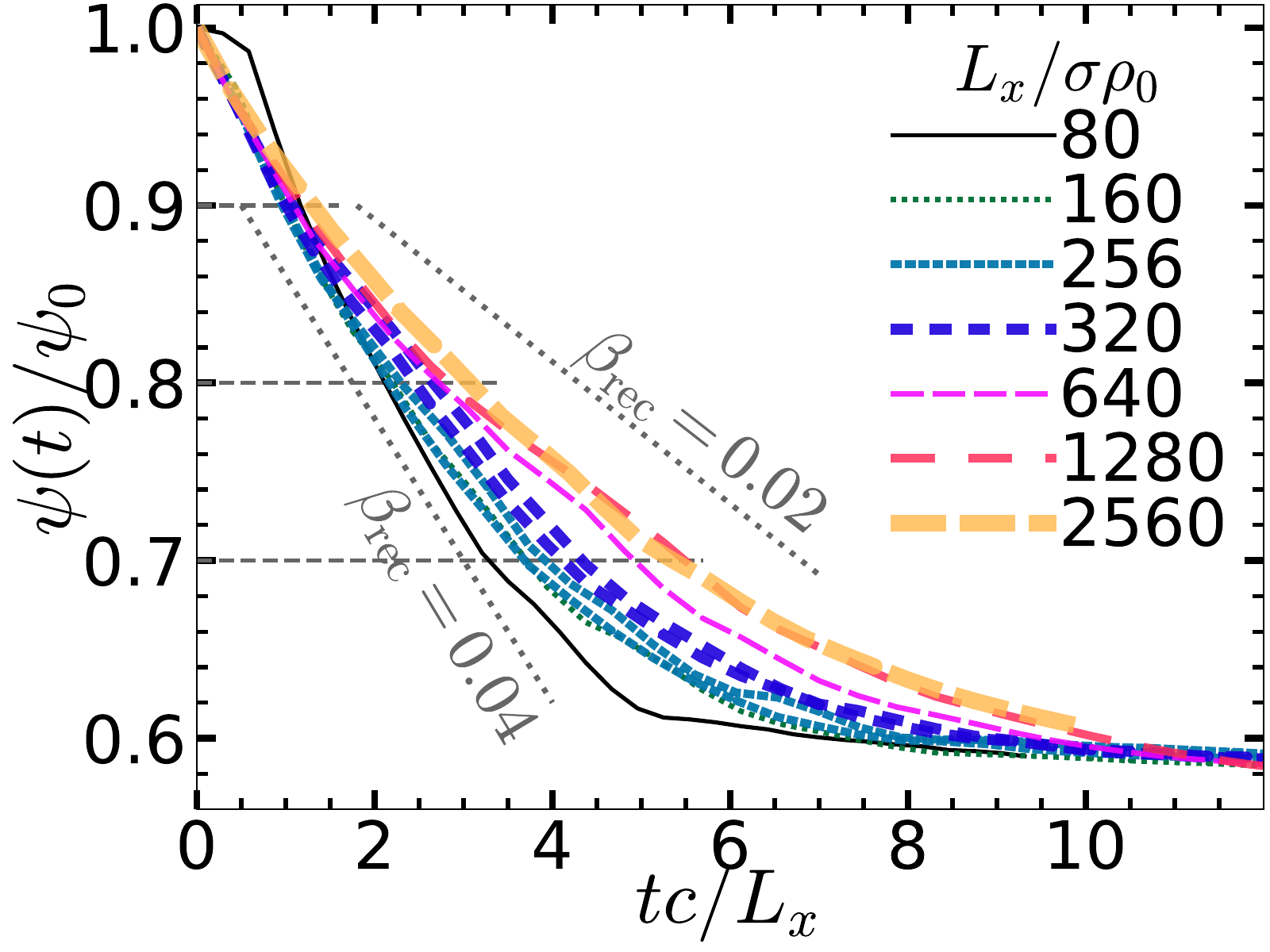}
}
\caption{ \label{fig:energyFluxVsLx2d}
For different system sizes $L_x$:
(left) transverse magnetic energy versus time, with an inset showing~$L_x=1280\sigma\rho_0$ alone, and (right) unreconnected flux versus time.
The horizontal dashed lines indicate the periods over which the average reconnection rate is given in Fig.~\ref{fig:reconRateAndTotalDissipationVsLx2d},
and slopes corresponding to $\beta_{\rm rec}=0.04$ and~0.02 (cf.~\S\ref{sec:unreconnectedFlux}) are shown as grey dotted lines.
}
\end{figure}

Before discussing how~$L_x$ affects typical reconnection behaviour, we note that stochastic evolution of the plasmoid hierarchy affects energy conversion and NTPA, especially over short timescales.
For example, in the $L_x=1280\sigma\rho_0$ simulation, the magnetic energy 
depletion slows (compared with earlier times) between~$t\approx 3L_x/c$ and~$t \approx 5L_x/c$, so that $U_{Bt}(t)$ is distinctly higher at $t=5L_x/c$ than in all the other simulations (Fig.~\ref{fig:energyFluxVsLx2d} left, inset). 
At~$t\approx 3L_x/c$, the lower layer in the simulation domain for $L_x=1280\sigma\rho_0$ has two large,
roughly equal-sized plasmoids, separated by~$\approx L_x/2$
(while the upper layer already has only one final plasmoid at this time). 
In principle, one of the plasmoids could either move left to merge with the 
other, or (since the simulation is periodic in~$x$) it could move right.
Thus they reach a ``Buridan's Ass'' metastable state that seems to slow reconnection evolution for a short time.
However, by $4.4L_x/c$, the two plasmoids have 
decided which way to move, and by $4.8L_x/c$ they have
started to merge, causing a burst of magnetic energy depletion; by the time the merger is nearly complete at $6.1L_x/c$, $U_{Bt}(t)$ has ``caught up'' to all the other simulations.
Over short times, the reconnection dynamics thus depends on the detailed behaviour of large plasmoids, which has an element of randomness;
however, the random plasmoid behaviour does not alter the long-term behaviour of reconnection.

\begin{figure}
\centering
\fullplot{
\includegraphics*[width=0.49\textwidth]{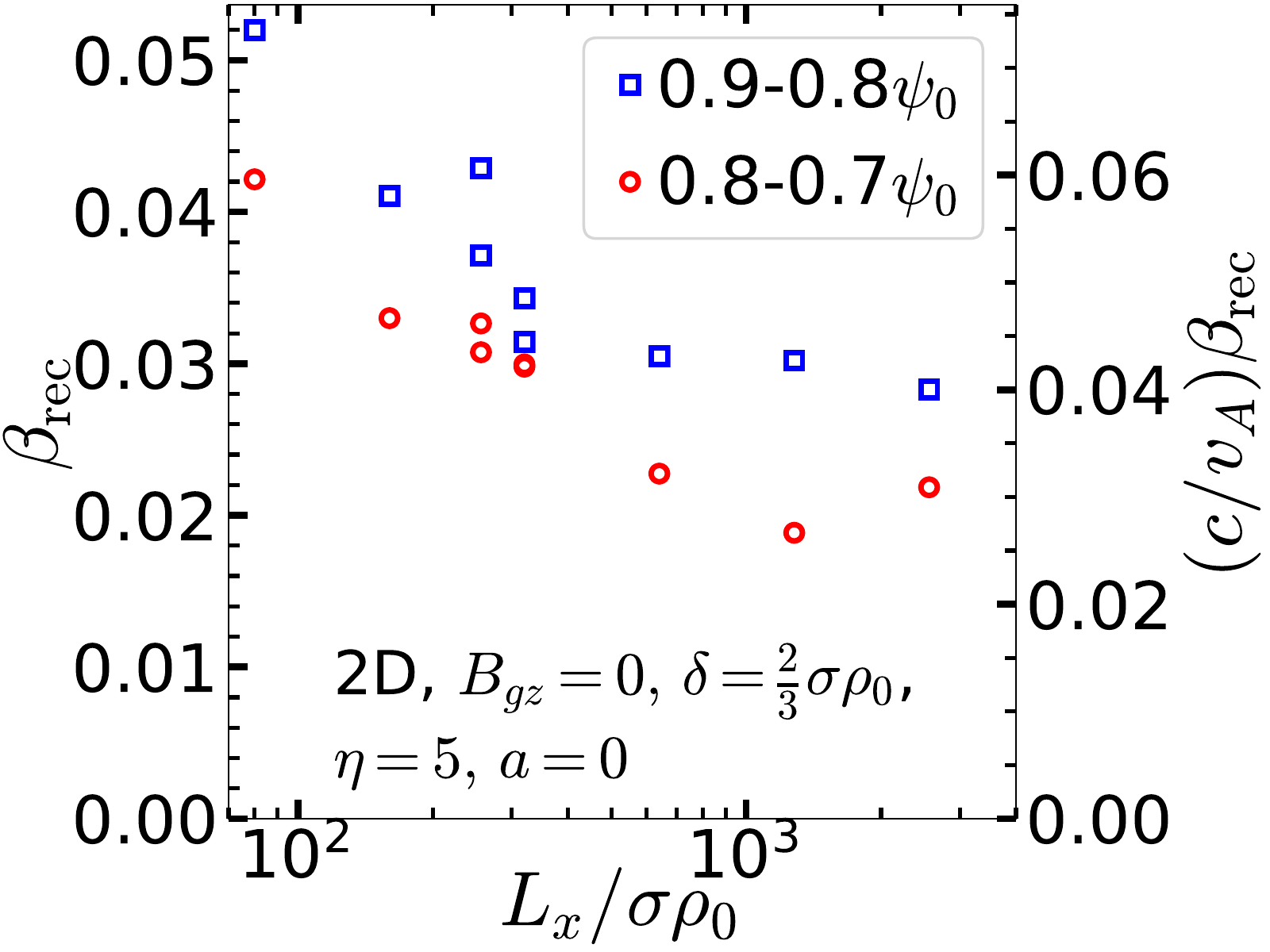}%
\hfill
\includegraphics*[width=0.49\textwidth]{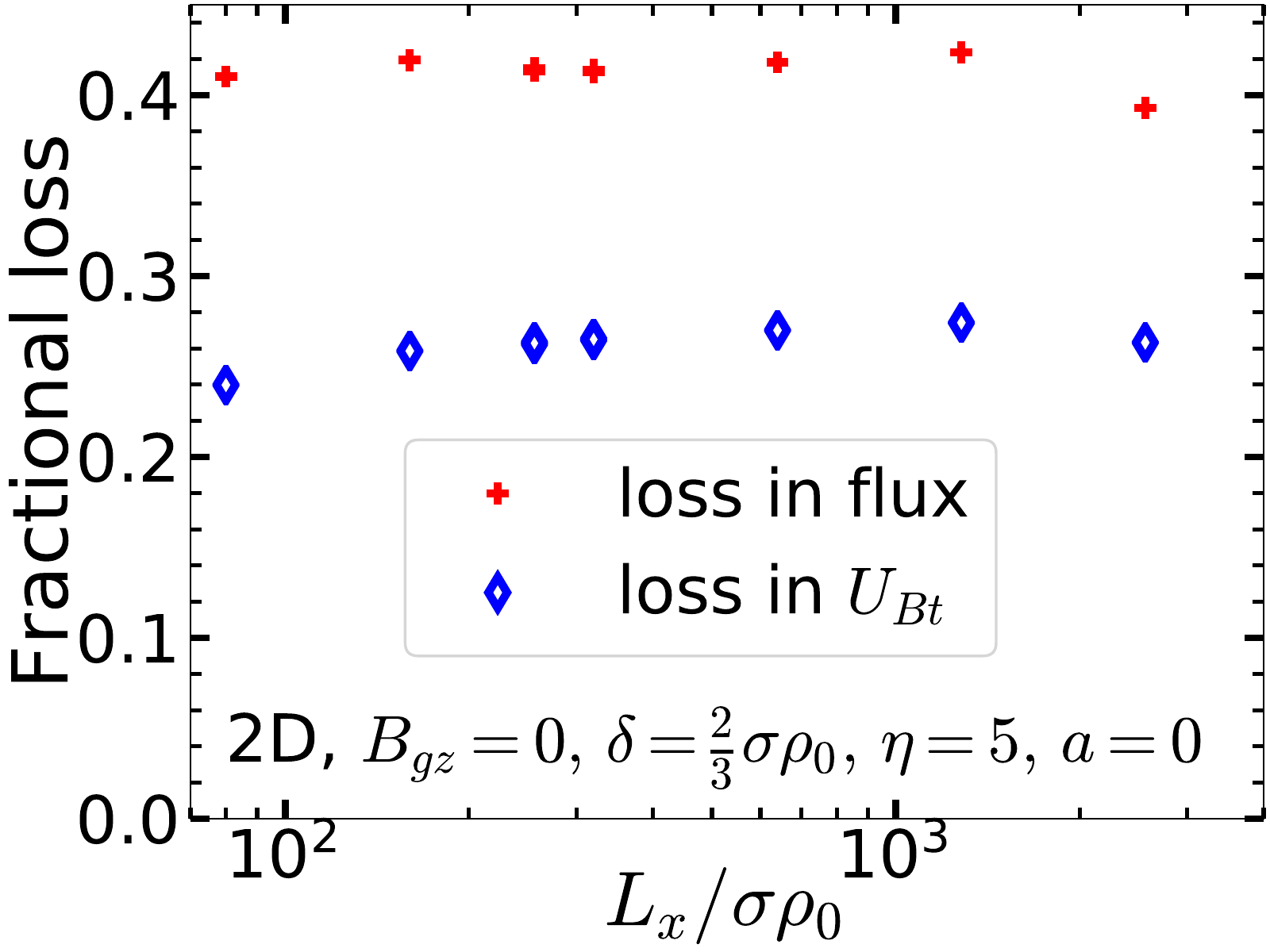}%
}
\caption{ \label{fig:reconRateAndTotalDissipationVsLx2d} 
(Left) The
average reconnection rate (normalized to $B_0 c$ on the left scale, and to $B_0 v_A$
on the right scale) versus system size $L_x$, over the period when the unreconnected flux falls from 0.9--0.8$\psi_0$ (blue squares), and also from 0.8-0.7$\psi_0$ (red circles)---i.e., between the horizontal dashed lines in Fig.~\ref{fig:energyFluxVsLx2d}(right).
(Right) The fraction of initial (transverse) magnetic field energy converted to plasma energy (blue diamonds) and of upstream flux depleted (red plusses) during reconnection versus
system size $L_x$ (for aspect ratio $L_y/L_x=2$, with two layers).
The largest system size, $L_x=1280\sigma\rho_0$ had not quite finished reconnection when the simulation halted around $10L_x/c$; it would have depleted slightly more energy and flux if given more time.
}
\end{figure}

Figure~\ref{fig:energyFluxVsLx2d} shows the evolution of transverse magnetic energy~$U_{Bt}(t)$ and unreconnected flux~$\psi(t)$ for a range of~$L_x$.
Although~$U_{Bt}(t)$ often decreases fitfully, $\psi(t)$ tends to decrease more smoothly.
Initially, $\psi(t)$ falls very rapidly as reconnection is dominated by the initial current sheet, and the rate of fall slows a little as the upstream plasma begins to dominate reconnection (the larger~$L_x/\delta$ is, the sooner this happens relative to the total duration of reconnection).
Eventually, the system reaches a stable magnetic configuration and $\psi(t)$ approaches a constant value;
therefore, the reconnection rate must decrease over time, eventually toward zero (in a closed system).
The reconnection rates over two different time intervals are shown in Fig.~\ref{fig:reconRateAndTotalDissipationVsLx2d}(left), namely for the two intervals $0.9 > \psi(t)/\psi_0 > 0.8$ and $0.8 > \psi(t)/\psi_0 > 0.7$, i.e., between the horizontal dashed lines in Fig.~\ref{fig:energyFluxVsLx2d}(right).
As the system size~$L_x/\sigma\rho_0$ increases from~160 to~2560, $\beta_{\rm rec}$ (cf.~\S\ref{sec:unreconnectedFlux}) falls from~$\approx 0.04$ to~$\approx 0.03$ in the earlier interval, and from~$\beta_{\rm rec}\approx 0.03$ to~$\beta_{\rm rec}\approx 0.02$ in the later interval.
However, $\beta_{\rm rec}$ stabilizes within the uncertainty of measurement by $L_x\gtrsim 640\sigma\rho_0$.

Reconnection continues until approximately~$0.27U_{B0}$ and~$0.42\psi_0$ have been depleted; the fractions of lost energy and flux are practically independent of~$L_x$ (Fig.~\ref{fig:reconRateAndTotalDissipationVsLx2d}, right), although we note that they are specific to the fixed aspect ratio~$L_y/L_x=2$.
The largest simulation, $L_x=2560\sigma\rho_0$ shows slightly less loss in~$U_{Bt}$ and~$\psi$, likely because it had not quite finished reconnecting when it halted around $10L_x/c$ (cf.~Fig.~\ref{fig:energyFluxVsLx2d}).

Importantly, the \emph{upstream} magnetic energy and flux are both depleted by the same amount---42~per~cent.
However, of that~$0.42U_{B0}$ initially associated with magnetic field lines that eventually undergo reconnection, only~$0.27U_{B0}$ is converted to plasma energy; the remaining~$0.15 U_{B0}$ ends up stored in the ``reconnected'' magnetic field around and within plasmoids.

\begin{figure}
\centering
\fullplot{
\includegraphics*[width=0.325\textwidth]{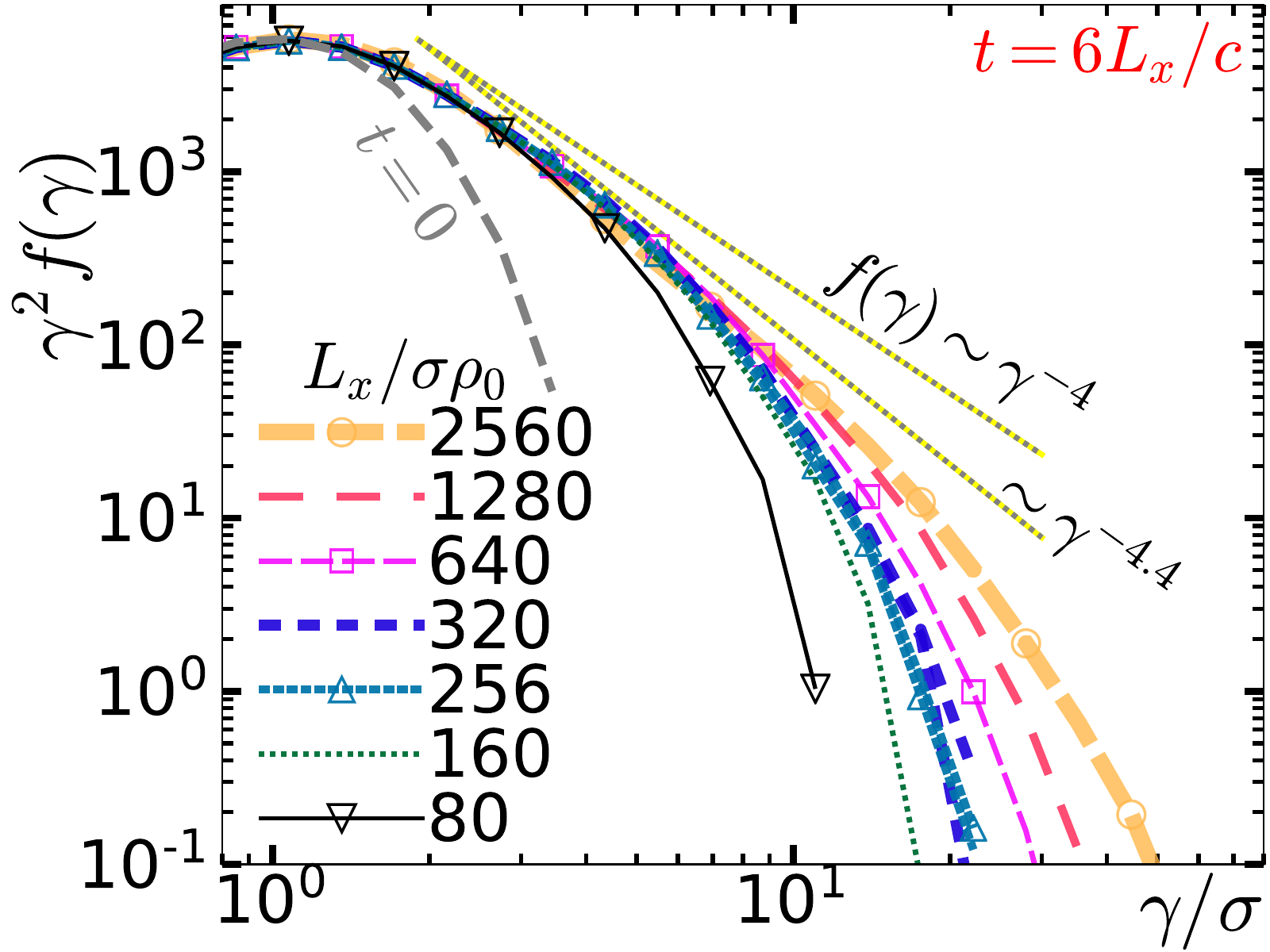}%
\hfill
\includegraphics*[width=0.325\textwidth]{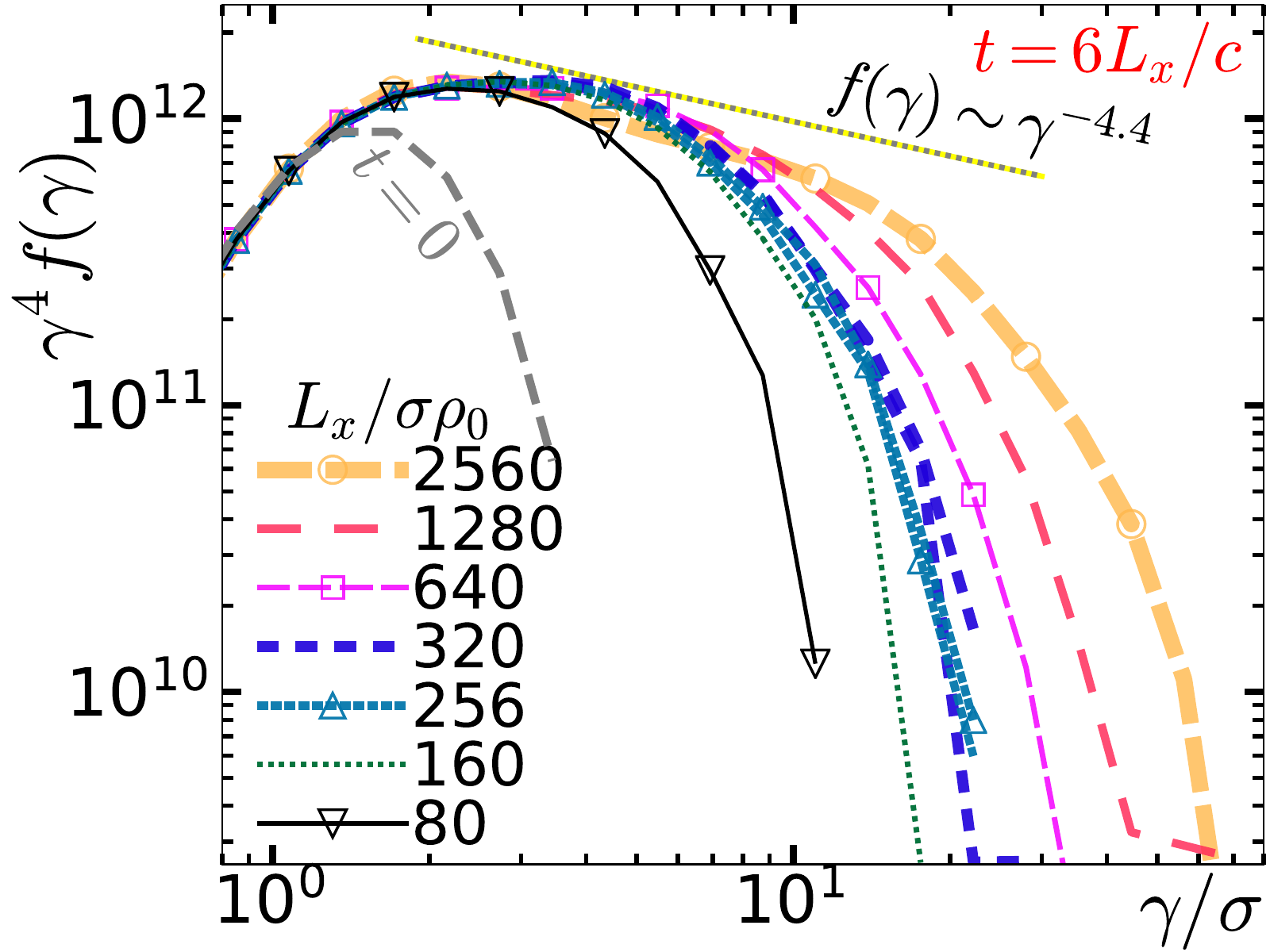}
\hfill
\includegraphics*[width=0.325\textwidth]{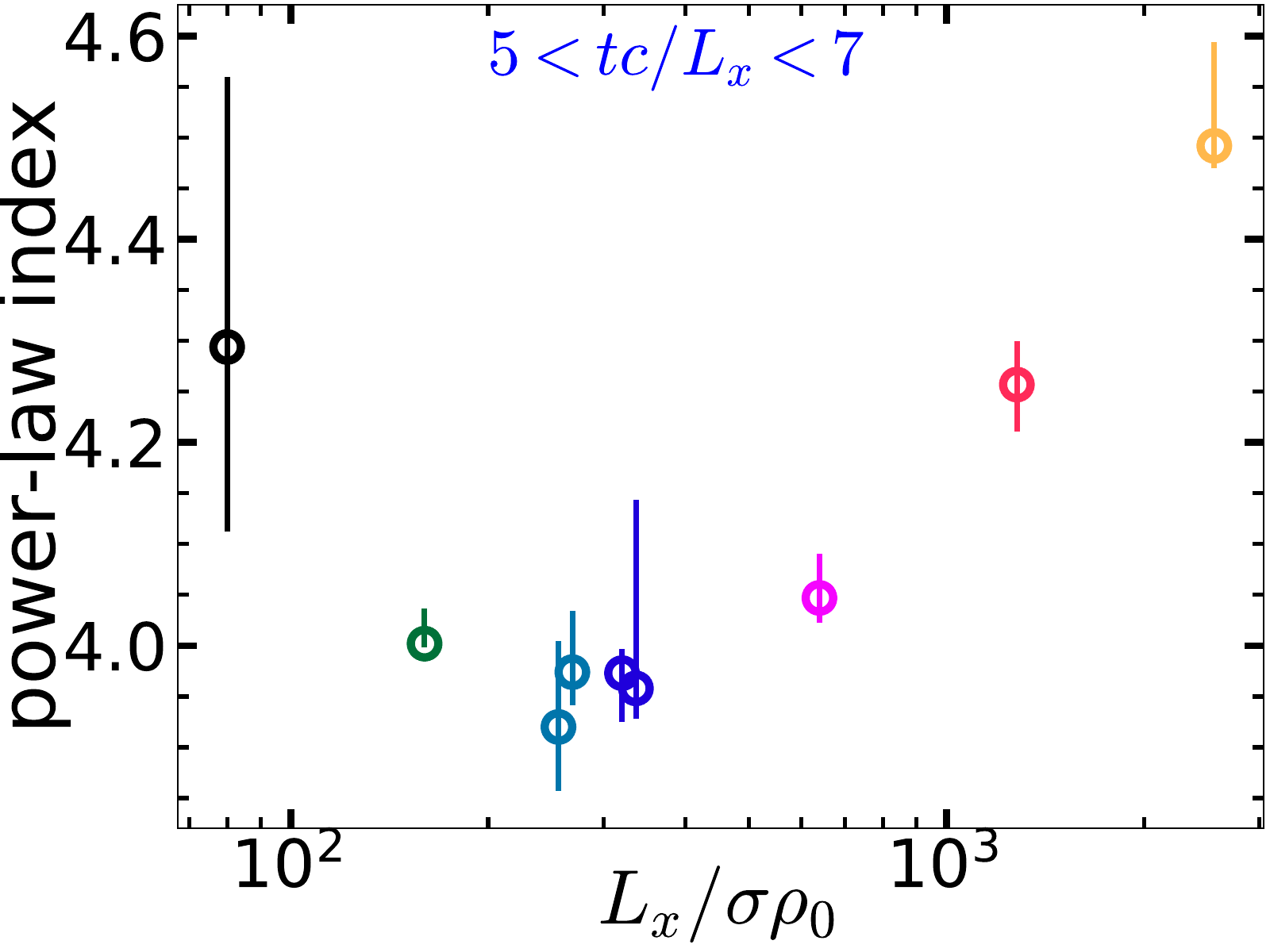}
}
\caption{ \label{fig:spectraVsLx2d}
For different system sizes $L_x$:
(left) electron energy spectra $f(\gamma)$, compensated by $\gamma^2$,
at time $t=6L_x/c$, with power-law indices of~4 and~4.4 indicated by dotted grey-on-yellow lines,
(middle) $\gamma^{4} f(\gamma)$ at $t=6L_x/c$, to enhance differences, 
(right) the median power-law index during the time period 
$5 < tc/L_x < 7$, with ``error'' bars containing 68~per~cent of the values fitted to spectra within this period.
For $L_x/\sigma\rho_0=$256 and~320, results from two simulations are shown (slightly offset in $L_x$ for clarity in the right panel).
}
\end{figure}

Turning our attention to NTPA, 
Fig.~\ref{fig:spectraVsLx2d} shows the electron energy spectra~$f(\gamma)$ at $t=6L_x/c$ for simulations with different~$L_x$ [the left panel shows $\gamma^2 f(\gamma)$, and the middle shows $\gamma^4 f(\gamma)$ to enhance small differences].
Aside from the smallest, all simulations have a 
clear high-energy nonthermal power-law $f(\gamma)\sim \gamma^{-p}$; 
for the largest simulation, it extends over almost a decade in energy. 
At~$t=6L_x/c$, the power-law index~$p$ steepens slightly as $L_x/\sigma\rho_0$ increases from~160 to~2560, from~$p\approx 4.0$ to~$p\approx 4.5$ (Fig.~\ref{fig:spectraVsLx2d}, right---cf.~\S\ref{sec:NTPAfit}). 
This steepening may be caused by a different acceleration mechanism that operates over long times (hence only in very large simulations), as we will discuss presently.

\begin{figure}
\centering
\fullplot{
\includegraphics*[width=0.325\textwidth]{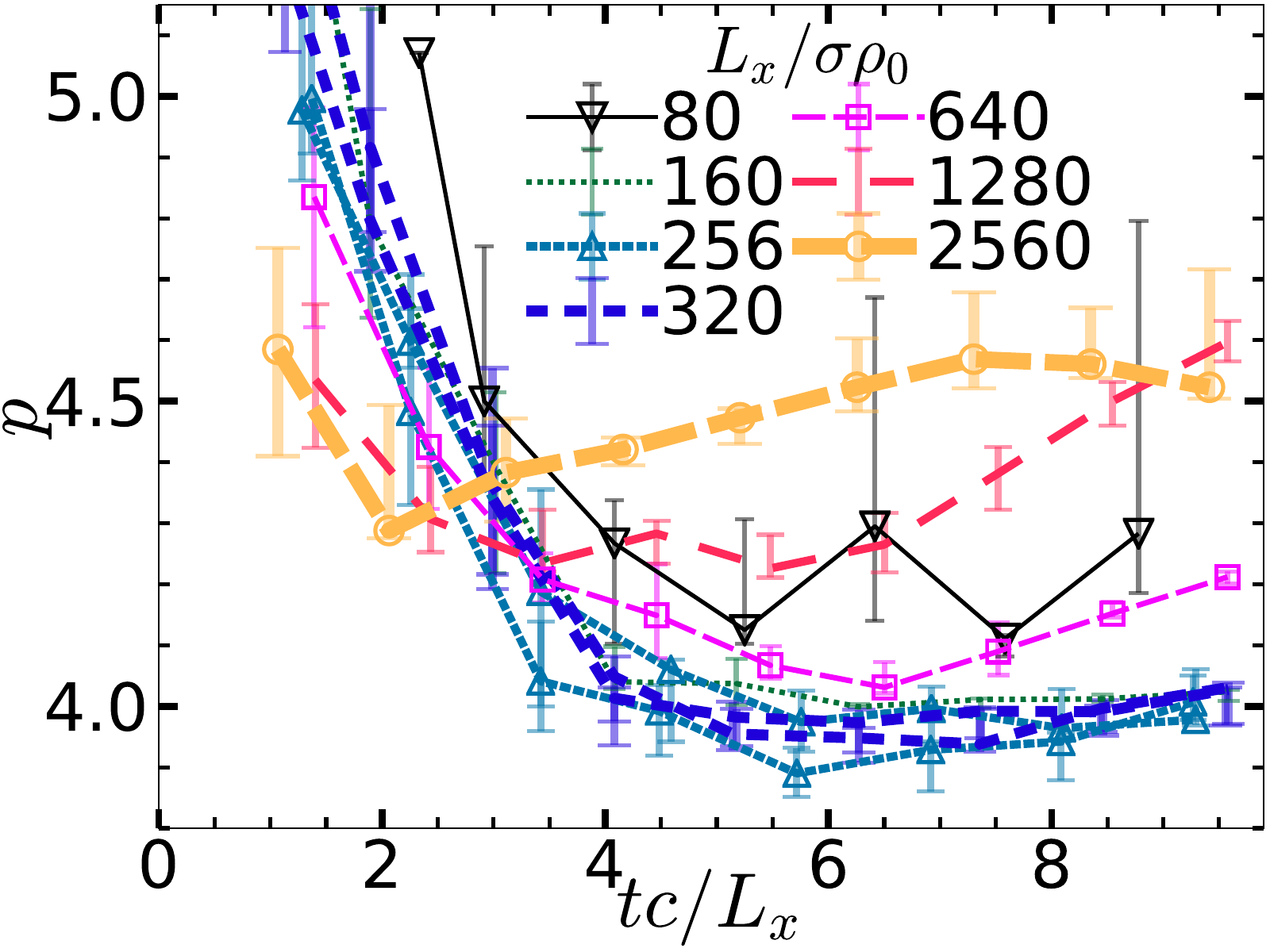}%
\hfill
\includegraphics*[width=0.325\textwidth]{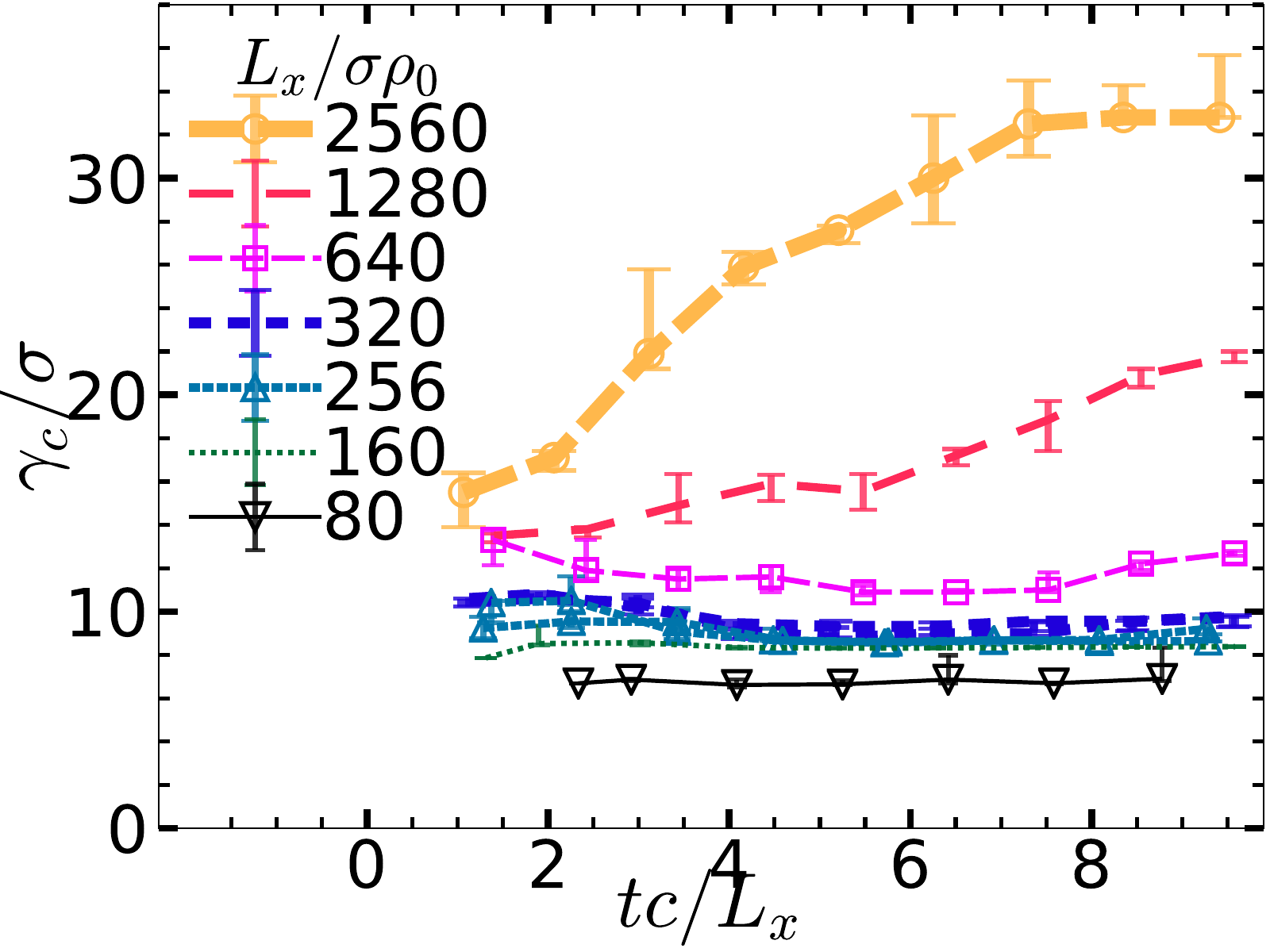}%
\hfill
\includegraphics*[width=0.325\textwidth]{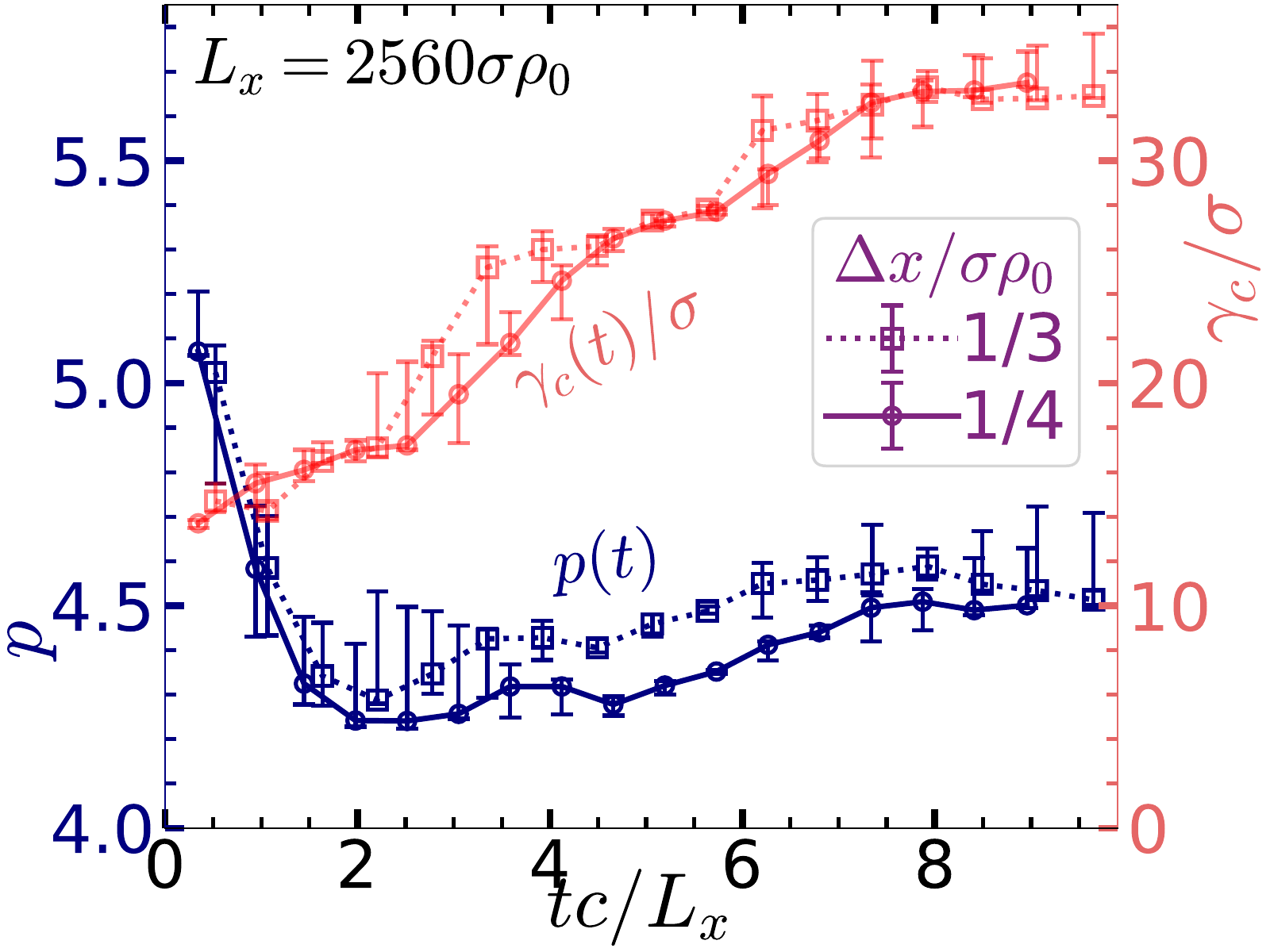}%
}
\caption{ \label{fig:cutoffVsLx2d}
For different system sizes~$L_x$, the fitted values for the 
power-law index~$p$ (left) and high-energy cutoff~$\gamma_c$ (middle) versus time.
(Right) For $L_x=2560\sigma\rho_0$, $p(t)$ (blue, left axis) and $\gamma_c(t)$ (red, right axis) are essentially the same for 
two different resolutions, $\Delta x/\sigma\rho_0 = 1/3$ and $1/4$,
even though for this large system size, $\Delta x/\sigma\rho_0=1/3$ 
resulted in unphysical total energy growth of~1~per~cent, while $\Delta x/\sigma\rho_0 =1/4$ saw a more satisfactory~0.05~per~cent increase.
Error bars indicate uncertainty, encompassing the middle 68~per~cent of values measured over the associated time interval.
}
\end{figure}

The time evolution of power-law index~$p$ and high-energy cutoff~$\gamma_c$ (cf.~\S\ref{sec:NTPAfit}), for all the~$L_x$, are shown in Fig.~\ref{fig:cutoffVsLx2d} (left and middle panels).
The initial development of the power law, as the first particles reach high energies, results in~$p(t)$ rapidly falling (i.e., the power law hardens/flattens) to approach an index between~3.9 and~4.5.
From its minimum value, the slope appears to steepen slowly over time, mostly notably for simulations with $L_x\geq 640\sigma\rho_0$, all of which show $p(t)$ rising by about~0.2 from its minimum. 

Correspondingly, the cutoff~$\gamma_c(t)$ grows rapidly as the power law develops, attaining after only 1--2$L_x/c$ a value ranging from~$\gamma_c=6\sigma$ for $L_x=80\sigma\rho_0$ to~$\gamma_c=15\sigma$ for $L_x=2560\sigma\rho_0$.
For~$L_x\geq 1280\sigma\rho_0$, $\gamma_c(t)$ continues to grow over time, and even the case $L_x=640\sigma\rho_0$ experiences some limited growth after about~$t=7L_x/c$.
Thus at late time~$t=9L_x/c$, $\gamma_c$ ranges 
from~$6\sigma$ to~$33\sigma$ as $L_x$ increases by a factor of 32, consistent with $\gamma_c \sim \sqrt{L_x}$;
importantly, although~$\gamma_c$ increases with~$L_x$, the increase is clearly sublinear.
This is consistent with very rapid (extreme) acceleration up to~$\gamma_c\sim 4\sigma$ \citep{Werner_etal-2016,Uzdensky-2020arxiv}, at which energy particles become trapped in plasmoids and cease extreme acceleration, but undergo much slower acceleration inside plasmoids \citep{Petropoulou_Sironi-2018,Hakobyan_etal-2021,Uzdensky-2020arxiv}.
Interestingly, the slow growth in~$\gamma_c(t)$ seems to correspond to the slow steepening in~$p(t)$, suggesting that the slower acceleration mechanism may yield a steeper power law.
The increase in~$p$ with~$L_x$ seen in Fig.~\ref{fig:spectraVsLx2d}(right) may simply reflect the increasing influence of the slower acceleration mechanism after longer times (accessible in large systems).

Very large reconnection simulations sometimes suffer at late times from slow-growing instabilities.
The largest case, $L_x=2560\sigma\rho_0$, exhibited potentially troubling energy nonconservation, with the total energy growing by almost~1~per~cent between $t=7L_x/c$ and $t=10L_x/c$.
We find that this typically results in unphysical heating of the lowest-energy particles but does not affect the high-energy spectrum.
To rule out this possibility, we compare the simulation with fiducial resolution $\Delta x=\sigma\rho_0/3$ to one with higher resolution, $\Delta x=\sigma\rho_0/4$ in Fig.~\ref{fig:cutoffVsLx2d}(right); both show essentially similar evolution of~$p(t)$ and~$\gamma_c(t)$, although the higher-resolution case suffered only a~0.05~per~cent increase in total energy.

In summary, the system-size dependence of 2D reconnection with $\sigma_h=1$ is weak but not entirely insignificant for $L_x \gtrsim 160\sigma\rho_0$.
The reconnection rate (normalized to $cB_0$) is $\beta_{\rm rec}\approx 0.03$ or $(c/v_A)\beta_{\rm rec}\approx 0.04$.
During reconnection, some particles are very rapidly accelerated into a power-law energy spectrum $\sim \gamma^{-p}$ with $p\approx 3.9$--$4.4$, extending up to a cutoff around $\gamma_c \gtrsim 6\sigma$.
This rapid acceleration presumably ends when particles become trapped in plasmoids.
Over time, a much slower acceleration mechanism continues to accelerate particles trapped in plasmoids \citep{Petropoulou_Sironi-2018,Hakobyan_etal-2021}, resulting in a slight steepening of~$p$ (up to~$\approx 4.6$ in our largest simulation) and slow but significant growth in $\gamma_c$ (up to~$\approx 33\sigma$ in our largest simulation).

\subsection{2D reconnection: guide magnetic field}
\label{sec:Bz2d}

We now investigate the effect of adding an initial
uniform guide magnetic field $B_{gz}\hat{\boldsymbol{z}}$, considering
$B_{gz}/B_0 \in \{$0, 0.25, 0.5, 0.75, 1, 1.5, 2, 3, 4$\}$.
All simulations presented in this subsection are 2D 
with $L_x=1280\sigma\rho_0$,
starting with an initial current sheet half-thickness
$\delta=(2/3)\sigma\rho_0$ ($\eta=5$, $\beta_d=0.3$), and zero perturbation ($a=0$).

\begin{figure}
\centering
\fullplot{
\includegraphics*[width=0.49\textwidth]{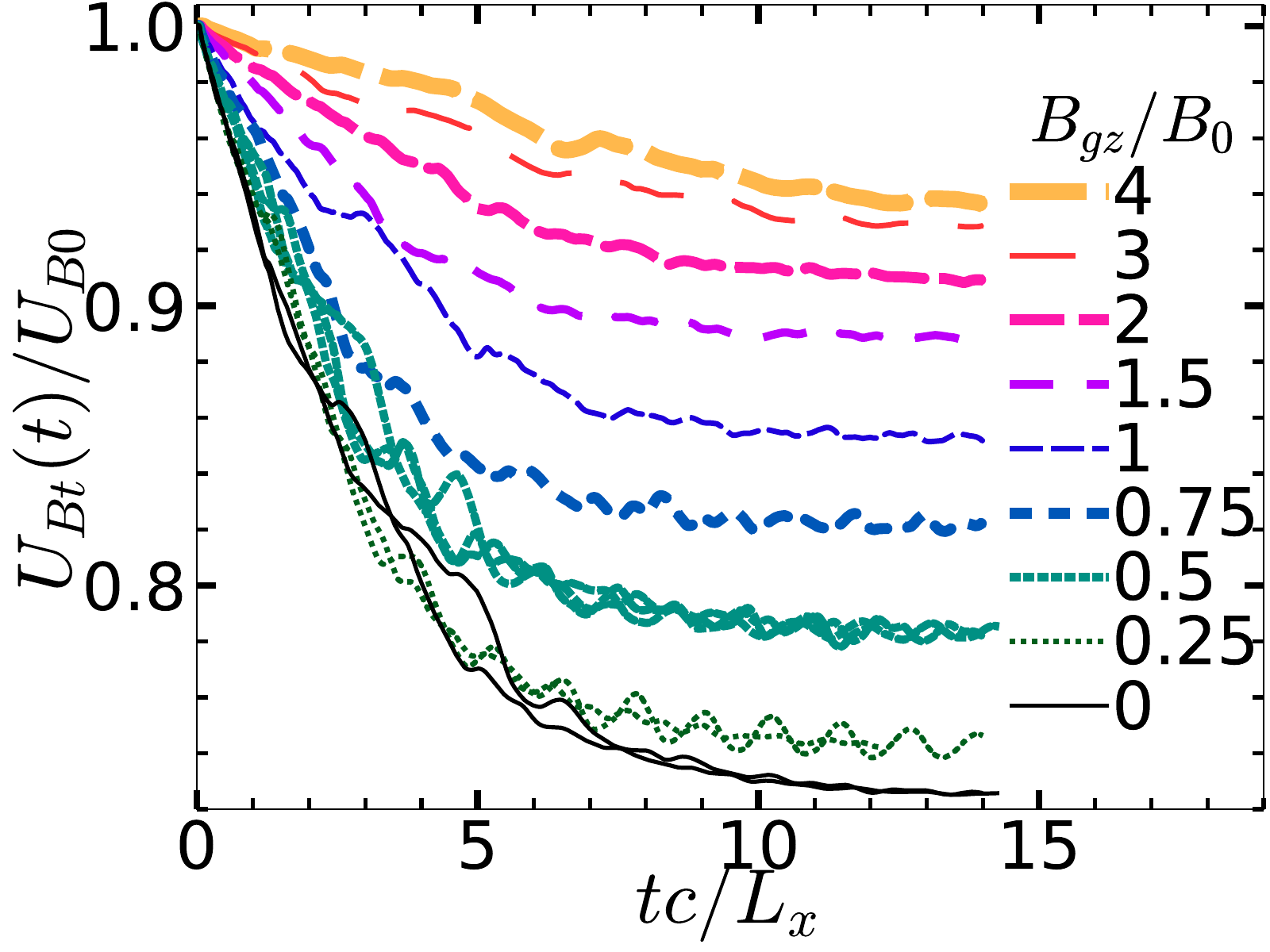}%
\hfill
\includegraphics*[width=0.49\textwidth]{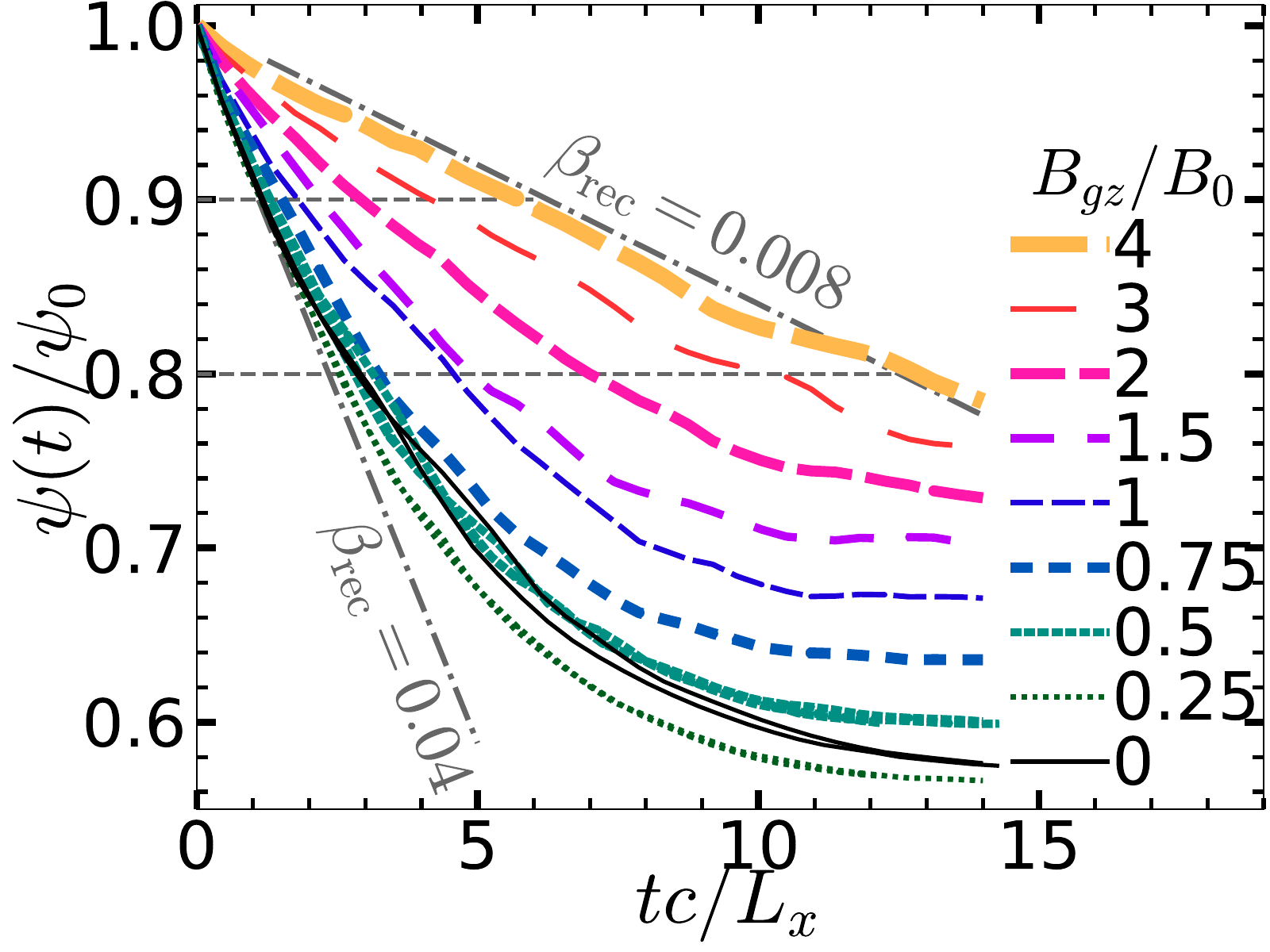}%
}
\caption{ \label{fig:energyFluxVsBgz2d}
For several guide field strengths $B_{gz}$:
(left) transverse magnetic energy versus time,
and (right) unreconnected flux versus time, with dot-dashed lines showing
the slopes corresponding to $\beta_{\rm rec}=$0.04 and~0.008 (cf.~\S\ref{sec:unreconnectedFlux}).
To estimate stochastic variability, two simulations are shown for
$B_{gz}=0$, two for $B_{gz}=0.25B_0$, and three for $B_{gz}=0.5B_0$.
}
\end{figure}

A weak guide field $B_{gz} \lesssim 0.25 B_0$ has very little effect on the magnetic energy evolution.
A strong guide field, however, slows reconnection and inhibits overall magnetic
energy conversion, as shown in Fig.~\ref{fig:energyFluxVsBgz2d}.
The left panel shows the transverse magnetic energy $U_{Bt}$ versus time, for simulations with $B_{gz}/B_0$ ranging from~0 to~4;
as $B_{gz}$ increases above $\simeq 0.5B_0$, $U_{Bt}(t)$ falls more
slowly, and ultimately decreases by a smaller amount.
Correspondingly (cf. Fig.~\ref{fig:energyFluxVsBgz2d}, right), the
unreconnected flux decreases more slowly, and less flux is reconnected
overall.

This slowing of reconnection has been previously attributed to
the decrease in the effective component ($x$-projection) of the Alfv\'{e}n velocity
\citep{Liu_etal-2014,Liu_etal-2015,Werner_Uzdensky-2017}.
The upstream Alfv\'{e}n velocity, including~$B_0$ and~$B_{gz}$, is 
$v_A=c\sqrt{(B_0^2+B_{gz}^2)/(B_0^2+B_{gz}^2 + 4\upi h)}$,
where $h$ is the plasma enthalpy density ($h=4n_b\theta_b m_e c^2$ in the ultrarelativistic limit).
The projection of $v_A/c$ along~$B_0$ is therefore
\begin{eqnarray} \label{eq:vAx}
  \frac{v_{A,x}}{c} &=& 
  \sqrt{ \frac{B_0^2}{B_0^2 + B_{gz}^2}} \frac{v_A}{c}
  =
  \sqrt{ \frac{B_0^2}{B_0^2 + B_{gz}^2 + 4\upi h}} 
  =
  \sqrt{ \frac{B_0^2}{B_0^2 + 4\upi h_{\rm eff}}} 
\end{eqnarray}
where, following \citet{Werner_Uzdensky-2017}, we define 
$h_{\rm eff}\equiv h+B_{gz}^2/4\upi$ to be an effective enthalpy density including a contribution from the guide field.
Then, defining $\sigma_{h,\rm eff}\equiv B_0^2/(4\upi h_{\rm eff})$, we have $v_{A,x} = (\sigma_{h,\rm eff}^{-1}+1)^{-1/2}$, which reduces to
$v_{A}/c = (\sigma_h^{-1}+1)^{-1/2}$ for~$B_{gz}=0$.

\begin{figure}
\centering
\fullplot{
\includegraphics*[width=0.49\textwidth]{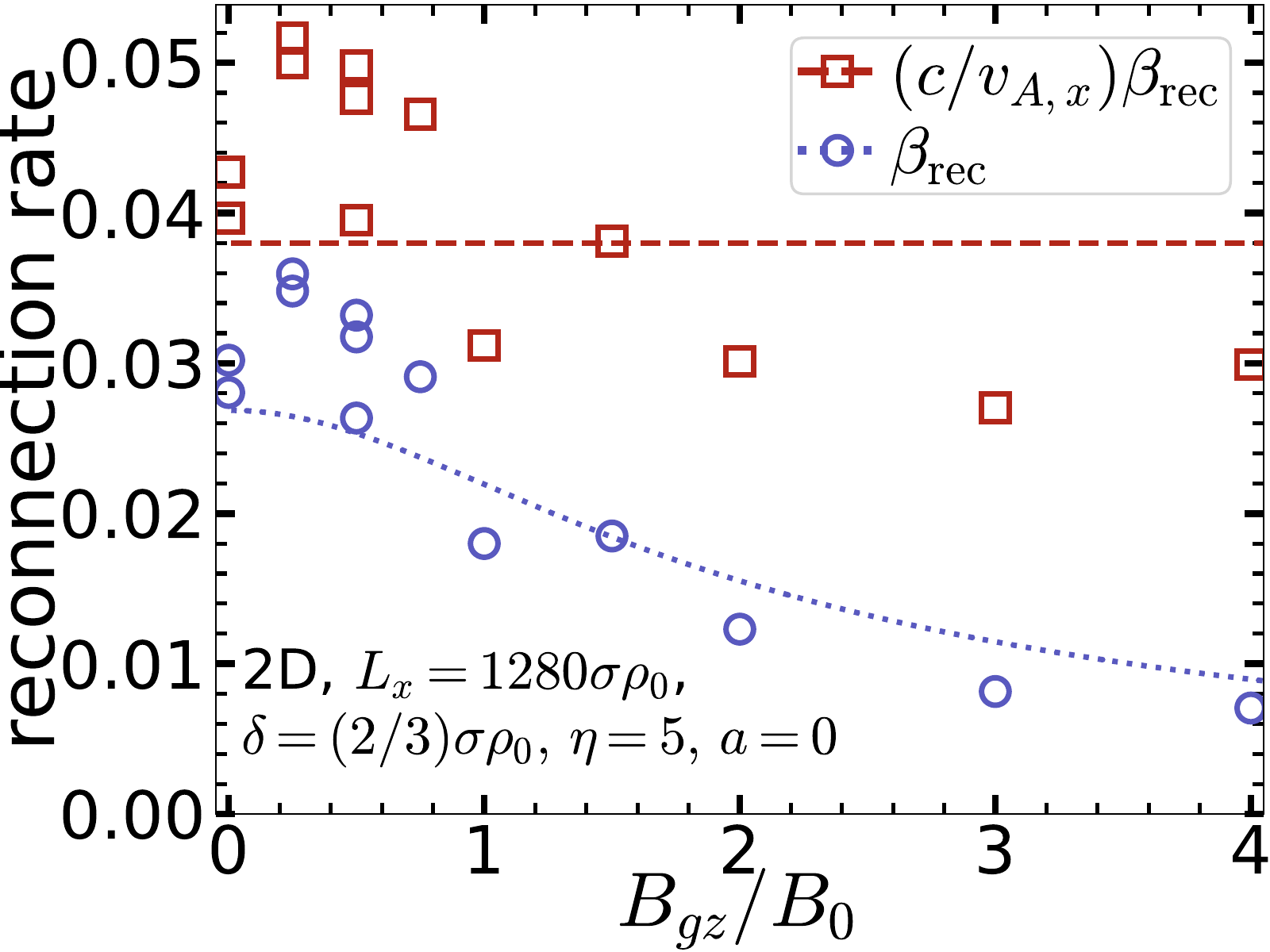}%
\hfill
\includegraphics*[width=0.49\textwidth]{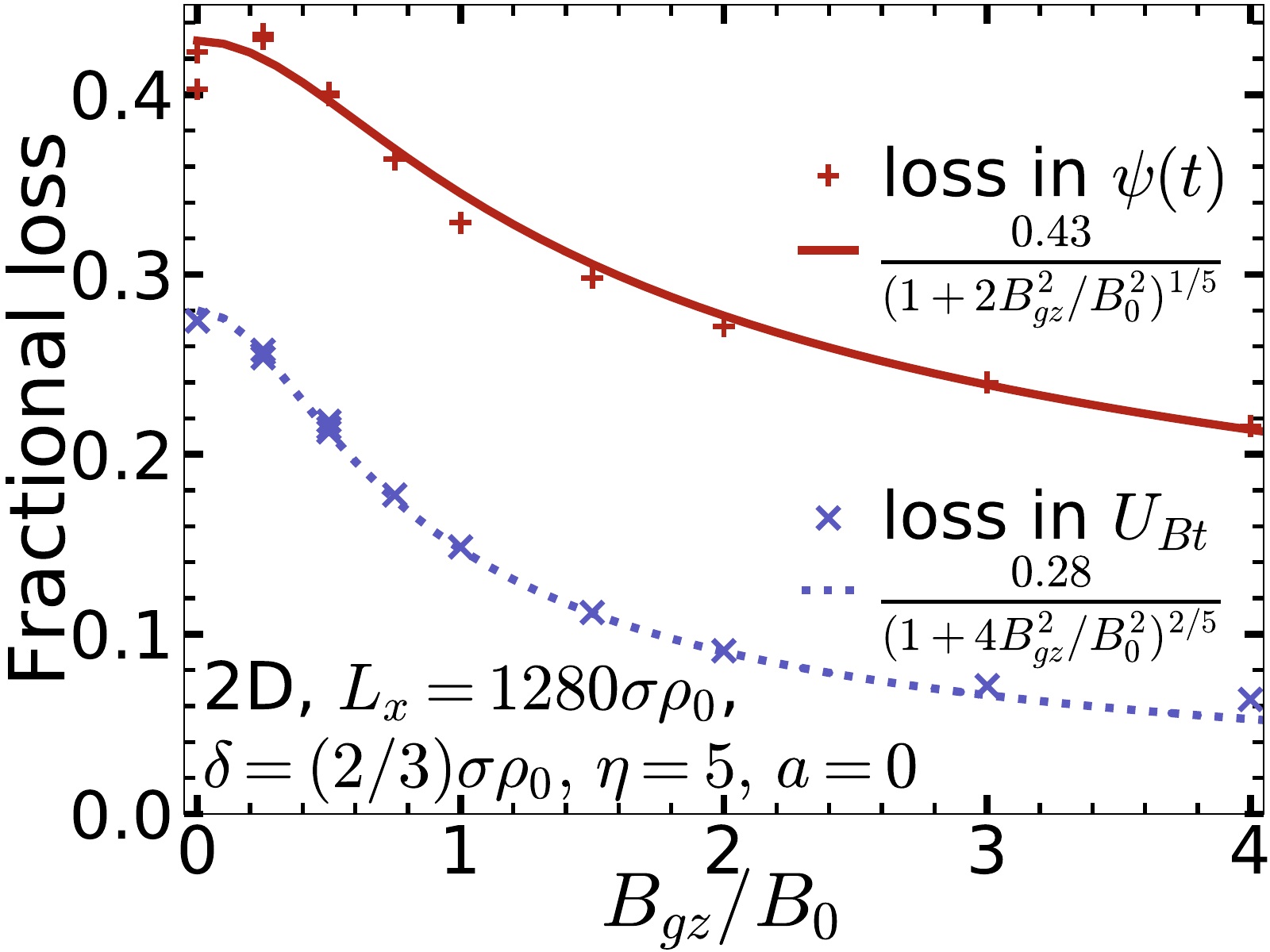}%
}
\caption{ \label{fig:reconRateAndTotalDissipationVsBgz2d}
(Left) Reconnection rate versus $B_{gz}$: 
blue circles show the reconnection
rate $\beta_{\rm rec}$ (normalized to $B_0 c$), and red squares show $(c/v_{A,x})\beta_{\rm rec}$---cf.~Eq.~(\ref{eq:vAx}).
The blue dotted line shows what~$\beta_{\rm rec}$ would be if $(c/v_{A,x})\beta_{\rm rec}$ were constant at~0.038 (the red dashed line).
The reconnection rates are averages over the period during which~$\psi(t)$  
falls from $0.9\psi_0$ to $0.8\psi_0$ (i.e., between the horizontal
grey, dashed lines in Fig.~\ref{fig:energyFluxVsBgz2d}, right).
(Right) The fractional amount of transverse magnetic energy $U_{Bt}$ (blue x's) and upstream magnetic flux (red plusses) lost during reconnection, versus $B_{gz}$, along with empirical fits for both.
}
\end{figure}

Figure~\ref{fig:reconRateAndTotalDissipationVsBgz2d} quantifies the reconnection rates (left panel) and amount of magnetic energy and upstream flux depleted (right).
The left panel shows the
reconnection rates,
averaged over the time interval during which~$\psi(t)$
falls from~$0.9\psi_0$ to~$0.8\psi_0$.
The reconnection
rate, normalized to $B_0 c$ (blue circles) falls from roughly 0.03--0.04 for 
$B_{gz}\lesssim 0.75B_0$ to around 0.01 for $B_{gz} = 4 B_0$.
The red squares show
reconnection rates normalized to $B_0 v_{A,x}$.
For $B_{gz}/B_0\gtrsim 1$, $(c/v_{A,x})\beta_{\rm rec}$ is relatively constant around 0.03; however, $(c/v_{A,x})\beta_{\rm rec}$ is closer to~0.04 or even~0.05 for $B_{gz}/B_0 < 1$.
As a compromise, Fig.~\ref{fig:reconRateAndTotalDissipationVsBgz2d} shows a dashed line at~$(c/v_{A,x})\beta_{\rm rec}=0.038$; the dotted line shows the corresponding $\beta_{\rm rec}$. 

According to Eq.~(\ref{eq:vAx}), the guide field should not significantly
suppress reconnection for $B_{gz}/B_0 \ll \sigma_h^{-1/2} = 1$.
We note that a much stronger guide field would be needed to suppress nonrelativistic reconnection (where $\sigma_h \ll 1$).
On the other hand, 
our previous work \citep{Werner_Uzdensky-2017} considered $\sigma_h \simeq 25$, and correspondingly 
observed that even $B_{gz} = B_0/2$ significantly suppressed reconnection, slowing the reconnection rate and reducing the overall magnetic depletion
(although $B_{gz}=B_0/4$ had only a weak effect).

Figure~\ref{fig:reconRateAndTotalDissipationVsBgz2d}(right) shows that
the total fraction of $U_{Bt}$ converted to $U_{\rm plasma}$ decreases with stronger guide field---from around 27~per~cent for $B_{gz}=0$ to only about 6~per~cent for $B_{gz}=4 B_0$; similarly, the amount of flux reconnected decreases from 42~per~cent to around 21~per~cent.
We find that the fractional loss in~$\psi$ is fit fairly well by $0.86(L_x/L_y)(1+2B_{gz}^2/B_0^2)^{-1/5}$, 
and the loss in~$U_{Bt}$ by $0.56(L_x/L_y)(1+4B_{gz}^2/B_0^2)^{-2/5}$;
these functional forms are meant for convenient comparison with future results and we offer no justification for them.

Even a weak guide appears to enhance the reconnection rate slightly.
However, because stochastic variability can yield different reconnection rates, more work is needed to determine whether this trend is
statistically significant.

With stronger guide field, reconnection ends (or reaches a stable configuration) with a smaller amount amount of magnetic energy and flux depletion.
This may be
caused by the increased pressure of the guide magnetic field in plasmoids.
The guide magnetic field pressure resists compression with adiabatic index~$\Gamma=2$, compared with only~$\Gamma=4/3$ for relativistic plasma alone [e.g., compressing a flux tube with uniform magnetic field from volume~$V_i$ to~$V_f$ while preserving the magnetic flux inside requires work~$\propto (V_i/V_f)^{\Gamma-1}-1$ where~$\Gamma=2$; whereas compressing the plasma in the tube requires work~$\propto (V_i/V_f)^{4/3-1}-1$]. 
As a result, for the same amount of (transverse magnetic field) flux reconnected, the major plasmoid will be larger when the guide field is stronger.
Therefore, the growth of the major plasmoid reaches the system-size scale earlier (in terms of amount of flux reconnected, not time) and shuts down reconnection.

\begin{figure}
\centering
\fullplot{
\includegraphics*[width=0.49\textwidth]{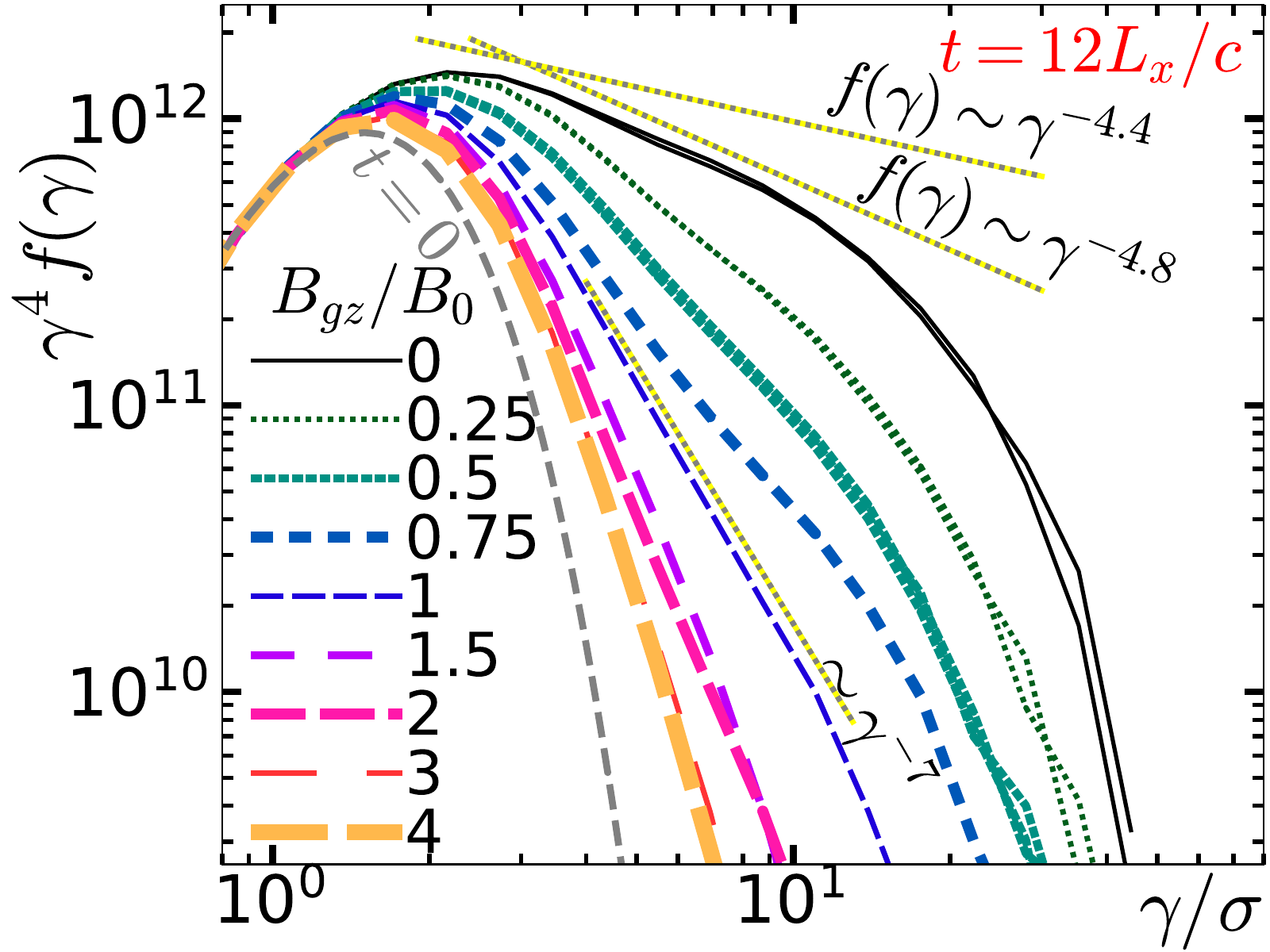}%
\hfill
\includegraphics*[width=0.49\textwidth]{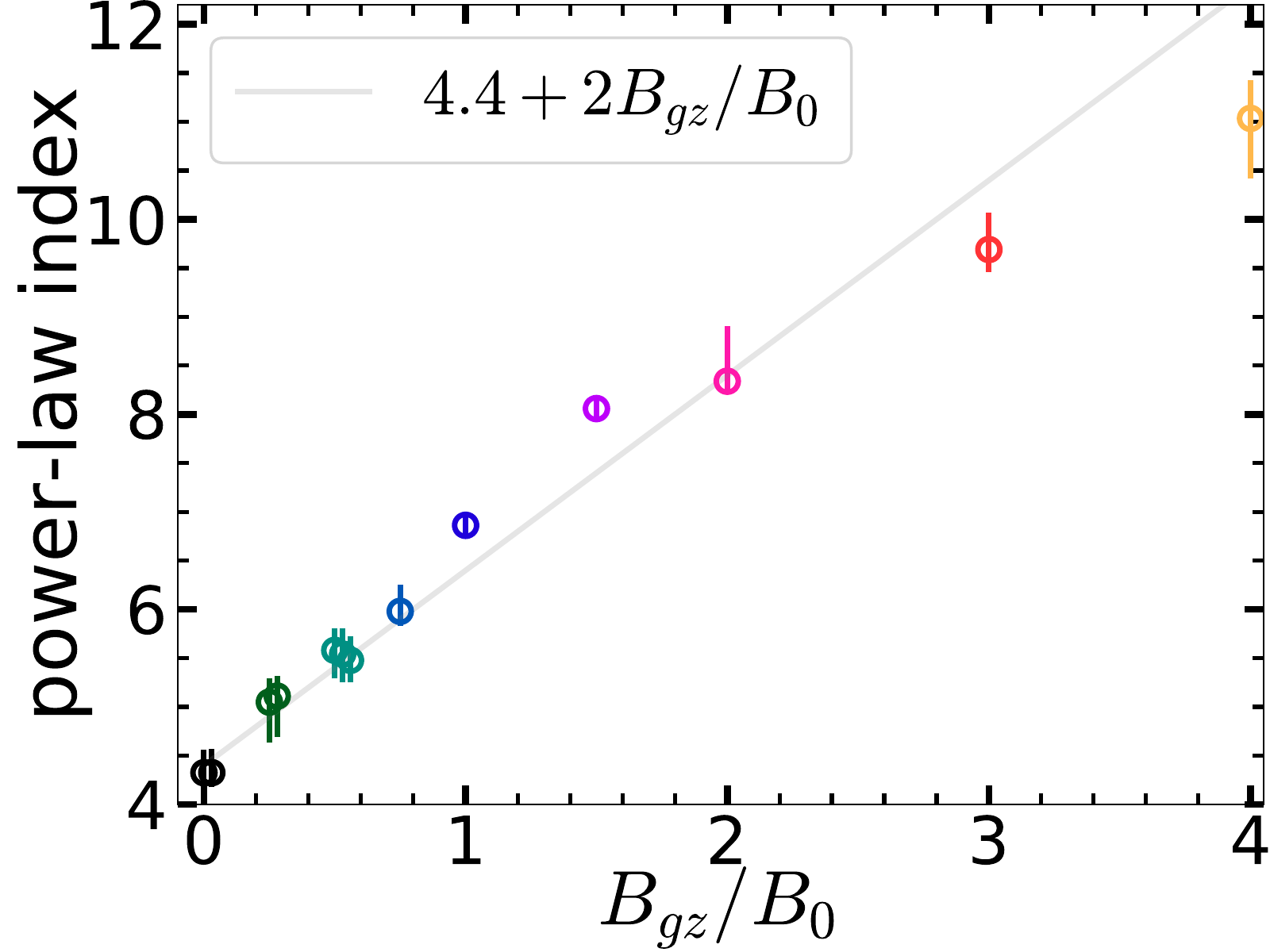}%
}
\caption{ \label{fig:ntpaVsBgz2d}
(Left) Particle energy spectra $f(\gamma)$ at $t=12L_x/c$, for different
imposed guide magnetic fields, show that increasing guide field 
inhibits NTPA, resulting in steeper power laws.  Slopes with power-law indices of 4.4, 4.8, and 7 are shown for comparison.  (Right) The steepening 
is quantified by power-law indices fit to the high-energy
part of $f(\gamma)$, with error bars encompassing the middle 68~per~cent
of fitted indices over times from 4 to~10$L_x/c$.  The line $4.4+2B_{gz}/B_0$ is shown as a guide, but it is important to remember that precise measurement of steep power laws is very difficult, and in fact the cases with $B_{gz}/B_0 \geq 1.5$ are not very different from the initial Maxwellian at $t=0$ (shown in dashed grey).
}
\end{figure}

Perhaps not surprisingly, the slowing of reconnection inhibits NTPA,
as shown in Fig.~\ref{fig:ntpaVsBgz2d},
which displays the final particle energy spectra (left panel)
and the fitted power-law indices (right).
Overall, less energy is transferred from the magnetic field to the plasma,
and at any given well-above-average energy, stronger guide field results in
a smaller fraction of particles at that energy, hence NTPA is less
efficient.
As $B_{gz}$ increases, the power law
steepens from $p\simeq 4$ to $p\gtrsim 10$ at $B_{gz}=4B_0$.
The line $p\approx 4.4 + 2B_{gz}/B_0$ very roughly captures this trend, although it is important to remember that the uncertainty in measuring steep power laws ($p\gtrsim 4$) can be quite high (our fitting method returns consistent values characterizing the spectra, but fundamentally one should question what it means to measure a steep power law, e.g.,~$\sim \gamma^{-8}$ as $\gamma$ varies over just a half decade).
This general trend was also observed for $\sigma_h=25$ reconnection in \citet{Werner_Uzdensky-2017}, where the steepening was fairly well described by 
$p=1.9+0.7\sigma_{h,\rm eff}^{-1/2}$; however, this formula provides a very poor fit for the $\sigma_h=1$ simulations, perhaps because we are in the regime of very steep power laws that cannot be reliably measured at higher energies without orders of magnitude more simulated particles.

\section{3D reconnection with moderate magnetization}
\label{sec:3d}

Despite many similarities between 2D and 3D in the $\sigma_h=1$ regime (especially in NTPA), current sheet evolution in 3D sometimes shows substantially different behaviour.
Reconnection still begins fairly rapidly in 3D, perhaps urged on (at very early times) by the thin tearing-unstable current sheet of the initial state, but the rate of magnetic energy conversion at early times is only about half that in 2D.
However, while 2D systems approach a non-reconnecting, relaxed steady state after 5--10$L_x/c$, 3D systems continue to convert magnetic energy at a slower rate (by an order of magnitude or more) for many tens of~$L_x/c$.
During 2D reconnection, part of the upstream magnetic energy is transferred to the plasma, and part is pumped into ``reconnected fields''---i.e., into plasmoids---where it remains in magnetic form \citep[see, e.g.,][]{Sironi_etal-2015}.
In 3D reconnection, plasma is also energized, but upstream magnetic energy pumped into plasmoids (or flux ropes) is subsequently converted to plasma energy \citep[rather than building up as in 2D; the decay of 3D flux ropes was also observed for $\sigma_h\gg 1$ in][]{Guo_etal-2020arxiv}.
Thus, while 2D reconnection simulations (in a closed periodic box) end up in a relaxed steady state with substantial magnetic field stored in one large plasmoid (for each initial current sheet), 3D reconnection results in a thickened, roughly uniform, much less structured, turbulent current layer with greatly diminished magnetic field.  This thickened layer continues slowly to convert magnetic energy to plasma energy; ultimately more magnetic energy can be converted in 3D than in 2D.
Despite a lower reconnection rate in 3D (hence lower reconnection electric field), NTPA remains robust.

While 2D reconnection can be bursty due to stochastic creation, movement, and merging of plasmoids, these events tend to be random blips in a predetermined course of evolution toward a unique final state.  In contrast, 3D reconnection exhibits greater variability, with stochastic events leading to a thicker turbulent layer in different ways.
In particular, we have often observed the following behaviour.
Because of RDKI, the current layer sometimes develops large-amplitude kinking in $z$, resulting in the layer dramatically folding over on itself like a breaking wave.  This tends to deplete very rapidly almost all the magnetic energy within the original amplitude of oscillation, resulting in a thick turbulent layer.
However, this behaviour is not inevitable; sometimes the kink amplitude grows but stops short of folding over on itself, resulting in much slower energy conversion and layer growth.
It remains unclear---because it might require running simulations for hundreds or thousands of $L_x/c$---whether a simulation that does not undergo ``layer-folding'' will ever convert as much energy or thicken the layer as much as one that does.

In the following subsection, we present an overview of differences between 2D and 3D current sheet evolution, using our largest simulations.  To study the effect of ``3D-ness'' we consider four configurations that are identical but for different values $L_z$.  The two smallest $L_z$ behave similarly (i.e., like 2D reconnection), whereas the two largest exhibit 3D effects.
Following the overview (\S\ref{sec:overview3d}), which will describe magnetic energy conversion and NTPA as well as plasma and field evolution, 
we include a brief discussion of increased stochastic variability, or sensitivity to initial conditions, in 3D (\S\ref{sec:variability3d}).
Then, using an extensive set of slightly smaller simulations, we will specifically explore the dependence of energy conversion and NTPA on the initial current sheet configuration
(\S\ref{sec:pertAndEta3d} and~\S\ref{sec:RDKIamp}), on the aspect ratio $L_z/L_x$ (i.e., the ``3D-ness,'' \S\ref{sec:Lz3d}), and finally on guide magnetic field (\S\ref{sec:Bz3d}).

We note that the 3D generalization of a plasmoid or magnetic island that forms in 2D reconnection is a flux rope.  However, we continue to use the term ``plasmoid'' to refer to flux ropes or any concentrations of plasma in 3D reconnection, both for simplicity and because we have not yet thoroughly investigated these objects to determine whether they truly have the magnetic structure of flux ropes.

\subsection{Overview of 3D reconnection with $\sigma_h=1$, and comparison with 2D}
\label{sec:overview3d}

In this section we begin with an overview of
basic 3D current sheet behaviour
for one specific configuration, namely:
$\sigma_h=1$, zero guide field ($B_{gz}=0$), zero initial perturbation ($a=0$), an initially-thin current sheet $\delta=(2/3)\sigma \rho_0$ ($\eta=5$), and $L_x=512\sigma \rho_0 = L_y/2$ (the largest $L_x$ of any 3D simulations in this paper). 
We will show results from simulations with $L_z/L_x=0$ (a 2D simulation),
1/32, 1/8, and 1; we will show in particular
that $L_z=L_x/32$ behaves essentially like the 2D simulation, while
the two largest simulations are similar to each other but quite different from the smaller two, presumably because of 3D effects that appear for sufficiently large $L_z$.

\begin{figure}
\centering
\fullplot{
\raisebox{0.315\textwidth}{\large (a)}%
\includegraphics*[width=0.455\textwidth]{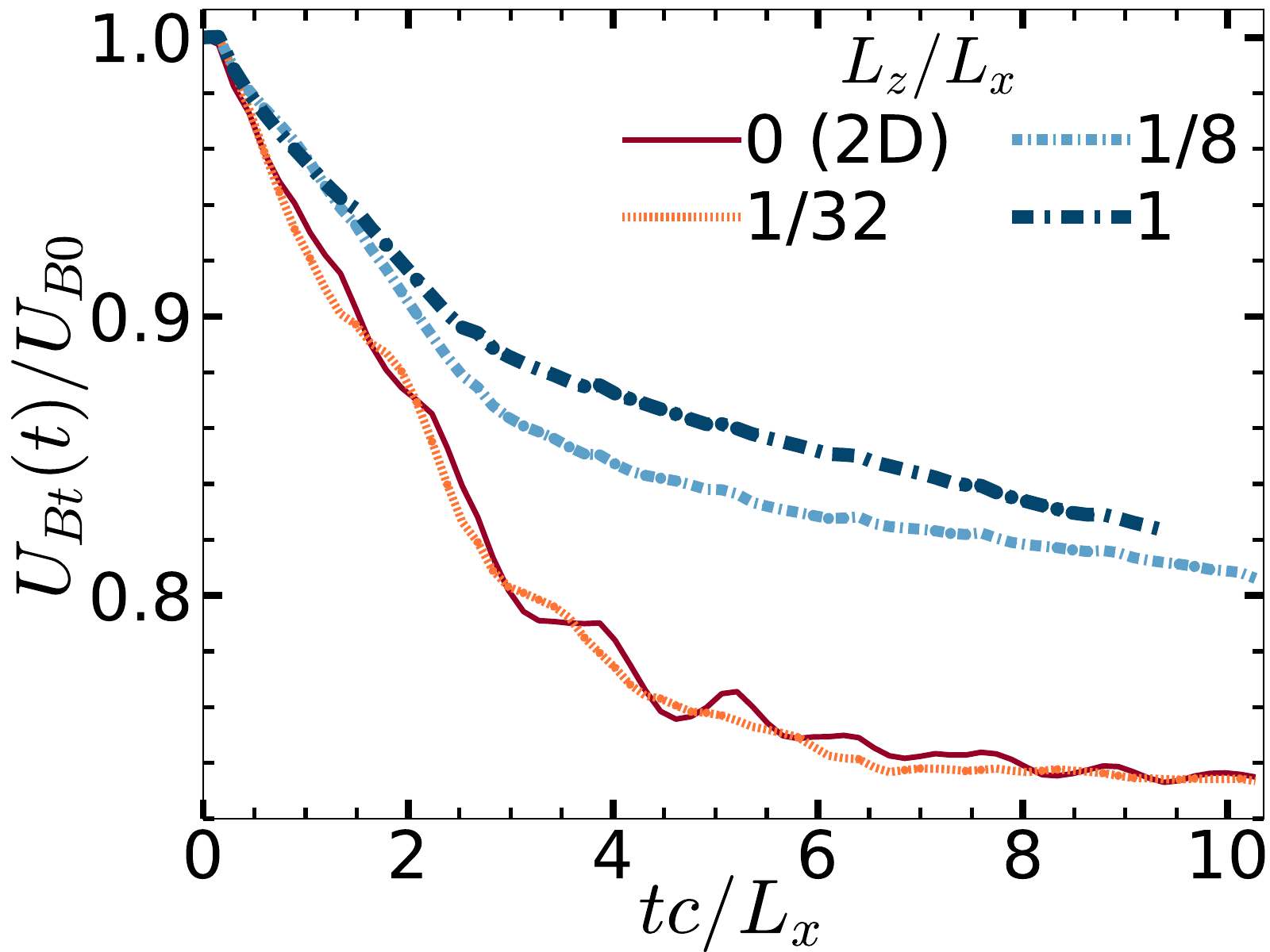}%
\hfill
\raisebox{0.315\textwidth}{\large (b)}%
\includegraphics*[width=0.455\textwidth]{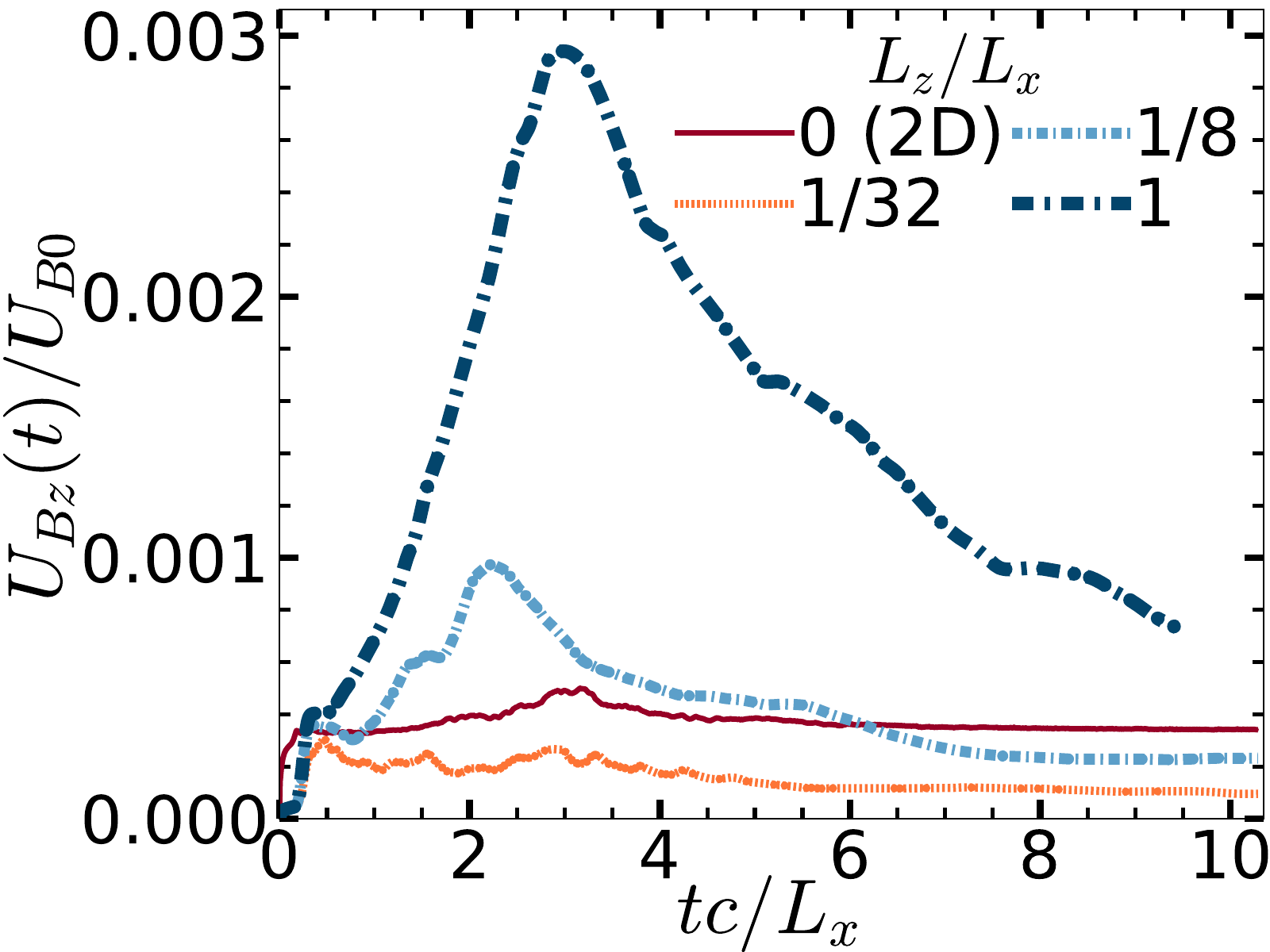}\\
\raisebox{0.315\textwidth}{\large (c)}%
\includegraphics*[width=0.455\textwidth]{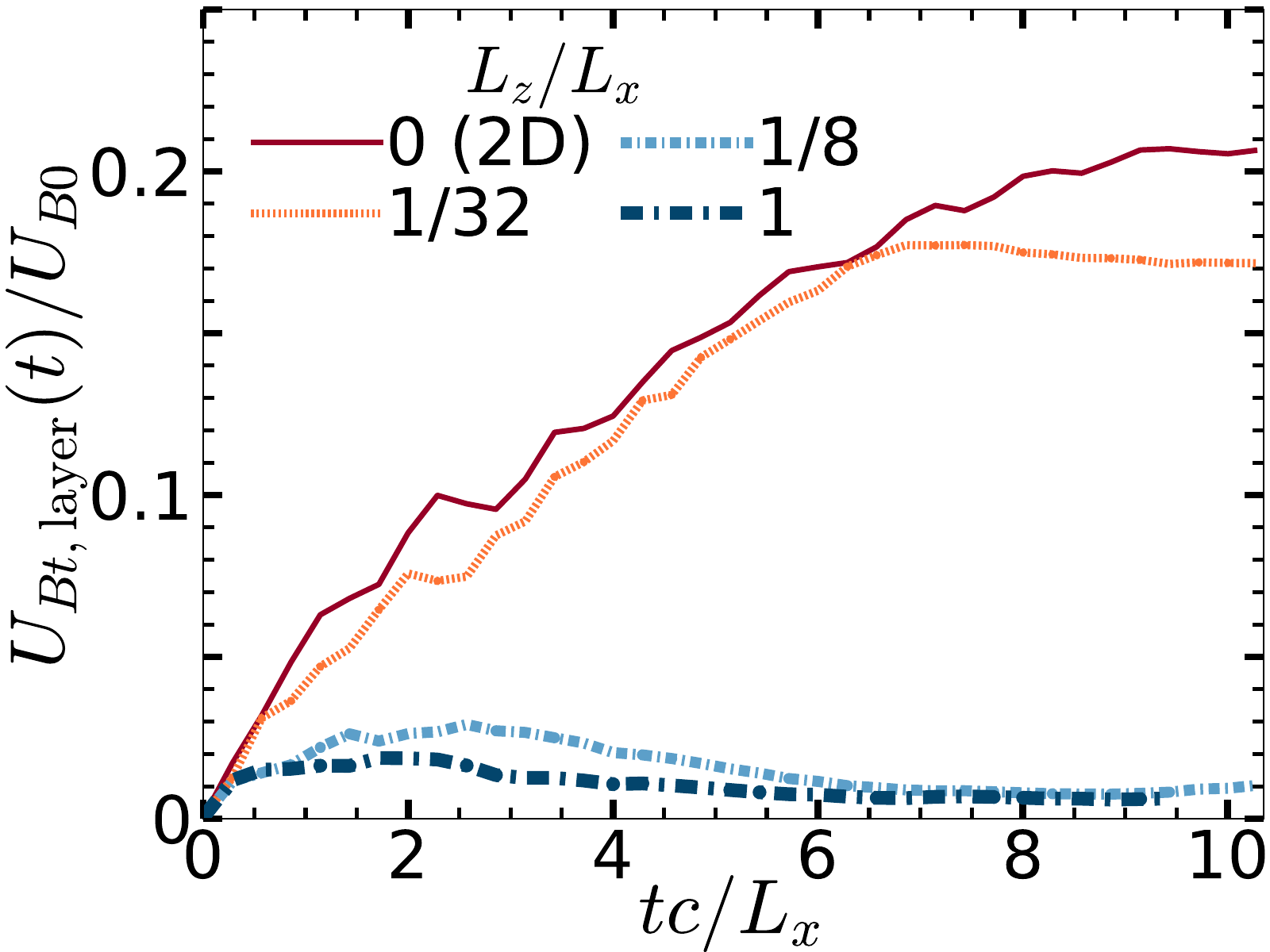}%
\hfill
\raisebox{0.315\textwidth}{\large (d)}%
\includegraphics*[width=0.455\textwidth]{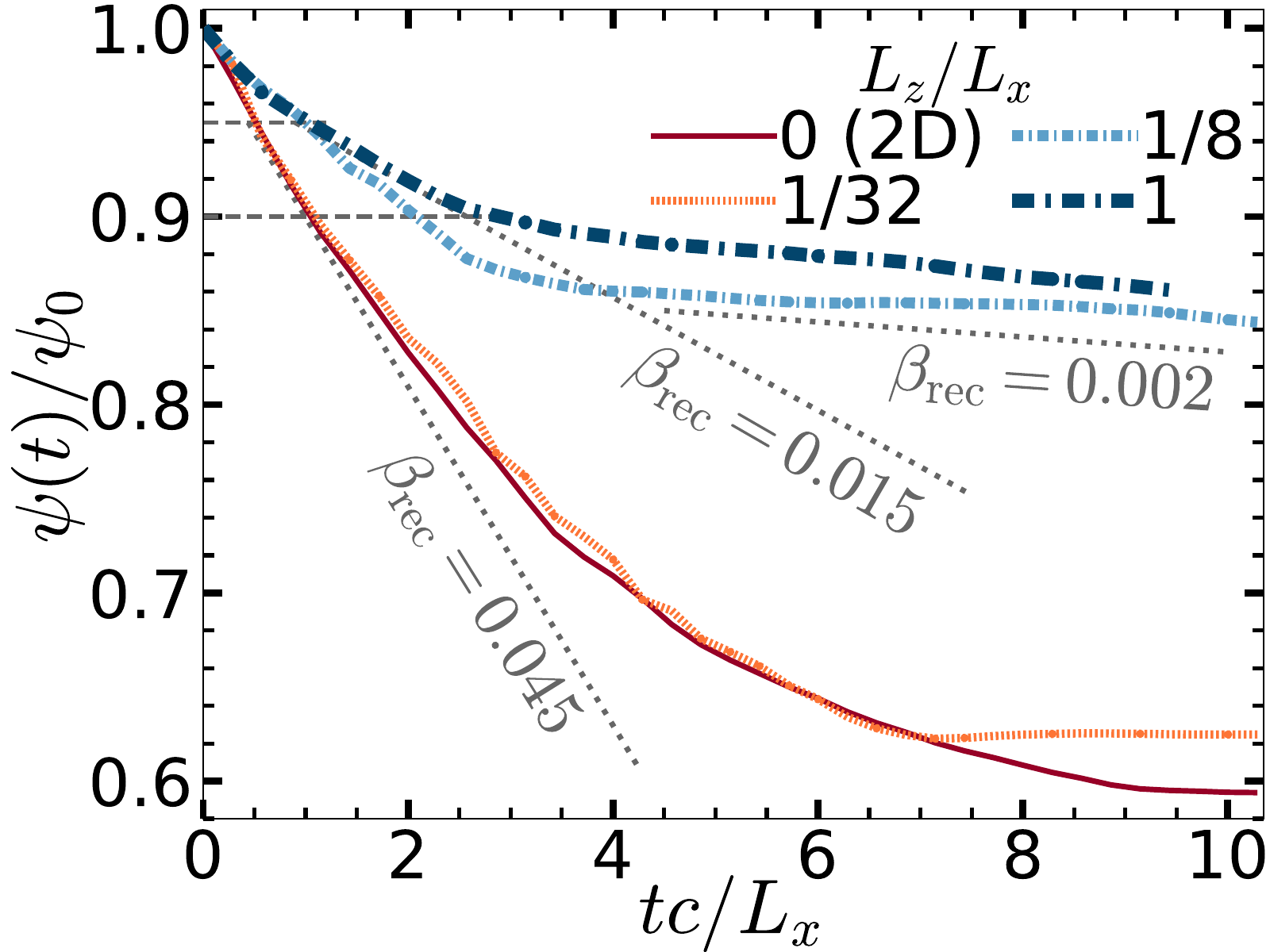}\\
}
\caption{ \label{fig:energyVsTime2D3D}
Energy in various magnetic field components versus time for 4 different $L_z/L_x$, normalized to
$U_{B0} \approx \int dV \, B_0^2/8\upi $: 
(a) $U_{Bt} = \int dV \, (B_x^2+B_y^2)/8\upi$,
(b) $U_{Bz} = \int dV \, B_z^2/8\upi$, 
and (c) $U_{Bt,\rm layer}$, the magnetic energy in the layer (cf.~\S\ref{sec:terminology}).  
The electric field energy (not shown) is always less than $0.003 U_{B0}$.
Finally (d), the unreconnected magnetic flux $\psi(t)$;
dotted grey lines indicate dimensionless reconnection rates,
$(cB_0L_z)^{-1} (d\psi /dt)$, of 0.045 and 0.015, which are roughly
the rates of nearly 2D and 3D simulations, respectively, between the grey 
dashed lines, and also 0.002, roughly the 3D rate at later times.
}
\end{figure}

Before looking at the evolution of 2D and 3D current sheets,
it is helpful to see how global energies and unreconnected flux vary in time.
We begin by looking at global field energies versus time, normalized to initial magnetic energy $U_{B0}$, in Fig.~\ref{fig:energyVsTime2D3D}(a)-(c).
In these simulations without guide field, 
almost all the field energy is contained in $U_{Bt}$, the energy in the transverse magnetic field
$\boldsymbol{B}_t = B_x \hat{\boldsymbol{x}} + B_y \hat{\boldsymbol{y}}$.
In all cases $U_{Bz}(t) < 0.003U_{B0}$,
and the energy in the electric field $\boldsymbol{E}$ (not shown) 
also remains below $0.003 U_{B0}$. 
The $B_z$ and $E$ field components can nevertheless be substantial within the (the small volume of) the layer.
Further investigation of $B_z$ is left for future work, but
we note that there should be no $B_z$ at all in a 2D MHD description \citep[for pair plasma with no initial guide field,][]{Zenitani_etal-2009}, while 2D PIC simulations do develop non-zero $B_z$ \citep{Zenitani_Hesse-2008}.

Figure~\ref{fig:energyVsTime2D3D}(a) shows $U_{Bt}$ decreasing over time in all cases as it is converted to plasma energy.
The simulations with larger $L_z/L_x=1/8,1$ show similar magnetic depletion, and those with smaller $L_z/L_x=0,1/32$ closely resemble each other; in this and other diagnostics we will see that the former ($L_z/L_x=1/8,1$) exhibit 3D behaviour, while the latter are essentially~2D.
We find that magnetic energy depletes
roughly half as fast in 3D as in 2D, at early times 
up to $\sim 4L_x/c$ (Fig.~\ref{fig:energyVsTime2D3D}a; see also table~\ref{tab:finalEnergyAndFlux}, which compares energy changes and reconnection rates).
While the 2D simulations have mostly finished reconnection by $t\sim 4L_x/c$ (approaching a final state a few $L_x/c$ later), the 3D simulations embark upon a second stage of prolonged slow energy conversion \citep[cf.][]{Liu_etal-2011}, converting magnetic energy to plasma energy an order of magnitude still more slowly for many tens of $L_x/c$.
Were these simulations to run many times longer (which they did not, because of prohibitive computational cost),
the 3D simulations would eventually convert more magnetic energy than the essentially 2D simulations; we will see in~\S\ref{sec:Lz3d}, using smaller simulations ($L_x=341\sigma\rho_0$), that 3D simulations convert more magnetic energy over $\sim 50 L_x/c$ than 2D simulations.

\begin{table}
\centering
\begin{tabular}{lccccc}
    & $-\frac{\Delta U_{Bt,\rm up}}{U_{B0}}$ 
    & $\frac{\Delta U_{Bt,\rm layer}}{U_{B0}}$ 
  & $-\frac{\Delta U_{Bt}}{U_{B0}} \approx 
      \frac{\Delta U_{\rm plasma}}{U_{B0}}$
    & $-\frac{\Delta \psi}{\psi_0}$
    & $\beta_{\rm rec,0.95-0.90}$
    \\
  nearly 2D $\left(\frac{L_z}{L_x}=0,\frac{1}{32}\right)$: & 
     0.44--0.47 & 0.17--0.20 & 0.27 & 0.37--0.41 & 0.043--0.047
   \\
  fully 3D $\left(\frac{L_z}{L_x}=\frac{1}{8}, 1\right)$:  & 
     0.18--0.21 & 0.01 & 0.18--0.20 & 0.14-0.16 & 0.01--0.02
\end{tabular}
\caption{\label{tab:finalEnergyAndFlux}  
The fractional loss 
in transverse magnetic
field energy $U_{Bt}$ in the entire simulation 
(nearly equal to the gain in plasma energy 
$U_{\rm plasma}$), divided into upstream and layer
regions ($U_{Bt,\rm up}+U_{Bt,\rm layer}=U_{Bt}$), and
the loss in unreconnected magnetic flux $\psi$,
over the first $10L_x/c$,
for simulations with $L_x=512\sigma\rho_0$ and $B_{gz}=0$.
Changes are shown normalized to initial magnetic energy $U_{B0}$ and 
initial unreconnected flux $\psi_0$.
The fully 3D simulations would continue to
deplete magnetic energy if run longer (cf.~\S\ref{sec:Lz3d}).
Also shown is $\beta_{\rm rec}$ (cf.~\S\ref{sec:unreconnectedFlux}), 
averaged over the time during which~$\psi(t)$
drops from~0.95$\psi_0$ to~0.90$\psi_0$.
}
\end{table}

More magnetic energy conversion is possible in 3D  because the magnetic energy pumped into plasmoids can be depleted in 3D, whereas in 2D plasmoids remain stable.
To demonstrate how little magnetic energy remains stored in plasmoids in 3D,
we separately measure the transverse magnetic energy in the upstream and in the layer regions (cf.~\S\ref{sec:terminology}):
$U_{Bt} = U_{Bt,\rm up} + U_{Bt,\rm layer}$.
With no guide field and no initial perturbation, the initial magnetic energy 
$U_{B0}=U_{Bt0}$ is entirely in unreconnected transverse field.
During magnetic reconnection in 2D, part of this gets converted into plasma energy $U_{\rm plasma}$, and part into $U_{Bt,\rm layer}$ in plasmoids.
Over the first $10L_x/c$, the (nearly) 2D simulations take the initial upstream magnetic energy $U_{B0}$ and convert about $\Delta U_{\rm plasma}\approx 0.27U_{B0}$ to plasma energy while building up a comparable amount, $\Delta U_{Bt,\rm layer}\approx 0.20_{B0}$, in plasmoids, where it remains trapped.
Since the energized particles are also trapped in plasmoids, we estimate that the magnetic energy in plasmoids is about 2/3 of the plasma energy in plasmoids.
Over the same time, fully 3D simulations convert $\Delta U_{\rm plasma}\approx 0.2 U_{B0}$ to plasma energy, but magnetic energy in the layer quickly decays, leaving a strikingly smaller upper bound on the amount of energy, $\Delta U_{Bt,\rm layer}\approx 0.01 U_{B0}$, that could be in plasmoids 
(see Fig.~\ref{fig:energyVsTime2D3D}c and table~\ref{tab:finalEnergyAndFlux}).
Although 3D simulations deplete upstream energy $U_{Bt,\rm up}$ more slowly, they thus convert it more completely to plasma energy.

Along with magnetic energy, we can also measure the unreconnected magnetic flux, which decreases over time, faster in 2D than in 3D.  
In 3D, as in 2D, some of the upstream flux is promptly reconnected, ending up in plasmoids, as in 2D.
Unlike in 2D, however, in 3D some upstream flux may be promptly annihilated, and also (only in 3D) reconnected flux is subsequently annihilated as plasmoids or flux rope structures become unstable and decay.
It is nontrivial to measure how much upstream flux is annihilated promptly without first undergoing reconnection, but we can conclude from the relatively small amount of magnetic energy in the layer that ultimately, in 3D, most of the lost upstream flux is annihilated (versus being stored permanently in plasmoids).
Figure~\ref{fig:energyVsTime2D3D}(d) shows the unreconnected flux $\psi(t)$ diminishing as reconnection occurs for the four different values of $L_z/L_x$.
Table~\ref{tab:finalEnergyAndFlux} shows that up to~$t=10L_x/c$, the
nearly 2D simulations lose 40 per~cent of the initial unreconnected flux, and fully 3D simulations lose only about 15 per~cent (though, over time, they would reconnect and/or annihilate more upstream flux, whereas the 2D simulations have essentially finished).
The reconnection rate $\beta_{\rm rec}$ at early times, measured between $\psi(t)=0.95\psi_0$ and $\psi(t)=0.90\psi_0$ (the horizontal dashed lines in Fig.~\ref{fig:energyVsTime2D3D}d) and normalized to~$B_0 c$, is about~0.045 for the nearly~2D simulations, and~0.015 for 3D (see table~\ref{tab:finalEnergyAndFlux}); later, around~$t=8L_x/c$, $\beta_{\rm rec}$ is on the order of $\sim 0.002$ for the 3D simulations (and fluctuating around zero for~2D).

\begin{figure}
\centering
\fullplot{
\includegraphics*[width=\textwidth]{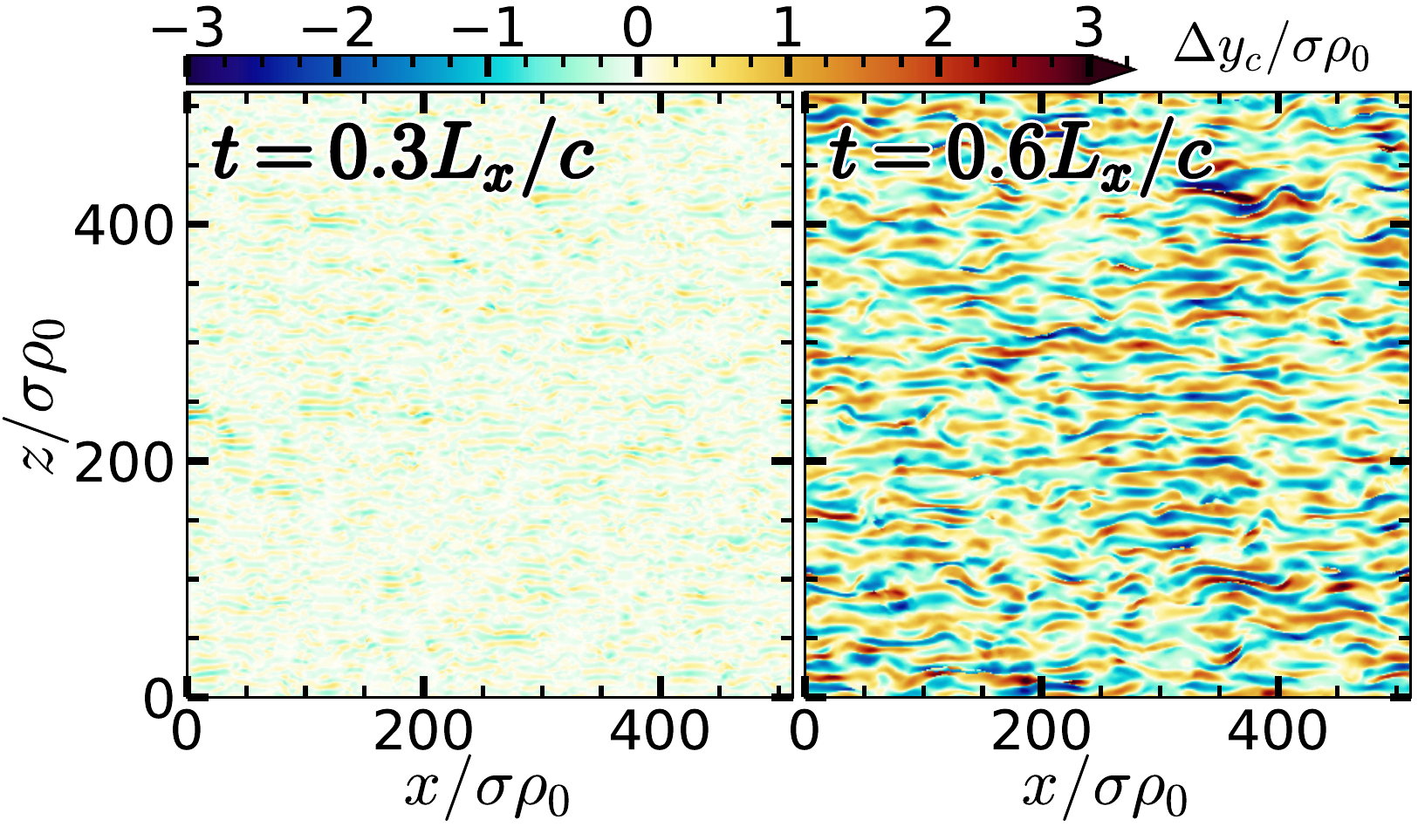}%
}
\caption{ \label{fig:layerOffsetVsXz}
The displacement $\Delta y_c(x,z,t)$ of the current sheet central surface
from the original midplane (cf.~\S\ref{sec:sheetCenter}), for
$L_z=L_x$ at two different times, $t=0.3L_x/c$ (left) and $0.6 L_x/c$ (right),
shows increasing rippling due to the RDKI.
}
\end{figure}

Having shown significant differences (as well as qualitative similarities) between 2D and 3D energy and upstream flux depletion, we will describe the evolution of the current sheet.  First, we will describe the kinking of the sheet, which cannot occur in 2D at all; then we will spend some time describing 3D behaviour that resembles 2D reconnection.

Perhaps the most expected (by now) new effect in 3D
is the kinking of the current sheet in the $z$ direction due to the RDKI \citep{Zenitani_Hoshino-2007,Zenitani_Hoshino-2008}.
Simulating the $z$ dimension enables the RDKI, which can occur because of the relative movement of electrons and positrons in the $z$-direction 
and causes
rippling of the sheet. 
Figure~\ref{fig:layerOffsetVsXz} shows (for the case with $L_z=L_x$) a current sheet rippling due to RDKI with an amplitude $\Delta y \approx \sigma\rho_0$ at $t=0.3L_x/c$, growing to $3\sigma\rho_0$ at $t=0.6L_x/c$ [for comparison, the initial current sheet thickness was $\delta=(2/3)\sigma \rho_0$].
It had originally been thought that RDKI---at least in cases with weak guide field---might out-compete the tearing instability, thereby suppressing reconnection \citep{Zenitani_Hoshino-2008}.
However, later, larger 3D simulations (in the high-$\sigma_h$ regime) found that, despite active RDKI, reconnection ultimately does not seem to be inhibited, even with zero 
guide field \citep{Sironi_Spitkovsky-2014,Guo_etal-2014,Guo_etal-2015,Werner_Uzdensky-2017}.

Indeed, 3D simulations do exhibit behaviour that, at least locally, shares many characteristics with 2D reconnection.
Although energy is clearly transferred from magnetic fields to plasma,
determining whether and where reconnection occurs is nontrivial in 3D.
We tentatively suggest, based on our 3D simulations, that reconnection occurs in patches within the layer---patches that may be the analogues of elementary current sheets in 2D reconnection (i.e., the smallest inter-plasmoid current sheets, which are not further broken up by secondary tearing).
It is possible that reconnection is slower (overall) in 3D because the area of actively-reconnecting regions (patches) is smaller, while the local reconnection rate remains as high as in 2D \citep[such patchiness with locally high reconnection rates has been observed for $\sigma_h\sim 1$ by][]{Yin_etal-2008}.
Our results are consistent with this hypothesis, but so far we have been unable to measure local reconnection rates with enough certainty to conclude this definitively; we leave a dedicated investigation of this problem to future studies.
However, we offer evidence that reconnection does in fact occur (though in patches), and that the patches evolve over time, and continue to exhibit signatures of reconnection even at late times during the ``slow 3D reconnection stage'' (e.g., $t>4L_x/c$).
Specifically, we will see that actively-reconnecting regions are areas
where the current sheet is thin, that they exhibit outflows (roughly) in the $\pm x$ directions, and that they are the areas with strongest parallel (to $\boldsymbol{B}$) electric fields and $\boldsymbol{E}\bcdot\boldsymbol{J}>0$.

\begin{figure}
\centering
\fullplot{
\includegraphics*[width=\textwidth]{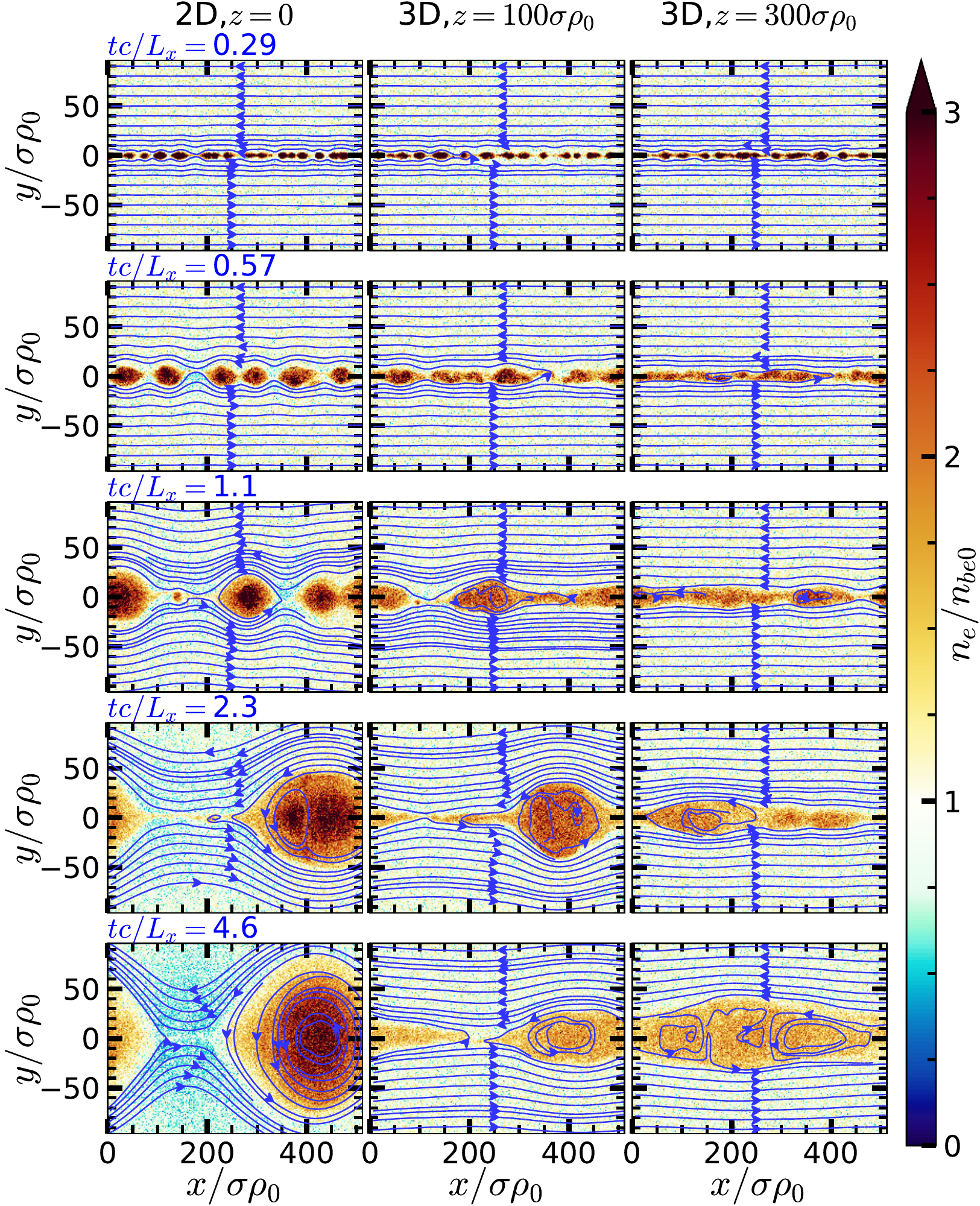}%
}
\caption{ \label{fig:neVsXy2D3D}
A time sequence of electron density $n_e$ in the $x$-$y$ reconnection plane, in nearly 2D ($L_z=L_x/32$, left column) and 3D ($L_z=L_x$, middle and right) simulations for $tc/L_x=0.3, 0.6, 1.1, 2.3, 4.6$.  
Blue lines follow magnetic field lines, ignoring $B_z$ (with a higher density of lines shown closer to the midplane).
The left column shows  $z=0$ (which looks fairly similar to all other $z$ values for this nearly 2D case); the middle column shows $z=100 \sigma \rho_0\approx 0.2 L_z$ and the right column shows $z=300 \sigma \rho_0 \approx 0.6L_z$.
}
\end{figure}

Figure~\ref{fig:neVsXy2D3D} shows the electron density $n_e$, normalized to $n_{be0}=n_{b0}/2$, at five different times ($tc/L_x=0.3$, 0.6, 1.1, 2.3, and 4.6) in $x$-$y$ slices,
at $z=0$ for the nearly 2D simulation with $L_z=L_x/32$, and at $z=100\sigma\rho_0$ and $z=300\sigma\rho_0$ for $L_z=L_x$.
A sample of magnetic field lines (traced while ignoring $B_z$) shows that $n_{e} \lesssim n_{be0}$ in the upstream region, where flux is unreconnected (cf.~\S\ref{sec:unreconnectedFlux}), while $n_e > n_{be0}$ in most of the layer.
Therefore, looking at density is a good if rough way to track the gross evolution of the shape of the layer.

For $L_z=L_x/32$, the picture is practically the same as in 2D:
the initial current sheet tears, small plasmoids form and merge into larger plasmoids, until there is just one plasmoid (and a monster one at that) in the layer.  Between plasmoids, the layer is thin at X-points (really, $X$-lines extended in the $z$ direction) where reconnection takes place.
For $L_z=L_x$, the picture is initially very similar to that in 2D, when viewed in an $x$-$y$ slice.
For example, at $t=0.3L_x/c$, a similar number of plasmoids is visible
(although a slice at any value of $z$ looks qualitatively similar, these structures do not appear to extend uniformly in $z$; i.e., at early times we do not observe long flux ropes of length $L_z$).
Soon (by $t=0.6L_x/c$), however, the 3D layer shows general thickening without the clear plasmoid structures so familiar in 2D. 
At some rare places (e.g., $x\approx z \approx 100 \sigma\rho_0$, $tc/L_x=1.1$--$2.3$), the layer remains very thin, as at X-points in 2D.
In the vicinity of these thin places we still see some ballooning of the layer, resembling a larger plasmoid, but smaller (in $y$) than in 2D, and with much less distinct density structure.
After $2.3L_x/c$, most of the layer is thick, but not as thick as plasmoids in 2D.

The $L_z=L_x/32$ simulation is fairly uniform in $z$, consistent with its
behaviour resembling 2D reconnection.
The 3D simulation, however, is very nonuniform, as we will see in a series
of plots showing various fields at time $t=2.3L_x/c$, at the centre of the layer, i.e., at $y_c(x,z,t)$ such that $B_x(x,y_c,z,t)=0$.

\begin{figure}
\centering
\fullplot{
\raisebox{0.285\textwidth}{a.}%
\includegraphics*[width=0.30\textwidth]{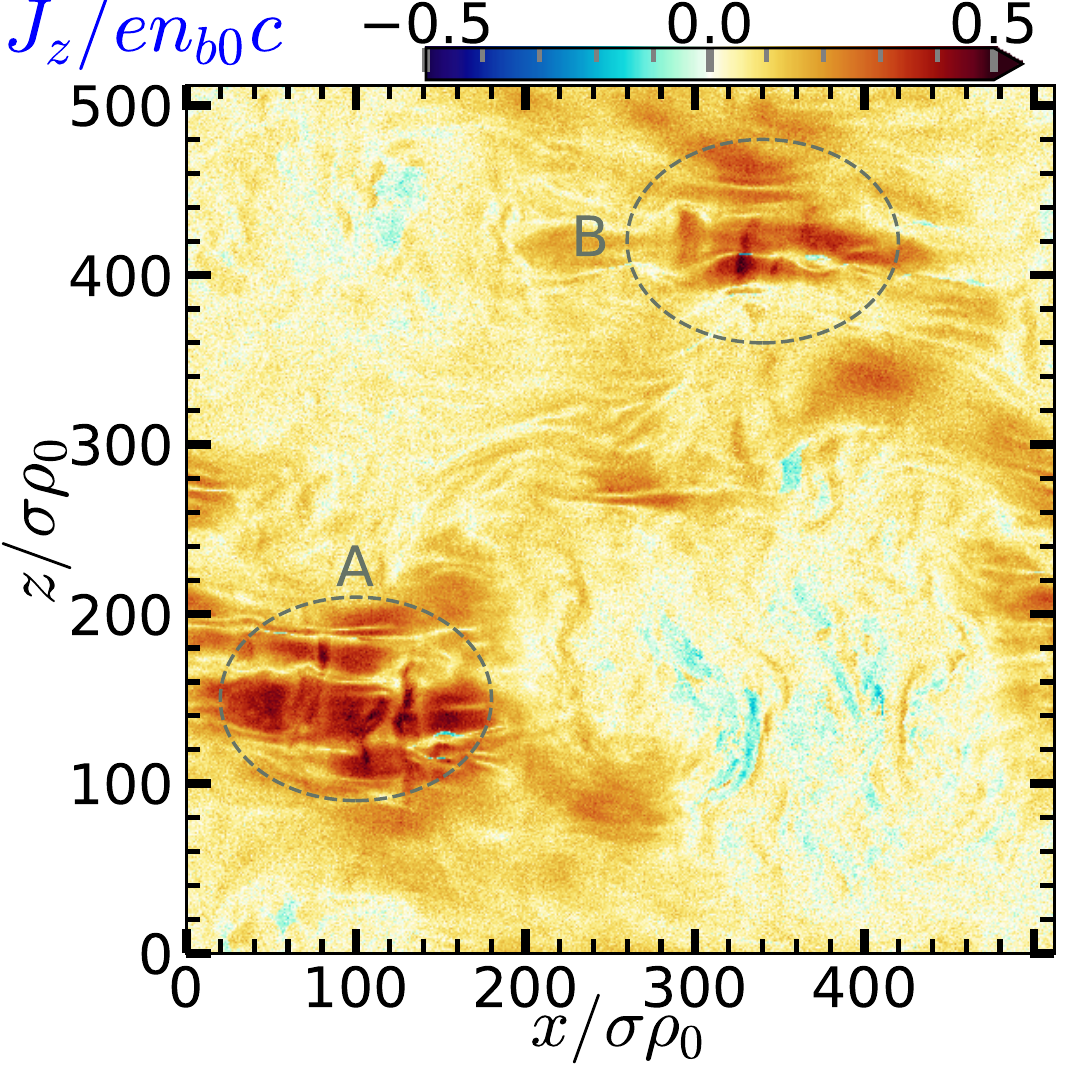}\hfill
\raisebox{0.285\textwidth}{c.}%
\includegraphics*[width=0.30\textwidth]{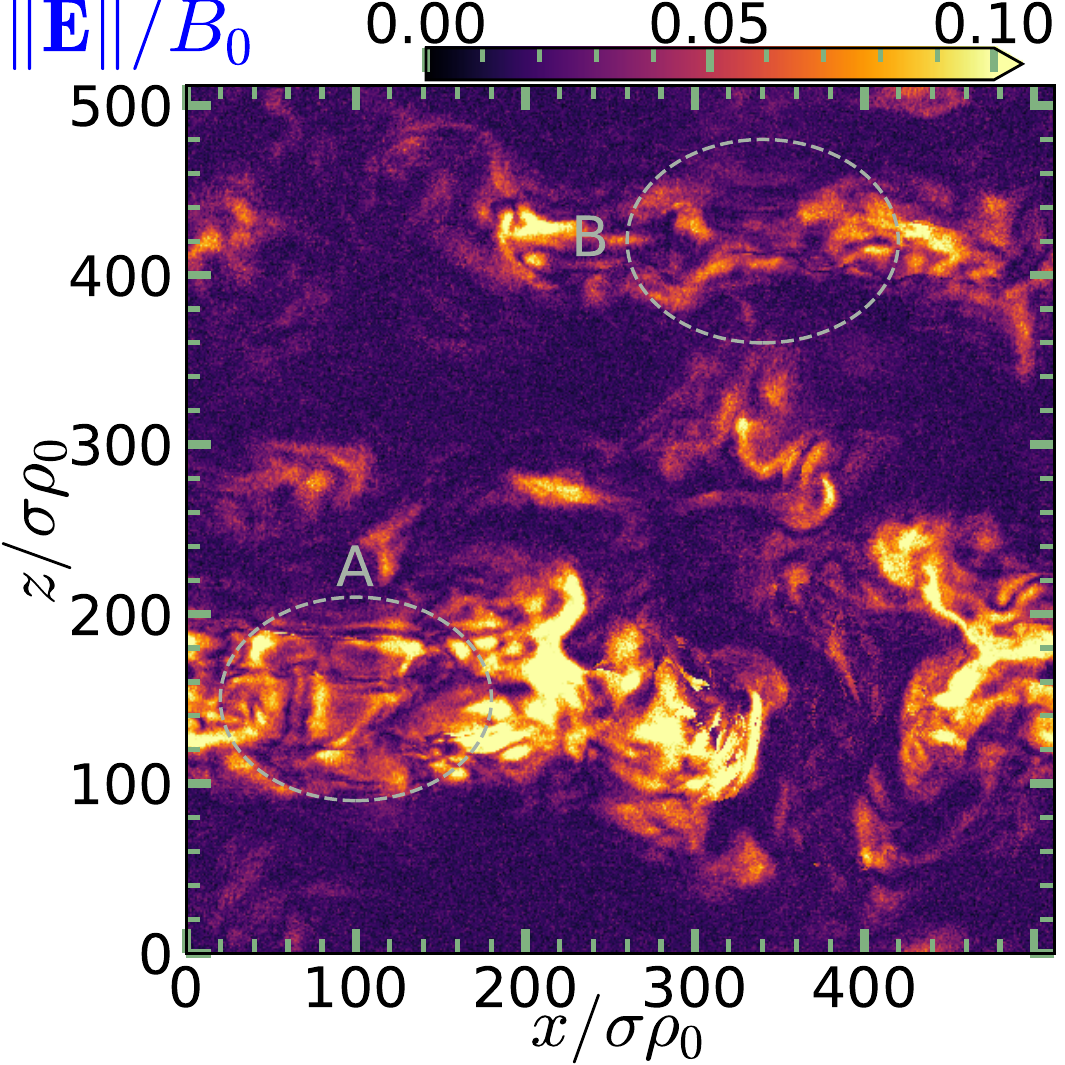}\hfill
\raisebox{0.285\textwidth}{e.}%
\includegraphics*[width=0.30\textwidth]{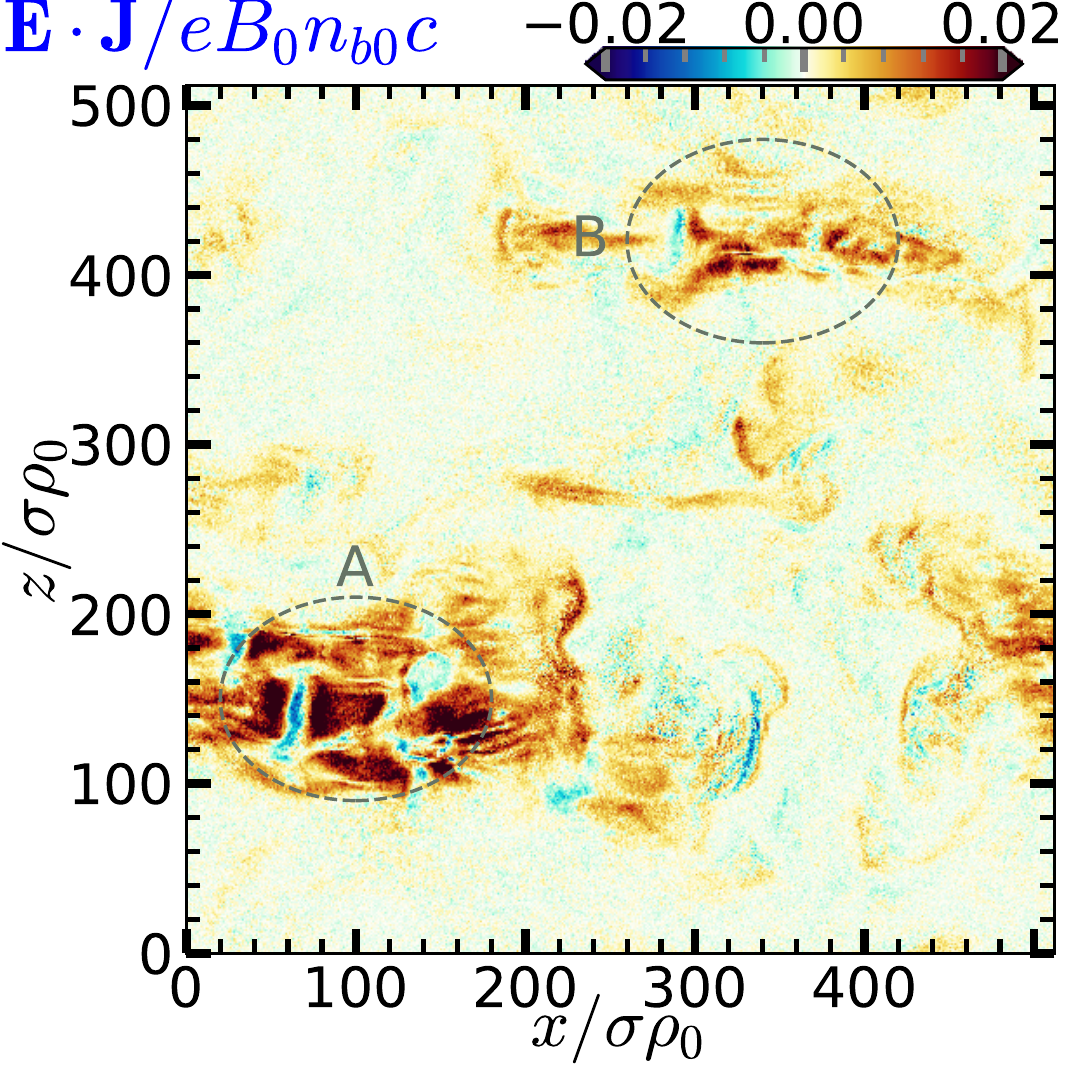}\\
\raisebox{0.285\textwidth}{b.}%
\includegraphics*[width=0.30\textwidth]{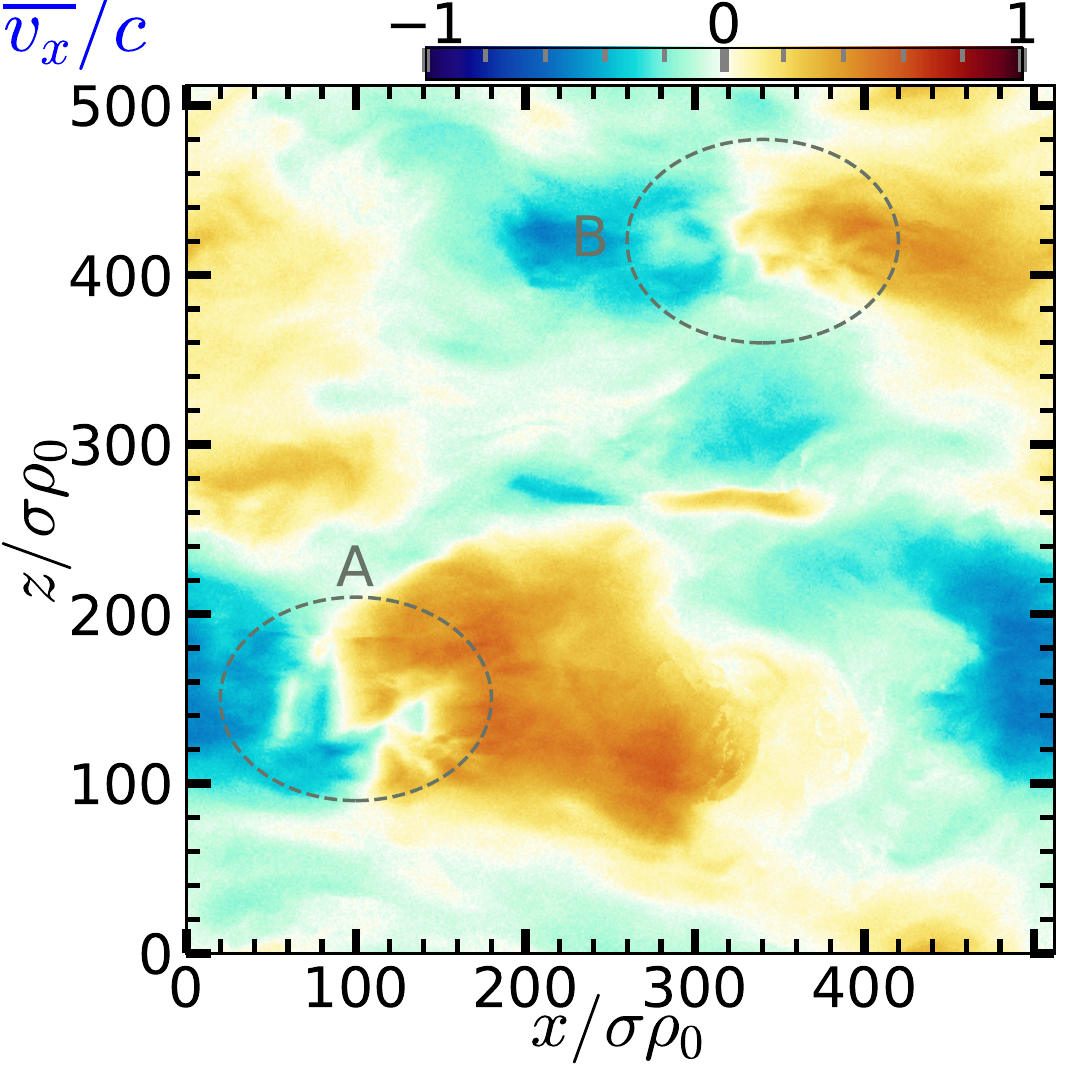}\hfill
\raisebox{0.285\textwidth}{d.}%
\includegraphics*[width=0.30\textwidth]{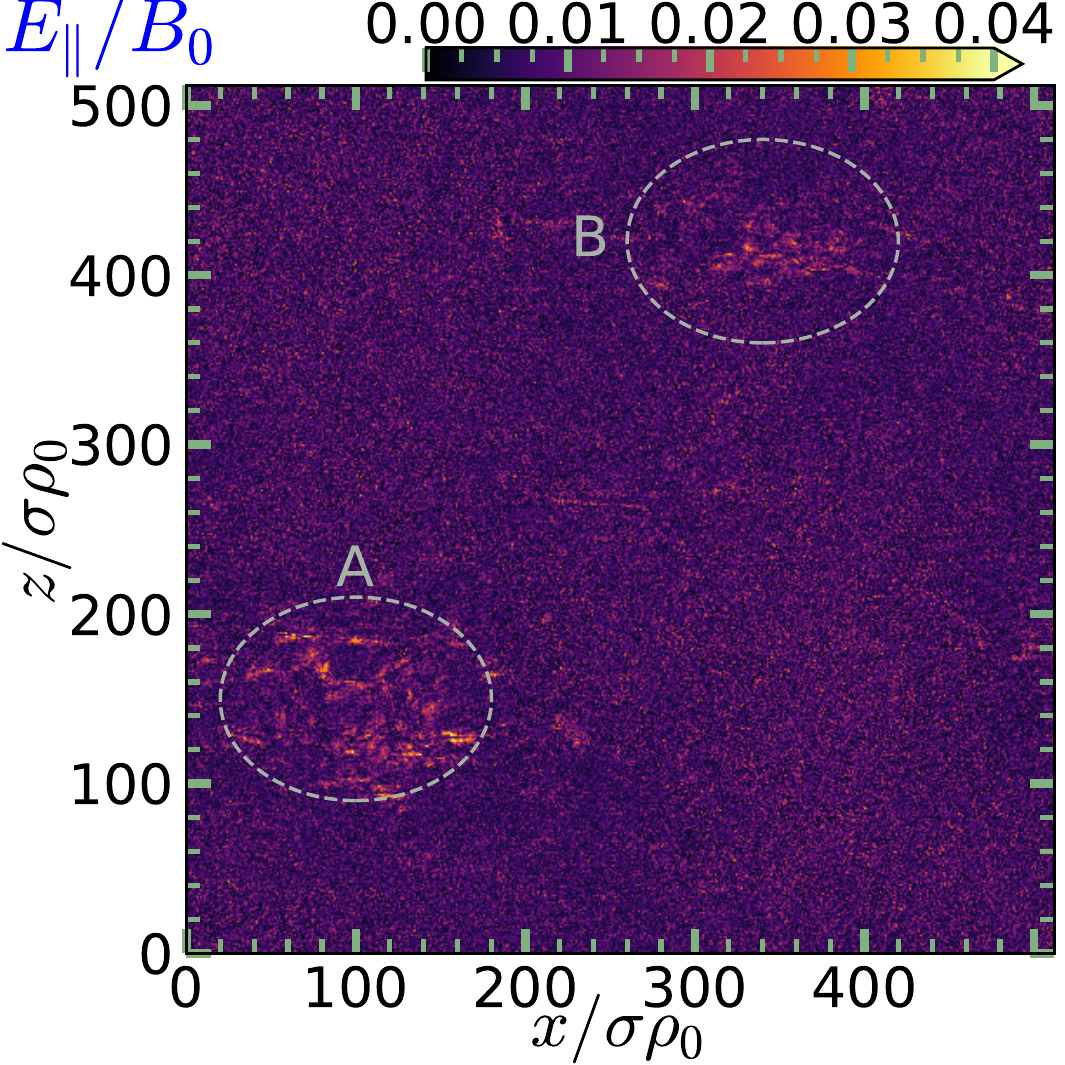}\hfill
\raisebox{0.285\textwidth}{f.}%
\includegraphics*[width=0.30\textwidth]{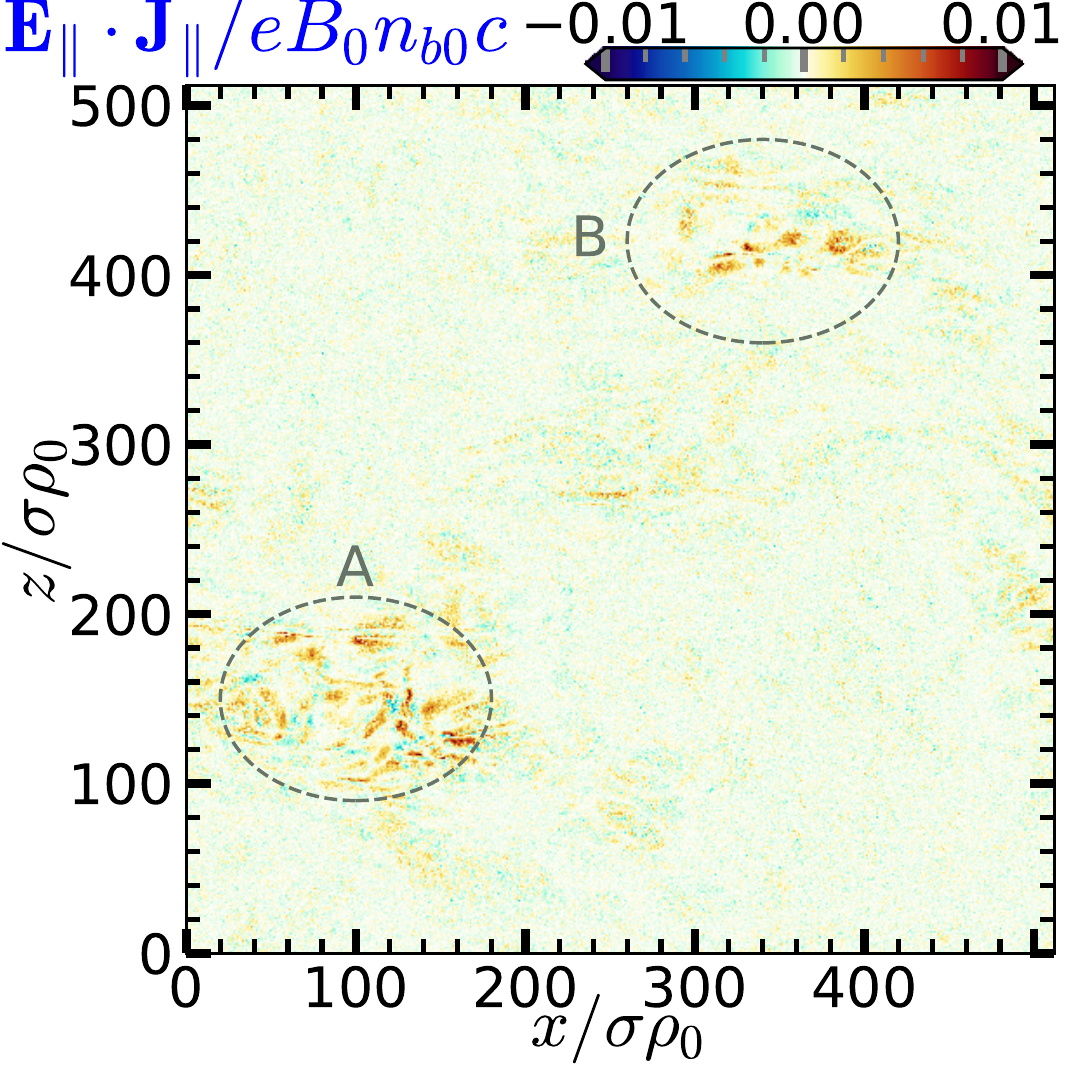}%
}
\caption{ \label{fig:layerXz3dn6144}
Field quantities versus $x$ and $z$ at the layer central surface, i.e., at $y=y_c(x,z,t)$ such that $B_x(x,y_c,z,t)=0$ (cf.~\S\ref{sec:sheetCenter}).
$A$ and $B$ indicate---roughly---examples of actively-reconnecting regions.
(a) The current density in the $z$-direction, $J_z$, is highest in reconnecting regions, where the layer is thin. 
(b) $\overline{v_x}=(v_{e,x} + v_{i,x})/c$, the average of electron and positron $x$-velocities, shows strong outflows from reconnection regions.
(c) The electric field, $\|\boldsymbol{E}\|$ is largest in reconnecting regions and also in strong outflows.
(d) Although small, $E_\|$, the magnitude of the component of $\boldsymbol{E}$ parallel to the local $\boldsymbol{B}$, is stronger in reconnecting regions.
(e) $\boldsymbol{E}\bcdot \boldsymbol{J}$ indicates dominant power transfer from fields to plasma, mostly in reconnecting and outflow regions.
(f) $\boldsymbol{E}_\|\bcdot\boldsymbol{J}_\|$ is positive and strongest in reconnecting regions, indicating significant plasma energization by parallel electric fields (or electric fields where $B\approx 0$).
All plotted quantities were averaged over $y_c-\sigma\rho_0 < y < y_c+\sigma\rho_0$.
}
\end{figure}

Figure~\ref{fig:layerXz3dn6144} shows various quantities at the layer central surface, $y_c(x,z,t)$ (cf.~\S\ref{sec:sheetCenter}), versus $x$ and $z$, for the $L_z=L_x$ case, at $t=2.3L_x/c$.
We focus particularly on region $A$, a patch of size (at this time) $\simeq 100\sigma\rho_0$ located around $x\approx z \approx 100\sigma\rho_0$, which we believe to be an actively-reconnecting region; we saw in Fig.~\ref{fig:neVsXy2D3D} that it is a site where the layer is near its thinnest. 
Panel (a) shows the current density $J_z$ in the $z$-direction, which is indeed most prominent in this reconnecting region, consistent with the layer being thin.  (We note that current sheets
are locally not necessarily parallel to $z$, but are nearly parallel to $z$ in many places.)
Panel (b) shows strong outflows from this region in the $\pm x$ directions,
as one would expect from reconnection; similarly, region~$B$, around $x\approx 320\sigma\rho_0$ and $z\approx 400\sigma\rho_0$ with size (at this time) similar to patch $A$, has a high current density and strong outflows.
There are also smaller reconnecting regions---e.g., at $x\approx 260\sigma\rho_0$ and $z\approx 280\sigma\rho_0$.
The ``line'' between the outflows thus appears to be a stagnation line, which is commonly associated with an X-line in the classic 2D picture of (symmetric) reconnection.
Panel (c) shows $\|\boldsymbol{E}\|$, which is sizeable ($\sim 0.1 B_0$) over a significant region, especially in the neighbourhood of large, strong outflows.
However, (panel d) the parallel electric field $|E_\||$ (parallel to the local
$\boldsymbol{B}$) is largest only in a region much closer to the stagnation line, as expected.
Panel (e) shows that $\boldsymbol{E}\bcdot \boldsymbol{J}$ has more strongly-positive areas than negative, indicating energy transfer from electromagnetic fields to the plasma;
this transfer occurs around X-lines and extended outflows, but as panel (f) shows, the region where $E_\| J_\|$ is most positive is limited to areas closer to the largest stagnation lines (in regions $A$ and $B$).
Figure~\ref{fig:layerXz3dn6144} thus suggests that reconnection is
occurring in patches (such as regions $A$ and $B$), perhaps not with such neat geometry as in 2D, but exhibiting basic signatures, such as a thin layer with high current density, strong outflows, particle energization, and indeed particle energization by parallel electric fields or by electric fields in a region of weak magnetic field.

Very early in time, the layer exhibits many small patches of active reconnection.  At any point in time, a patch tends to have roughly the same size in $x$ and $z$ (the $x$-extent can be seen in Fig.~\ref{fig:neVsXy2D3D}).
For example, patches have size $\sim 20\sigma\rho_0$ at $t=0.3L_x/c$.
It is perhaps therefore not surprising that the simulation with $L_z=L_x/32=16\sigma\rho_0$---which is smaller than the patch size even at this early time---behaves like 2D reconnection.
Over time (for $L_z=L_x$), the number of reconnecting patches decreases rapidly, but the surviving patches grow in size, in both $x$ and $z$.
By $2.3L_x/c$, the 3D simulation has 2 major patches (regions $A$ and $B$), with perhaps a few much smaller, probably less active ones.
By $4.6L_x/c$, only region $A$ remains, although it seems to have shrunk in the $z$ direction, perhaps spreading its current density a little wider in $x$.
Importantly, region $A$ does continue to exhibit signs of active reconnection at these late-stage times, when the global reconnection rate is relatively slow in 3D.
We leave a systematic investigation of the behaviour and statistics of these patches to future work.

\begin{figure}
\centering
\fullplot{
\includegraphics*[width=\textwidth]{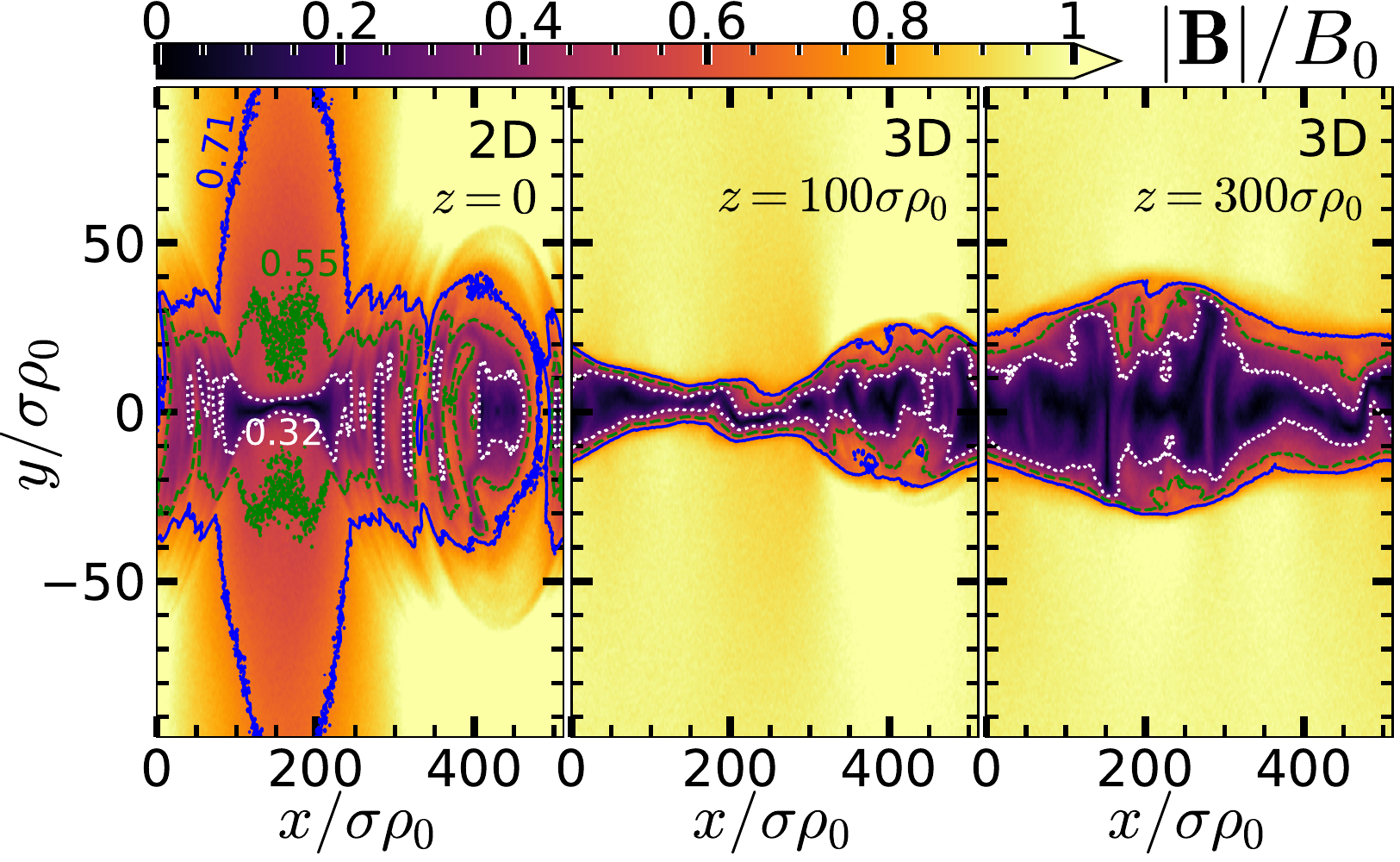}%
}
\caption{ \label{fig:BvsXy2D3D}
The magnetic field strength $|\boldsymbol{B}|$ at $t=4.6L_x/c$ in the same $z$-planes shown in Fig.~\ref{fig:neVsXy2D3D}; (left) nearly 2D, $z=0$, (middle) 3D, $z=100\sigma\rho_0$, and (right) 3D, $z=300\sigma\rho_0$.
Contours are drawn for $B/B_0=\sqrt{0.1}\approx 0.32$ (dotted white), $\sqrt{0.3}\approx 0.55$ (dashed green), and $\sqrt{0.5}\approx 0.71$ (solid blue).
}
\end{figure}

Although the above-described patches in 3D locally resemble elementary reconnecting current sheets in 2D, important differences emerge away from these patches.
As previously mentioned, we do not see the long-term storage of magnetic energy in highly-structured plasmoids as in 2D;
for example, in 3D reconnection we do not see clear signatures of persistent, growing structures extended across the simulation in the $z$-direction (as we would expect flux ropes to be), e.g., in Fig.~\ref{fig:layerXz3dn6144}.
In (nearly) 2D, the magnetic field is somewhat diminished far upstream of major X-lines, e.g., to $\approx 0.7 B_0$ more than $100\sigma\rho_0$ upstream at $t=4.6L_x/c$; at the same time, the magnetic field rarely
falls below $0.3B_0$ anywhere (Fig.~\ref{fig:BvsXy2D3D}, left).
In 2D, plasmoids continue to grow roughly up to the size of the simulation,
when the magnetic configuration prevents further reconnection.
In 3D, however, the current layer becomes a somewhat thickened turbulent region with relatively sharp boundaries (where most current flows) inside which the magnetic field is generally diminished.
The magnetic field is practically full strength outside of the sharp boundaries, while being less than $0.3 B_0$ in a large fraction of the volume within the boundaries (Fig.~\ref{fig:BvsXy2D3D}, middle and right).

The evolution of current sheets in 3D is complicated, and demands much more study.
For now, however, we will move on to consider one of the most important consequences of current sheet evolution and resulting energy conversion, namely NTPA.

\begin{figure}
\centering
\fullplot{
\includegraphics*[width=0.49\textwidth]{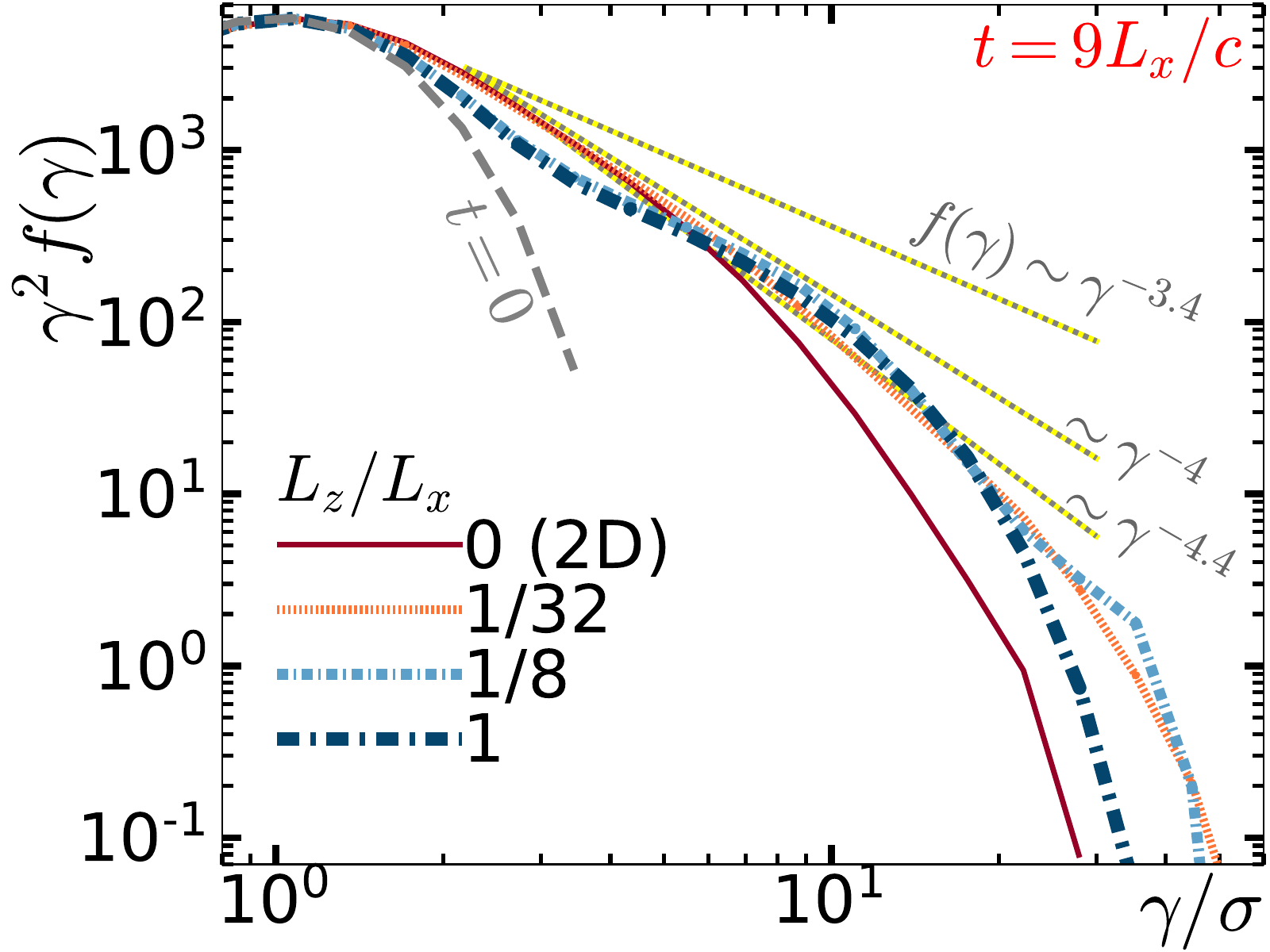}%
\hfill
\includegraphics*[width=0.49\textwidth]{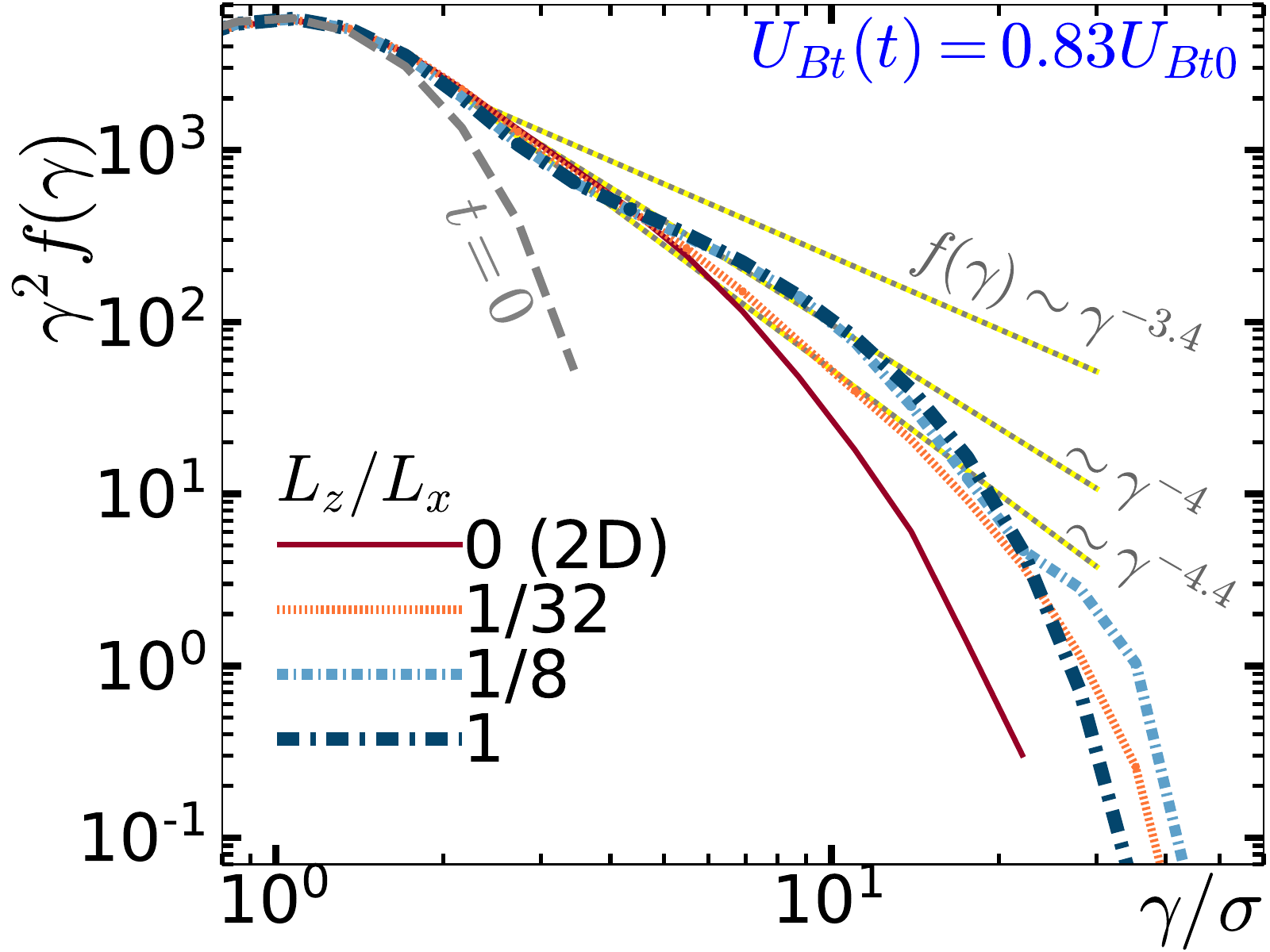}%
}
\caption{ \label{fig:ntpa2D3D}
Electron energy spectra $f(\gamma )$, compensated by $\gamma^2$, where particle energy is $\gamma m_e c^2$, for $L_z/L_x=0$, 1/32, 1/8, 1, (left) at the same time, $t=9 L_x/c$ and (right) when the same amount of magnetic energy ($\simeq 17$ per~cent) has been depleted.  Dotted grey-on-yellow 
lines indicate a range of power-law slopes $f(\gamma)\sim \gamma^{-3.4}$, $\gamma^{-4}$, and $\gamma^{-4.4}$.
}
\end{figure}

Reconnection-driven NTPA is a promising candidate to explain, e.g., high-energy X-ray and gamma-ray emission from a variety of astrophysical sources.
It has been previously suggested that NTPA in 3D might be suppressed by the RDKI interfering with reconnection \citep{Zenitani_Hoshino-2007, Zenitani_Hoshino-2008}.
And since the reconnection electric field responsible for accelerating particles is proportional to the reconnection rate, and we have shown here that the global reconnection rate is slower in 3D than in 2D, we have another reason to suspect that NTPA might be considerably less efficient in 3D.
However, that is emphatically not the case here.
Figure~\ref{fig:ntpa2D3D} shows electron energy spectra~$f(\gamma)$ (compensated by~$\gamma^2$)---compared both at the same time ($t=9 L_x/c$) and after the same amount of magnetic energy has been given to particles ($\Delta U_{\rm plasma} \approx 0.17 \: U_{B0}$).
Remarkably, all the spectra are pretty similar, exhibiting high-energy nonthermal behaviour with a power-law slope $\approx -4$ [i.e., $f(\gamma)\sim \gamma^{-4}$].
However, though the difference may be small, 3D reconnection yields more particles with very high energies than 2D; in addition, since ultimately 3D reconnection can convert more magnetic energy to plasma energy than 2D, 3D reconnection may accelerate more particles overall (not just at the very highest energies).

\begin{figure}
\centering
\fullplot{
\includegraphics*[width=0.49\textwidth]{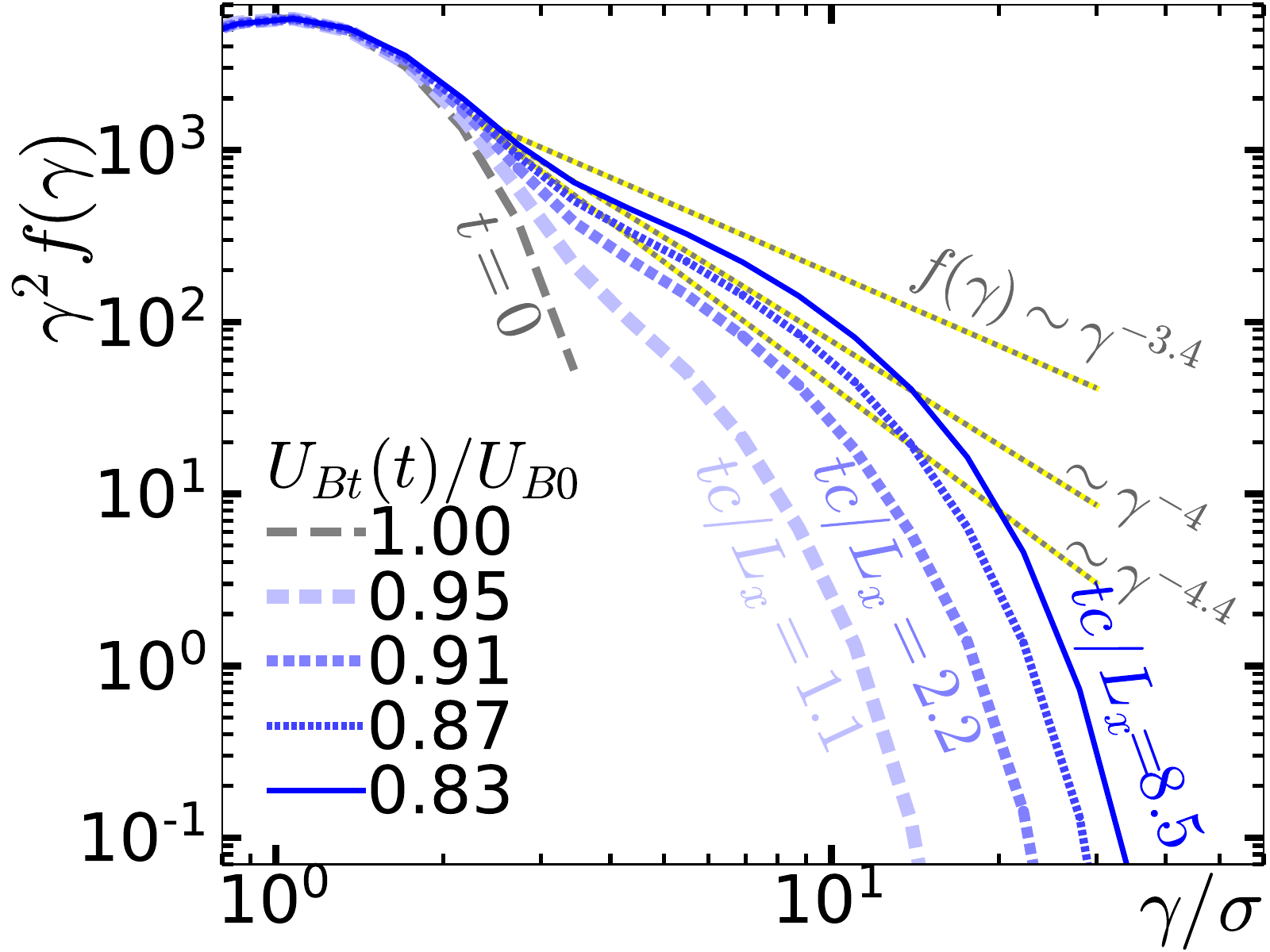}%
}
\caption{ \label{fig:ntpaVsTime}
Electron energy spectra $f(\gamma)$, compensated by $\gamma^2$, at different times for the 3D case $L_z/L_x=1$: the spectra are shown at times $tc/L_x=1.1$, 2.2, 4.3, and 8.6, when fractions 0.05, 0.09, 0.13, and 0.17 of the initial magnetic energy $U_{B0}$ have been depleted.
Power-law indices of 3.4, 4, and 4.4 are shown for reference, as in
Fig.~\ref{fig:ntpa2D3D}.
Even though reconnection is much slower after $4L_x/c$ than before, significant NTPA occurs between $tc/L_x=$4.3 and 8.6.
}
\end{figure}

NTPA continues to occur (in 3D) even during the second stage of reconnection, when the global rates of reconnection and magnetic energy conversion are smaller by an order of magnitude compared with the first stage.
Figure~\ref{fig:ntpaVsTime} shows the spectra for $L_z=L_x$ at four different times, $tc/L_x=1.1, 2.3, 4.3, 8.6$, chosen so that between successive times, $U_{\rm plasma}$ increases by equal amounts, namely $0.04 U_{B0}$---i.e., $\Delta U_{\rm plasma}/U_{B0} = 0.05, 0.09, 0.13, 0.17$.
Even after $t\simeq 3L_x/c$, when the magnetic energy is at 89~per~cent and after which the reconnection rate drops significantly, we see that $f(\gamma)$ continues to increase at high energies.
Indeed, the increase in $f(\gamma)$ preserves the power-law slope at high energies; the high-energy slope of $f(\gamma)$ does not decrease after $3L_x/c$, nor does the high-energy cutoff decrease (if anything, the opposite occurs, possibly indicating re-acceleration of already-energetic particles that have returned coincidentally to actively reconnecting regions).

It is possible that NTPA in 3D is at least as efficient as in 2D, despite the slower (global) reconnection rate in 3D, because in 3D the local reconnection rate is as high as in 2D within actively-reconnecting patches.
Thus NTPA may be occurring in a similar manner in 2D and in 3D, but in a relatively smaller region in 3D.
This result is consistent with \citet{Yin_etal-2008};
the global reconnection rate drops because there are fewer/smaller patches, not because local reconnection rates drop.

Furthermore, the stark contrast in plasmoid evolution in 2D and 3D might promote more efficient NTPA in 3D.
In 2D, ballooning plasmoids expand deeply into the upstream, and can contain higher magnetic field than the upstream region; the magnetic fields of large plasmoids can trap high-energy particles \citep[with Larmor radii a large fraction of the plasmoid size;][]{Sironi_etal-2016,Uzdensky-2020arxiv}.
In 3D, however, any structures (flux ropes) formed from reconnected magnetic field decay \citep[see also][]{Zhou_etal-2020};
high-energy particles may be confined to the layer by the upstream magnetic fields, but probably not by distinct sub-structures within the current layer.  
The lack of trapping may result in more particle
acceleration in 3D; whereas a particle trapped in a plasmoid does not get accelerated by reconnection \citep[but could be accelerated by a different, slower mechanism;][]{Petropoulou_Sironi-2018}, untrapped particles could potentially pass multiple times through reconnecting layers, thus experiencing a sort of multi-stage acceleration \citep{Dahlin_etal-2015,Zhang_etal-2021arxiv}.

It is worth estimating, from a Hillas criterion perspective \citep{Hillas-1984}, the
energy gain of a particle in a typical reconnection electric field
$E_{\rm rec}\sim 0.04B_0$, traversing an entire patch of active reconnection.  
By $t=2.3L_x/c$, there is a patch (region $A$ in Fig.~\ref{fig:layerXz3dn6144}) in the simulation with $L_z=L_x$ that is about 
$\ell\sim 100\sigma\rho_0$ in size.
Thus we estimate an energy gain $eE_{\rm rec}\ell\sim 4 \sigma m_e c^2$, which is roughly consistent with the high-energy cutoff
\citep[a similar result was given for elementary current sheets in 2D, for high-$\sigma_h$ reconnection;][]{Werner_etal-2016}.
We do not suggest this proves that all particles are being accelerated 
directly by the reconnection electric field; the acceleration mechanism
is beyond the scope of this paper and remains under debate \citep[see, e.g.,][]{Drake_etal-2006,Nalewajko_etal-2015,Ball_etal-2018,Ball_etal-2019,Guo_etal-2019,Sironi_Beloborodov-2020,Kilian_etal-2020,Guo_etal-2020arxiv,Uzdensky-2020arxiv,Zhang_etal-2021arxiv}, but at least for 2D and 3D reconnection with weak guide field, above-average-energy particles probably have gyroradii larger than the layer thickness and will experience more acceleration by the motional electric field of plasmoids, 
$\boldsymbol{E} = -(1/c)\boldsymbol{v}\times \boldsymbol{B}$
\citep{Guo_etal-2019,Kilian_etal-2020,Guo_etal-2020arxiv,Uzdensky-2020arxiv}.
We offer this estimate simply as a way to compare system size to
particle energies.

Thus we have seen both striking similarities and striking differences between 2D and 3D reconnection.  Both convert significant amounts of magnetic energy to plasma energy, but the conversion rate is slower in 3D.  Many of the signatures of reconnection at 2D X-points (or X-lines) are apparent in 3D simulations, but what happens in reconnection outflows is very different.  While 2D simulations beget a growing, highly-structured hierarchy of distinct plasmoids containing energized particles and reconnected magnetic field, 3D simulations only form small plasmoid-like features, which subsequently decay, allowing magnetic field energy in the layer to be converted to plasma energy.  As a result, 3D simulations ultimately convert more magnetic energy than in 2D.
The decay of plasmoids in 3D means that the reconnected flux in those plasmoids is annihilated, while in 2D, flux is almost perfectly conserved, such that all the lost upstream flux ends up as reconnected flux inside plasmoids.  In 3D, the lost upstream flux is mostly annihilated (and it is yet to be determined what fraction is directly annihilated and what fraction is first reconnected and later annihilated).
Interestingly, perhaps the strongest similarity between 2D and 3D reconnection is NTPA; the spectrum of high-energy particles is very similar in both cases, especially if compared at times when the depleted magnetic energy is the same.
With this overview finished, we will continue to examine differences between 2D and 3D reconnection in more depth.

\subsection{3D reconnection: stochastic variability}
\label{sec:variability3d}

An important difference between 2D and 3D current sheet simulations is the greater stochastic variability in 3D.
When initializing the simulations, the initial electric and magnetic fields are precisely determined.  However, the particle positions and velocities representing the plasma distribution in Monte Carlo fashion are chosen randomly (according to a precisely-given distribution).  Thus two ``identical'' simulations will not be exactly the same, and the initial microscopic randomness can lead to variability at macroscopic scales.

In 2D reconnection, two ``identical'' simulations will yield different plasmoid formation, motion, and merging; this can affect the global reconnection and energy conversion rates.
Smaller simulations often show greater stochastic variation in global quantities such as energy versus time, because there are fewer plasmoids overall, increasing the influence of erratic individual plasmoids.
We have often observed that the process of plasmoid merging (especially in simulations without initial perturbation; cf.~\S\ref{sec:pertAndEta2d}) results in two monstrously large plasmoids in a layer spaced apart by nearly $L_x/2$.  These two plasmoids will ultimately merge; but by symmetry due to periodic boundary conditions, one plasmoid can either move left or right to merge with the other.
This symmetry is always eventually broken to yield a lower-energy magnetic configuration, but its breaking can be ultimately traced to the initial microscopic randomness.
As these plasmoids decide which way to go for the final merger, the magnetic energy often stagnates at a roughly constant level for a short time before rapidly declining during the final plasmoid merger (see, e.g., Fig.~\ref{fig:energyFluxVsLx2d}, $L_x/\sigma\rho_0=640$, 1280, and 2560, around 3--5$L_x/c$).
However, due to randomness, sometimes the final two plasmoids are not created so symmetrically, and the magnetic energy decline continues smoothly.
Thus we often describe reconnection as ``bursty''; in radiative reconnection, where high-energy particles cool rapidly, e.g. via inverse Compton radiation, this burstiness is visible even in the spectrum of high-energy particles \citep{Werner_etal-2019}.

Thus initial random particle positions and velocities lead to macroscopic variability in 2D at system-size scales.
However, although significant, this stochastic variability is a sequence of random short detours on a path to essentially the same final state.  In fact, \S\ref{sec:pertAndEta2d} showed that varying the initial current sheet configuration (varying more than just different random velocities from the same distribution) still had little effect on the overall evolution of reconnection.
Looking at coarse time scales of $O(L_x/c)$, the overall rate and amount of energy depletion are fairly independent of initial conditions (for the same upstream conditions).

\begin{figure}
\centering
\fullplot{
\includegraphics*[width=0.325\textwidth]{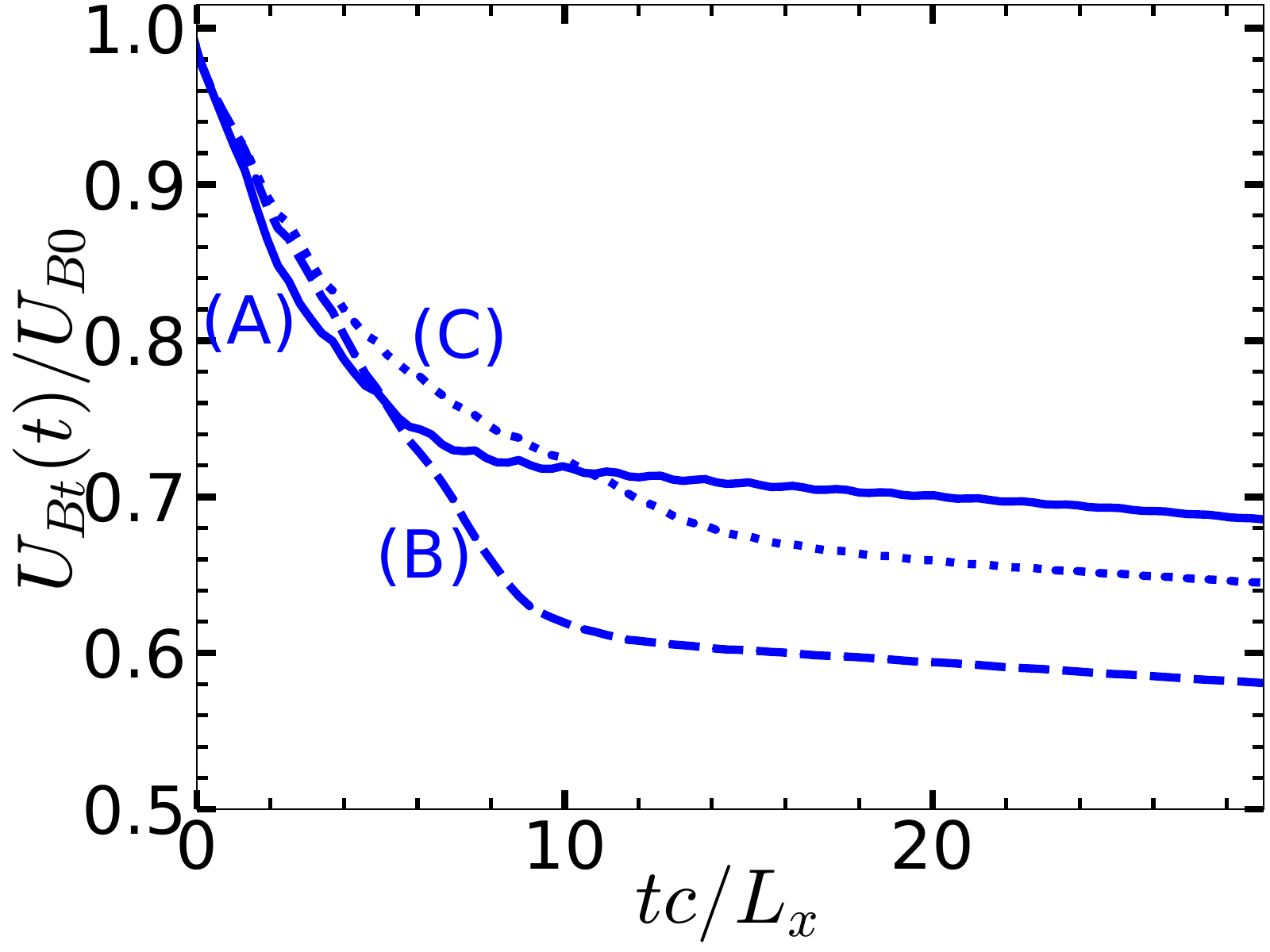}%
\hfill
\includegraphics*[width=0.325\textwidth]{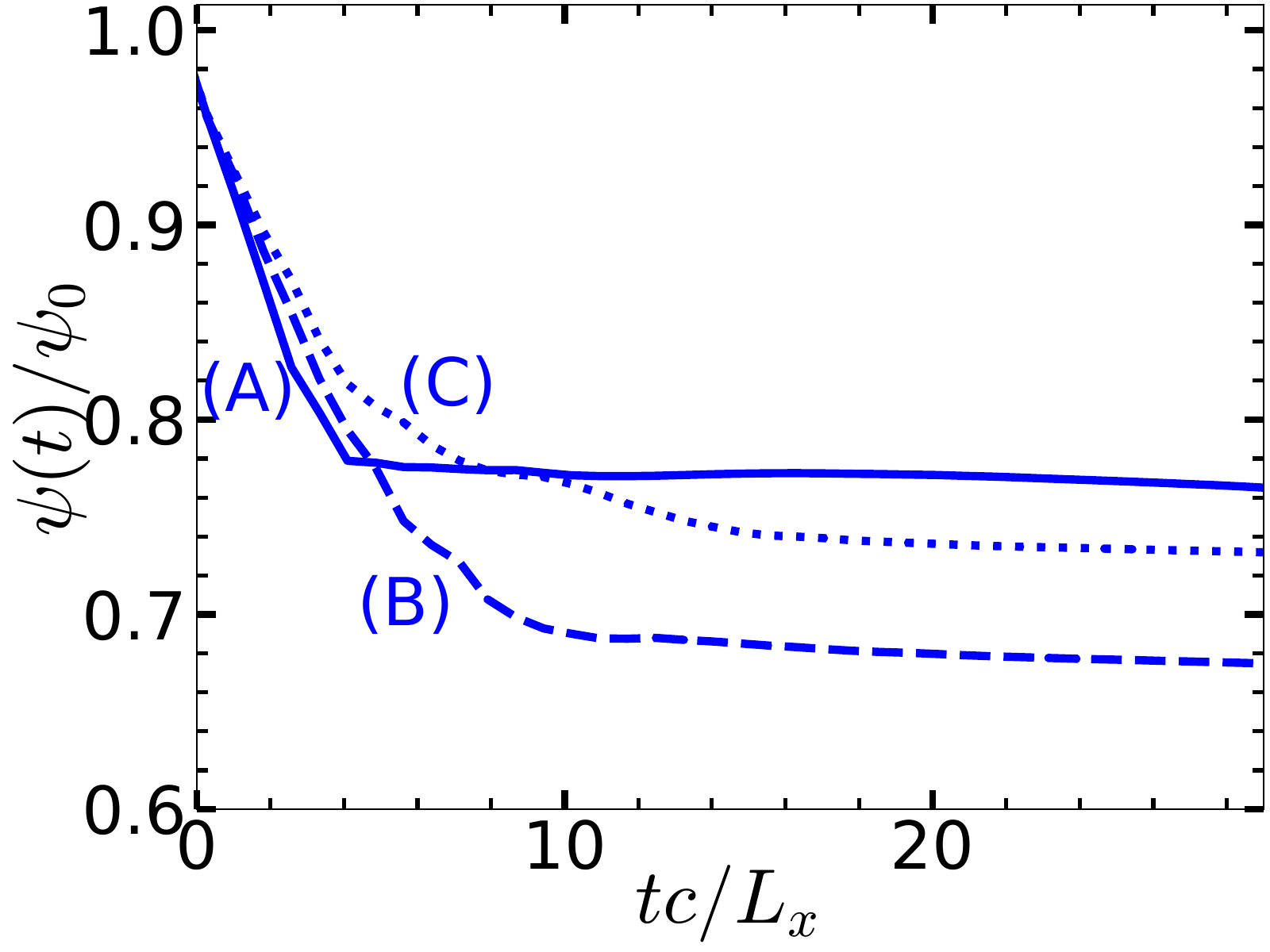}%
\hfill
\includegraphics*[width=0.325\textwidth]{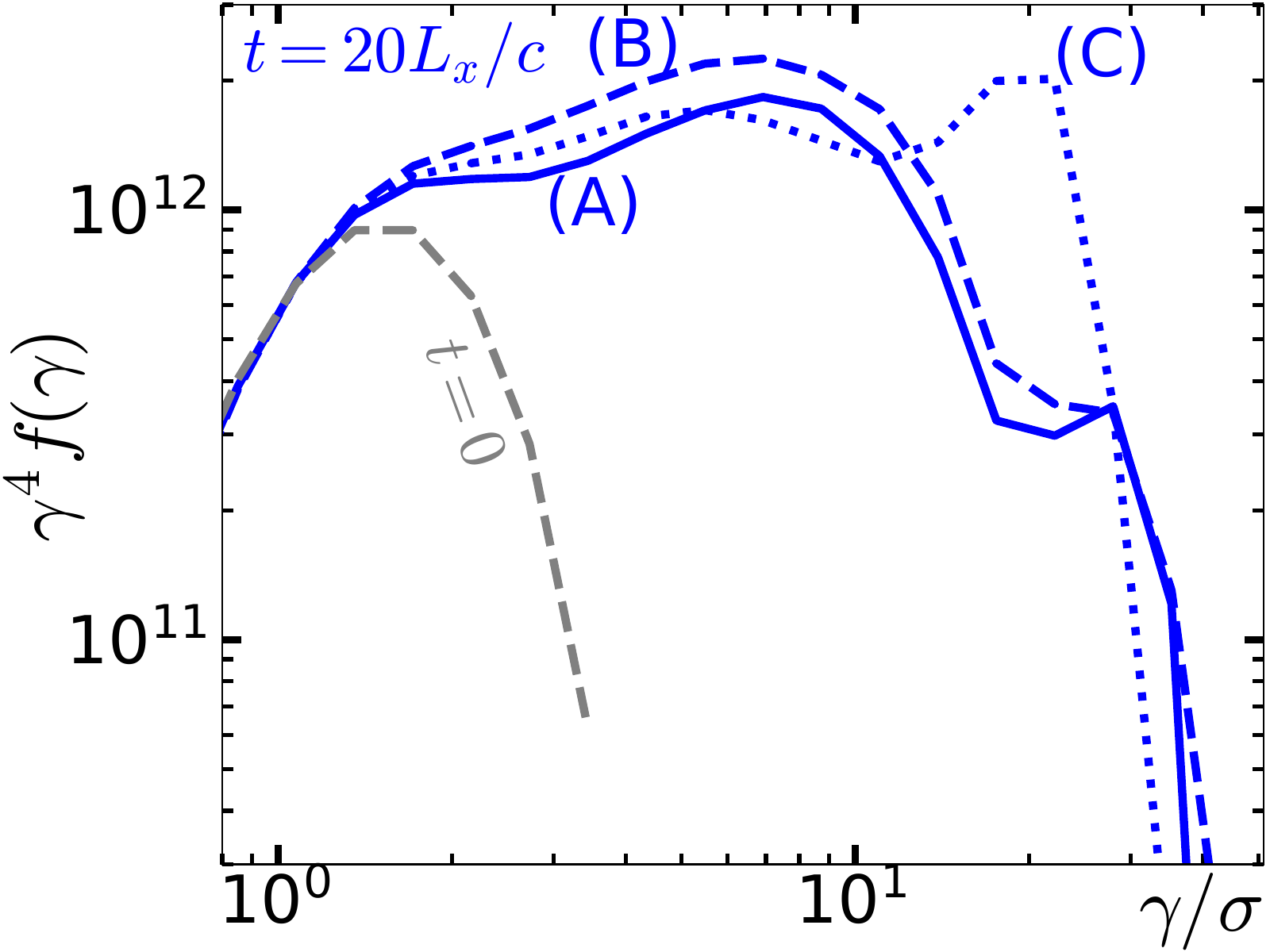}%
}
\caption{ \label{fig:variability3d}
3D reconnection  exhibits significant stochastic variability, as
shown by three simulations (A, B, and~C) with $L_x=256\sigma\rho_0$ and $L_z=L_x/4$, with
otherwise identical parameters, but
different initial random particle positions and velocities.
Panels show:
transverse magnetic energy (left) and unreconnected flux (middle) versus time, and
(right) particle energy spectra at $20L_x/c$, compensated by 
$\gamma^4$.
}
\end{figure}

In contrast, in a 3D reconnection simulation, random initialization can lead to significant macroscopic differences that persist for long times---at least as long as we have been able to simulate.
While a detailed study of random variation in large ensembles of ``identical'' simulations is beyond the scope of this paper, we present here an anecdote of high variability in 3D in a case with $L_x=256 \sigma\rho_0$ and $L_z=L_x/4$, which we ran three times with ``identical'' set-ups ($B_{gz}=0$, $\eta=5$, $\beta_d=0.3$, $a=0$).
The three simulations ultimately exhibit very different magnetic energy and flux evolution (Fig.~\ref{fig:variability3d}) even over scales of tens of $L_x/c$; this
variation is much greater than we see in 2D (although not perfectly comparable, Figs.~\ref{fig:perturb2dLx320}, \ref{fig:eta2dLx1280}, and \ref{fig:energyFluxVsLx2d} place a rough upper bound on variability in 2D;
also see Fig.~\ref{fig:energyAndFluxVsPert3d}).
Nevertheless, despite these differences, the particle energy spectra are similar despite different energy evolution (Fig.~\ref{fig:variability3d}, right), with the exception of one case exhibiting a bump above $\gamma \gtrsim 10\sigma$.
At early times, the evolutions of magnetic energy and flux (Fig.~\ref{fig:variability3d}, left and middle) are reasonably similar, but they diverge after $\simeq 5L_x/c$.
Simulation (A) starts off with the highest magnetic conversion rate but ends up converting the least amount (30~per~cent) of magnetic energy after $30L_x/c$.
Simulation (B) rapidly converts about 40~per~cent of the magnetic energy within $10L_x/c$, and then continues to convert energy at a much slower rate for at least the next $20L_x/c$.
And simulation (C) exhibits a medium-slow stage of energy conversion between~5 and~15$L_x/c$, followed by a stagnant or very slow final stage similar to the others.

One mechanism that can result in large random variation in 3D is the following.  We will show in~\S\ref{sec:RDKIamp} that in some cases RDKI causes the layer to kink to very large amplitudes (comparable to the kink wavelength).
At some amplitude, the instability becomes highly non-linear, resulting in severe distortion of the current sheet, somewhat resembling breaking waves; when this happens, magnetic energy is rapidly depleted, leaving a much-thickened layer that is much more stable against tearing and RDKI.
This process rapidly transforms the layer, putting the simulation in a very different state.
Some initial configurations (e.g, with higher $\beta_d$, as we will see in~\S\ref{sec:RDKIamp}) seem to make this transformation more likely, but there appears to be a large range of conditions under which it is possible but not inevitable.
Thus two ``identical'' simulations, one that suffers this transformation and one that does not, may evolve very differently over long times (e.g., $50L_x/c$).

In light of this, we try to be careful in this study about not drawing conclusions from small differences between individual 3D simulations. 

\subsection{3D reconnection: dependence on initial current sheet configuration}
\label{sec:pertAndEta3d}

We showed in~\S\ref{sec:pertAndEta2d} that an initial magnetic perturbation barely affects energy conversion rates and NTPA in 2D reconnection, despite some differences in the evolution of the plasmoid hierarchy \citep[but see][and our comparison to it at the end of~\S\ref{sec:pertAndEta2d}]{Ball_etal-2018}.
Similarly, other parameters describing the initial current sheet can vary over a wide range without much affecting 2D reconnection.
Here, we explore a range of different current sheet configurations, as in~\S\ref{sec:pertAndEta2d}, but now in 3D, all the while keeping the upstream/background plasma the same.  
In the substudies (3D-a,b,c) below, we will vary the same parameters as in~\S\ref{sec:pertAndEta2d}, (2D-a,b,c).
However, the range of varied parameters will be much more limited because of the increased cost of simulation in 3D; for instance, we examine only two different perturbation strengths in (3D-a) and two different values of $\eta$ in (3D-b).
As in~\S\ref{sec:pertAndEta2d}, we set $B_{gz}=0$.
We will see that the initial current sheet has more long-term influence in 3D than in 2D.

\begin{figure}
\centering
\fullplot{
\includegraphics*[width=0.325\textwidth]{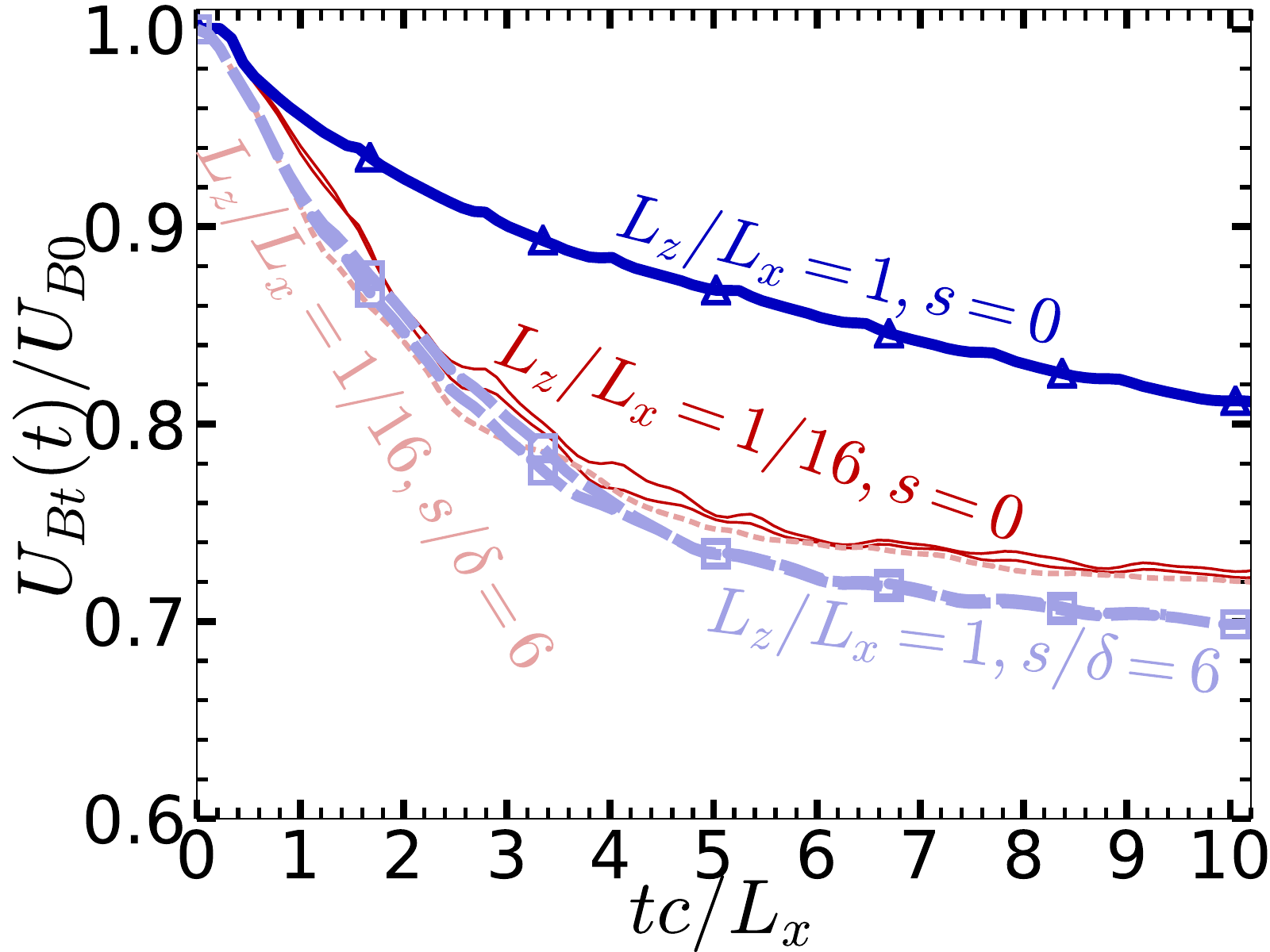}%
\hspace{0.05in}%
\includegraphics*[width=0.325\textwidth]{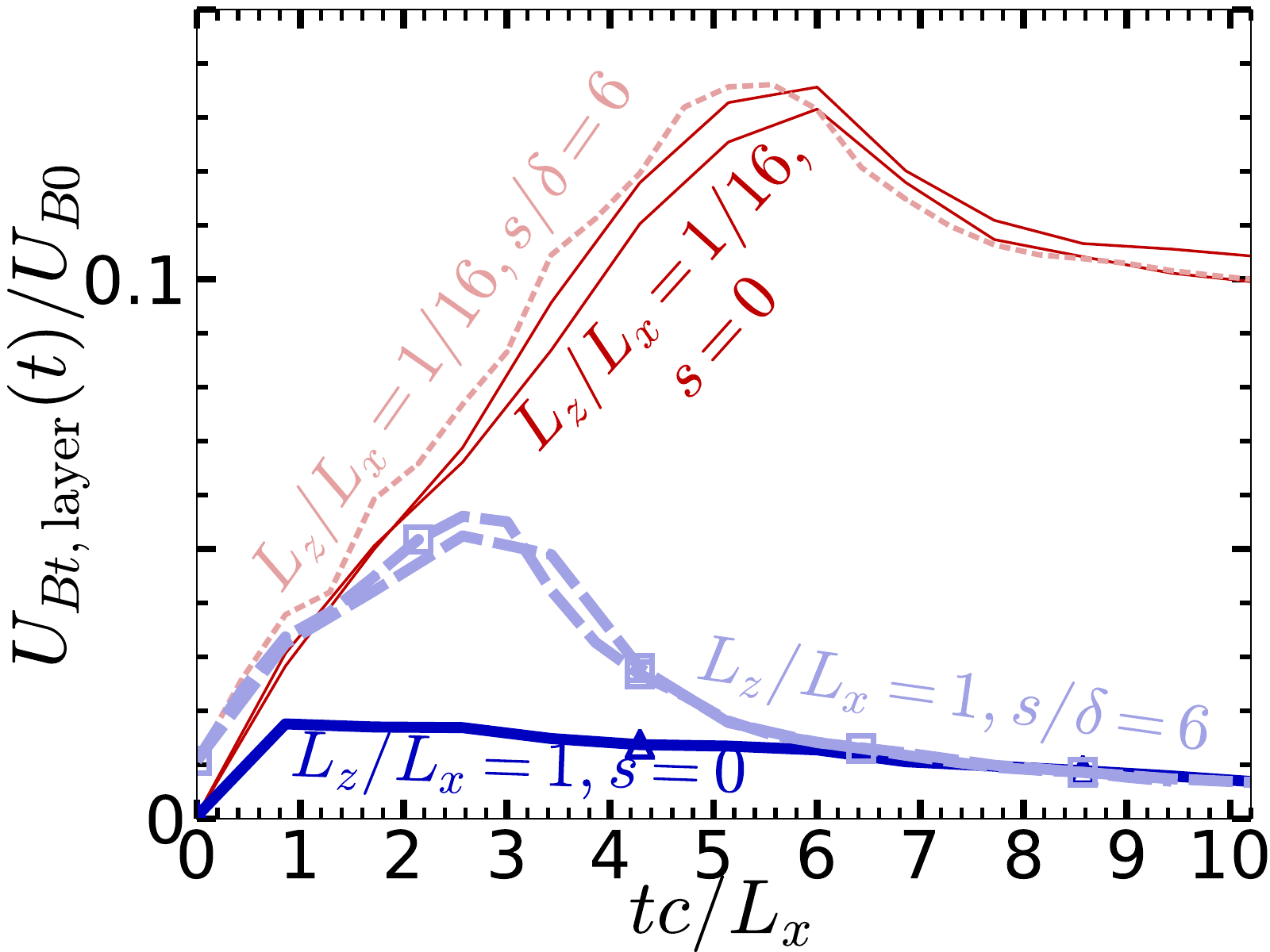}%
\hspace{0.05in}%
\includegraphics*[width=0.325\textwidth]{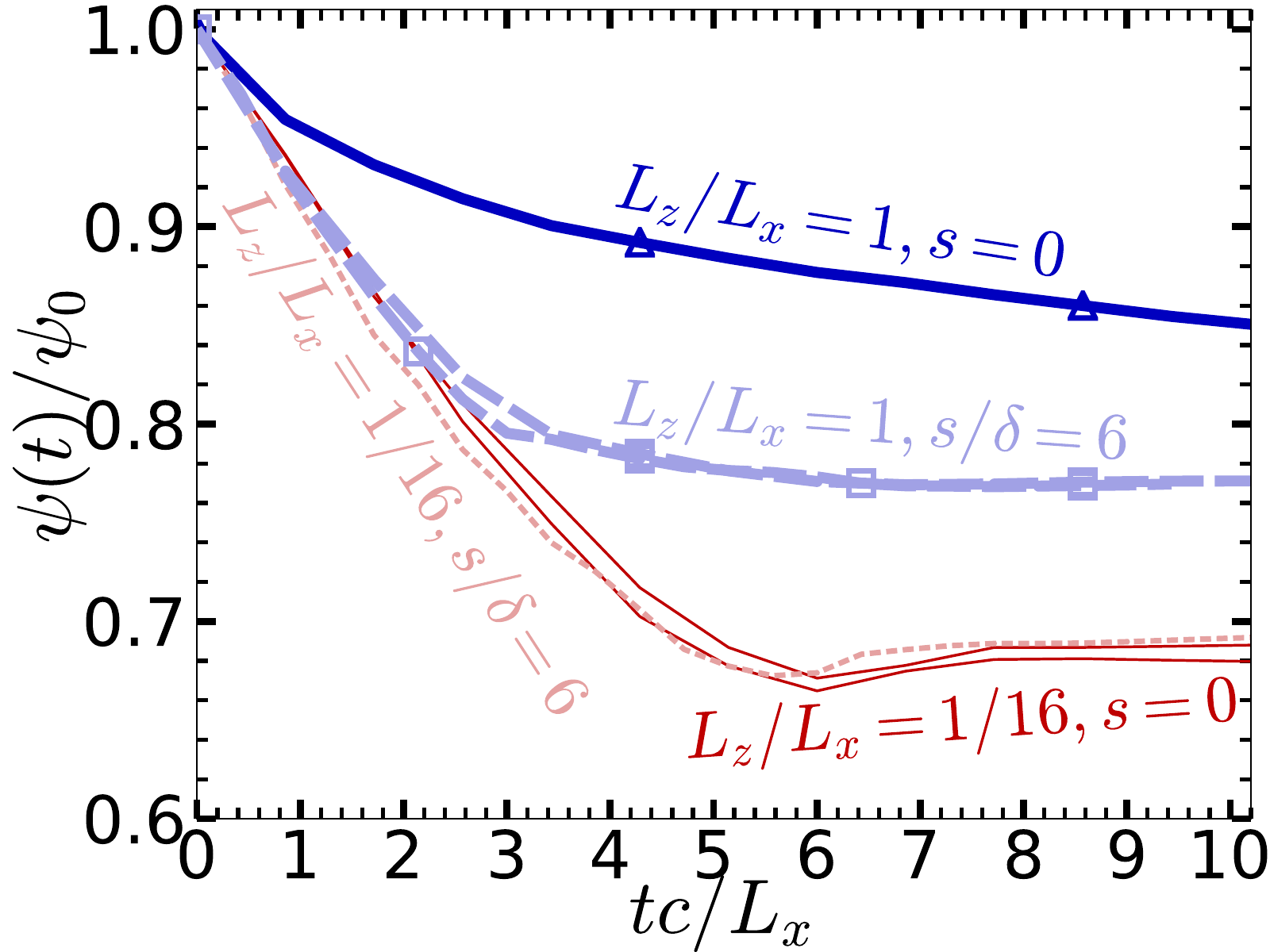}%
}
\caption{ \label{fig:energyAndFluxVsPert3d}
In 3D a perturbation (which is uniform in $z$) makes simulations behave more like 2D simulations, especially at early times.
These plots show energies and flux versus time for four simulation configurations: 
with perturbation ($s/\delta=6$, dashed lines) and without ($s=0$, solid), for 3D ($L_z=L_x$, blue) and nearly 2D ($L_z/L_x=1/16$, red) simulations.
All have $L_x=341\sigma\rho_0$, $\eta=5$, $\beta_d=0.3$, $B_{gz}=0$.
For two of the configurations---$L_z/L_x=1$, $s=0$ and $L_z=1/16,s/\delta=6$---two runs are shown (identical except for initially randomized particles).
The transverse magnetic energy, $U_{Bt}(t)/U_{B0}$ (left), is very similar for the 3D simulation with perturbation and both 2D simulations, but the 3D simulation without perturbation converts magnetic energy much more slowly.
The magnetic energy in the layer, $U_{Bt,\rm layer}(t)/U_{B0}$ (middle), however, shows that the 3D case with perturbation initially resembles the 2D simulations, but after~$3L_x/c$ depletes the ``reconnected'' magnetic field energy in the layer.
Similarly, the unreconnected magnetic flux $\psi(t)/\psi_0$ (right) shows that the 3D case with perturbation lies between the 2D cases and the 3D case without perturbation.
}
\end{figure}

\emph{(3D-a) Varying the initial perturbation.} 
In this substudy we will see that, in 3D, 
an initial magnetic field perturbation (uniform in $z$) significantly alters the subsequent evolution---unlike in 2D, where the initial perturbation is relatively unimportant, as shown in~\S\ref{sec:pertAndEta2d}(2D-a).
Perhaps the easiest way to describe the effect of a perturbation in 3D is that (in simulations with no guide magnetic field) the perturbation tends to make reconnection more like 2D reconnection.
Figure~\ref{fig:energyAndFluxVsPert3d} shows energies and
flux versus time, for perturbed and unperturbed initial current sheets, in a fully 
3D ($L_z=L_x$) simulation and---for comparison---in a nearly 2D ($L_z=L_x/16$) simulation---all with $L_x=341\sigma\rho_0$, $\eta=5$, $\beta_d=0.3$, and $B_{gz}=0$.
We consider only one non-zero perturbation strength, $a=12$, for which the initial magnetic separatrix height extends to about $s=6\delta = 4\sigma\rho_0\approx L_x/85$, beyond the current sheet but still very far from $L_y/4$.
To give an idea of stochastic variability, we show two simulation runs for the nearly 2D case without perturbation, and two for the 3D case with perturbation.
As expected from the 2D simulations in~\S\ref{sec:pertAndEta2d}(2D-a), the nearly 2D cases are almost identical in all three panels---the total transverse magnetic energy $U_{Bt}(t)$, the magnetic energy in the layer $U_{Bt,\rm layer}(t)$, and the upstream flux $\psi(t)$; the perturbation has no significant effect on energy or flux evolution in 2D.
In contrast, the 3D cases with $s=0$ and $s/\delta=6$ differ markedly; the $s=0$ case initially converts magnetic energy more slowly, with a correspondingly slower decline in unreconnected flux.
Interestingly, the early energy evolution in 3D
\emph{with} perturbation ($s/\delta=6$)
is very similar to the nearly 2D cases (\emph{with} and \emph{without} perturbation).
After$\:\sim 3L_x/c$, the initially-perturbed 3D simulation starts to exhibit slower energy conversion (compared with earlier times or with the 2D case at times up to $t\lesssim 6L_x/c$), as shown in Fig.~\ref{fig:energyAndFluxVsPert3d}(left).
Whereas the \emph{unperturbed} 3D case never has much magnetic energy $U_{Bt,\rm layer}$ in the layer,
the \emph{perturbed} 3D simulation builds up energy in $U_{Bt,\rm layer}$ until~$3L_x/c$, just as in 2D (Fig.~\ref{fig:energyAndFluxVsPert3d}, middle).
After~$3L_x/c$, the perturbed 3D case continues to deplete magnetic energy by converting \emph{upstream} magnetic field energy to plasma energy and also by converting~$U_{Bt,\rm layer}$ to plasma energy.
The amount of unreconnected flux $\psi(t)$ in the perturbed 3D case also follows the 2D cases up until~$3L_x/c$; after that the decline of $\psi$ slows substantially in 3D but continues apace in 2D for another $~3L_x/c$.
Although not shown here, the total magnetic energy continues to fall after~$10L_x/c$ in both 3D cases; for example it falls to~67~per~cent of its initial value after~$25L_x/c$ for the perturbed simulation and after~$50L_x/c$ for the unperturbed 3D case (and continues to drop further). 
The uniform-in-$z$ initial perturbation thus appears to kick-start reconnection in 2D mode, creating long flux ropes that, being more uniform in $z$, are more stable;  eventually 3D effects appear, however, and the plasmoid structures decay along with their associated magnetic fields.

\begin{figure}
\centering
\fullplot{
\includegraphics*[width=0.49\textwidth]{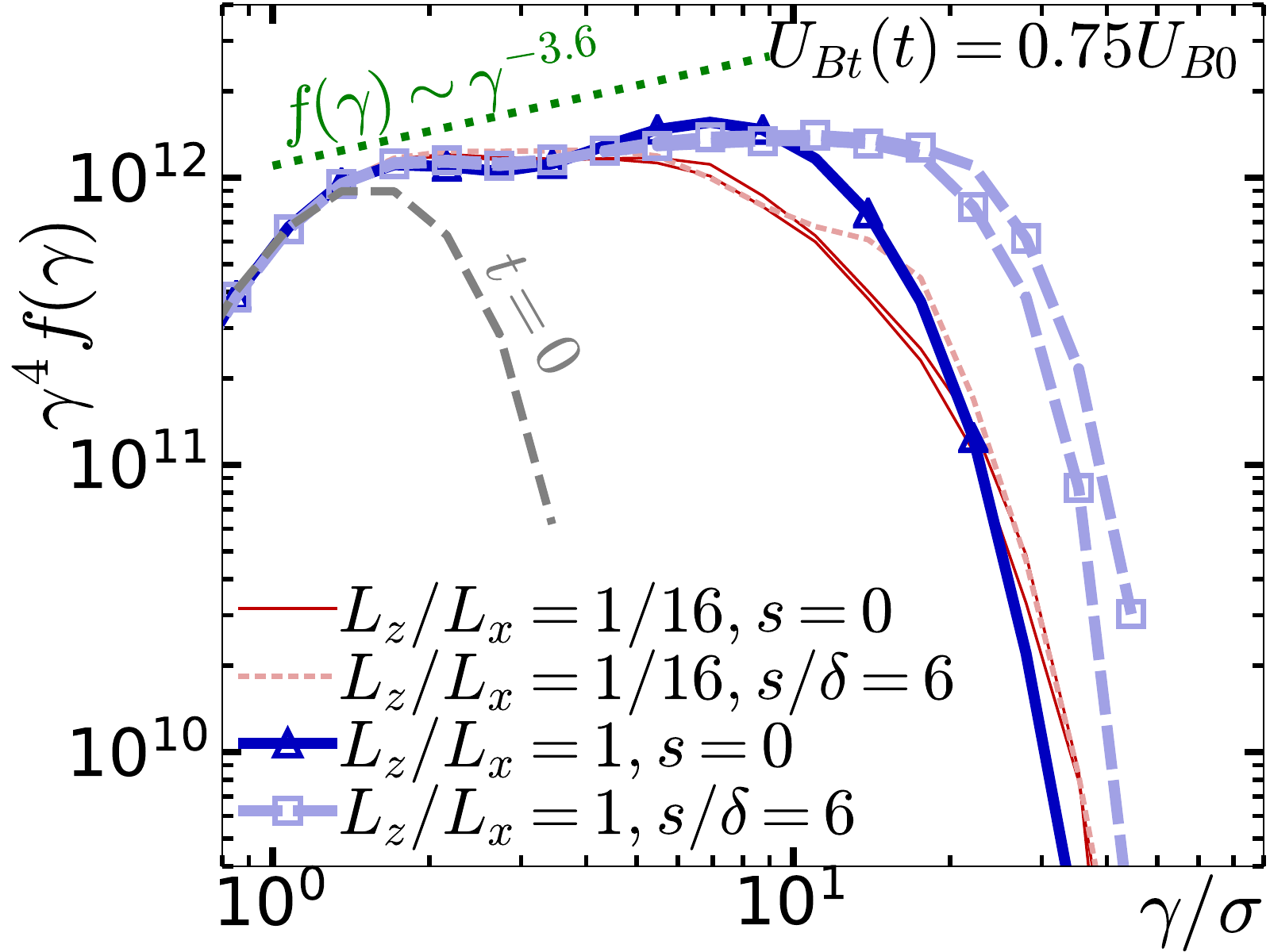}%
\hfill
\includegraphics*[width=0.49\textwidth]{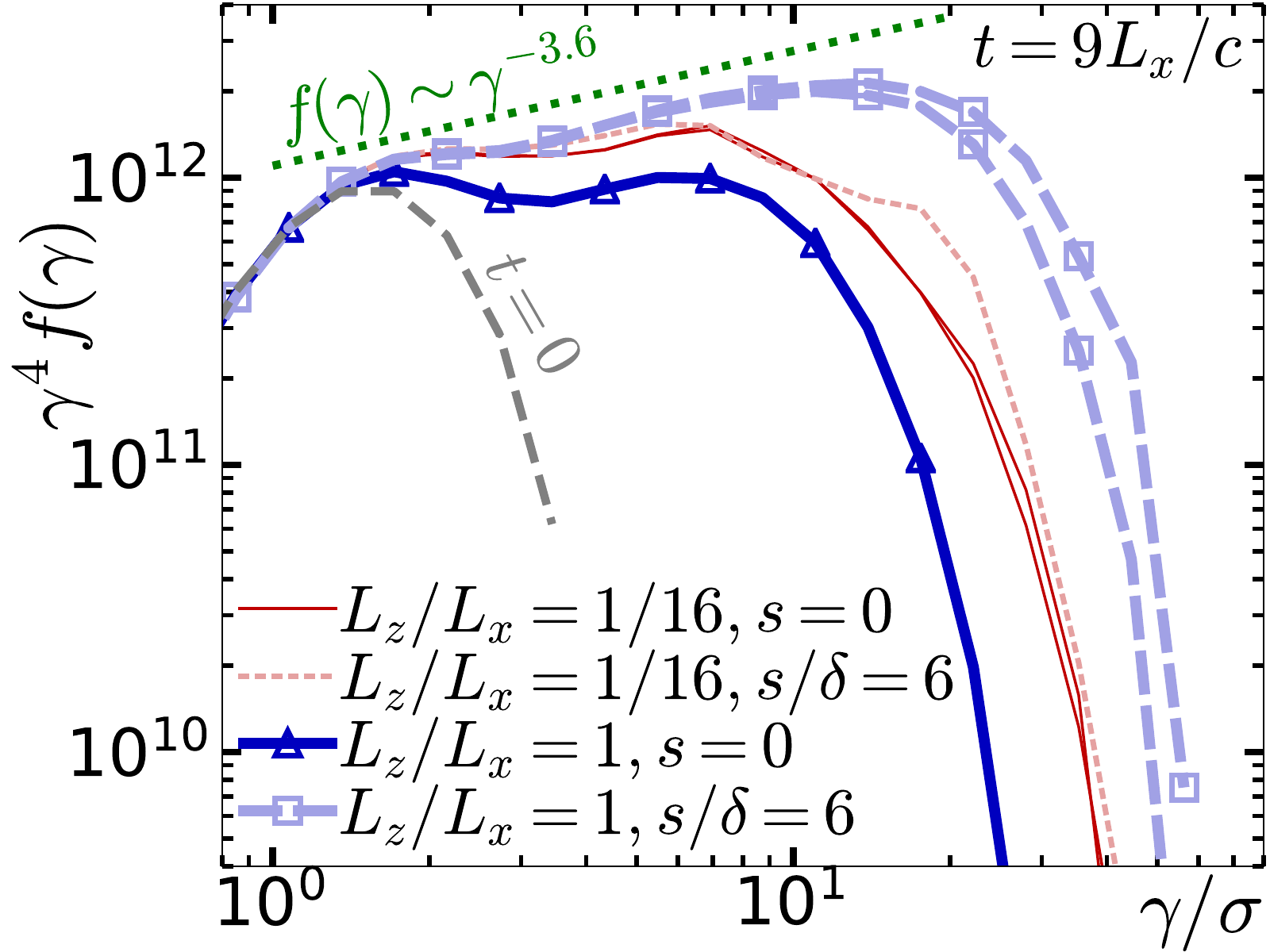}%
}
\caption{ \label{fig:ntpaVsPert3d}
Electron energy spectra (compensated by $\gamma^4$) 
for the same (perturbed/unperturbed, 2D/3D) simulations as in Fig.~\ref{fig:energyAndFluxVsPert3d}, 
are all fairly similar at the time
when~$U_{Bt}/U_{B0}=0.75$ (left)---the differences at high energy would hardly be remarkable were it not for the $\gamma^4$.
Spectra are also shown (right)
at $t= 9L_x/c$, at which time simulations that have converted more magnetic field energy have more high-energy particles.  
For comparison, the grey/dashed line shows the initial spectrum at $t=0$ in all simulations;
and, a power-law segment with index $p=3.6$ (green/dotted) demonstrates a ten~per~cent (i.e., small) difference from $p=4$ (which would be horizontal).
We note, since Fig.~\ref{fig:energyAndFluxVsPert3d} shows only the first $10L_x/c$, that the 3D case with $s=0$ does not reach $U_{Bt}(t)=0.75U_{B0}$ until
about $t=20L_x/c$, while all the other simulations convert $0.25U_{B0}$ in about $5L_x/c$.
}
\end{figure}

The electron energy spectra are fairly similar for all cases,
as shown in Fig.~\ref{fig:ntpaVsPert3d}, especially when comparing spectra at the same amount of total energy depleted ($U_{Bt}=0.75U_{B0}$).
The 3D simulations have perhaps slightly more efficient NTPA than the nearly 2D simulations (e.g., they have more particles with $\gamma \approx 10\sigma$), but this enhancement might be attributable to stochastic variation. 
The high-energy spectrum for the unperturbed 3D simulation grows more
slowly (than either the perturbed 3D simulation or 2D simulations), consistent with the slower energy conversion, and this is reflected in less NTPA at~$t=9L_x/c$ relative to the other cases.
Similarly, the perturbed 3D simulations have converted more magnetic energy to plasma energy at $t=9L_x/c$ than the other cases, and they correspondingly show more particles at high energies.
Differences in the high-energy spectra are magnified by plotting $\gamma^4f(\gamma)$; although the perturbed 3D cases appear to show significantly more NTPA than the 2D or unperturbed 3D cases, the difference is less than a 10~per~cent change in power-law index $p\approx 4$.

To avoid the early similarities between 2D and 3D with an initial perturbation, simulations discussed in the rest of~\S\ref{sec:3d} will have zero initial perturbation.

\emph{(3D-b) Varying current sheet density $\eta$ with fixed $\beta_d=0.3$; i.e., varying $\delta/\sigma\rho_0$ for fixed $\delta/\rho_d=2.1$}.  
Varying the overdensity $\eta=n_{d0}/n_{b0}$, and fixing $\beta_d=0.3$ while maintaining the Harris equilibrium, 
was seen to have relatively little effect on 2D reconnection in large systems, as long as $\eta \gg 0.1$ (see~\S\ref{sec:pertAndEta2d})---with the notable exception that decreasing $\eta$ (increasing $\delta/\sigma\rho_0$) merely delayed reconnection onset.
As noted in~\S\ref{sec:pertAndEta2d} (also see table~\ref{tab:etaEffect}), the initial gyroradius $\rho_d$ and skin depth $d_{ed}$ of the current sheet plasma, as well as its half-thickness $\delta$, all vary as $\eta^{-1}$ when the upstream $n_{b0}$ and $\theta_{b}$ are kept constant (hence associated time scales also vary as $\eta^{-1}$, explaining the onset delay).
As $\eta$ varies, the thickness $\delta$ thus varies relative to upstream plasma scales, but not relative to the drifting plasma scales.

\begin{figure}
\centering
\fullplot{
\includegraphics*[width=0.325\textwidth]{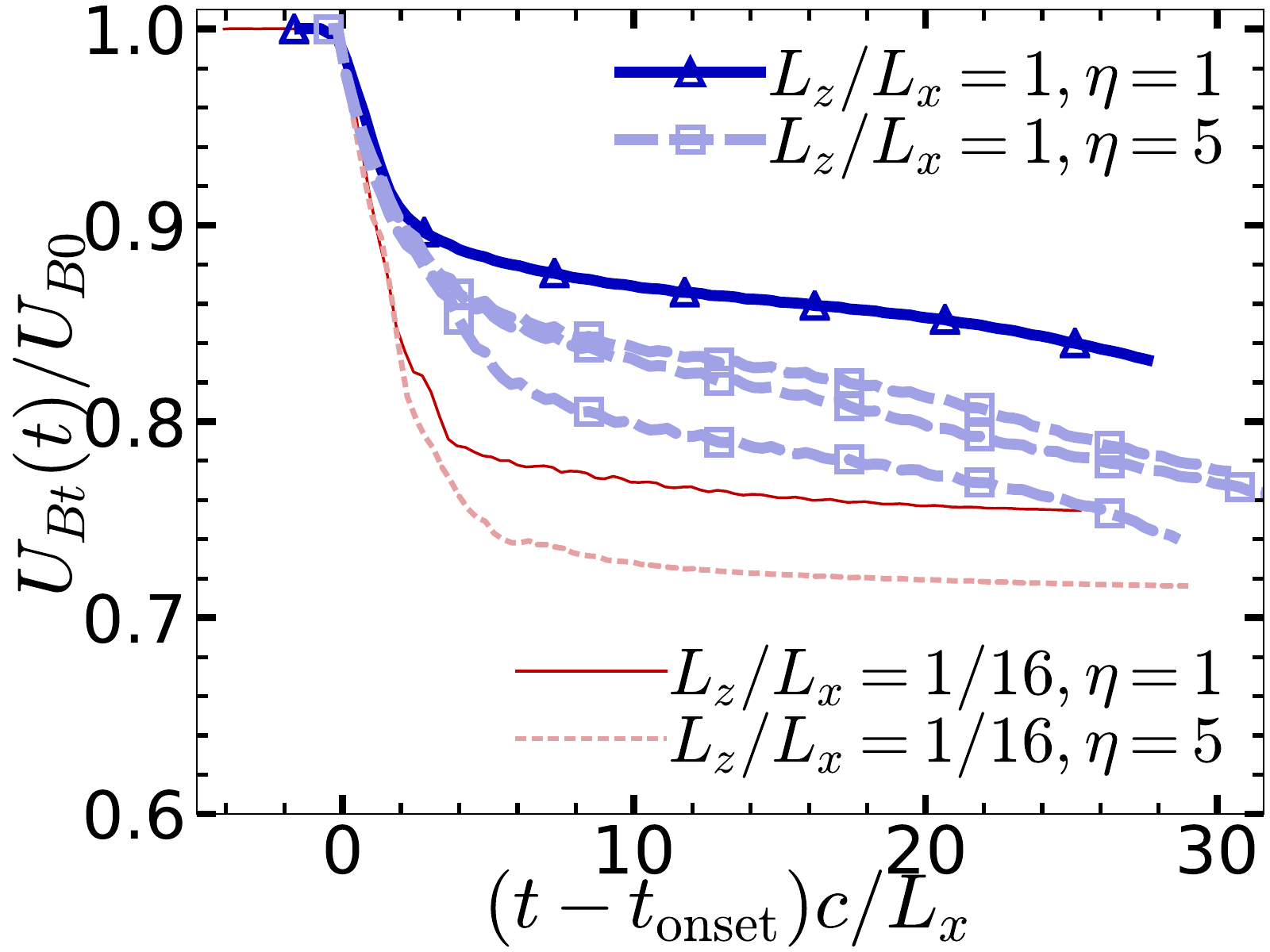}%
\hspace{0.05in}%
\includegraphics*[width=0.325\textwidth]{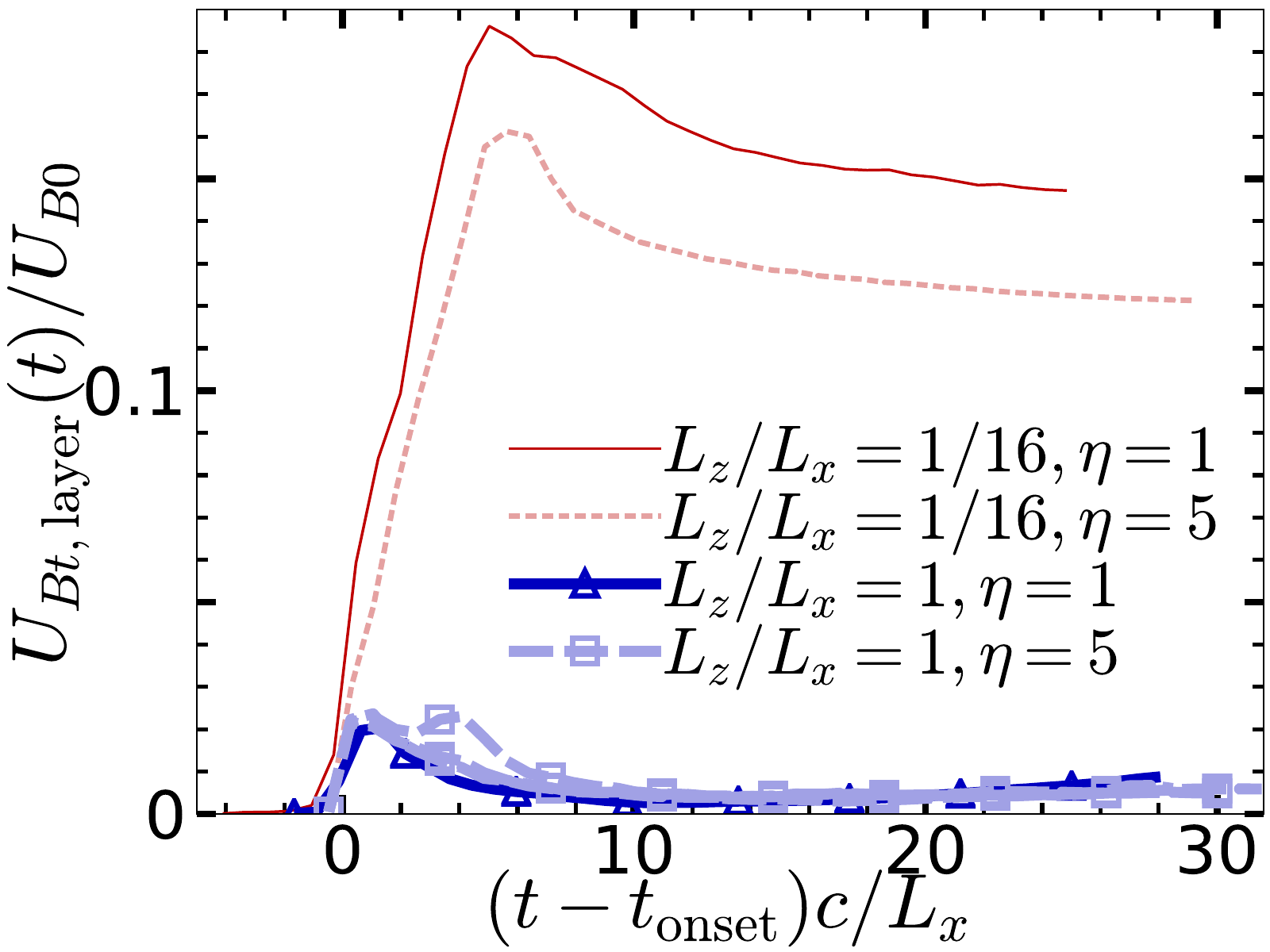}%
\hspace{0.05in}%
\includegraphics*[width=0.325\textwidth]{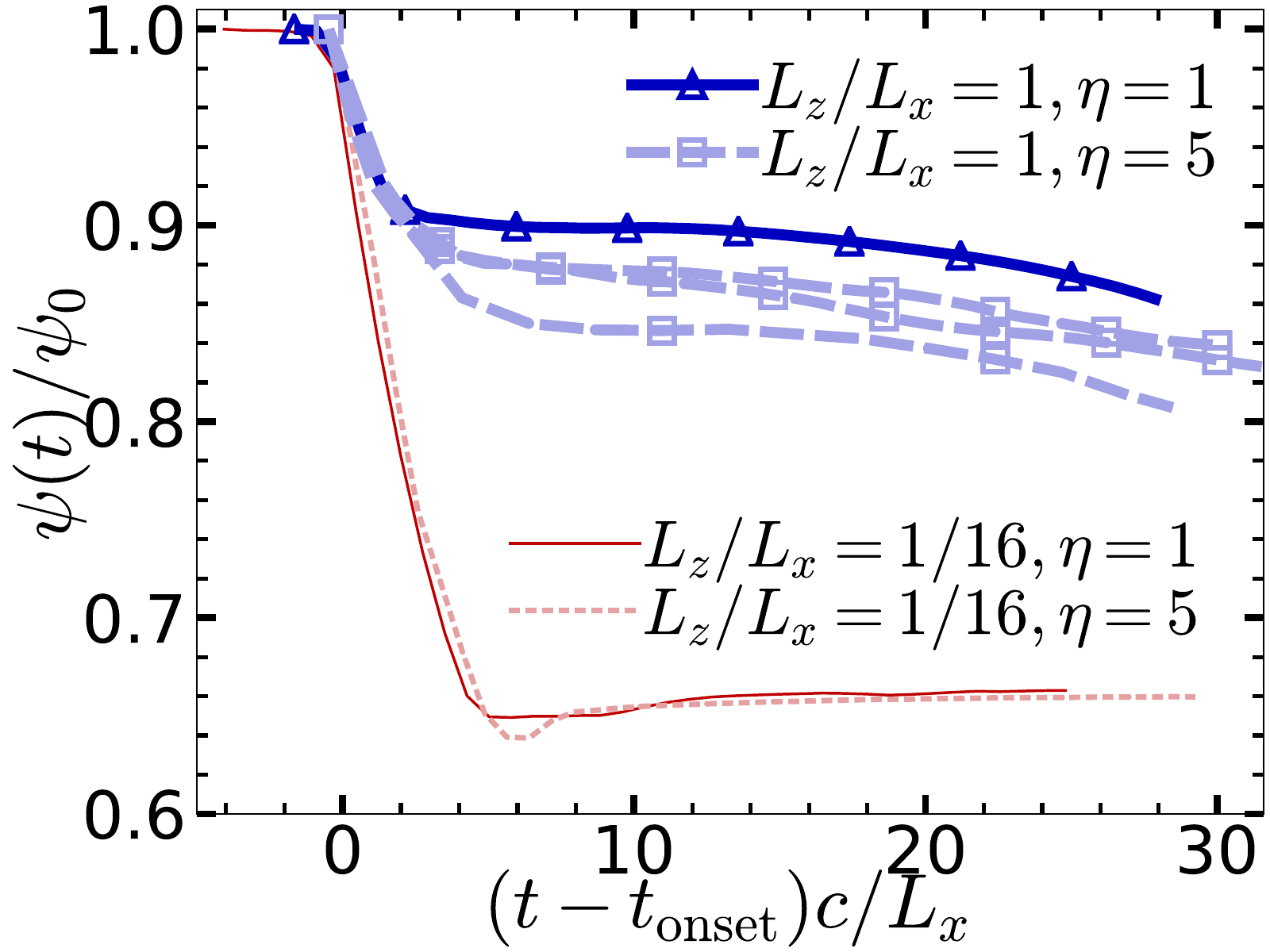}%
}
\caption{ \label{fig:energyAndFluxVsEta3d}
As in 2D, varying overdensity $\eta$ in 3D does not dramatically affect the overall energy and flux evolution (differences shown here are attributable to finite system size).
These plots show energies and flux versus time for four simulation configurations: 
$\eta=1$ (solid lines) and $\eta=5$ (dashed),  
3D ($L_z/L_x=1$, blue) and nearly 2D ($L_z/L_x=1/16$, red).
All have $L_x=256\sigma\rho_0$, $\beta_d=0.3$, $B_{gz}=0$, and $a=0$.
Three different runs are shown for the 3D, $\eta=5$ case.
The magnetic energy evolution $U_{Bt}(t)$ (left) shows slower energy conversion in 3D than in 2D as expected; but the difference between $\eta=1$ and $\eta=5$ in 3D is comparable to the difference in 2D (also it is comparable to the stochastic variation among the three 3D, $\eta=5$ cases).
The difference in 2D vanishes in larger systems (see~\S\ref{sec:pertAndEta2d}, 2D-b).
The magnetic energy in the layer, $U_{Bt,\rm layer}(t)$ (middle), evolves similarly in all the 3D cases, very unlike the 2D cases, which store significant energy in ``reconnected'' field. 
Similarly, the unreconnected flux $\psi(t)$ (right) is similar for all 3D cases, which differ greatly from the nearly identical 2D cases.
}
\end{figure}

\begin{figure}
\centering
\fullplot{
\includegraphics*[width=0.49\textwidth]{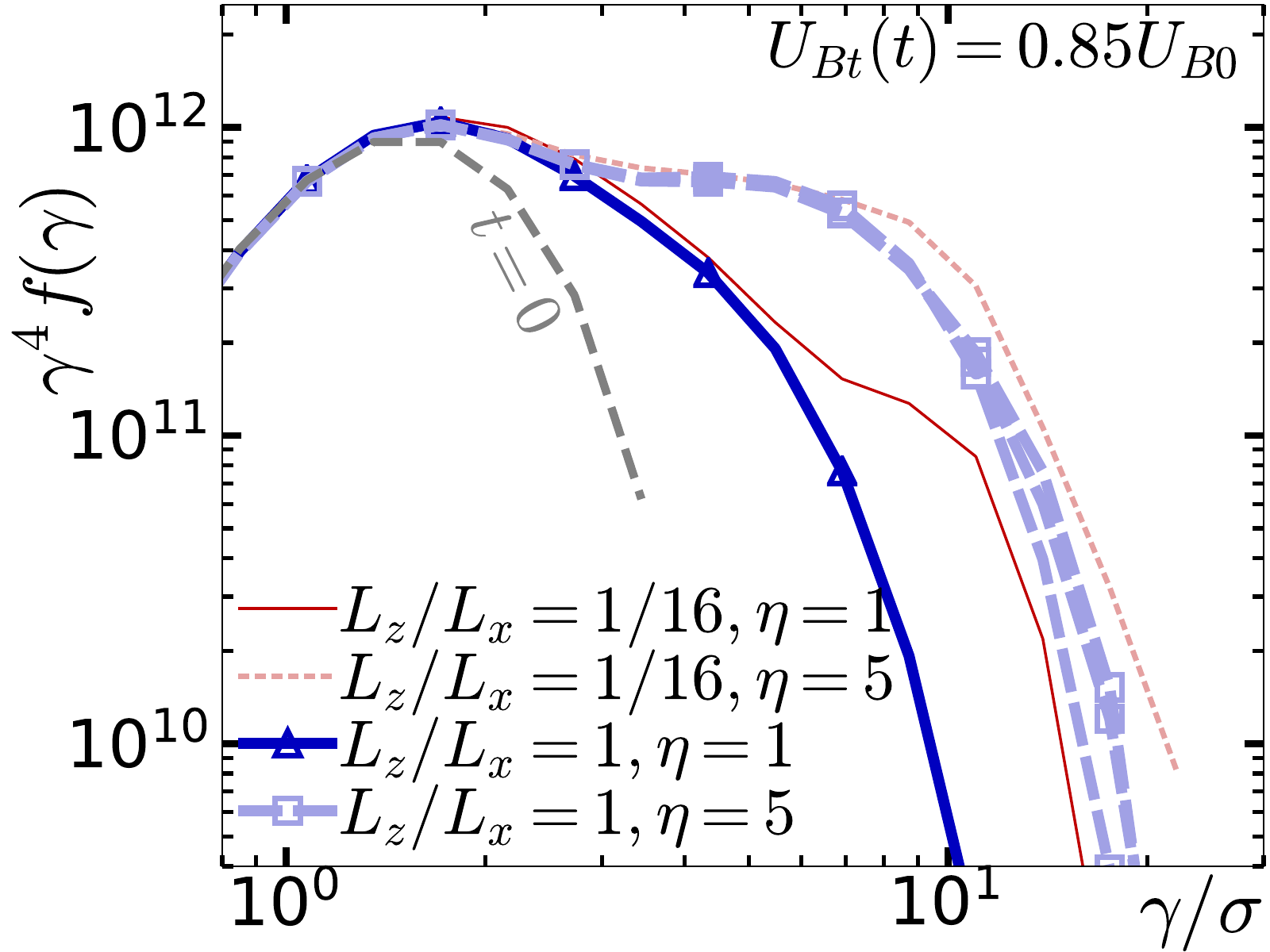}%
\hfill
\includegraphics*[width=0.49\textwidth]{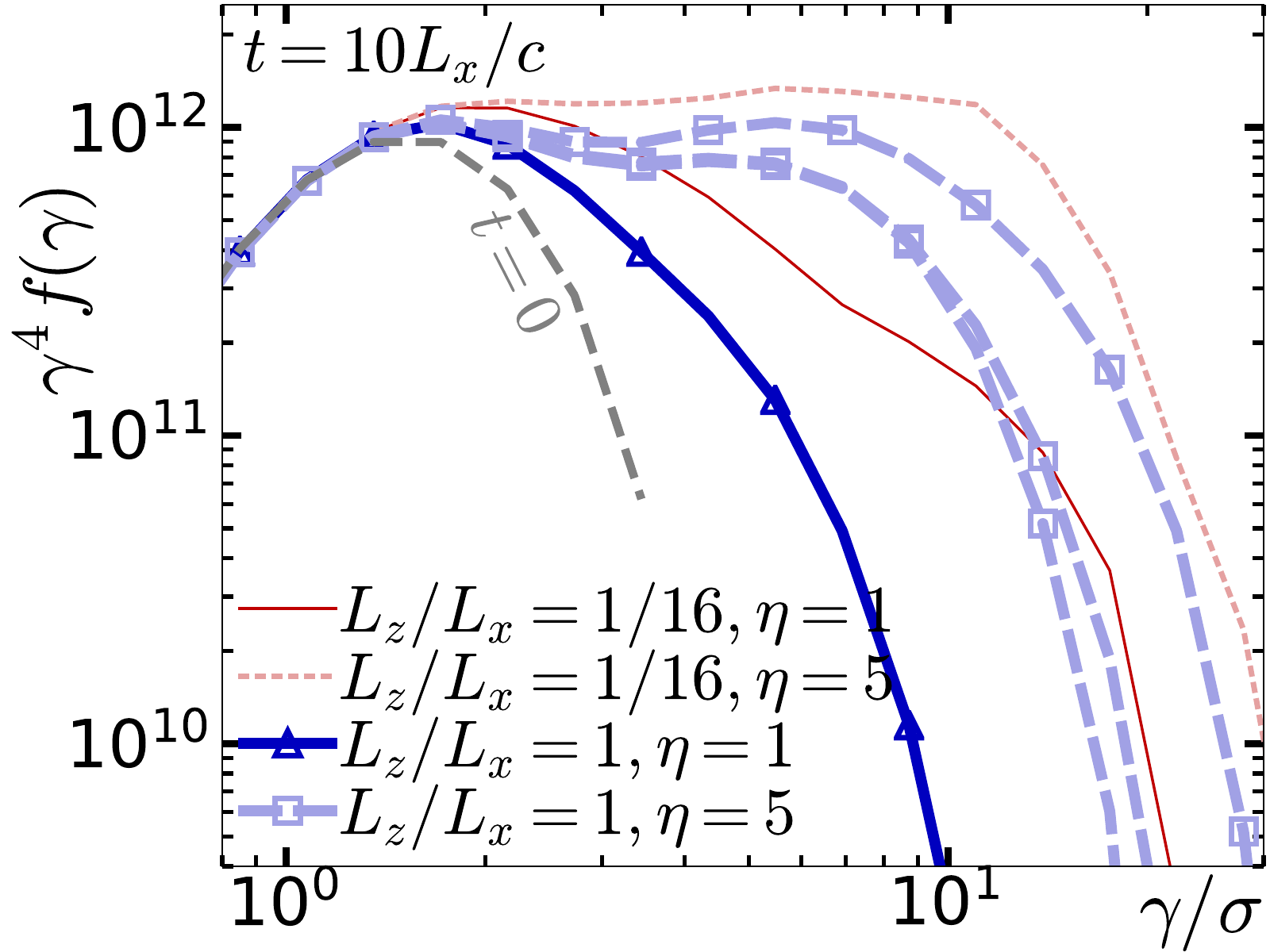}%
}
\caption{ \label{fig:ntpaVsEta3d}
Varying overdensity $\eta$ does not have any more effect on NTPA in 3D than it has in 2D; and in 2D, the effect vanishes for larger system sizes (see~\S\ref{sec:pertAndEta2d}, 2D-b).
These plots show the electron energy spectra (compensated by $\gamma^4$) 
for the same simulations ($\eta\in \{1,5\}$, 2D/3D) as in Fig.~\ref{fig:energyAndFluxVsEta3d}, at the time
when~$U_{Bt}(t)/U_{B0}=0.85$ (left), and 
at $t=10L_x/c$ (right).  The grey/dashed line shows the initial spectrum at $t=0$ in all simulations.  
}
\end{figure}

Here we examine overdensities $\eta=1$ and $\eta=5$, in 3D ($L_z=L_x$) simulations with $L_x=256 \sigma\rho_0$, $\beta_d=0.3$, $B_{gz}=0$, and zero initial perturbation. 
The choice of zero initial perturbation differs from~\S\ref{sec:pertAndEta2d}(2D-b), and is motivated by (3D-a), which showed that an initial perturbation can suppress 3D effects, at least temporarily.
Figures~\ref{fig:energyAndFluxVsEta3d} and~\ref{fig:ntpaVsEta3d} compare
energy and flux versus time, and resulting NTPA, for 3D simulations with
dense [$\eta=5$, $\delta=(2/3)\sigma\rho_0$, $\theta_d=0.10\sigma$] 
and less dense [$\eta=1$, $\delta=(10/3)\sigma\rho_0$, $\theta_d=0.52\sigma$] initial
current sheets; for comparison, nearly 2D simulations are shown for the same parameters as the 3D simulations, except $L_z=L_x/16$.
To estimate stochastic variability in 3D, we show the results of three simulations with $\eta=5$ and $L_z=L_x$ (but just one run each for $\eta=1$ and both nearly-2D cases).

The total transverse magnetic energy (Fig.~\ref{fig:energyAndFluxVsEta3d}, left) shows that energy conversion is generally slower in 3D than in 2D, as expected (e.g., averaged over the first 10$L_x/c$).  Also, energy conversion is slower for $\eta=1$ than for $\eta=5$.
However, the difference between the $\eta=1$ and $\eta=5$ cases is roughly the same in 3D as it is in 2D; and we saw in~\S\ref{sec:pertAndEta2d} that in 2D this difference vanishes for larger systems.
Therefore, any effect of $\eta$ could be entirely caused by the small system size; there is no indication that $\eta$ has a different effect in 3D than in 2D.

Most telling, the magnetic energy in the layer
(Fig.~\ref{fig:energyAndFluxVsEta3d}, middle) is nearly independent
of $\eta$ in 3D; i.e., varying $\eta$ does not appear to enhance or inhibit 3D effects significantly.
Although we did not explore the effect of $\eta$ 
on 3D simulations with an initial perturbation, we expect that a perturbation would make 3D simulations more closely resemble 2D simulations, which have at most a weak dependence on $\eta$.  

The decay of upstream flux $\psi(t)$ (Fig.~\ref{fig:energyAndFluxVsEta3d}, right) also indicates no $\eta$-dependence in 3D beyond what exists in 2D.  
The 2D cases have very similar $\psi(t)$; $\psi(t)$ is also similar for the 3D cases, which exhibit slower decay than the 2D cases (e.g., averaged over the first~$10L_x/c$).
As with total magnetic energy, the decay of upstream flux continues over much longer times than in 2D, at a slower pace than at early times.

Although it appears that $\eta=1$ current sheets result in more 
efficient NTPA (Fig.~\ref{fig:ntpaVsEta3d}), we note again that 
the difference in 3D between $\eta=1$ and $\eta=5$ is almost the same as the difference in 2D (a factor of~$\sim 2$ in particle energy).
This is especially clear when comparing particle energy spectra at the same amount of magnetic energy converted (namely, when $U_{Bt}/U_{B0}=0.85$).
In fact, for each $\eta$, the spectra from corresponding 2D and 3D simulations are nearly the same, aside from a small burst of particles accelerated up to $\gamma\simeq 10\sigma$ for the 2D, $\eta=1$ case; however, this burst is probably the result of some randomness that would be averaged out in a larger simulation---moreover, the ``burst'' is less than a factor of 2 in energy, and $f(\gamma)$ is already very low by such high energies.
Thus we conclude that the effects of $\eta$ are no different in 3D than in 2D; and we note that the effects of $\eta$ diminish in 2D for larger systems (\S\ref{sec:pertAndEta2d}, 2D-b).

To summarize: considering energy and particle acceleration, we detect no
differences between $\eta=1$ and $\eta=5$ that are clearly linked to 3D effects.  However, an ensemble of larger 3D simulations may be needed to rule out significant 3D effects of $\eta$.

\emph{(3D-c) Varying $\beta_d\propto \eta^{-1}$; i.e., varying $\delta/\rho_d$ for fixed $\delta/\sigma\rho_0=(10/3)\sigma\rho_0$.}
When we vary $\eta\propto \beta_d^{-1}$, 
we see dramatic differences in the evolution of 3D current sheets, although NTPA will turn out to be similar in all cases.
As described in \S\ref{sec:pertAndEta2d}(2D-c), changing $\beta_d$ and $\eta$ together, to maintain a constant product $\eta \beta_d=0.3$, keeps $\delta=\sigma\rho_0/(\eta \beta_d)$ constant (with respect to upstream plasma scales), while $\delta/\rho_d=(2/3)(\beta_d^{-2}-1)^{1/2}$ varies.
This allows us to study the evolution of different current sheet configurations while ensuring that the initial current sheet is always well resolved ($\delta/\Delta x = 10$) and of constant size relative to the simulation box [$(L_y/4)/\delta = 39$].

We consider cases that, in large 2D simulations in~\S\ref{sec:pertAndEta2d}(2D-c), yielded essentially similar results;
in 3D we use smaller systems, $L_z=L_x=256 \sigma\rho_0$, 
but as in (2D-c), $B_{gz}=0$ and there is no initial perturbation.
We scan over
$\beta_d \in \{0.075$, 0.15, 0.3, 0.6$\}$, which implies $\eta \in \{4$, 2, 1, 0.5$\}$, respectively; 
corresponding parameters (see table~\ref{tab:etaEffect}) are
$\delta/\sigma\rho_0=10/3$,
$\delta/\rho_d \in \{$8.9, 4.4, 2.1, 0.89$\}$, and
$\theta_d/\theta_b \in\{$0.50, 1.0, 2.1, 5.0$\}$.
For comparison, we also show 2D simulations with identical set-ups except for $L_z=0$.

\begin{figure}
\centering
\fullplot{
\includegraphics*[width=0.325\textwidth]{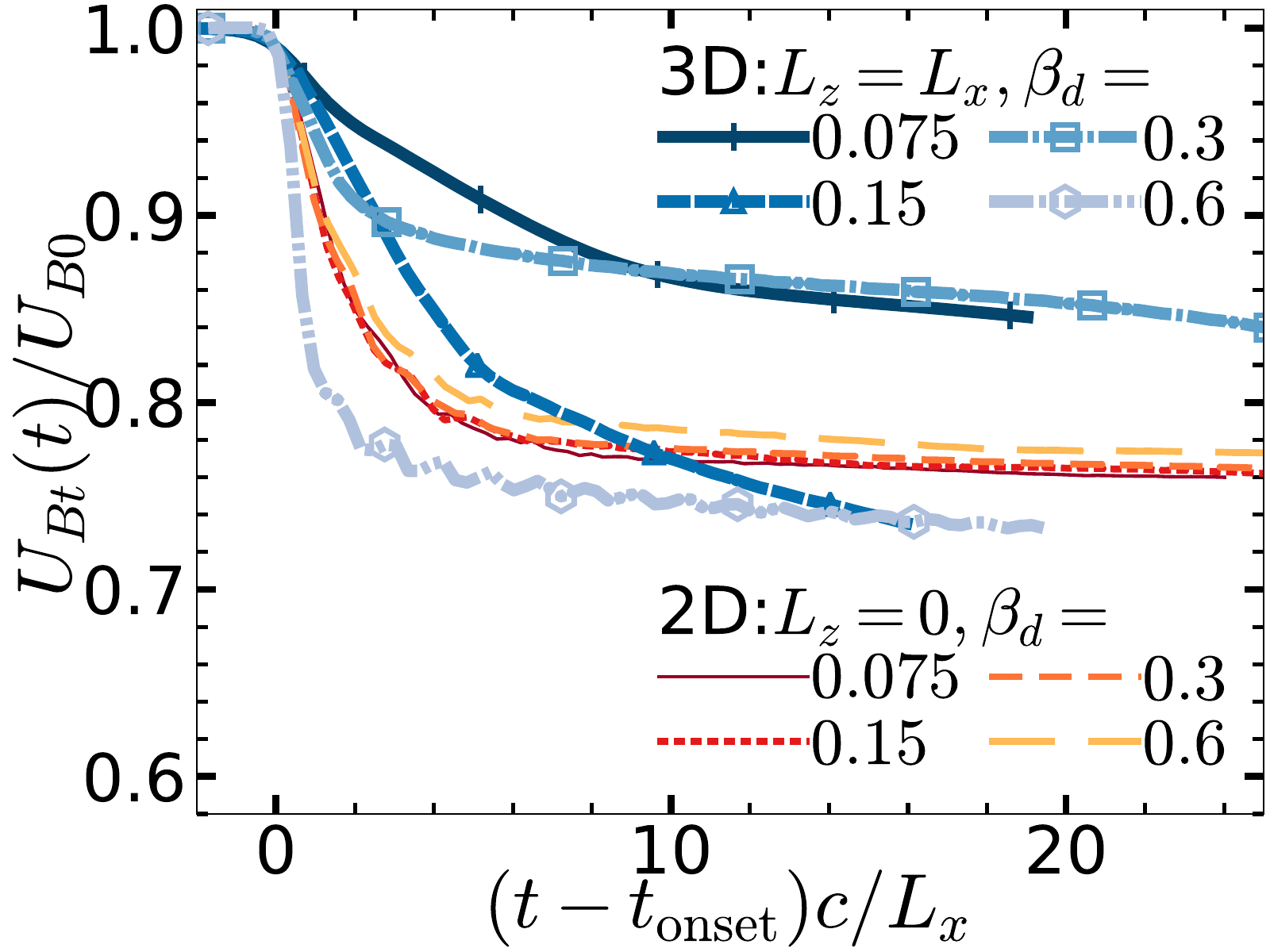}%
\hspace{0.05in}%
\includegraphics*[width=0.325\textwidth]{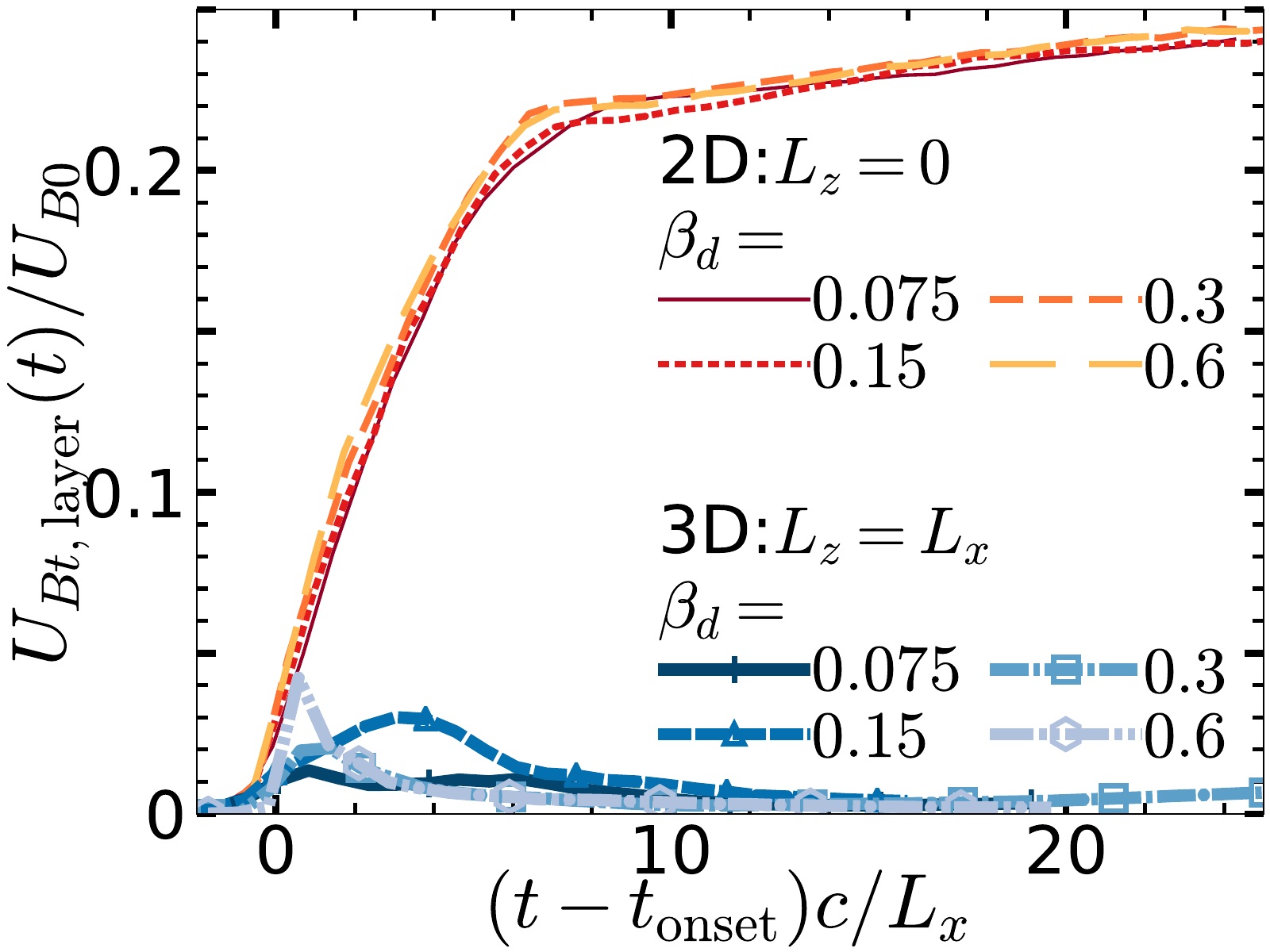}%
\hspace{0.05in}%
\includegraphics*[width=0.325\textwidth]{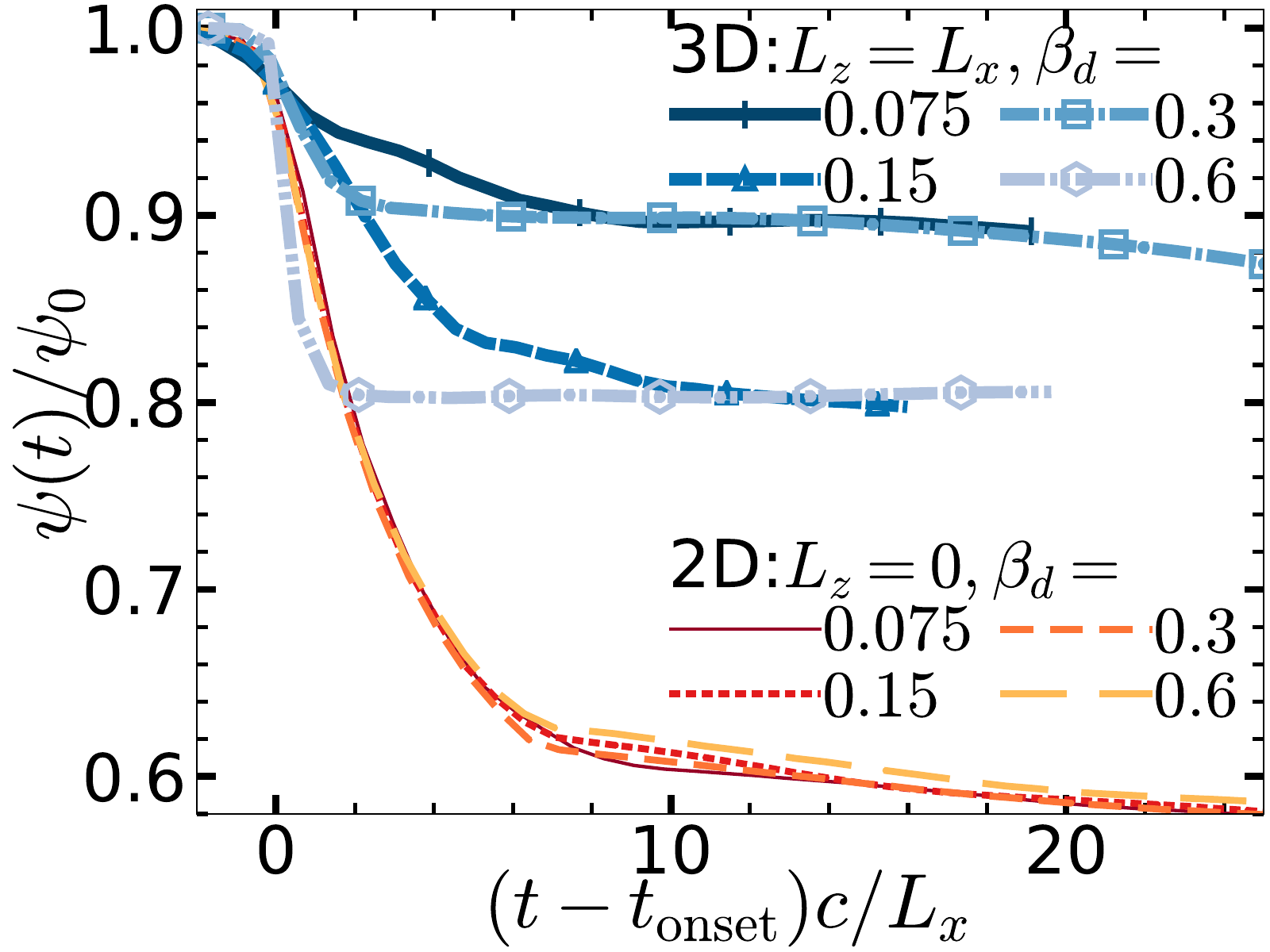}%
}
\caption{ \label{fig:energyAndFluxVsBetad3d}
Varying $\beta_d$ while varying $\eta$ (to hold $\eta\beta_d=0.3$ constant) dramatically alters reconnection in 3D, but not in 2D.
These plots show energies and flux versus time, for eight simulations: 
$\beta_d\in\{$0.075,0.15,0.3,0.6$\}$ (hence $\eta\in\{4$,2,1,0.5$\}$), for both $L_z=L_x$ (3D) and $L_z=0$ (2D).
All simulations have $L_x=256\sigma\rho_0$, $B_{gz}=0$, and $a=0$.
The magnetic energy, $U_{Bt}(t)$ (left), is very similar for all 2D cases, but the 3D cases differ substantially in terms of rates and amounts of magnetic energy conversion.
The magnetic energy in the layer, $U_{Bt,\rm layer}(t)$ (middle), is again similar for all 2D cases; the 3D cases all exhibit the characteristic depletion of reconnected field energy, but differ by small amounts at early times.
The unreconnected flux $\psi(t)$ (right) qualitatively resembles the magnetic energy, except that the 2D simulations reconnect relatively more flux.
}
\end{figure}

Looking at overall characteristics, we notice first (in Fig.~\ref{fig:energyAndFluxVsBetad3d}, left) that the amount and rate of magnetic energy conversion is almost identical for the 2D simulations, but varies drastically with $\beta_d$ in 3D.
Correcting for the difference in reconnection onset times, we see that over the first few~$L_x/c$ after onset, the rate of magnetic energy conversion generally increases with~$\beta_d$.
For $\beta_d=0.6$, the rate is even faster than in 2D; for $\beta_d=0.3$ and $\beta_d=0.15$, it is a little slower than in 2D; and for $\beta_d=0.075$, the initial rate is the slowest of all.
After a few~$L_x/c$, all 3D cases exhibit a slower stage of energy conversion;
however, 
the behaviour becomes more complicated, depending non-monotonically on~$\beta_d$.  
(It might not be useful to try to ascribe a direct $\beta_d$-dependence to the later time evolution.  The dependence could involve an element of randomness, and it could depend on the state of the system after early evolution that did strongly depend on $\beta_d$.  For example, it often happens that a simulation that converts more energy at early times will convert energy more slowly at later times---for example, comparing $\beta_d=0.6$ and $\beta_d=0.15$, or comparing $\beta_d=0.3$ and $\beta_d=0.075$.)
The~$\beta_d=0.6$ (3D) case drastically slows down after~$2L_x/c$, almost as in 2D, except that, being 3D, it can convert somewhat more magnetic energy than in 2D.
The $\beta_d=0.3$ case also slows fairly drastically (though not quite as much as $\beta_d=0.6$) and as a result converts much less energy over~$20L_x/c$.
In contrast, the $\beta_d=0.15$ case slows a bit, but maintains a more moderate magnetic energy conversion rate, and catches up to the $\beta_d=0.6$ case in terms of total magnetic energy conversion after~$\simeq 15L_x/c$.
The $\beta_d=0.075$ case goes from slow to slower, and, like $\beta_d=0.3$, does not convert as much energy over~$20L_x/c$ as the other cases.

The magnetic energy in the layer (Fig.~\ref{fig:energyAndFluxVsBetad3d}, middle) is again almost identical for all the 2D cases.
In 3D, however, we see some differences.
The $\beta_d=0.6$ case, which has such rapid energy conversion at very early times, shows a 2D-like increase for~$\simeq 1L_x/c$, before decaying in a characteristically 3D manner.
The $\beta_d=0.15$ case shows a smaller, slower rise before decaying, while the other cases ($\beta_d=$0.075, 0.3) never have much magnetic energy in the layer.
Despite the differences, all 3D cases strongly contrast with the 2D cases, which store significant amounts of magnetic energy in the layer.  Thus 3D cases can convert more magnetic energy overall even while the upstream flux is less depleted.

The upstream magnetic flux evolution (Fig.~\ref{fig:energyAndFluxVsBetad3d}, right) is consistent with the overall magnetic energy depletion.  As usual, more upstream flux must be reconnected in 2D simulations to convert the same amount of magnetic energy, because in 2D much more of the upstream magnetic energy ends up stored in plasmoids.  As we discussed above, in 2D, flux is very nearly conserved, with an increase in reconnected flux corresponding precisely to the decrease in upstream flux~$\psi$.  In 3D, however, the lost upstream flux is eventually mostly annihilated (although some of it may first be reconnected before being subsequently annihilated).

\begin{figure}
\centering
\fullplot{
\includegraphics*[width=0.49\textwidth]{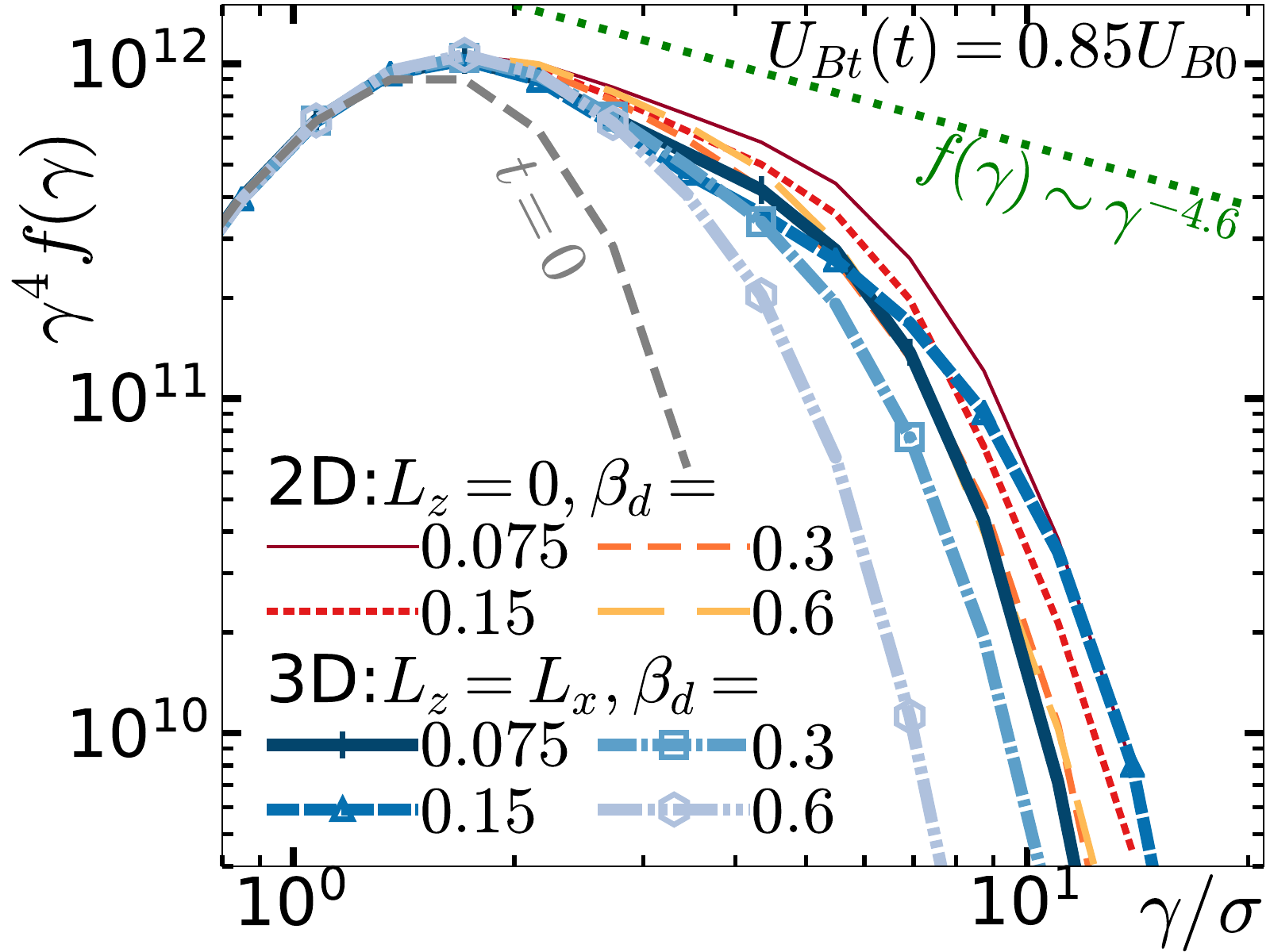}%
\hfill
\includegraphics*[width=0.49\textwidth]{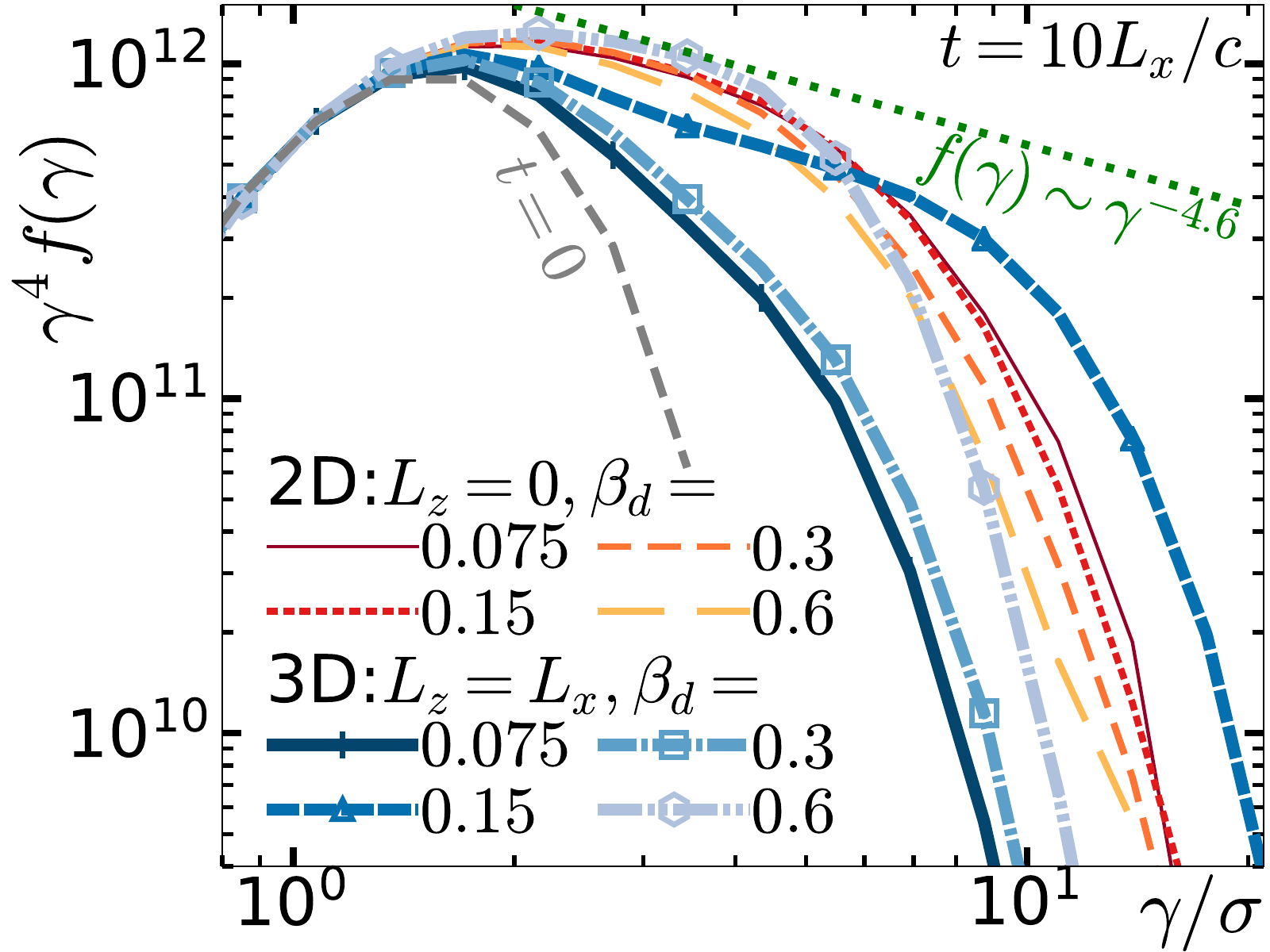}%
}
\caption{ \label{fig:ntpaVsBetad3d}
Although varying $\beta_d$ with $\eta$ (holding $\eta\beta_d$ constant) dramatically affects energy conversion versus time, NTPA is less affected, especially when compared at the same $U_{Bt}/U_{B0}$.
These plots show the electron energy spectra (compensated by $\gamma^4$) 
for the same simulations ($\beta_d\in \{$0.075, 0.15, 0.3, 0.6$\}$, 2D/3D) as in Fig.~\ref{fig:energyAndFluxVsBetad3d} at the time 
when~$U_{Bt}/U_{B0}=0.85$ (left), and 
at $t=10L_x/c$ (right).  For comparison, the grey/dashed line shows the initial spectrum at $t=0$; and, a power-law segment with $p=4.6$ (green/dotted) demonstrates a 15~per~cent (i.e., small) difference from $p=4$ (which would be horizontal).
}
\end{figure}

Particle energy spectra are shown in Fig.~\ref{fig:ntpaVsBetad3d}---both 
at times when 15~per~cent of the initial magnetic energy has been lost (left panel),
and at the same time, $t=10\: L_x/c$ (right).
All cases exhibit fairly similar NTPA, although when the spectra
are compensated by~$\gamma^4$, differences are quite apparent.
The trend is non-monotonic with the input parameters $\beta_d$ and $\eta$, which is not surprising given the non-monotonic dependence of energy conversion on these parameters.
There is evidently a good deal of stochastic variation in these cases, and
more work needs to be done to settle this issue.
However, given that the time over which 15~per~cent of the magnetic energy is converted differs by a factor of~$\simeq 10$ for 3D simulations with $\beta_d=0.6$ and $\beta_d=0.075$, the energy spectra are remarkably similar.

The substantial differences in energy and flux depletion (in 3D simulations with $\beta_d=0.6$ and $\beta_d=0.075$) are very intriguing, and worthy of discussion in a separate section,~\S\ref{sec:RDKIamp}.
Before that, we briefly summarize the results of this subsection.

\emph{Summary of 3D-a,b,c.} 
In contrast to 2D, we have seen here and in~\S\ref{sec:variability3d}  (and will see in more detail in~\S\ref{sec:RDKIamp}) that current sheets in 3D display a greater variety of behaviours, and that the overall evolution is less inevitable, than in 2D.  Rather, the current sheet evolution can take very different paths, depending on initial conditions and stochastic variability.  For example, simulations with identical upstream conditions can convert rather different amounts of magnetic energy over~$\sim 20L_x/c$, whereas 2D simulations with a wide variety of initial current sheets convert nearly the same magnetic energy when averaged over~$\sim 2L_x/c$ time windows, despite some random variability on shorter timescales (e.g., Figs.~\ref{fig:perturb2dLx320}, \ref{fig:eta2dLx1280}, \ref{fig:betad2dLx1280}, \ref{fig:energyFluxVsLx2d}).

An initial magnetic field perturbation, uniform in the $z$ dimension, triggers time evolution resembling 2D reconnection at early times; however, at later times when 2D reconnection would slow and approach a final stable magnetic configuration with substantial magnetic energy in plasmoids, 3D simulations generally deplete the magnetic energy in the layer (as well as in the upstream).
The effect of varying the overdensity $\eta=n_{d0}/n_{b0}$ for fixed $\beta_d$ (hence varying $\delta/\sigma\rho_0$ but constant $\delta/\rho_d=2.1$) is less clear, hard to distinguish amid stochastic variability and moderate system size, but our simulations did not show any effect of $\eta$ in 3D beyond what is observed in 2D (which is very minimal in large systems).
In contrast, varying $\beta_d\propto \eta^{-1}$ (hence varying $\delta/\rho_d$ while keeping $\delta/\sigma\rho_0=10/3$ fixed) had a pronounced effect on current sheet evolution, with larger $\beta_d$ leading to faster initial magnetic energy conversion and more conversion over $20L_x/c$---and we will explore this next.

Overall, global NTPA is surprisingly consistent given the variety of behaviours evolving from different initial current sheet configurations.  This is especially true if particle spectra are compared at times when the same amount of magnetic energy has been converted to plasma energy.

\subsection{3D reconnection: the effect of varying $\beta_d$ on RDKI amplitude}
\label{sec:RDKIamp}

The linear and nonlinear stages of RDKI growth were investigated in detail in \cite{Zenitani_Hoshino-2007,Zenitani_Hoshino-2008}, both in theory and in 2D and 3D simulation, including the effect of RDKI on NTPA.  However, by exploring a range of initial parameters, specifically a range of initial drift speeds, $\beta_d c$, we find that the evolution from the linear to the nonlinear stage can have varied and often important consequences for later evolution.

Varying $\beta_d\propto \eta^{-1}$, as in~\S\ref{sec:pertAndEta3d}(3D-c), to keep $\eta\beta_d=0.3$ constant, hence fixing $\delta/\sigma\rho_0=10/3$, leads to dramatic differences in current sheet evolution, including differences in the evolution of global quantities such as magnetic energy. 
Here, we show these differences in more detail, and attribute them to the maximum amplitude to which RDKI can grow before becoming nonlinearly saturated.  In the following, $\Delta y_c$ is the amplitude of the rippling of the current sheet central surface, $y_c(x,z)$ (see~\S\ref{sec:sheetCenter}).

We start by looking at the same 3D simulations run for~\S\ref{sec:pertAndEta3d}(3D-c), with common parameters $L_z=L_x=256\sigma\rho_0$, $\eta\beta_d=0.3$, $\delta/\sigma\rho_0=10/3$, $B_{gz}=0$, and zero initial perturbation. 
We focus on the two extreme cases: $\beta_d=0.6$ ($\eta=0.5$, $\delta/\rho_d=0.89$), which shows very rapid, substantial magnetic energy conversion, followed by very slow, limited energy conversion; and $\beta_d=0.075$ ($\eta=4$, $\delta/\rho_d=8.9$), which shows slower energy conversion and less conversion overall after $20L_x/c$ (see table~\ref{tab:etaEffect} for values of other relevant parameters).

\begin{figure}
\centering
\fullplot{
\includegraphics*[width=\textwidth]{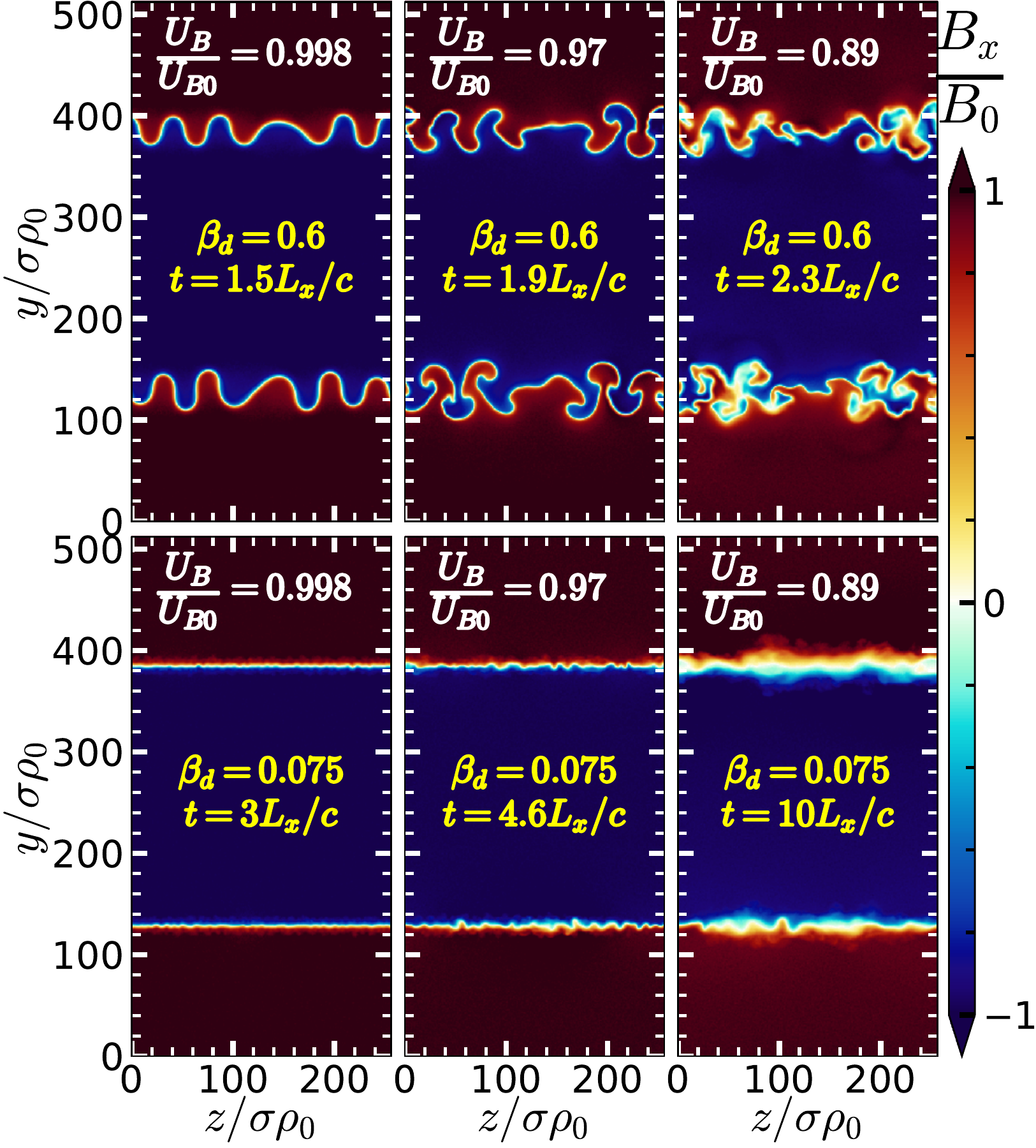}%
} 
\caption{ \label{fig:kinkAmpVsBetad3d}
$B_x(y,z)$ in the $x=0$ plane for 3D ($L_z=L_x$) for two cases, (top row) $\beta_d=0.6$, $\eta=0.5$ and (bottom row) $\beta_d=0.075$, $\eta=4$, each at three times, when the total magnetic energy $U_B$ has fallen to fractions~0.998, 0.97, and 0.89 of its initial value $U_{B0}$.
For large~$\beta_d$ (top) the kink wavelength is much longer than the current sheet thickness, and the sheet ripples with large amplitude $\Delta y_c$ while converting very little magnetic energy to plasma energy, until nonlinear development causes the sheet to fold over on itself, rapidly converting magnetic energy~$0.08U_{B0}$ over a time~$0.4L_x/c$. 
For small~$\beta_d$ (bottom) the kink wavelength is short, and~$\Delta y_c$ does not increase much beyond the original sheet thickness, resulting in slower magnetic energy conversion ($0.08U_{B0}$ over $5.4L_x/c$).
}
\end{figure}

The reason for these differences is illustrated in Fig.~\ref{fig:kinkAmpVsBetad3d}, which shows $B_x(0,y,z)$ in the $x=0$ plane at the three different times when 0.2, 3, and 11~per~cent of the initial magnetic energy has been converted to plasma energy, for the $\beta_d=0.6$ case (top) and $\beta_d=0.075$ (bottom).
For $\beta_d=0.6$, the layer kinks with long wavelength ($\lambda_z \gg \delta$) and the RDKI amplitude $\Delta y_c$ can grow to a relatively large value, $\Delta y_{c*} \sim\lambda_z$, before becoming highly nonlinear;
the nonlinear development causes the highly-distorted layer to fold over on itself, rapidly depleting magnetic energy in a layer of thickness~$\sim \lambda_z \gg \delta$.
It is this process, and not 2D-like reconnection (perhaps surprisingly, given that the energy conversion versus time resembles that in 2D reconnection), that results in rapid energy dissipation at early times.
In contrast, for $\beta_d=0.075$ (Fig.~\ref{fig:kinkAmpVsBetad3d}, bottom), the initial sheet kinks with short wavelength ($\lambda_z \lesssim \delta$); since $\Delta y_c$ cannot grow much beyond $\lambda_z$, the instability saturates before it can distort the sheet much (compared with $\delta$).
In this case, the nonlinear RDKI can deplete magnetic energy only within a layer of thickness $\sim \lambda_z \lesssim \delta$.

\citet{Zenitani_Hoshino-2007} derived the most unstable wavelength for RDKI in relativistic plasma \citep[using two-fluid theory with an assumption that the current-carrying particles execute Speiser orbits within a sublayer of half-thickness $\sqrt{\rho_d \delta}$, and ignoring the background plasma; an improved approximation for $\beta_d\gtrsim 0.8$ can be found in][]{Hoshino-2020arxiv}: $\lambda_{z,\rm RDKI} = 16\upi \gamma_d \beta_d^2 \delta = (32 \upi/3)\beta_d \rho_d$, which can be significantly larger than $\rho_d$.
For $\beta_d=0.6$, $\lambda_{z,\rm RDKI}\approx 23\delta = 75 \sigma\rho_0$ (in Fig.~\ref{fig:kinkAmpVsBetad3d}, we see $\lambda_z\approx 50 \sigma\rho_0$).
For $\beta_d=0.075$, $\lambda_{z,\rm RDKI}\approx 0.3 \delta = \sigma\rho_0$ (in Fig.~\ref{fig:kinkAmpVsBetad3d}, we estimate very roughly, $\lambda_z\sim 7 \sigma \rho_0$).
While the measured instability wavelength does not precisely match the
theory, the qualitative behaviour is in agreement, and in any case, there
is a strong correlation between $\lambda_z/\delta$ and the rate and amount of magnetic energy conversion.

In both these simulations we find that the maximum $\Delta y_c$ is on the order of half the wavelength in the $z$-direction.
For the $\beta_d=0.6$ case, the kink grows roughly to a maximum $\Delta y_{c*} \approx 20\sigma\rho_0 \approx 0.4 \lambda_{z}$; for the $\beta_d=0.075$ case, $\Delta y_{c*} \approx 3\sigma\rho_0 \approx 0.4 \lambda_z$.
Given the uncertainty in these measurements, we might as well say $\Delta y_{c*} \approx \lambda_z/2$ for simplicity.
We note that the largest kink amplitudes (for $\beta_d=0.6$) are a fraction $\Delta y_{c*}/(L_y/4)\lesssim 0.2$ of the global system; this fraction is sizeable, but not so close to 1 that we need to worry about the lower and upper current layers interacting.

We hypothesize that in 3D, RDKI at some wavelength $\lambda_{z}$ grows linearly, without substantial magnetic energy conversion or NTPA due to RDKI, until it nears a critical amplitude, $\Delta y_{c*} \sim 0.5 \lambda_{z}$.  
If the kink mode grows beyond $\Delta y_{c*}$, RDKI becomes highly nonlinear and the current layer severely distorts, folding over on itself and rapidly depleting magnetic energy within $|y| < \Delta y_{c*}$.
This yields a turbulent layer of thickness $\simeq 2\Delta y_{c*}$ with greatly diminished magnetic field.
After this time, magnetic energy continues to be converted to plasma energy, but at a much slower rate because of the much thicker layer.
Importantly, RDKI does not inevitably grow to $\Delta y_{c*}$; depending on the initial layer parameters and also on stochastic behaviour, it may not reach the nonlinear stage that severely distorts and rapidly transforms the initially-thin current sheet.

\begin{figure}
\centering
\fullplot{
\includegraphics*[width=\textwidth]{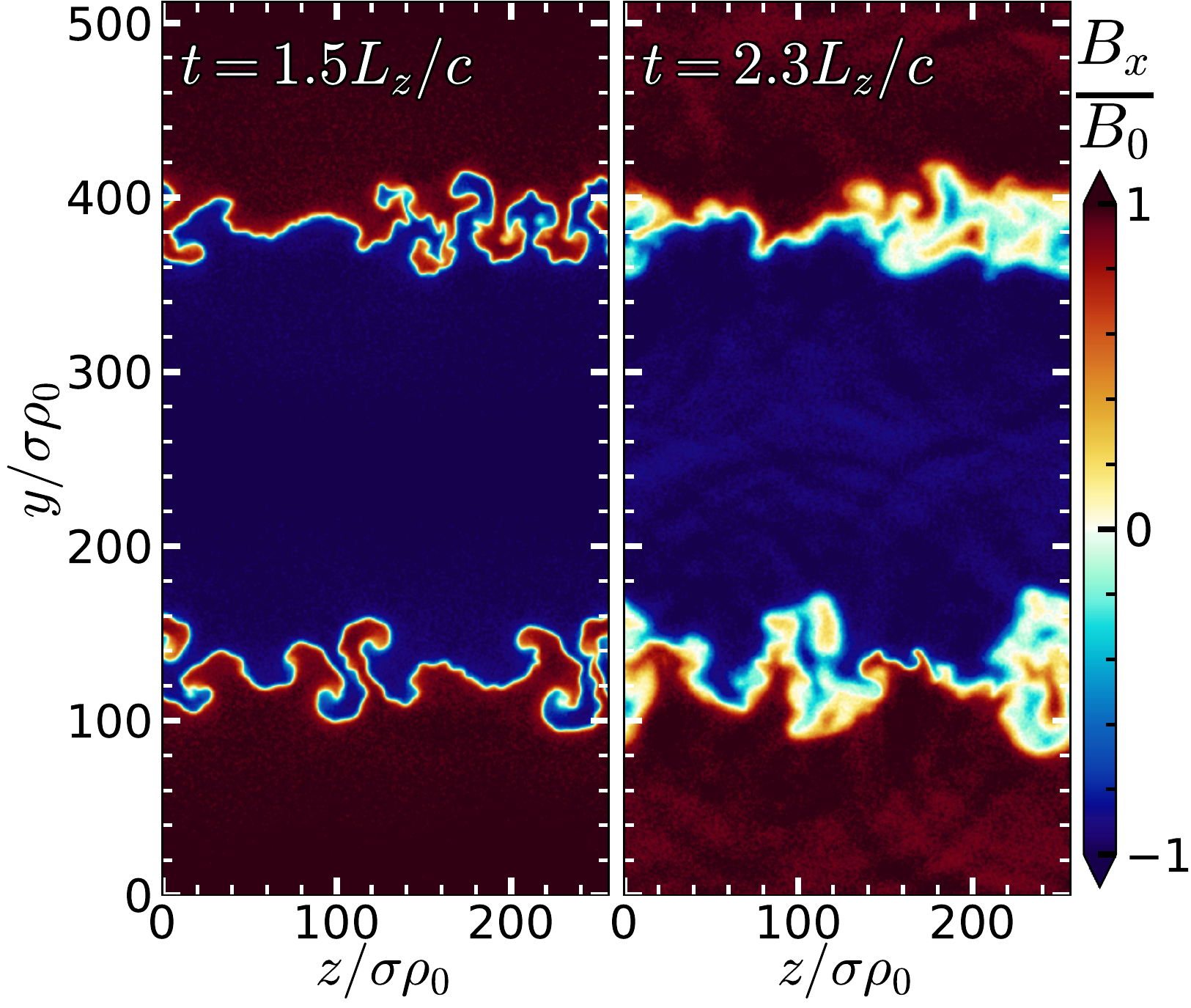}%
}
\caption{ \label{fig:kinkAmp2dyz}
$B_x(y,z)$ for 2D ($y$-$z$ plane) simulations  at two different times, $1.5 L_z/c$ and $2.3 L_z/c$, for an initial current sheet of thickness $\delta=(10/3)\sigma\rho_0$ with $\beta_d=0.6$ and $\eta=0.5$.
Without $x$-dependence, tearing and reconnection are forbidden, but RDKI is allowed.
The kink instability grows to large amplitudes, where the current layer becomes highly deformed (from the small-amplitude sinusoidal perturbation), folding over onto itself and depleting magnetic energy
(compare to Fig.~\ref{fig:kinkAmpVsBetad3d}).
}
\end{figure}

This phenomenon of extreme nonlinear kinking can occur even in a 2D simulation in the $y$-$z$ plane, where tearing (at least in the $x$-direction) is forbidden and reconnection cannot occur \citep{Zenitani_Hoshino-2007}.
Although a systematic examination of RDKI without reconnection in 2D$yz$ simulations is beyond the scope of this paper \citep[but see, e.g.,][]{Zenitani_Hoshino-2005a,Zenitani_Hoshino-2007,Zenitani_Hoshino-2008,Cerutti_etal-2014b},
we compare to one such 2D simulation, identical in set-up with $\beta_d=0.6$ and $\eta=0.5$ (but lacking any $x$-dependence).
There we see clear development of the kink instability growing to
large amplitudes and developing highly nonlinear behaviour in Fig.~\ref{fig:kinkAmp2dyz} (which looks similar to Fig.~\ref{fig:kinkAmpVsBetad3d}, for $\beta_d=0.6$).

\begin{figure}
\centering
\fullplot{
\begin{tabular}{cc}
\includegraphics*[width=0.48\textwidth]{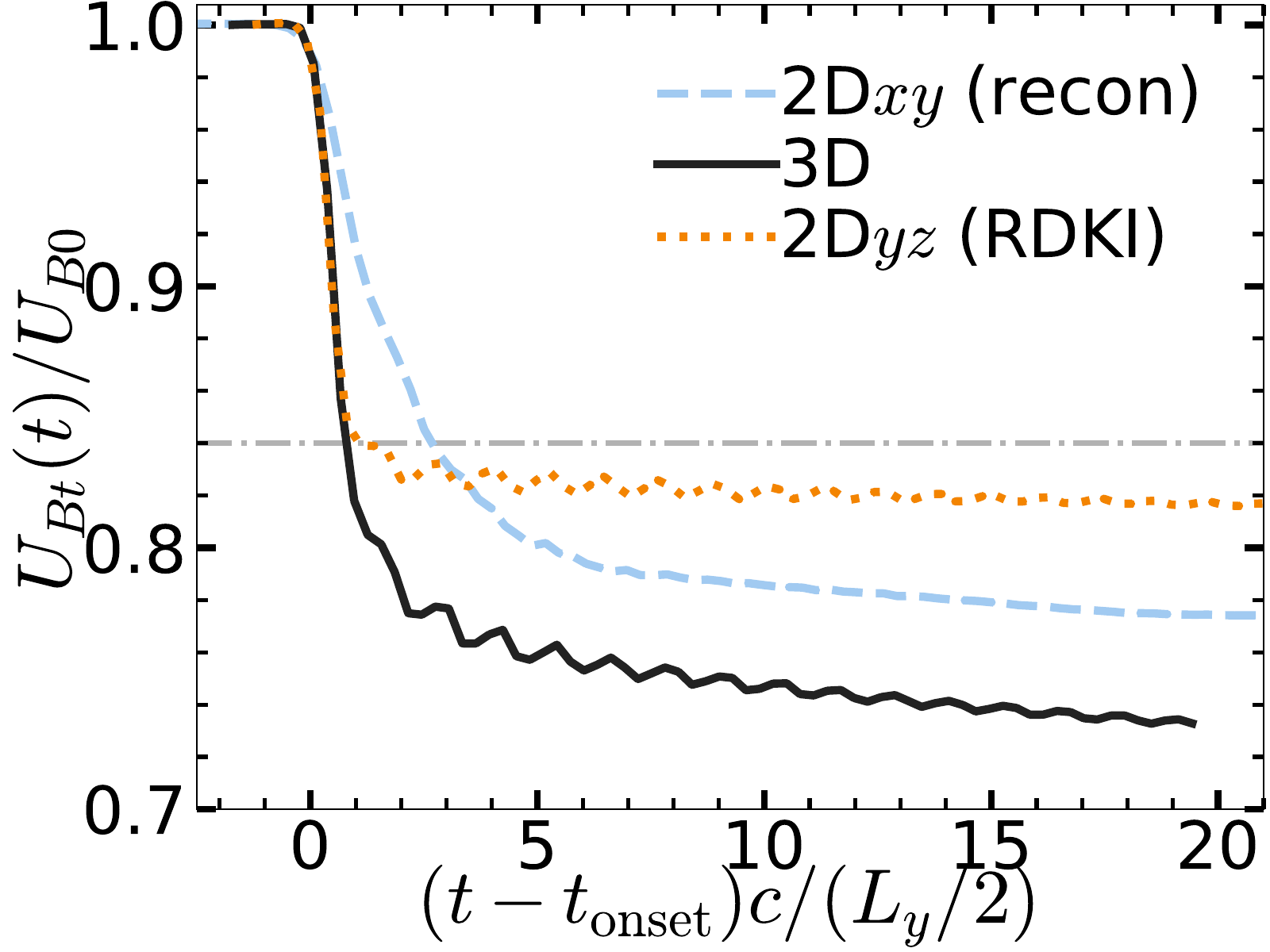}
\hfill &
\includegraphics*[width=0.48\textwidth]{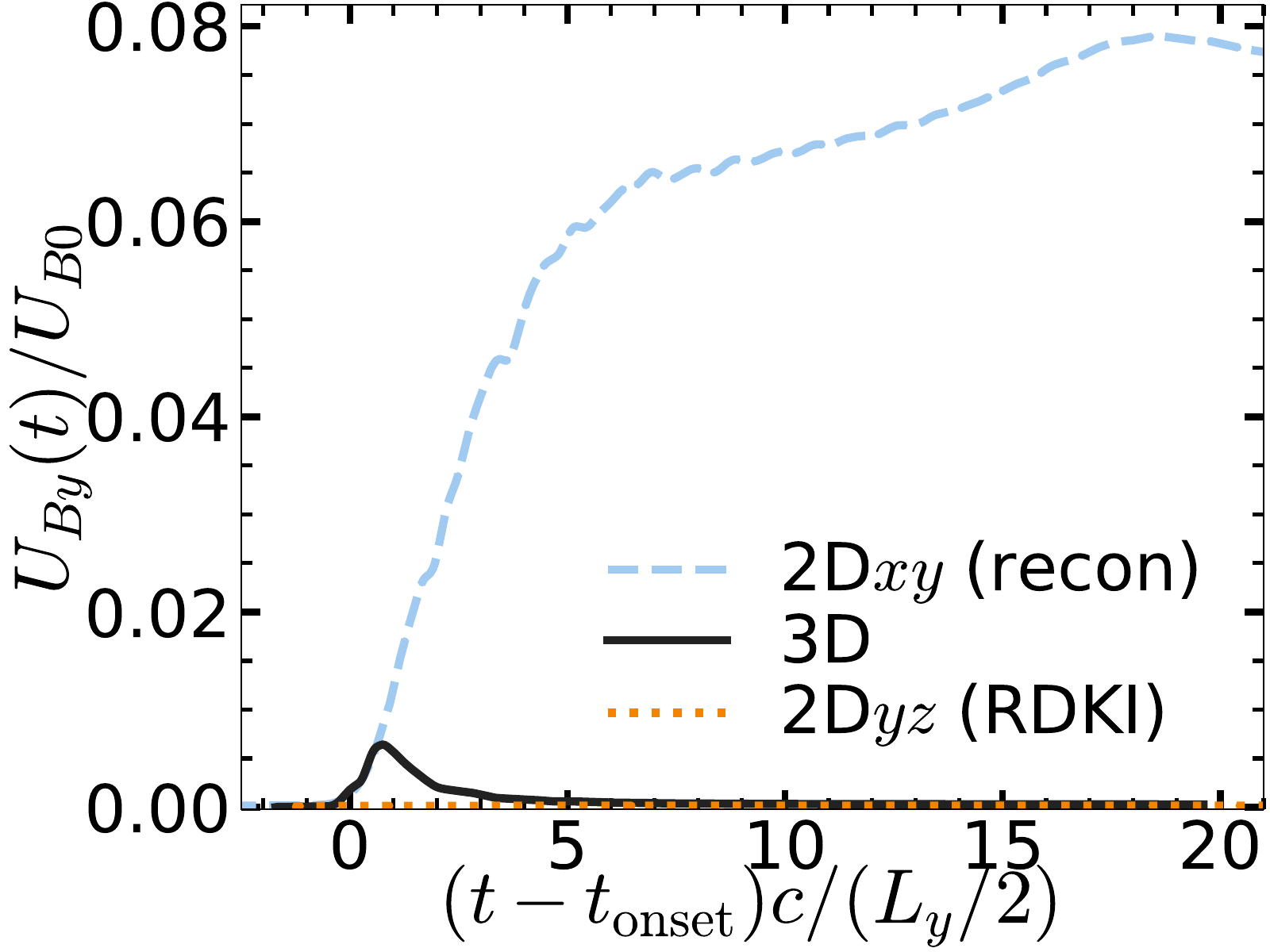}%
\\
\includegraphics*[width=0.48\textwidth]{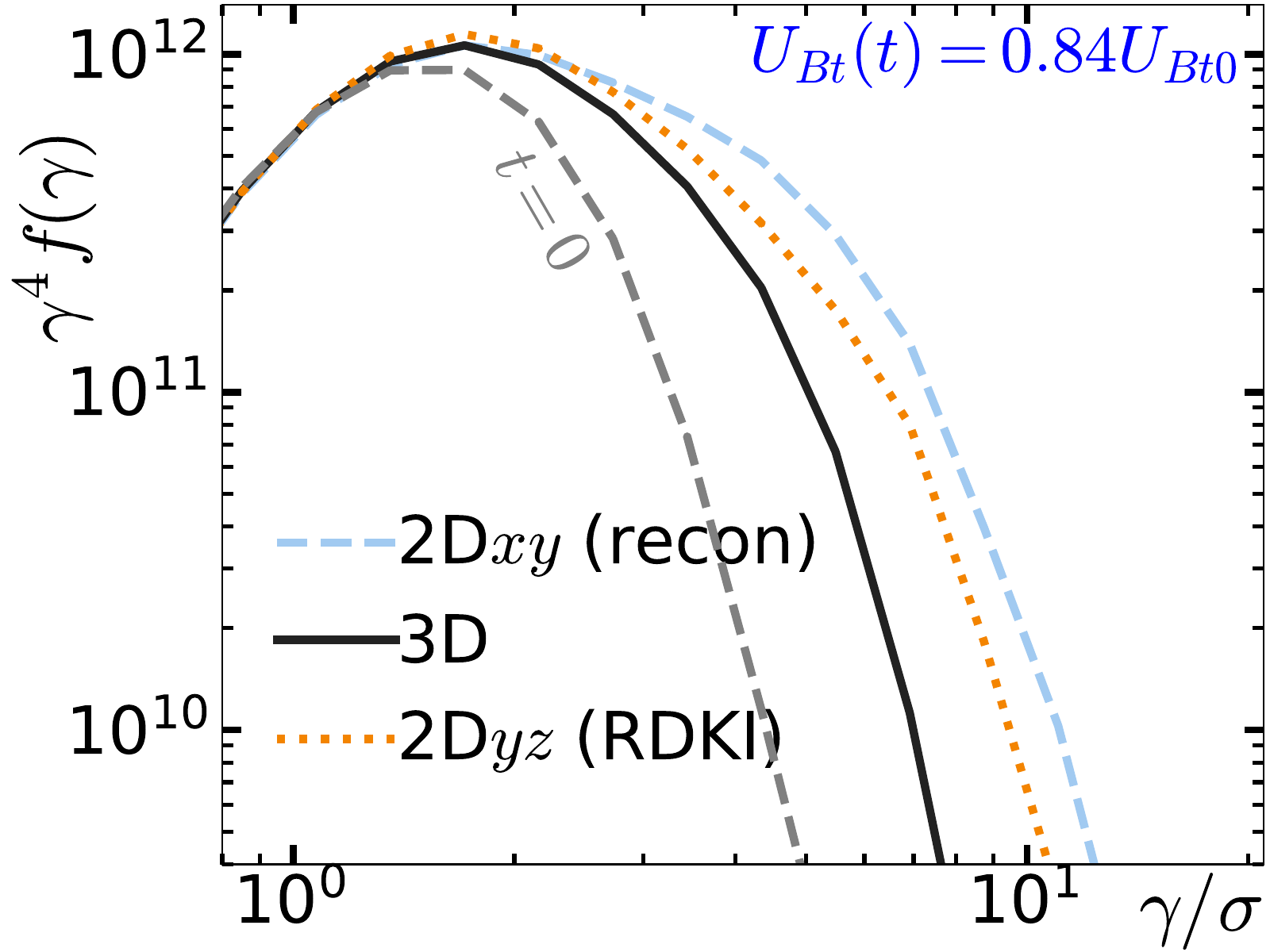}%
\hfill &
\includegraphics*[width=0.48\textwidth]{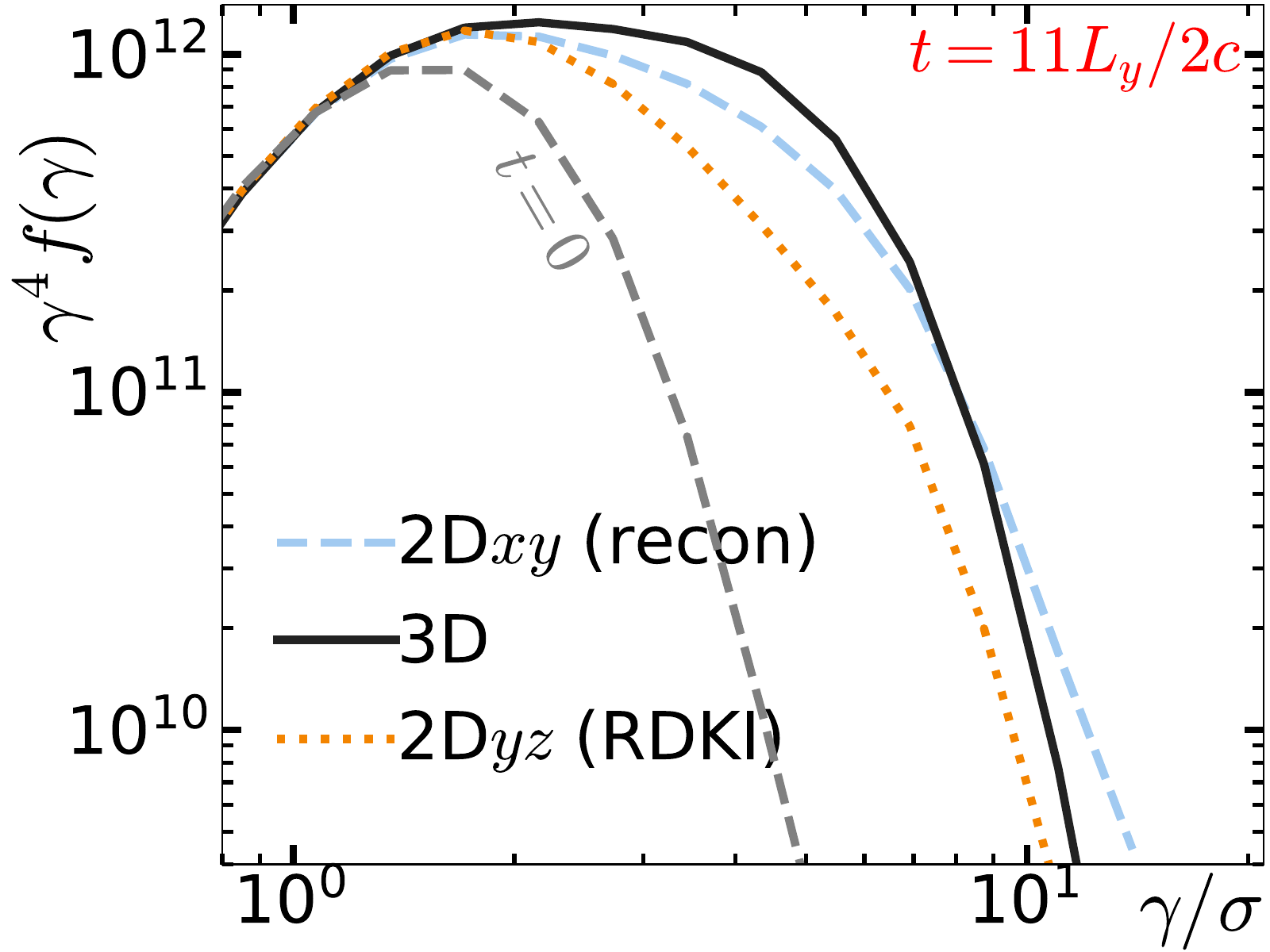}%
\end{tabular}
}
\caption{ \label{fig:energyAndNTPAbetadp6}
A comparison of energy conversion and NTPA for 2D$xy$, 3D, and 2D$yz$ simulations with the same set-up and size, except for unsimulated dimensions.  Upper left: (transverse) magnetic energy $U_{Bt}(t)$; upper right: energy $U_{By}(t)$, stored in $B_y$ field components, a rough indicator of the amount of magnetic energy stored in the layer; lower left: particle spectra (compensated by $\gamma^4$) when $U_{Bt}$ has fallen to $0.84U_{B0}$ (cf. the grey dash-dotted line in the upper left); lower right: spectra at $t=11.4\:(L_y/2)/c$.  The 2D$xy$ simulations exhibit familiar 2D reconnection but not RDKI; the 2D$yz$ simulations can undergo RDKI but not reconnection.
}
\end{figure}

Indeed, this 2D$yz$ configuration rapidly depletes magnetic energy, at least when it reaches the extreme nonlinear stage where the current layer is massively distorted.
For comparison, in Fig.~\ref{fig:energyAndNTPAbetadp6} 
we overplot results for the same set-up
[$\beta_d=0.6$, $\eta=0.5$, $\delta/\rho_d=0.89$, $\delta=(10/3)\sigma\rho_0$, $a=0$, $L_y=512\sigma\rho_0$]
but different dimensionalities: (1) 2D$xy$ with $L_x=L_y/2$ and $L_z=0$,
(2) 3D with $L_x=L_z=L_y/2$, and (3) 2D$yz$ with $L_z=L_y/2$ and $L_x=0$.
First, in panel (a), we see that magnetic energy is converted to particle energy more rapidly (presumably triggered by RDKI) in the 2D$yz$ and 3D simulations than by reconnection in 2D$xy$, but that all cases exhibit rapid magnetic energy conversion (a decline in $U_B$) of roughly similar amounts.  E.g., they all convert about 20--25 per~cent of $U_{B0}$ in less than 5$\:L_y/2c$ after onset and ultimately convert 20--27 per~cent of $U_{B0}$ within 20$\:L_y/2c$ of onset; the initial energy conversion rate is somewhat slower for the 2D$xy$ case than for the others, while the total energy converted to plasma energy is slightly higher for the 3D simulation (most notably for $t\gtrsim 5L_y/2c$).
Looking at the energy $U_{By}$ stored in $B_y$ field components ($U_{By}$ is a proxy for $U_{Bt,\rm layer}$ in the 2D$xy$ and 3D cases), we see (panel b) that the $2Dyz$ simulation has almost no $U_{By}$ (as expected), while the 3D simulation develops a small $U_{By}$ that decays, and the 2D$xy$ simulation stores $U_{By}$ permanently.
Thus the 2D$xy$ simulation sees a larger decrease in upstream magnetic energy than other cases, but some of that energy is transformed to ``reconnected field'' energy rather than being converted to plasma energy.

The similarities among these three cases are striking, especially considering that the 2D$xy$ simulation exhibits classic reconnection, which is forbidden in~2D$yz$.
The 2D$xy$ case conserves flux, while the 2D$yz$ simulation directly annihilates upstream flux.  We see clear suggestions of both these 2D behaviours in 3D---plasmoid formation and the growth of $U_{Bt,\rm layer}$ (as in the 2D$xy$ case) and rapid magnetic energy depletion due to nonlinear kinking of the current sheet (as in the 2D$yz$ case; see~\S\ref{sec:Lz3d} for more detail).  Thus in 3D, we expect that classic reconnection occurs to some extent, converting upstream flux to flux around plasmoids (as in 2D$xy$), and that some upstream flux is directly annihilated (as in 2D$yz$); furthermore, there is a unique 3D effect, namely the dissipation of plasmoids and annihilation of the associated ``reconnected'' flux.  Without much more extensive diagnostics, it is difficult to determine how much upstream flux is directly annihilated as opposed to reconnected and then annihilated, i.e., to determine precisely the extents to which the different 2D-like processes occur in 3D.

Interestingly, all these cases exhibit similar NTPA, despite the fact that very different mechanisms come into play in (at least) the 2D$xy$ and 2D$yz$ simulations.
Compensating $f(\gamma)$ by $\gamma^4$, however, we do see some differences.
If we compare the spectra at times when the magnetic energy has fallen by 16~per~cent (Fig.~\ref{fig:energyAndNTPAbetadp6}c), we find that NTPA is strongest for 2D$xy$, followed by 2D$yz$, and weakest for 3D; here, differences in $f(\gamma)$ reach an order of magnitude, but this is not as significant as it might seem, considering the steepness $f(\gamma)\sim \gamma^{-4}$---the spectra are separated by less than a factor of~2 in $\gamma$ (or energy).
Over time, the 3D case develops stronger NTPA, and the three cases look fairly similar; differences in $f(\gamma)$ are less than a factor of a few, which, again, is relatively little considering the steep slope.
At late times [e.g., around $11L_y/2c$, Fig.~\ref{fig:energyAndNTPAbetadp6}d], the 3D case has the hardest spectrum (shallowest slope).

Whereas \citet{Zenitani_Hoshino-2007,Zenitani_Hoshino-2008} also observed RDKI becoming nonlinear, those works observed little NTPA due to RDKI (for zero guide field)---in the linear stage, but also specifically in the nonlinear stage.
In contrast, when a long-wavelength RDKI mode ($\lambda_{z} \gg \delta$) reaches the nonlinear stage, we observe much more rapid magnetic energy declines than in the linear stage, and the nonlinear stage generates significant NTPA.

Although we must leave a systematic exploration of 2D$yz$ simulations to future work, we note that 2D$yz$ simulations with lower $\beta_d$ often do not show the violent nonlinear growth of the kink instability at large amplitudes (as in 3D; see Fig.~\ref{fig:kinkAmpVsBetad3d}).
In these cases, less magnetic energy is converted overall, although slow magnetic energy conversion can continue for tens of light-crossing times (as in 3D), and there is often no significant NTPA at all.

When looking at 3D simulations with different $\beta_d$, and comparing them to 2D reconnection, the similarly rapid magnetic energy conversion in 2D reconnection and 3D simulations with large $\beta_d$ (small $\delta/\rho_d$) might seem to suggest that these 3D simulations exhibit 2D-like reconnection.
This is (mostly) not the case here.  The 3D simulation with large $\beta_d=0.6$ has relatively long $\lambda_{z,\rm RDKI}$ and exhibits extreme nonlinear kinking that much more closely resembles the 2D$yz$ simulation;
it bears little resemblance to the 2D$xy$ simulation in terms of magnetic field structure.
While magnetic reconnection could be playing a role at small scales in the 3D simulation with $\beta_d=0.6$, its strong resemblance to the 2D$yz$ simulation suggests that reconnection may be unimportant at early times when magnetic energy conversion is dominated by the nonlinear RDKI.

If long-wavelength kink modes grow to large amplitudes and distort the current layer enough to suppress reconnection, in the process converting magnetic energy and producing NTPA, then we will need to understand how this extreme-kinking mechanism compares with large-scale reconnection (considering both 2D and 3D reconnection models).
If, as the $\beta_d=0.6$ simulation in this section suggests, extreme kinking results in similar NTPA and overall energy conversion (in terms of both rate and amount) as 2D reconnection, then this mechanism becomes a competing candidate to explain phenomena for which reconnection has been invoked as an explanation.
One important difference is that guide magnetic field is believed to curtail RDKI rather severely \citep{Zenitani_Hoshino-2008}.
Another important difference is that the layer distortion appears to be self-limiting in the following sense:
it depletes magnetic energy within a thickness $2\Delta y_c \sim \lambda_{z,RDKI}$, resulting in a much thicker layer no longer highly unstable to RDKI.
(The same might be said of the tearing instability, which would saturate when the plasmoids fed by an elementary reconnecting current sheet grow to a height in $y$ comparable to the elementary current sheet length in $x$; however, in a large reconnecting system, the plasmoids detach from the elementary current sheet before growing so large and coalesce with other plasmoids, thus always maintaining thin current sheets.  Consequently tearing and reconnection do not saturate until the plasmoids become comparable to the system size.)
For our 3D simulations, $\lambda_{z,\rm RDKI}$ happens to be a significant fraction of the system size, and so this mechanism yields dramatic conversion of magnetic to plasma energy.
However, if the most unstable wavelength $\lambda_{z,\rm RDKI}$ is independent of system size, then in the limit of astronomically-large systems, this mechanism may convert a negligible fraction of available magnetic energy, while suppressing reconnection.
In contrast, 2D reconnection is self-limiting (in a closed system) only when plasmoids grow to a size $\sim L_x$ that scales with the system size.  We believe reconnection would continue for an arbitrary time, given a large-enough system.
(This is perhaps a remarkable property of 2D reconnection---that the macroscopically-large plasmoids do not destroy the kinetic-scale current sheets where reconnection occurs.)

On the other hand, because the maximum kink amplitude depends on the wavelength, $\Delta y_{c*} \sim \lambda_{z}/2$, it might be a mistake to assume that the most prominent or influential wavelength corresponds to the RDKI mode with the highest growth rate.
The most unstable mode might saturate without thickening the layer much, while a more slowly-growing mode with a longer wavelength might eventually reach a much larger amplitude.
This possibility needs further study; however, we will show in the following section that some simulations (with $\beta_d=0.3$) initially kink on a scale $\lambda_z \ll L_z$, but nonetheless over time develop large-amplitude kinking on the scale of the system, $\lambda_z = L_z$.
Whether such behaviour can scale with system size remains an outstanding question.

If large-amplitude kink modes are not inevitable, as some of our simulations suggest, it may be that short-wavelength RDKI modes develop and rapidly saturate, before their amplitudes grow much past the current sheet thickness.
In this case, RDKI may have a negligible effect on reconnection;
although
it appears that reconnection in this regime is slow (though perhaps locally fast, as suggested in~\S\ref{sec:overview3d}), this slowness may not have anything to do with RDKI.

\subsection{3D reconnection: $L_z$-dependence}
\label{sec:Lz3d}

In this subsection, we systematically 
vary $L_z$
to explore when and how 3D effects arise with increasing $L_z$.
We consider $L_z/L_x\in \{$0, 1/32, 1/16, 3/32, 1/8, 1/4, 1/2, 1, 3/2$\}$
(where $L_z=0$ means 2D) for system size $L_x=341\sigma \rho_0$,
$\eta=5$, $\beta_d=0.3$, $\delta=(2/3) \sigma \rho_0$, $B_{gz}=0$, and
zero initial perturbation.

\begin{figure}
\centering
\fullplot{
\includegraphics*[width=0.325\textwidth]{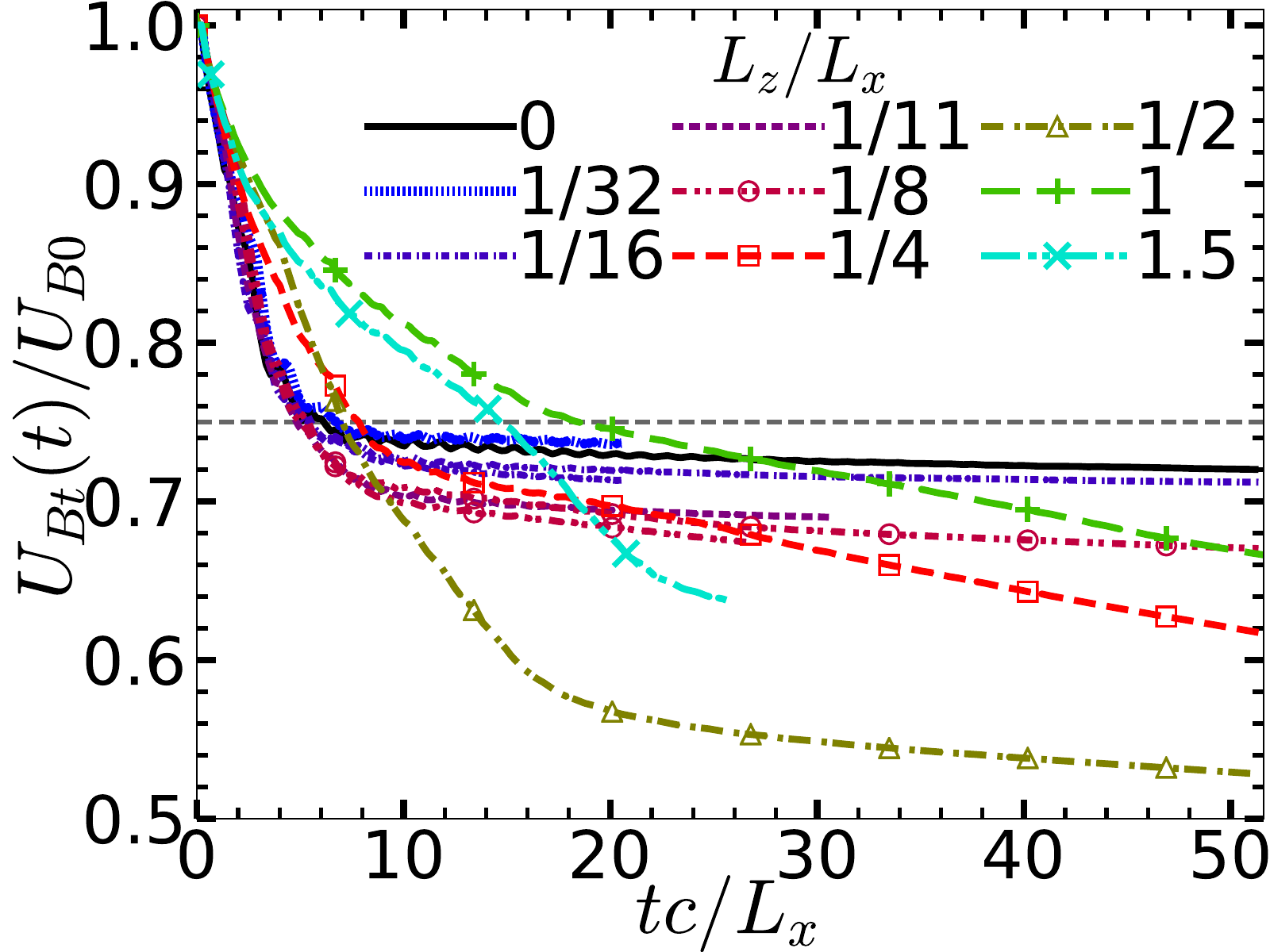}%
\hspace{0.05in}%
\includegraphics*[width=0.325\textwidth]{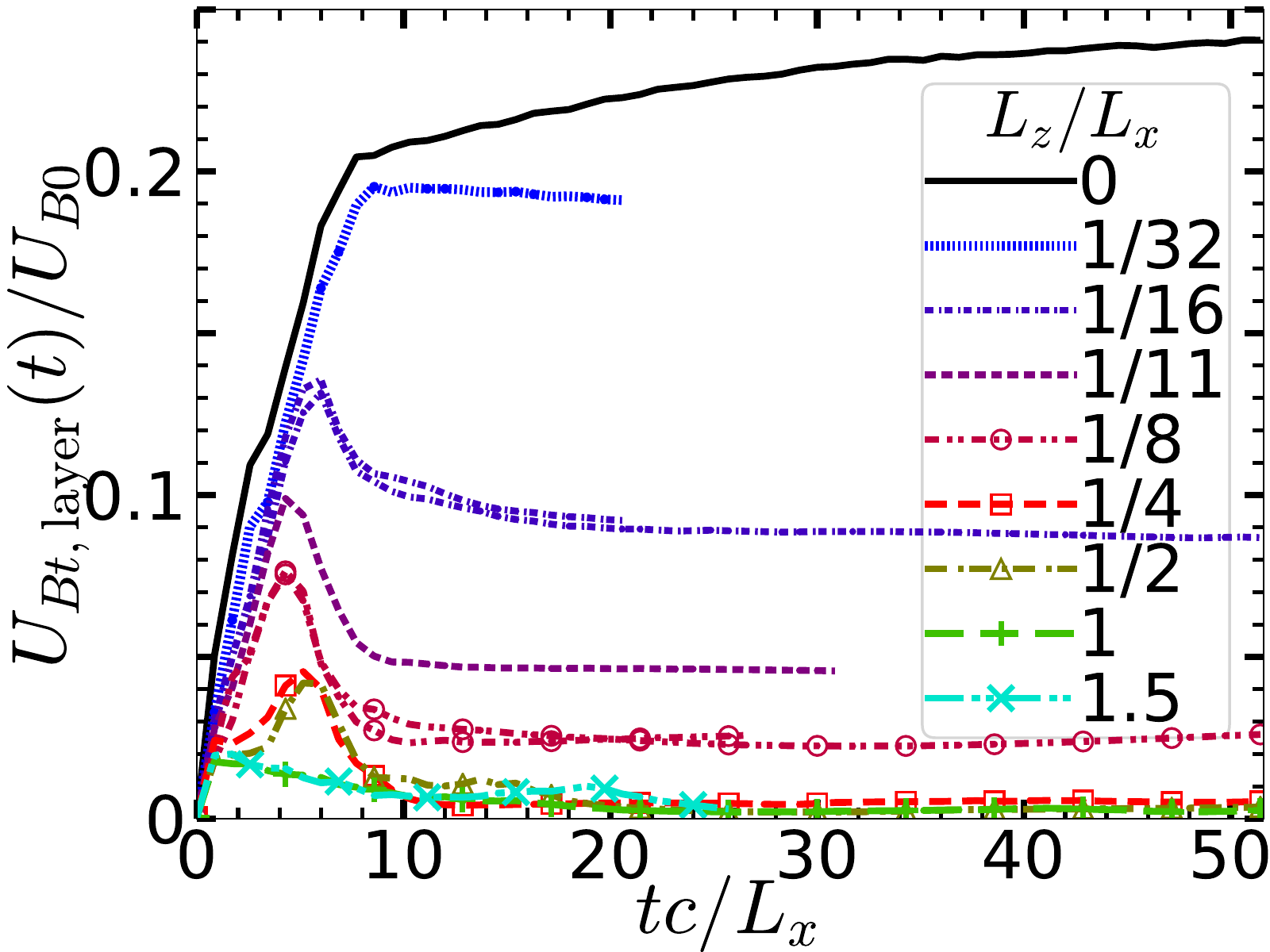}%
\hspace{0.05in}%
\includegraphics*[width=0.325\textwidth]{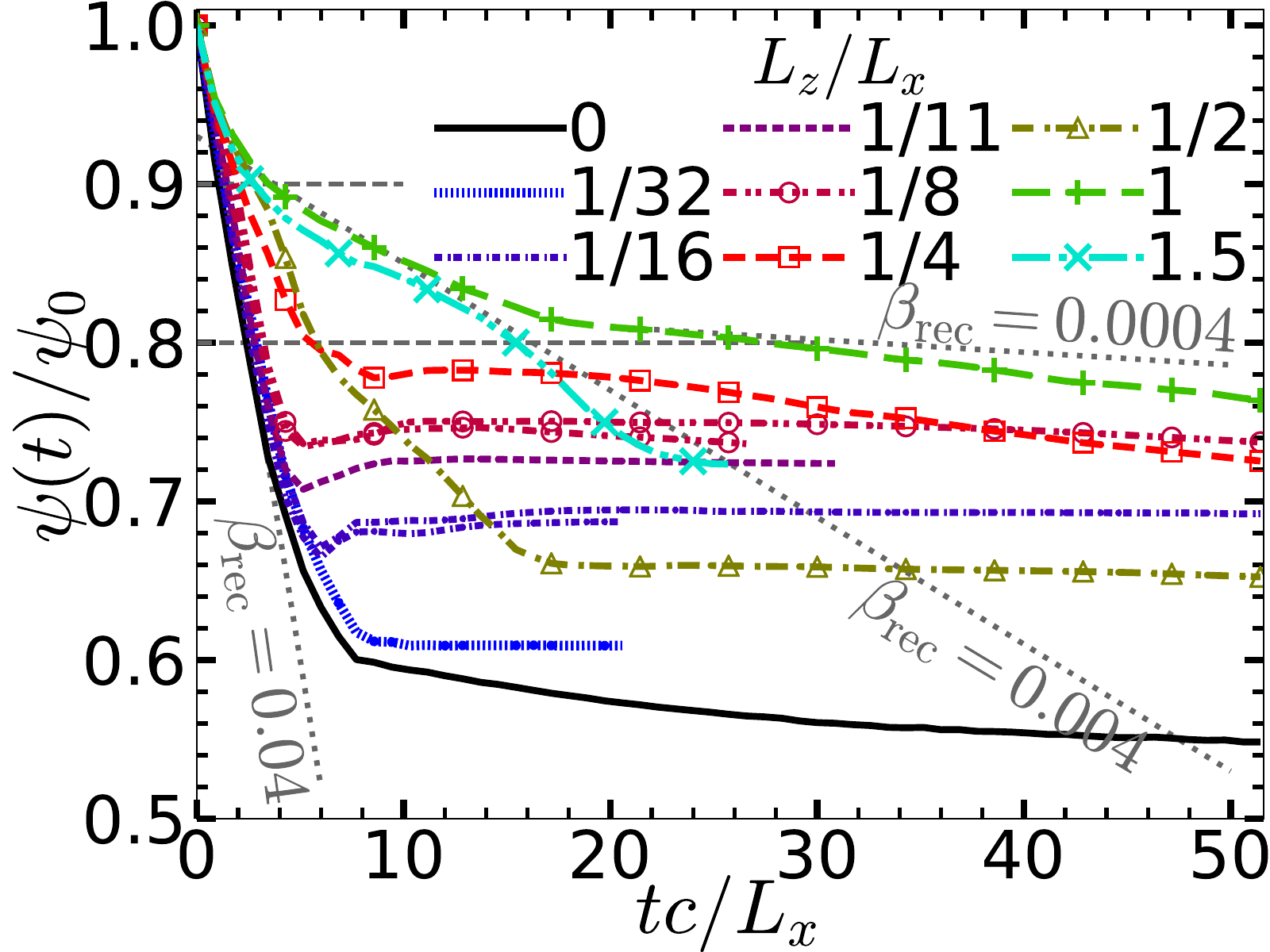}%
}
\caption{ \label{fig:energyAndFluxVsLz}
Total transverse magnetic energy $U_{Bt}(t)$ (left), magnetic energy in the layer $U_{Bt,\rm layer}(t)$ (middle), and unreconnected flux $\psi(t)$ (right),
for simulations with the same $L_x=341\sigma\rho_0$, $B_{gz}=0$, $\eta=5$, $\beta_d=0.3$, $a=0$, but varying $L_z/L_x$.
Cases $L_z/L_x=$1/16 and~1/8 each show two
simulations identical except for random initialization of particles.
In the plot of $U_{Bt}$ (left), the horizontal grey dashed line indicates $U_{Bt}=0.75U_{B0}$---NTPA spectra are shown in Fig.~\ref{fig:ntpaVsLz} when $U_{Bt}(t)$ crosses this line.
In the flux plot (right), the intersections of the two horizontal grey, dashed lines with $\psi(t)$ indicate where the average reconnection rate is measured for Fig.~\ref{fig:reconRateVsLz}; three grey, dotted lines show the slopes for constant reconnection rates~$\beta_{\rm rec}\in\{$0.04,0.004,0.0004$\}$ normalized to $cB_0$.
}
\end{figure}

We begin by looking at magnetic energy versus time in Fig.~\ref{fig:energyAndFluxVsLz} (we note that the total and transverse magnetic energy, $U_{B}$ and  $U_{Bt}$, are nearly the same; in all these cases,  $U_{Bz} < 0.004U_{B0}$).
In the left panel, we see that in 2D and nearly-2D (i.e., $L_z/L_x \ll 1$), $U_{Bt}(t)$ decays rapidly over $\sim 8 L_x/c$;
after this time, reconnection releases very little additional magnetic energy.
As $L_z$ increases, the initial rate of magnetic energy depletion
slows, but ultimately more magnetic energy is converted to 
particle energy---the case $L_z=L_x/2$ is an outlier, transferring
almost half the initial magnetic energy to particles (compared with
less than $0.3U_{B0}$ for 2D).
In these large-$L_z$ cases, we often see a stage of
relatively fast magnetic energy conversion ($\lesssim 5 L_x/c$),
followed by a later stage of much slower energy conversion.
Even after $50L_x/c$, slow energy conversion continues.

It is important to remember that, in the large-$L_z$ (3D) regime, stochastic variability can be considerable (cf.~\S\ref{sec:variability3d}).
Because of high computational costs, we were unable to run multiple simulations for each value of $L_z$; e.g., determining whether the dramatic energy release is caused by the special value of $L_z/L_x=1/2$ or is merely a statistical outlier, will have to be left to future studies.  In the meantime, we can have confidence only in clear, monotonic trends with $L_z$.
A possible reason for the increased stochastic variability in 3D was discussed in~\S\ref{sec:RDKIamp}, and we will explore this further at the end of this subsection.

The clearest signature of the effect of $L_z$ 
is the ``reconnected magnetic field energy''
(Fig.~\ref{fig:energyAndFluxVsLz}, middle), which (at any fixed time~$t$)
decreases nearly monotonically with $L_z/L_x$.
We interpret this as follows, as discussed previously 
in~\S\ref{sec:overview3d}.
In 2D, reconnection results in magnetic energy being stored in plasmoids; in the fully 2D simulation, the magnetic energy in the layer (predominantly in plasmoids) thus grows monotonically.
In 3D, reconnection can pump magnetic energy into plasmoids, but these magnetic structures are unstable (at least in the absence of appreciable guide field): they decay, converting their magnetic energy over time into plasma energy.
The almost-2D simulation with $L_z=L_x/32$ pumps nearly as much energy into plasmoids as the fully 2D simulation, but after $10L_x/c$ this energy
starts to decay slowly.
As $L_z/L_x$ becomes larger, the peak energy stored in plasmoids decreases.

We remind the reader that our measure of ``reconnected-field energy'' is the magnetic energy that is not in unreconnected field lines. It is thus an upper bound on the energy stored in structures formed from reconnected magnetic field; it could also contain, e.g., turbulent magnetic energy in the current layer.  

The unreconnected magnetic \emph{flux} (Fig.~\ref{fig:energyAndFluxVsLz}, right) shows the expected overall decay over time as flux is reconnected or annihilated.
At any fixed time~$t$, the remaining upstream flux tends to increase with $L_z/L_x$ (i.e., larger $L_z/L_x$ implies a slower reconnection rate)---with the striking exception of $L_z/L_x=1/2$, and the less striking exception of $L_z/L_x=3/2$.  Also, at very late time ($\simeq 40L_x/c$) we see the $L_z/L_x=1/4$ case overtaking $L_z/L_x=1/8$ in the amount of flux decay. 
This may represent a trend with increasing $L_z/L_x$ up to some minimally-3D value $L_z/L_x \sim 1/4$, above which results are independent of $L_z/L_x$ (but with high stochastic variability);
a statistical ensemble of simulations with large $L_z/L_x$ will be needed to distinguish the real trend.

\begin{figure}
\centering
\fullplot{
\includegraphics*[width=3in]{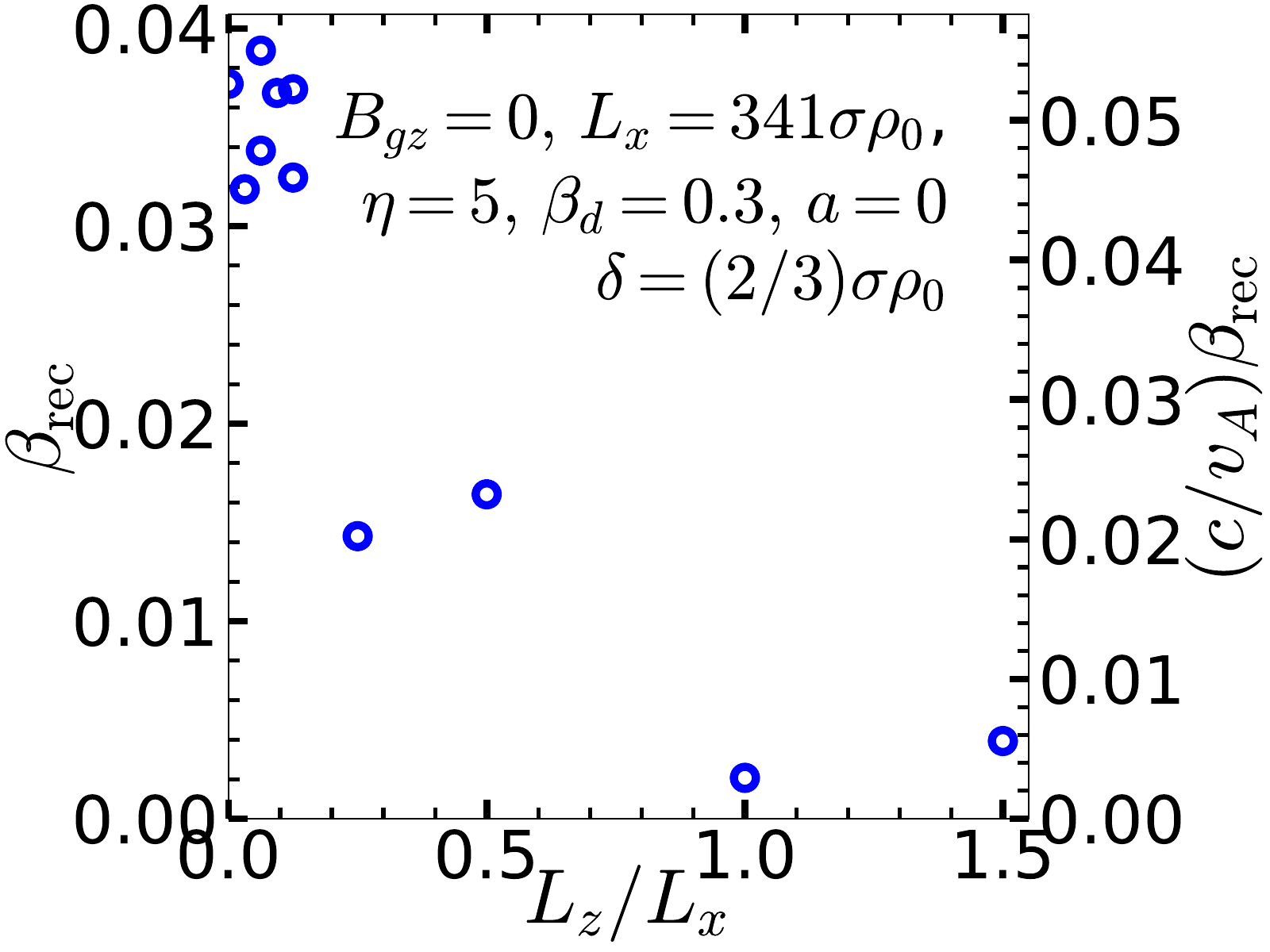}%
}
\caption{ \label{fig:reconRateVsLz}
Reconnection rates, normalized to $B_0 c$ on the left axis and to~$B_0v_A$ on the right (cf.~\S\ref{sec:unreconnectedFlux}), versus $L_z/L_x$, averaged over the period during which the upstream flux falls from~$0.9\psi_0$ to~$0.8\psi_0$ (between the horizontal grey dashed lines in Fig.~\ref{fig:energyAndFluxVsLz}, right),
for simulations with $L_x=341\sigma\rho_0$.
N.B. For 3D cases, the values of the instantaneous reconnection rate may vary significantly over the averaging time, and stochastic variability may also introduce significant uncertainty.
}
\end{figure}

Normalized to $B_0 c$, the reconnection rates $\beta_{\rm rec}$ (cf.~\S\ref{sec:unreconnectedFlux}) for all cases start out between 0.025--0.05, with larger $L_z/L_x$ already exhibiting slightly lower $\beta_{\rm rec}$; this is at very early times, probably dominated by the initial thin Harris sheet.
Within $10L_x/c$, the reconnection rates have all fallen below 0.01---because reconnection has finished in nearly-2D cases, and because 3D cases enter a stage of much slower reconnection. 
Figure~\ref{fig:reconRateVsLz} shows reconnection rates averaged
over the time for the unreconnected flux
to fall from $0.9\psi_0$ to $0.8\psi_0$, as a function of $L_z/L_x$ (we note that, for $L_z/L_x \leq 1/8$, $\psi(t_{0.8})=0.8\psi_0$ for $t_{0.8}<4L_x/c$).
For 3D simulations, the average rates should be used with caution, because the reconnection rate can change substantially during this time (cf. Fig.~\ref{fig:energyAndFluxVsLz}, right).
In any case, it shows clearly how much slower 3D reconnection can be.
At later times, the cases $L_z/L_x\geq 1/4$ eventually exhibit reconnection rates $\beta_{\rm rec}\sim 10^{-3}$ (this measurement does not necessarily guarantee that 2D reconnection is actually taking place; flux could be depleted by direct annihilation as well as 2D-like reconnection---cf.~\S\ref{sec:terminology},~\ref{sec:overview3d}).
It is notable that by $\gtrsim 30L_x/c$, e.g., the $L_z/L_x=1$ simulation has ``used up'' only about half the upstream flux as the 2D simulation, but has converted more magnetic energy to particle energy, because in 2D more magnetic flux and energy remains in plasmoids.

\begin{figure}
\centering
\fullplot{
\includegraphics*[width=0.49\textwidth]{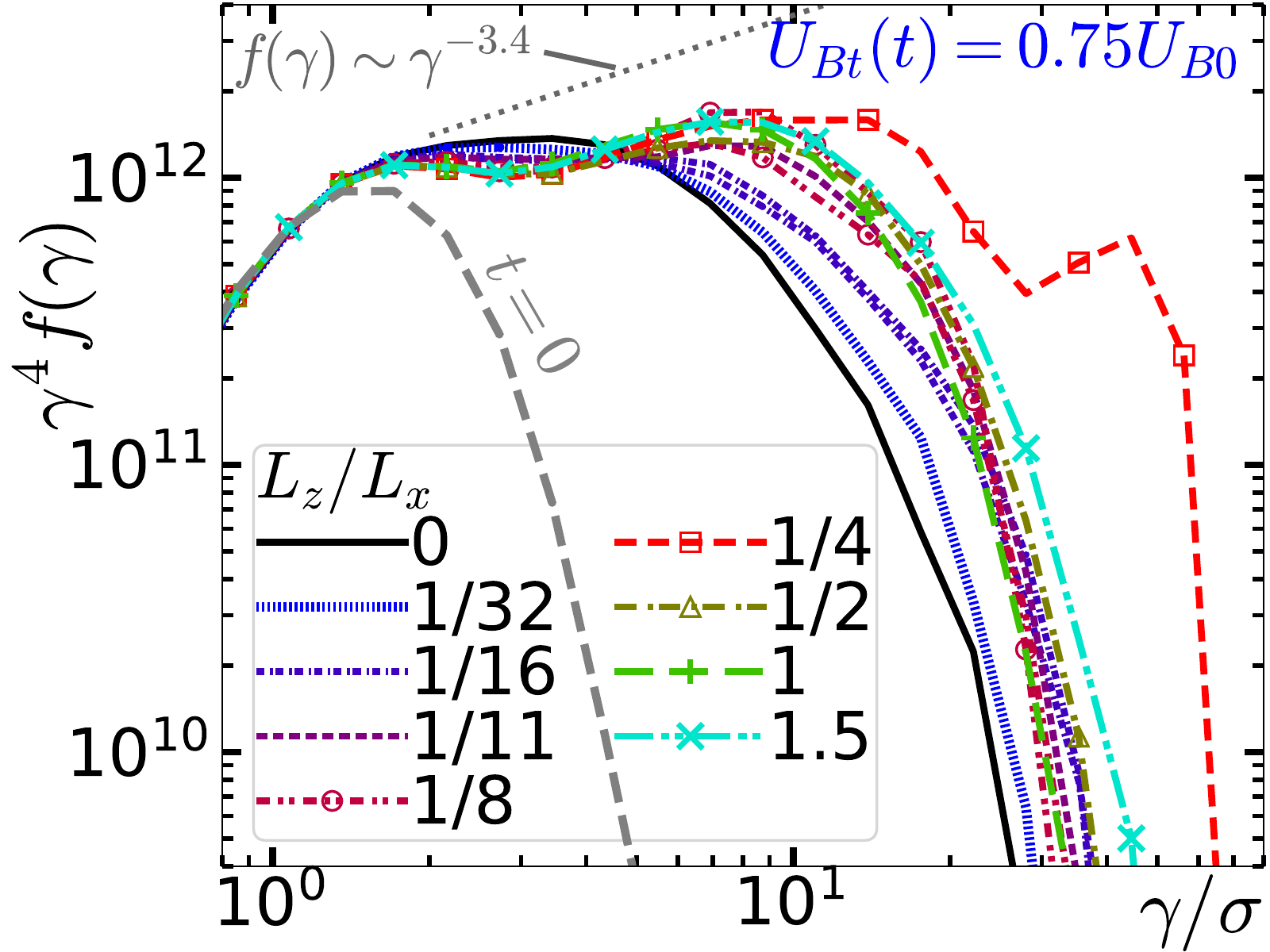}%
\hfill
\includegraphics*[width=0.49\textwidth]{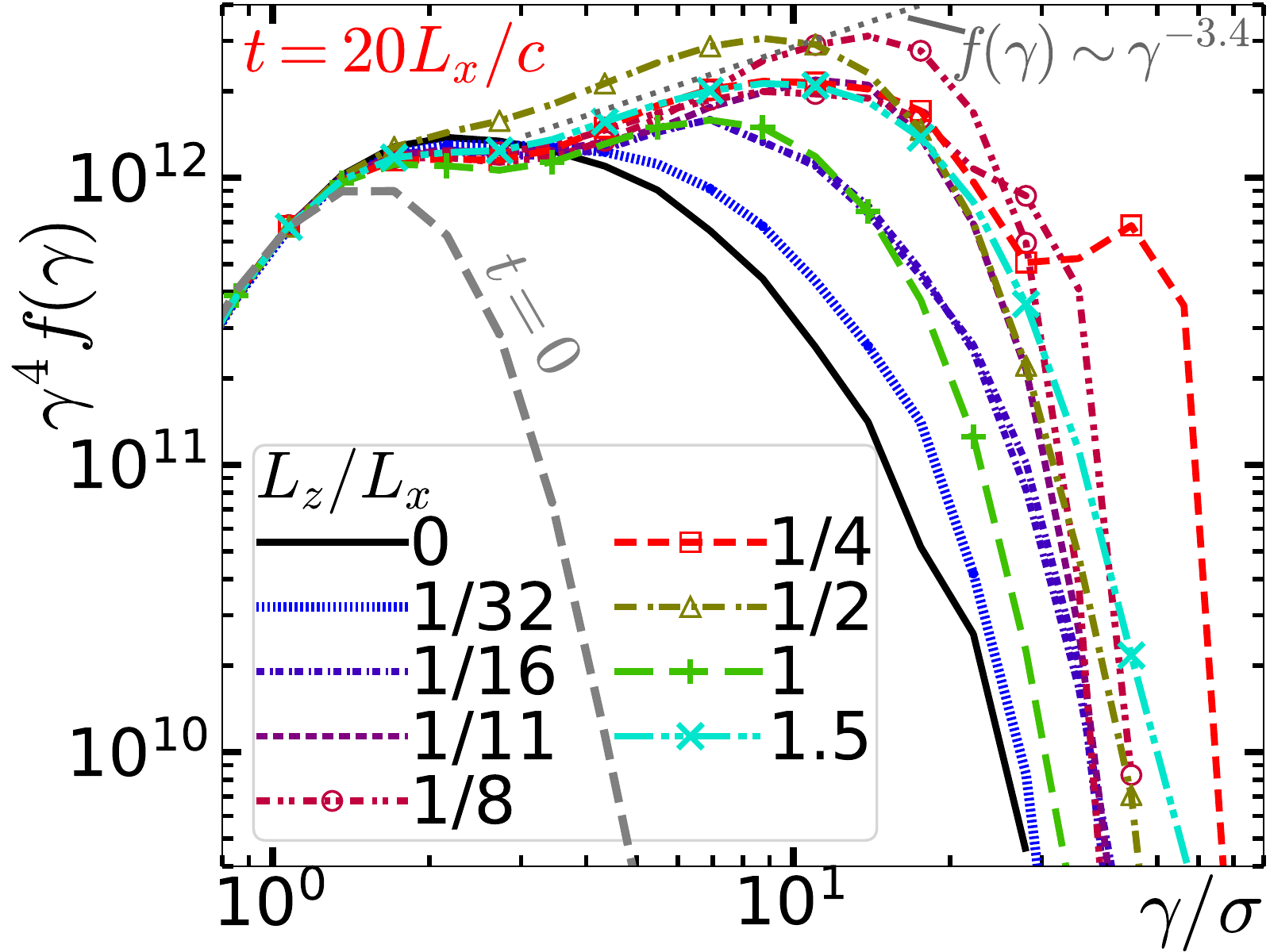}%
}
\caption{ \label{fig:ntpaVsLz}
Particle energy spectra (compensated by $\gamma^4$) for different
$L_z/L_x$, for $L_x=341\sigma\rho_0$, at the time when 25~per~cent of
the initial magnetic energy has been depleted (left---and cf. Fig.~\ref{fig:energyAndFluxVsLz}, left), and at 
time $t=20L_x/c$ (right).
The spectra are rather similar (especially on the left), except possibly for the $L_z=L_x/3$ case, but
there is a rough trend of increasing numbers of high-energy particles as
$L_z/L_x$ increases.
Dotted grey lines show the slope of the power law $f(\gamma)\sim \gamma^{-3.4}$; local power-law indices in the high-energy spectrum range from roughly~4.2 to~3.4.
}
\end{figure}

\begin{figure}
\centering
\fullplot{
\includegraphics*[width=0.49\textwidth]{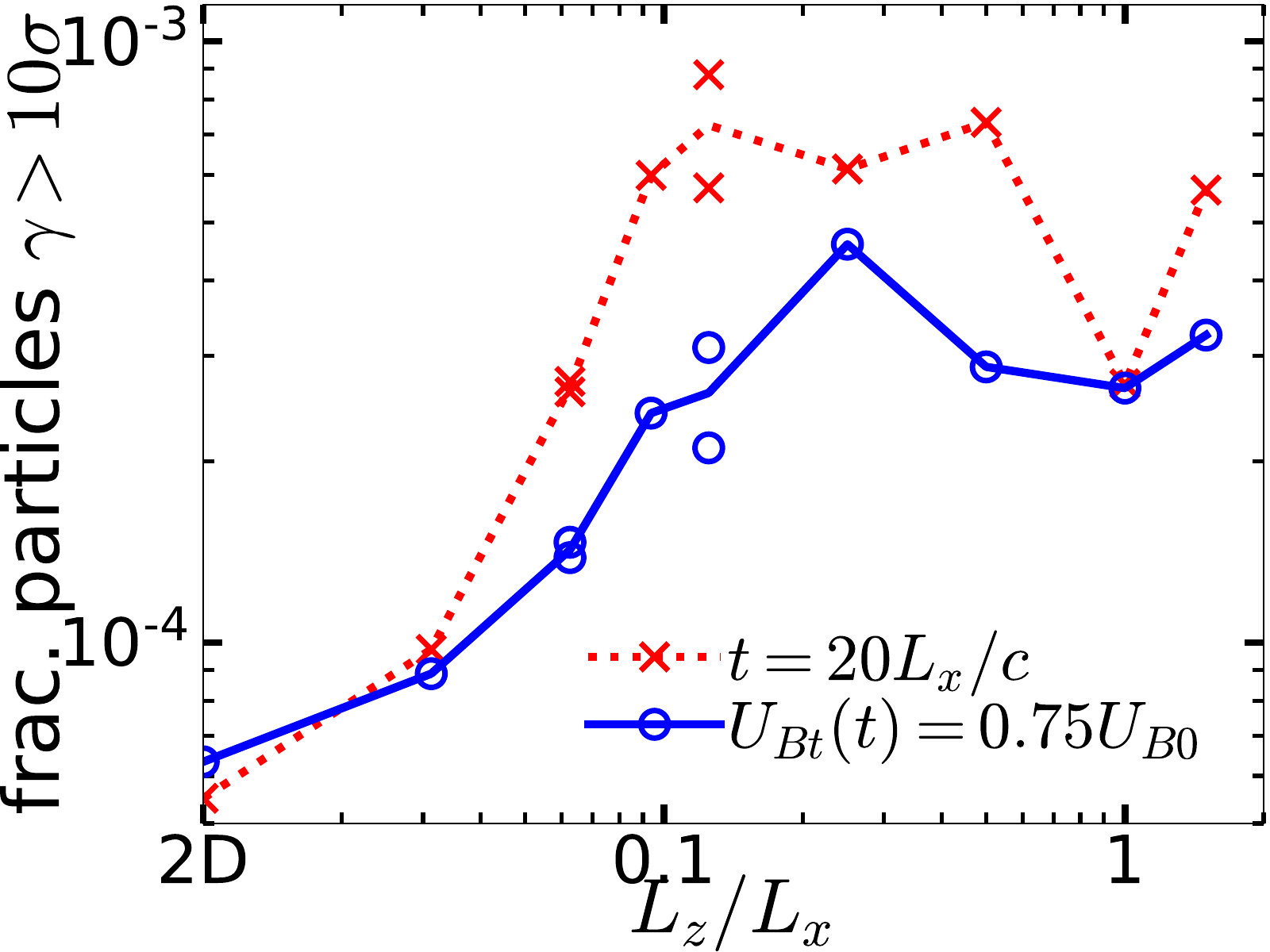}%
}
\caption{ \label{fig:ntpaFracVsLz}
The fraction of particles with Lorentz factor $\gamma>10\sigma$ for the spectra shown in Fig.~\ref{fig:ntpaVsLz}---blue circles for spectra at time $t$ such that $U_{Bt}(t)=0.75U_{B0}$, and
red x's for spectra at $t=20L_x/c$.
Although the trend with~$L_z/L_x$ appears fairly robust, the significance of these results must be carefully considered due to the steep power law.
For example, increasing the threshold from~$10\sigma$ to~$12\sigma$ lowers the fraction of threshold-exceeding particles by a factor of~4--6 (for all $L_z/L_x$).
}
\end{figure}

Figure~\ref{fig:ntpaVsLz}(left) shows
particle energy spectra at times when 25~per~cent of the initial
magnetic energy has been depleted---i.e., when $U_{Bt}(t)=0.75U_{B0}$; the times can be estimated from the intersection of $U_{Bt}(t)$ and the horizontal grey dashed line in Fig.~\ref{fig:energyAndFluxVsLz}(left), and range from $t=$5--8$L_x/c$ for $L_z/L_x\leq 0.5$, and $t=15$--$18L_x/c$ for $L_z/L_x\in\{1,1.5\}$.
We see first that all these
spectra are fairly similar, especially for $\gamma \lesssim 10\sigma$;
and even above $\gamma \gtrsim 10\sigma$, spectra vary by less than a
factor of 2 in energy, with the exception of $L_z/L_x=1/4$, which 
accelerates some particles to unusually high energies (we note that
a different case,
$L_z/L_x=1/2$, is the outlier in total magnetic energy depletion).
Second, however, there is a trend of increasing 
numbers of high-energy particles (say with $\gamma\simeq 10\sigma$)
as $L_z/L_x$ increases, at least for small $L_z/L_x$.
Looking at spectra at the same time, $20L_x/c$, we see differences
enhanced somewhat, with competing effects: larger $L_z/L_x$ yields
slightly more high-energy particles, but larger amounts of magnetic
energy conversion also yield more high-energy particles.

Although the somewhat short range of the high-energy power-law section prevents us from measuring precise power-law indices in a useful way, we can roughly characterize the local power-law indices as varying between~4.2 and~3.4 (in Fig.~\ref{fig:ntpaVsLz}, horizontal represents a power law $\gamma^{-4}$, and dotted grey lines show $\gamma^{-3.4}$).  Whereas simulations with $L_z/L_x\leq 1/8$ have power-law indices closer to~4, those with larger $L_z/L_x\geq 1/4$ appear to exhibit harder power laws with indices around 3.4 in the range 
$3 \lesssim \gamma/\sigma \lesssim 8$---however these same simulations also have slightly steeper power-law indices around 4.2 in $2 \lesssim \gamma \lesssim 3$.  
An alternative measure of NTPA efficiency is the fraction of particles accelerated beyond some high energy, e.g., $\gamma=10\sigma$.  Figure~\ref{fig:ntpaFracVsLz} shows that this fraction increases significantly with $L_z/L_x$ until saturating around $L_z/L_x\approx 0.1$; thus 3D simulations accelerate almost an order of magnitude more particles beyond $\gamma=10\sigma$.

Enhancement of electron NTPA in 3D, relative to 2D reconnection, has been previously observed in 
subrelativistic electron-ion plasma with $B_{gz}=0$ as well as with guide field $B_{gz}/B_0\leq 1.5$
\citep{Dahlin_etal-2015,Dahlin_etal-2017,Li_etal-2019}, and has just recently been seen in pair plasma with $\sigma_h=10$, $B_{gz}/B_0=0.1$ \citep{Zhang_etal-2021arxiv};
high-energy electrons were trapped by plasmoids in 2D, but not in 3D, thus allowing untrapped electrons to experience more acceleration in 3D.
In the subrelativistic case, there is some disagreement over whether the high-energy spectra in question are power laws or exponentials; a steep power law with index $p\gtrsim 4$ over a relatively small range in~$\gamma$ can look very much like an exponential decay in~$\gamma$.
An apparently dramatic $f_{3D}(\gamma)/f_{2D}(\gamma)=10$ \citep[measured at very high energies by][]{Dahlin_etal-2017} can be interpreted as particles gaining a less dramatic 20 or 30 per~cent more energy in 3D than in 2D.
In the more relativistic $\sigma_h=10$ case, the shallower/harder power law around $p\sim 2$ shows NTPA more clearly, and the highest-energy particles in~3D gain about 2--3 times more energy than in~2D \citep{Zhang_etal-2021arxiv}.
In our simulations we find, whether comparing 2D and 3D simulations at the same time or at the same $U_{Bt}/U_{Bt0}$, that usually $0.1 < f_{3D}(\gamma)/f_{2D}(\gamma) < 10$, and the ratio is often quite close to 1, especially when comparing the spectra at the same $U_{Bt}/U_{Bt0}$.
Although these spectra are power laws, they are steep power laws
[$f(\gamma)\sim \gamma^{-4}$] and so again even a substantial $f_{3D}/f_{2D}$ may imply only a rather modest increase in particle energy.
E.g., if $f_{2D}(\gamma)=a \gamma^{-4}$ for some constant $a$, and
$f_{3D}(\gamma) = 10 f_{2D}(\gamma)$, we can write
$f_{3D} \approx a (\gamma/1.78)^{-4}$---particles in 3D need only gain 78 per~cent more energy in 3D than in 2D to explain $f_{3D}/f_{2D}=10$.

Our results are not inconsistent with increased NTPA in 3D, but 
from our perspective, the differences in NTPA between 2D and 3D, compared at the same $U_{Bt}/U_{Bt0}$, are fairly insignificant, on the order of stochastic variation (regardless of guide field, as we shall see in~\S\ref{sec:Bz3d}).
3D simulations convert magnetic energy less rapidly than 2D simulations, and so the building-up of the high-energy part of $f(\gamma)$ occurs more slowly; 
however, as 3D simulations convert more magnetic energy to particle energy over long times (compared with 2D), they ultimately accelerate more particles than 2D simulations.
Therefore, NTPA is enhanced in 3D relative to 2D, but the shape of the high-energy power-law spectrum and cut-off (up to an overall normalization factor) does not differ significantly.

\begin{figure}
\centering
\fullplot{
\includegraphics*[width=0.65\textwidth]{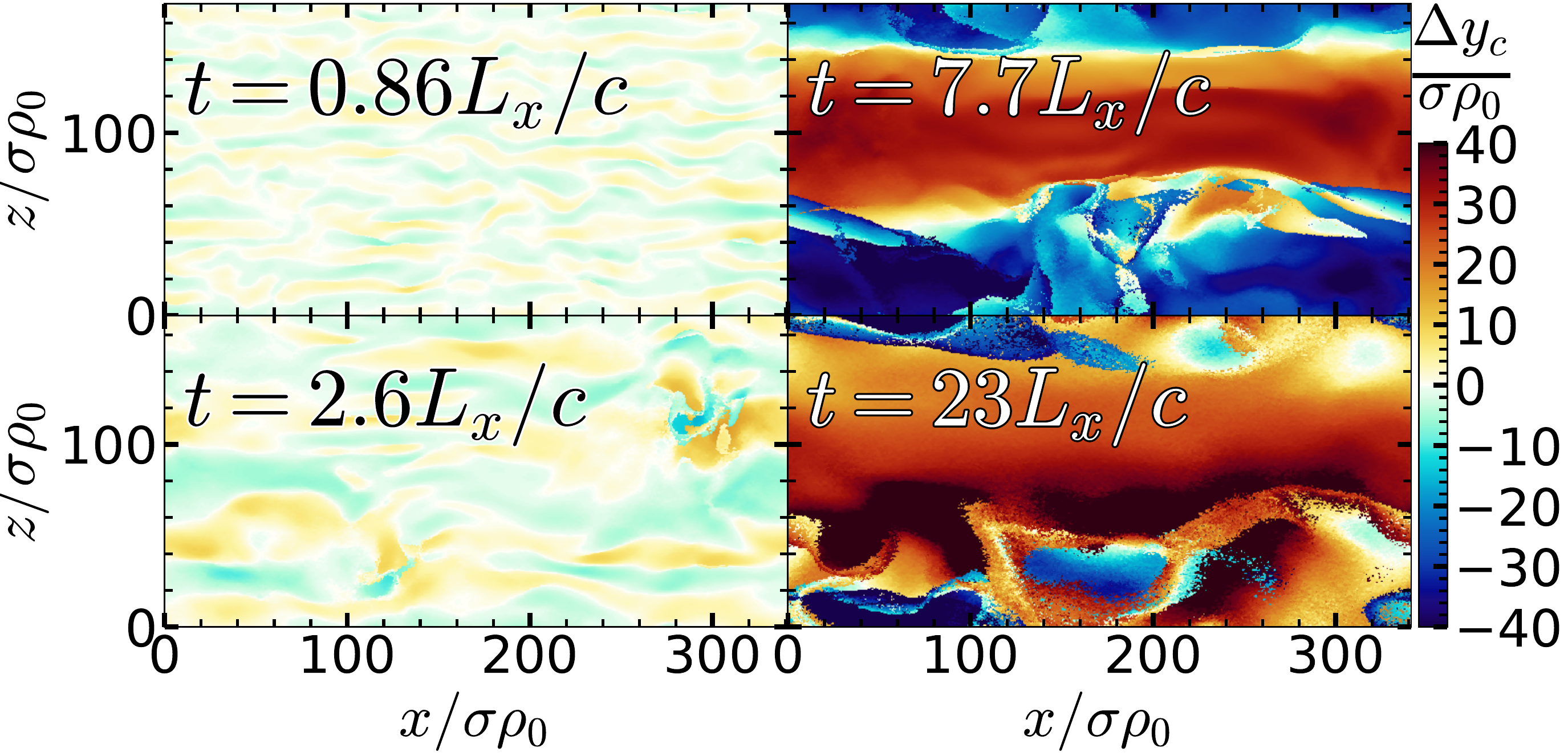}%
\hfill
\includegraphics*[width=0.33\textwidth]{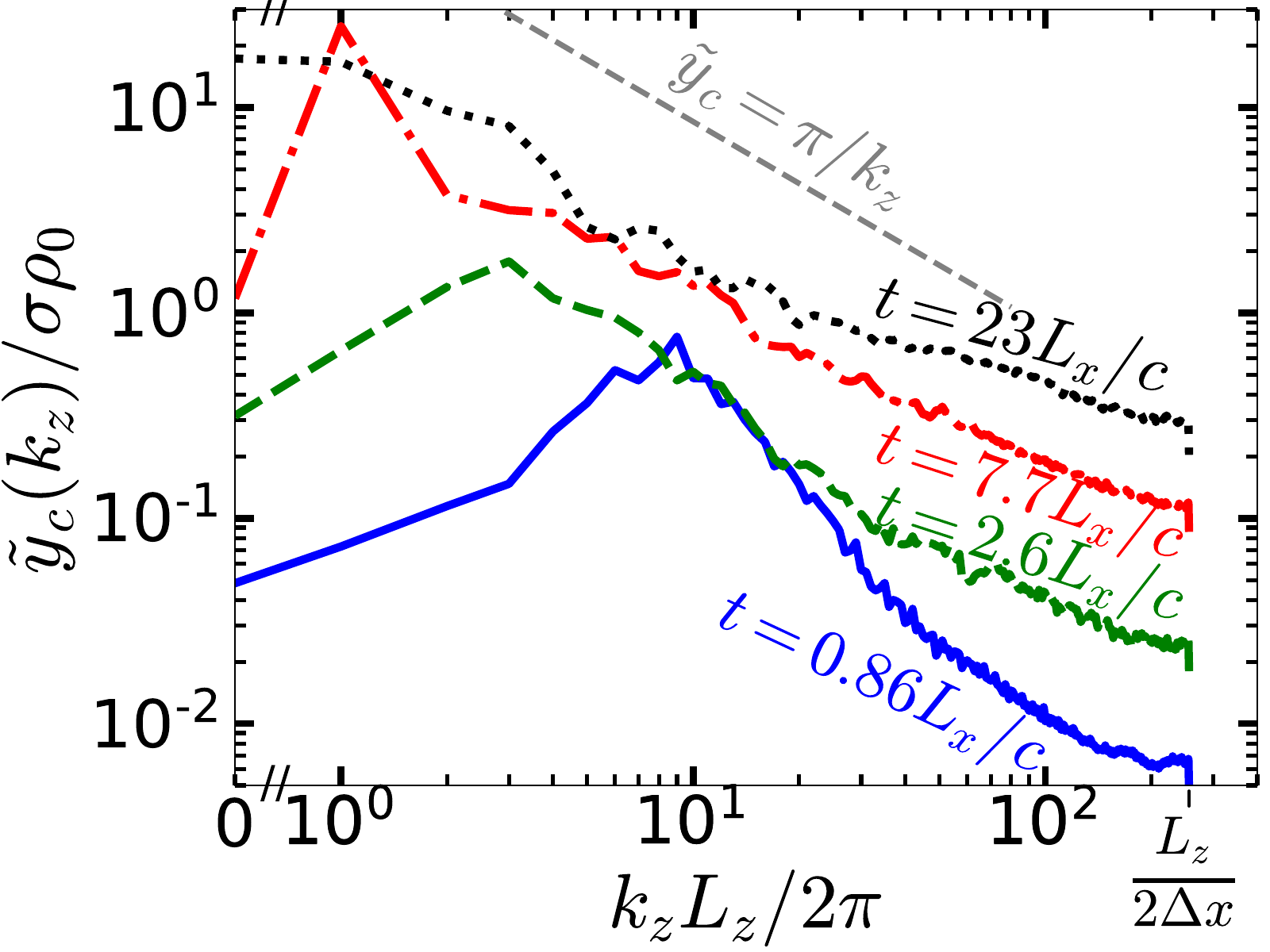}
}
\caption{ \label{fig:layerOffsetY}
(Left) The layer central surface displacement from the original midplane, $\Delta y_c(x,z)$ (cf.~\S\ref{sec:sheetCenter}), at four times, $tc/L_x\in \{$0.86, 2.6, 7.7, 23$\}$, shows low-amplitude short-wavelength RDKI modes dominating at early times, but at later times the dominant mode has $\lambda_z=L_z$.  At $7.7L_x/c$ (and thereafter), abrupt spatial variation in $\Delta y_c$ indicates the layer folding over on itself.
(Right) Fourier spectra magnitudes (cf.~\S\ref{sec:sheetCenter}) of the data shown on the left, averaged over all $x$ (after taking the magnitudes), show that over time, longer-wavelength RDKI modes grow to larger amplitudes.
}
\end{figure}

We have remarked several times on the stochastic variability in 3D reconnection (cf.~\S\ref{sec:variability3d},~\ref{sec:pertAndEta3d}) and have discussed a mechanism for large variability in~\S\ref{sec:RDKIamp}.
When an RDKI mode grows beyond an amplitude $\Delta y_{c*}$ on the order of its wavelength $\lambda_z$, it becomes extremely nonlinear, and the current layer folds over on itself and rapidly depletes magnetic energy within $|y|\lesssim \Delta y_{c*}$.
In Fig.~\ref{fig:energyAndFluxVsLz}, the simulation with $L_z/L_x=1/2$ experienced particularly large energy depletion, converting significantly more magnetic energy to plasma energy in~$20L_x/c$ than any other simulation did in~$50L_x/c$, including simulations with $L_z/L_x=1/4$ and~$1$.
Indeed, this simulation ($L_z=L_x/2$) developed a large-amplitude RDKI mode with the largest-possible wavelength, $\lambda_z=L_z$, and it became highly nonlinear, rapidly depleting magnetic energy in a thick layer.
Figure~\ref{fig:layerOffsetY}(left) shows the offset of the layer central surface (\S\ref{sec:sheetCenter}), 
$\Delta y_c(x,z,t) \equiv y_c(x,z,t)-y_c(x,z,0)$, 
at four times: $tc/L_x\in\{$0.9, 2.6, 7.7, 23$\}$;
then Fig.~\ref{fig:layerOffsetY}(right) shows the Fourier spectrum (in $z$, averaged over $x$; cf.~\S\ref{sec:sheetCenter}) of this quantity for each time.
At early times, $t=0.9L_x/c$, RDKI develops prominently at $\lambda_z\approx L_z/10 = 17 \sigma\rho_0$.
At later times, the most prominent wavelength gets longer; at $t=7.7L_x/c$, rippling with $\lambda_z=L_z$ is the dominant component.
Indeed, by this time, there are clear signs of nonlinear development and the layer folding over on itself (causing rapid magnetic energy conversion).
By $t=23L_x/c$, the layer has become generally thick and turbulent with weak magnetic field, and the long-wavelength rippling with $\lambda_z=L_z$ is no longer so apparent.

According to \citet{Zenitani_Hoshino-2007}, the wavelength of the fastest-growing (linear) RDKI mode is $\lambda_{z,\rm RDKI} = 16\upi \gamma_d \beta_d^2 \delta = 4.7\delta = 3\sigma\rho_0$.
However, this fastest-growing mode should saturate at an amplitude $\Delta y_{c*}\sim \lambda_{z,\rm RDKI}/2$ (cf.~\S\ref{sec:RDKIamp}), whereas longer-wavelength modes can grow (though possibly at a slower rate) to larger amplitudes.
Indeed, Fig.~\ref{fig:layerOffsetY} is very roughly consistent with longer-wavelength modes growing slower, but to larger amplitudes, $\tilde{y}_c(k_z) \sim k_z^{-1}$ (where $k_z=2\upi/\lambda_z$).
For some reason, in the simulation with $L_z=L_x/2=170\sigma\rho_0$, the mode with $\lambda_z=L_z$ grew sufficiently to enter highly-nonlinear development; presumably in simulations with other $L_z/L_x$, this did not happen.
Of course, for $L_z<L_x/2$, no mode can develop to such large amplitude; but the simulations with $L_z=L_x$ and $L_z=3L_x/2$ could have developed $\lambda_z=L_z/2$ as well as even longer-wavelength modes (and indeed, they did, to some extent---but for some reason the resulting magnetic energy conversion was not as dramatic).
We speculate that beyond some wavelength (in sufficiently large systems), long-wavelength RDKI modes will simply not have time to grow before being disrupted by reconnection or faster-growing, shorter-wavelength RDKI modes.
We leave it to future work to investigate what determines the nonlinear development of long-wavelength RDKI modes; we expect this to be a difficult task because we have already seen that stochastic variability can play a large role.

In summary, in 3D (i.e., for large $L_z/L_x$), with zero guide field and zero initial perturbation, reconnection and magnetic energy 
conversion occur more slowly (than in 2D) during the first active reconnection stage; for large $L_z/L_x$ almost all the 
released upstream/unreconnected magnetic energy is converted to particle energy
(whereas in 2D, roughly half of it is trapped in reconnected magnetic
field).
Whereas 2D simulations progress to a final state limited by the magnetic field configuration, with magnetic energy and flux trapped in plasmoids, 3D current sheet evolution does not have a clear endpoint (at least not before $50L_x/c$), but enters a long-lasting second stage in which gradual magnetic energy and upstream flux depletion continue at a slower pace.
RDKI modes can play a dramatic role in 3D reconnection---and perhaps it is even misleading to refer to ``3D reconnection'' in some cases where RDKI drives the most rapid magnetic energy conversion.
The most rapid and largest magnetic energy releases seem to be related to a long-wavelength RDKI mode developing nonlinearly when its amplitude reaches the order of its wavelength, causing extreme distortion of the current layer that quickly depletes magnetic energy within the mode amplitude.
Whether this happens at a given wavelength depends on $L_z$, but also has a random element, which explains the stochastic variability of 3D ``reconnection.''
If nonlinear RDKI does cause dramatic magnetic energy conversion, the resulting layer becomes thick and turbulent, and allows only much slower magnetic conversion thereafter, presumably because of its increased thickness.

Importantly, despite all the complications in 3D, with potentially very different mechanisms driving magnetic energy conversion to plasma energy, NTPA efficiency does not suffer in 3D and actually appears to
be slightly enhanced, although it may occur over somewhat longer times
because of the slower reconnection rate. 
The increase in the number of high-energy particles does depend on the rate of magnetic energy conversion, but for a fixed amount of energy gained by the plasma, $f(\gamma)$ is remarkably insensitive to $L_z$ (as well as to the initial current sheet configuration, as we saw previously).
However, in 3D, more magnetic energy can be converted to plasma energy than in 2D, and ultimately this leads to more particles at the highest (nonthermal) energies.

\subsection{3D reconnection: guide magnetic field}
\label{sec:Bz3d}

In this subsection we investigate 3D reconnection with different initial guide magnetic fields, $B_{gz}/B_0\in \{$0, 0.1, 0.25, 0.5, 0.75, 1$\}$,
using system size $L_z=L_x=256\sigma\rho_0$.
Other parameters are as usual: zero initial perturbation, $\eta=5$, $\beta_d=0.3$, $\delta=(2/3)\sigma\rho_0$.
The dependence of 3D reconnection on the guide magnetic field $B_{gz}$ is nontrivial and non-monotonic as a result of competing effects.
We find that the effect of guide magnetic field in 3D can be roughly summarized thus:
stronger $B_{gz}$ suppresses 3D effects, so that reconnection becomes more 2D-like;
this may be expected, because guide field suppresses 3D instabilities like RDKI \citep{Zenitani_Hoshino-2008}.
Stronger guide field promotes particle transport in $z$ (relative to perpendicular directions) by aligning $\boldsymbol{B}$ increasingly parallel to $z$, and this tends to even out variations in $z$ (relative to perpendicular directions).
Again, however, as with $B_{gz}=0$, we find that NTPA continues to be very similar in 2D and 3D for $B_{gz} > 0$.

In 2D, the added inertia of the guide field, being dragged along with the plasma, slows reconnection significantly when the guide field enthalpy density, $B_{gz}^2/4\upi$ becomes comparable to or larger than the upstream plasma enthalpy density, $h=4 \theta_b n_b m_e c^2$ (expressed here in the ultrarelativistic limit), i.e., when
$B_{gz}^2/B_0^2 \gtrsim 1/\sigma_h$ \citep[cf.~\S\ref{sec:Bz2d} and][]{Liu_etal-2015}.
The guide field enthalpy appears to slow reconnection in 3D as well.
However, because other 3D effects also slow reconnection, the fundamentally 2D effect of guide field enthalpy is not noticeable at low guide fields;
at low guide field, $B_{gz}/B_0 \lesssim 1/\sigma_h$, the guide field actually promotes reconnection by suppressing 3D effects that slow reconnection.

\begin{figure}
\centering
\fullplot{
\includegraphics*[width=0.49\textwidth]{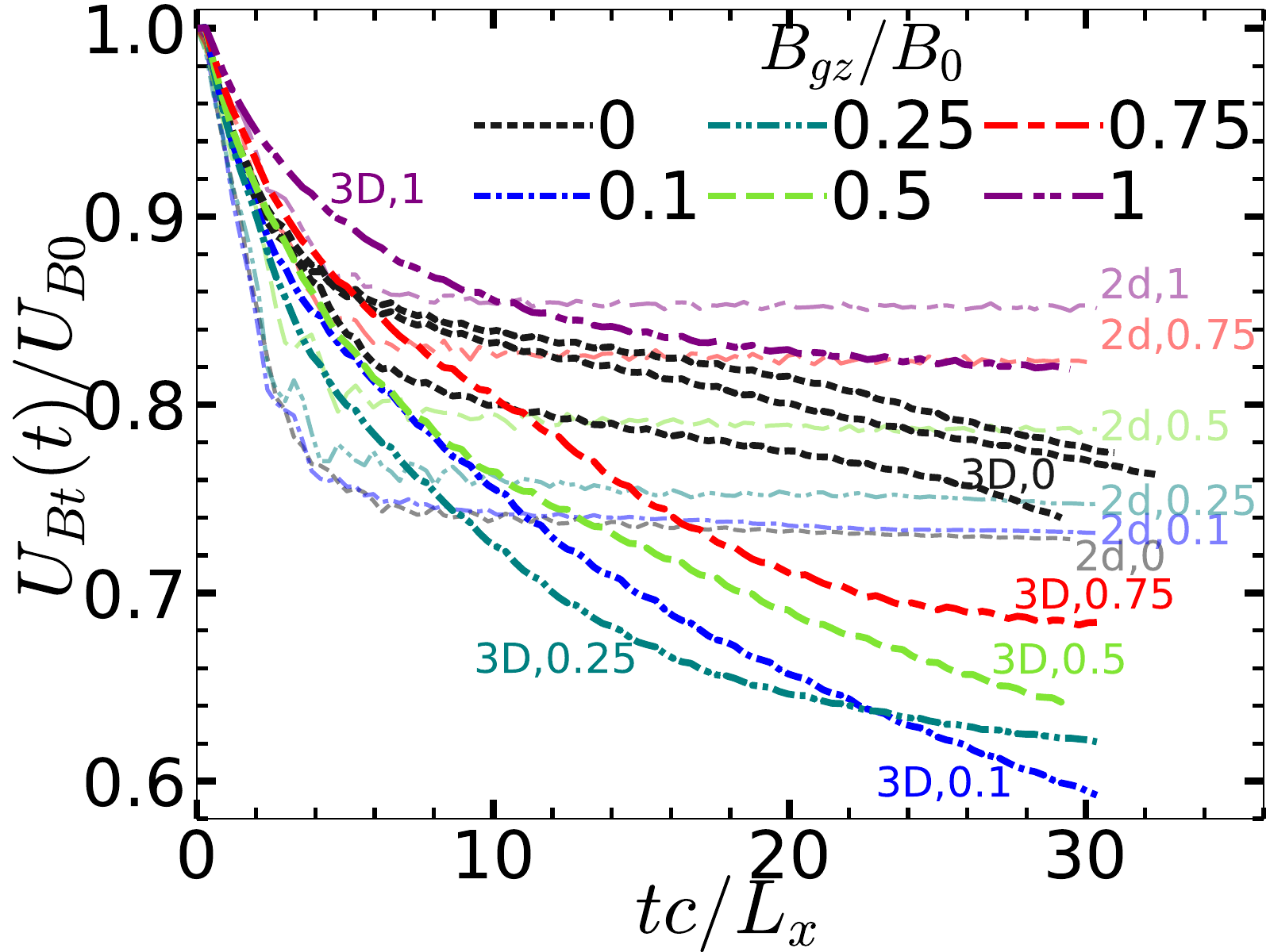}%
\hfill
\includegraphics*[width=0.49\textwidth]{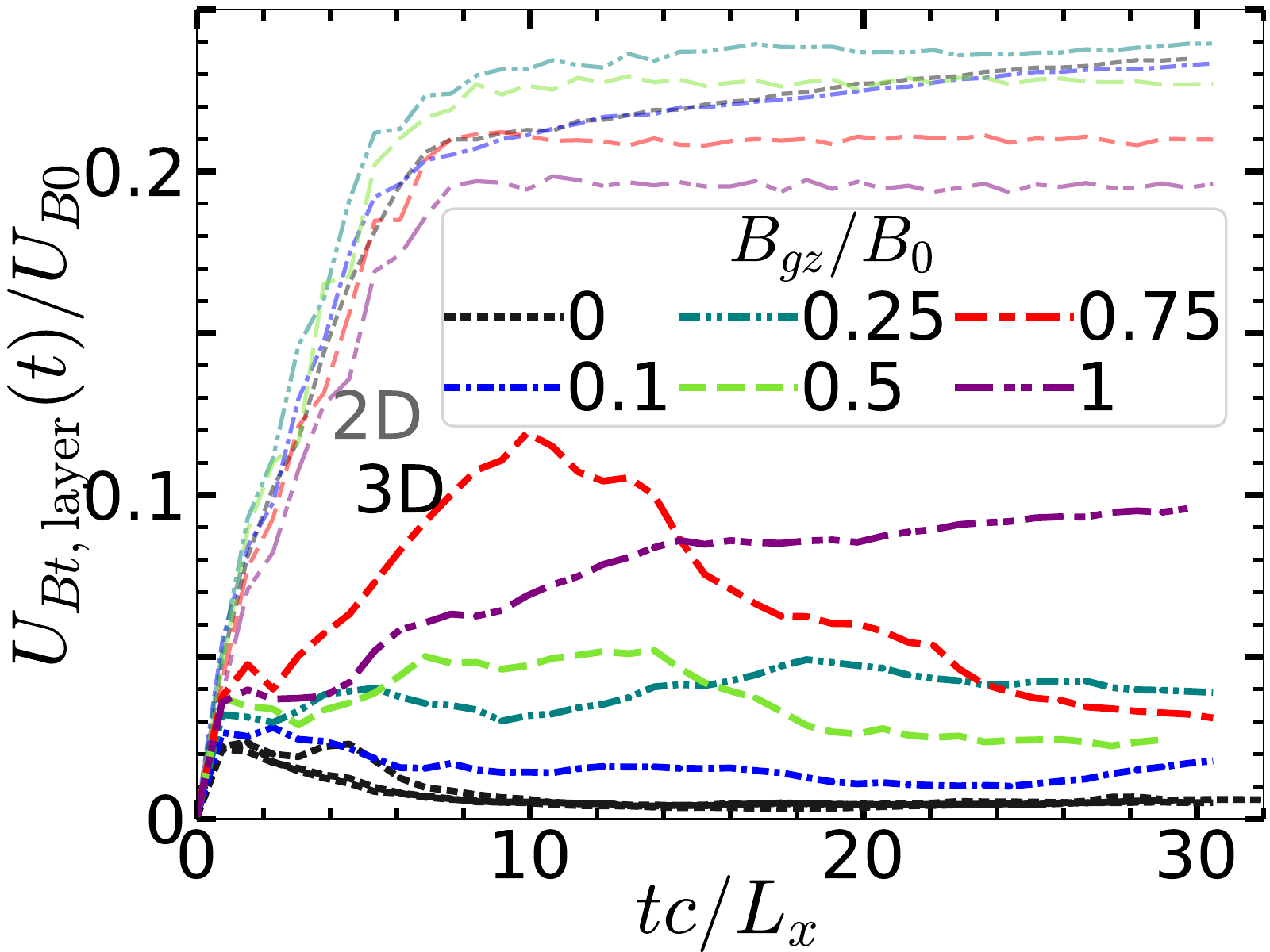}
}
\caption{ \label{fig:energyVsBgz3d}
Transverse magnetic energy $U_{Bt}(t)$ in the entire simulation (left) and $U_{Bt,\rm layer}(t)$, restricted to the layer (right), for both 2D (fainter, thinner lines) and 3D ($L_z=L_x$, thicker lines) simulations, for a range of
guide fields $B_{gz}$.  
Three 3D simulations are shown for $B_{gz}=0$.
On the left plot, lines are individually labelled
with dimensionality (``2d'' or ``3D'') and $B_{gz}/B_0$.
These simulations have size~$L_x=256\sigma\rho_0$.
}
\end{figure}

Figure~\ref{fig:energyVsBgz3d}(left) shows the time evolution of transverse magnetic energy $U_{Bt}(t)$ in 3D, for all six guide-field values (2D results are shown in fainter colours on the same graph, for comparison).
For $B_{gz}=0$, results from three 3D simulations are shown to give a rough estimate of stochastic variability, and all three exhibit relatively slow magnetic energy conversion.
Increasing $B_{gz}$ even to $B_{gz}=0.1B_0$ speeds up reconnection (presumably by suppressing 3D instabilities) and results in significantly more magnetic energy conversion over the first 20--30$L_x/c$; by $B_{gz}=B_0/4$, this trend saturates as the guide field enthalpy starts to suppress reconnection (this is more noticeable at later times).  Somewhere around $B_{gz}\gtrsim 0.75B_0$ (and clearly for $B_{gz}=B_0$), the guide field is suppressing reconnection as in 2D, so that for $B_{gz}=B_0$, 3D reconnection is slower than for $B_{gz}=0$ at all times.
We should keep in mind, however, that without guide field, stochastic variability can significantly affect system evolution, including energy dissipation (cf.~\S\ref{sec:variability3d}), and further work should investigate ensembles of simulations with different guide fields.

We argue that the enhancement of reconnection by moderate $B_{gz}$ is a consequence of reconnection becoming more 2D-like, stabilizing flux ropes and preventing 3D instabilities from interfering.
This is supported by Fig.~\ref{fig:energyVsBgz3d}(right),
which shows the magnetic energy in the layer versus time, $U_{Bt,\rm layer}(t)$.
In~\S\ref{sec:Lz3d} we showed (for $B_{gz}=0$) that for 2D simulations, $U_{Bt,\rm layer}(t)$
increases monotonically until it saturates at a significant fraction
of the initial transverse magnetic energy, $U_{Bt0}$; whereas, as 
$L_z/L_x$ is increased and 3D effects become important, $U_{Bt,\rm layer}$ increases only a little before decaying away.
Here we see the same thing in 3D: for $B_{gz}=0.1B_0$, $U_{Bt,\rm layer}(t)$ grows a little and then decays over time;
as $B_{gz}$ becomes stronger, 
$U_{Bt,\rm layer}(t)$ becomes more 2D-like, increasing to a larger value and (for $B_{gz}/B_0=1$) saturating at a roughly constant value (because the guide field stabilizes flux ropes containing the reconnected field).

\begin{figure}
\centering
\fullplot{
\includegraphics*[width=0.49\textwidth]{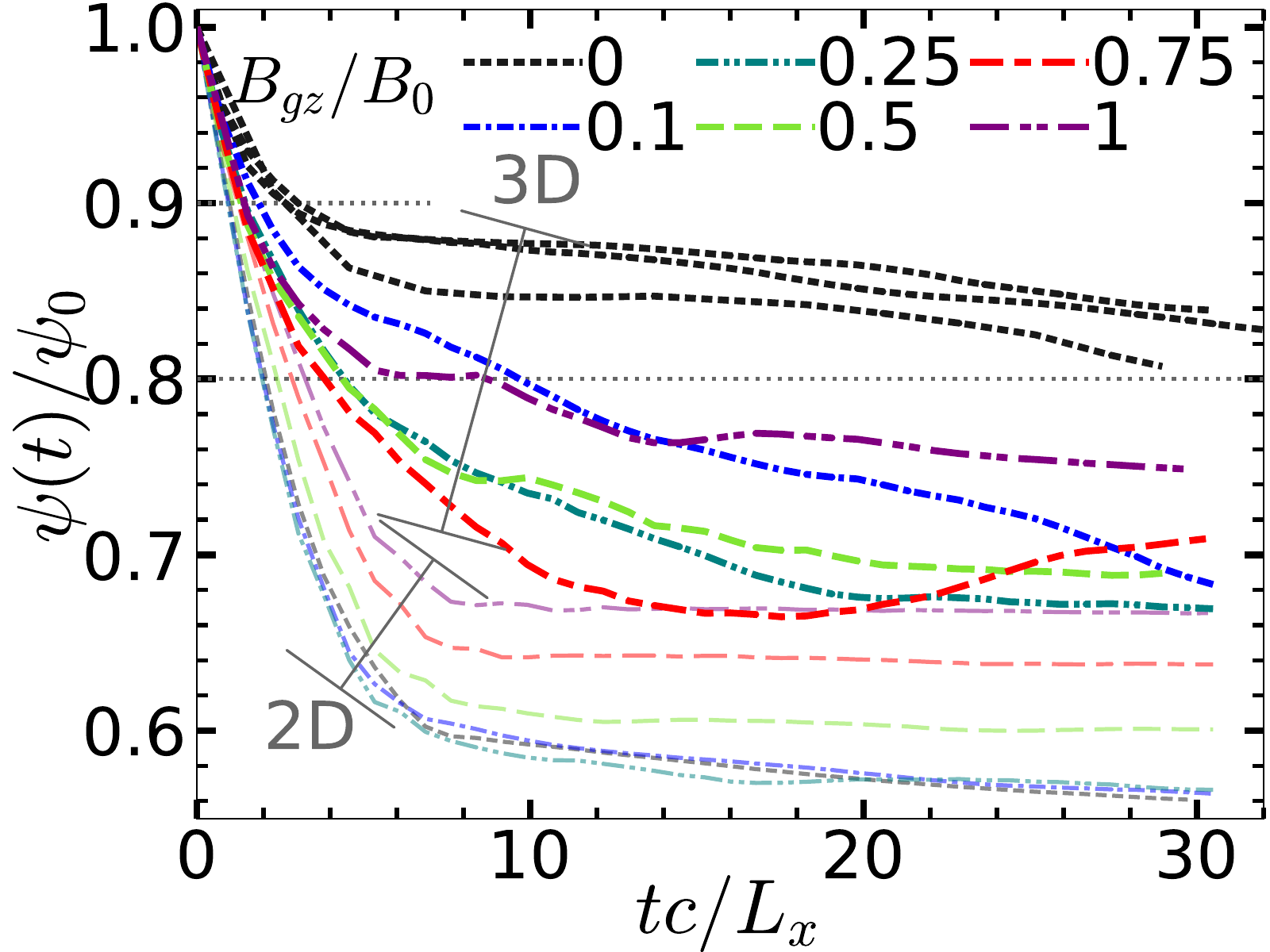}%
\hfill
\includegraphics*[width=0.49\textwidth]{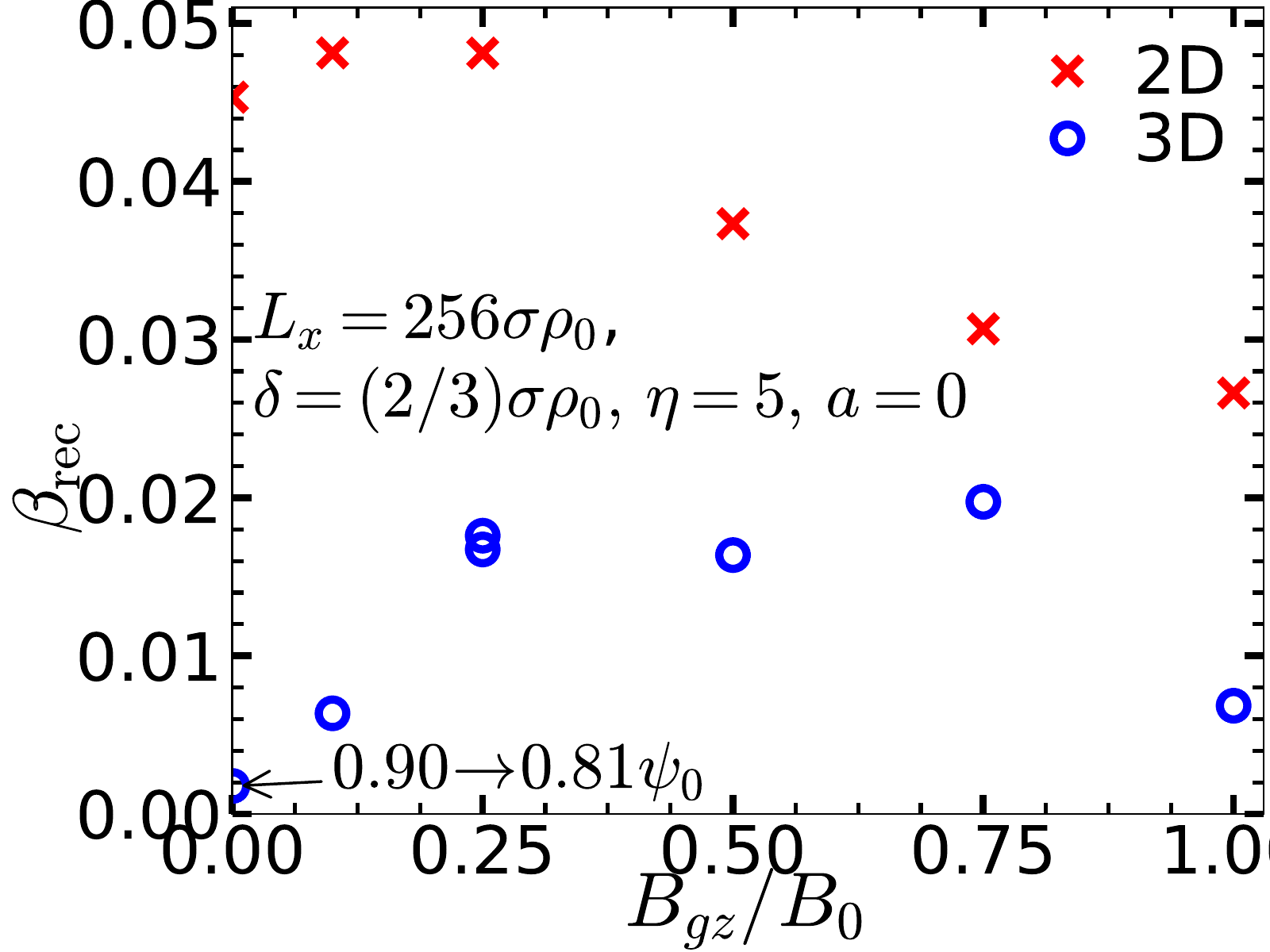}
}
\caption{ \label{fig:fluxAndReconRateVsBz2d3d}
(Left) Unreconnected flux $\psi(t)$ for the same simulations shown in Fig.~\ref{fig:energyVsBgz3d}, with horizontal grey, dotted lines indicating $0.9\psi_0$ and~$0.8\psi_0$.
(Right) Reconnection rates versus $B_{gz}/B_0$ for the simulations in the
left panel, averaged over the time for $\psi(t)$ to fall from~$0.9\psi_0$ to~$0.8\psi_0$ (except, since none of the three 3D, $B_{gz}=0$ cases reached~$0.8\psi_0$, we show~$\beta_{\rm rec}$ for just one of them, averaged between~$0.9\psi_0$ to~$0.81\psi_0$).
In 2D, increasing $B_{gz}$ reduces the reconnection rate; in 3D,
a small guide field increases the reconnection rate, but eventually, strong guide field will reduce the rate.
N.B. For 3D cases, the values of the instantaneous reconnection rate may vary significantly over the averaging time, and stochastic variability may also introduce significant uncertainty.
}
\end{figure}

The decay of upstream flux $\psi(t)$ in~3D, shown in Fig.~\ref{fig:fluxAndReconRateVsBz2d3d}(left), is consistent with this picture.  
$B_{gz}=0$ has the slowest reconnection rate ($\propto -d\psi/dt$), and, after $30L_x/c$, has the most remaining unreconnected flux upstream.
Intermediate guide fields yield faster reconnection and more ``reconnected'' flux, but by $B_{gz}/B_0=1$, reconnection is slowed as the guide field enthalpy becomes significant compared with the relativistic plasma enthalpy (as in 2D).

We now compare $U_{Bt}(t)$ and $\psi(t)$ for these 3D simulations with otherwise identical 2D simulations, for all six values of~$B_{gz}$.
Figure~\ref{fig:energyVsBgz3d}(left) shows~$U_{Bt}(t)$.
In 2D, increasing $B_{gz}$ monotonically slows magnetic energy conversion and results in less plasma energization overall.
In 3D, the initial depletion of $U_{Bt}$ is generally slower than in 2D, but plasma energization continues for a much longer time.
The $B_{gz}=0$ simulations reconnect so slowly in 3D that even after $30L_x/c$ they have converted less magnetic energy than their 2D counterparts.
For $0.1 \leq B_{gz}/B_0 \leq 0.5$, early-time conversion rates are slower in 3D, but after 20--30$L_x/c$, much more magnetic energy is converted in~3D than in~2D.
For $B_{gz}/B_0=0.75$, the 2D and 3D magnetic energy conversion rates are quite similar at early times, but again the 3D simulation eventually converts much more of $U_{Bt0}$ to $U_{\rm plasma}$.
Finally, for $B_{gz}/B_0=1$, the initial magnetic energy conversion rate in 3D is slower than in 2D (which, for reasons we do not yet understand, bucks the trend of becoming more 2D-like with strong guide field), but the 3D simulation ends up converting a roughly similar amount of energy as in 2D (because, like 2D simulations, it still stores a significant amount of energy in relatively stable flux ropes---cf.~Fig.~\ref{fig:energyVsBgz3d}, right).

The left panel of Fig.~\ref{fig:fluxAndReconRateVsBz2d3d} shows $\psi(t)$, again for 2D and 3D, for all six $B_{gz}$ values.
As $B_{gz}/B_0$ increases to $B_{gz}/B_0 \sim 0.75$, 2D reconnection rates ($\propto -d\psi/dt$) decrease, and 3D rates increase; correspondingly, the total amount of flux reconnected (over 30$L_x/c$) decreases in 2D and increases in 3D, until $\psi(t)$ becomes relatively similar in 2D and 3D for $B_{gz}/B_0\approx 0.75$.
This trend suggests that for $B_{gz}/B_0=1$, the 2D and 3D cases should be even more similar, but in fact (compared with $B_{gz}/B_0=0.75$) reconnection for $B_{gz}/B_0=1$ slows dramatically in 3D while it slows only a little in 2D.
We leave a full exploration of this to future work investigating 3D reconnection with stronger guide fields,
but we speculate that 3D reconnection may not reach the 2D limit until the purely 2D suppression (which begins for $B_{gz}/B_0 \gtrsim 1/\sigma_h$ and becomes stronger with higher $B_{gz}$) completely dominates over the suppression due to 3D effects.
At $B_{gz}/B_0=1$, we may be seeing the combined effects of guide-field suppression and 3D suppression of reconnection, while at some larger $B_{gz}$, we expect 3D effects to become weaker.
This is exemplified quantitatively in Fig.~\ref{fig:fluxAndReconRateVsBz2d3d}(right), where we graph the normalized reconnection rates, averaged over the period during which $\psi(t)$ falls from
$0.9\psi(0)$ to $0.8\psi(0)$.
The 2D reconnection rates fall with stronger $B_{gz}$, and the rate for $B_{gz}/B_0=1$ is roughly half of the maximum rate, which is realized for $B_{gz}/B_0 \lesssim 0.25$.
The 3D rate at $B_{gz}/B_0=0$ is an order of magnitude below the maximum 2D rate, but the 3D rate speeds up for intermediate $B_{gz}$ so that at $B_{gz}/B_0=0.75$, the 2D and 3D rates differ by less than a factor of 2.
As $B_{gz}/B_0$ increases from 0.75 to 1, the 2D rate drops smoothly, while the 3D rate drops more dramatically.

It appears likely that magnetic energy and flux would continue to decrease if we ran the 3D simulations for longer times.
However, we quantify the ``final'' amounts of remaining magnetic energy and flux after a long time, $t=30L_x/c$,
in table~\ref{tab:finalEnergyAndFluxVsBgz3d}---for both 2D and 3D simulations, for all $B_{gz}$.
In terms of these ``final'' values, we see that 2D and 3D reconnection
are most similar for $B_{gz}/B_0=1$, with $U_{Bt}$ differing by only 3 percentage points, and $\psi(t)$ by about 8 percentage points.
The table also shows that 3D simulations generally lose less upstream flux but convert more magnetic energy than 2D simulations,
because in 2D, energy in plasmoids is stable, whereas in 3D it is depleted (unless the guide field is sufficiently strong), as shown in Fig.~\ref{fig:energyVsBgz3d}(right). 

\begin{figure}
\centering
\fullplot{
\includegraphics*[width=0.49\textwidth]{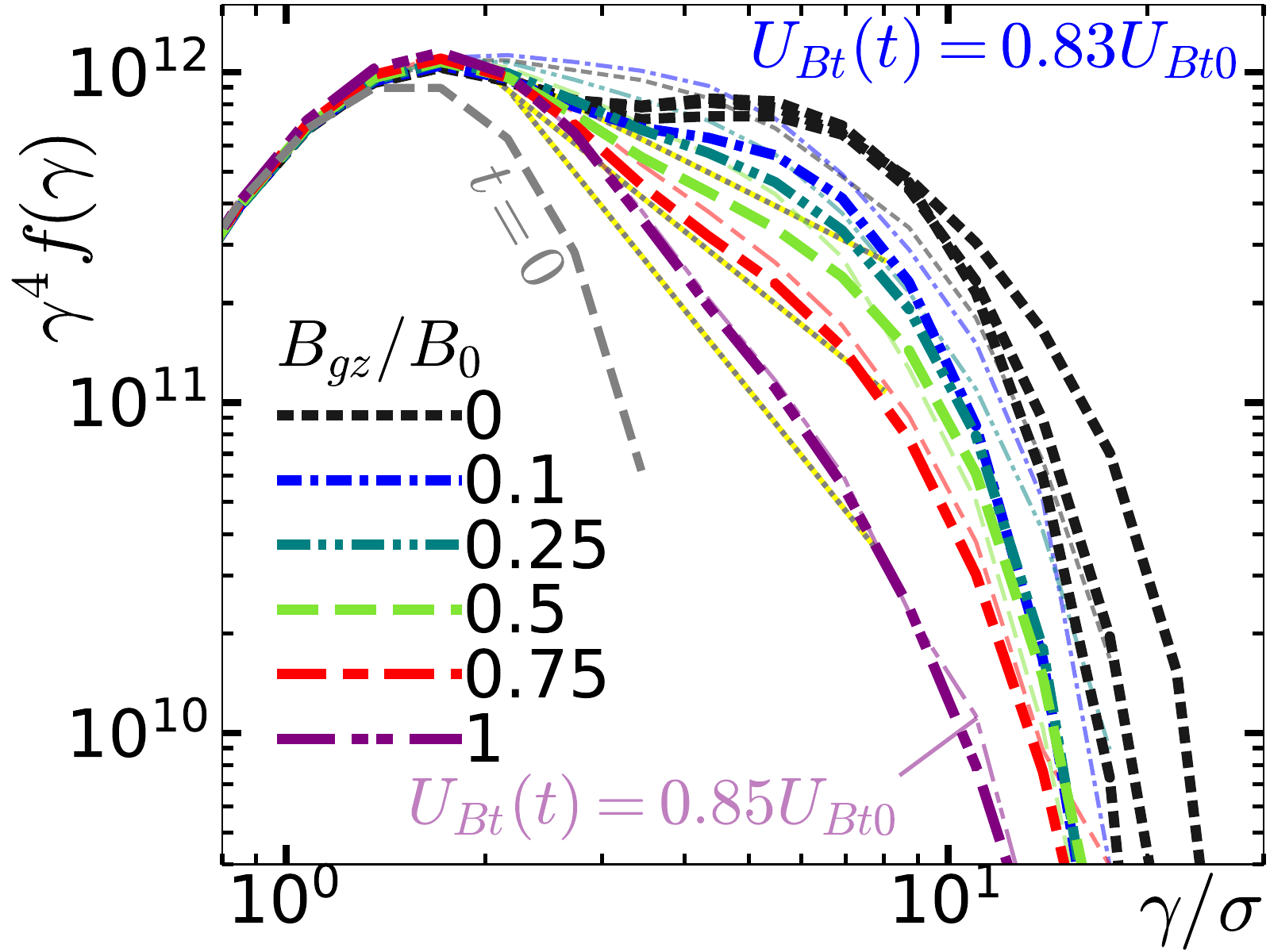}%
\hfill
\includegraphics*[width=0.49\textwidth]{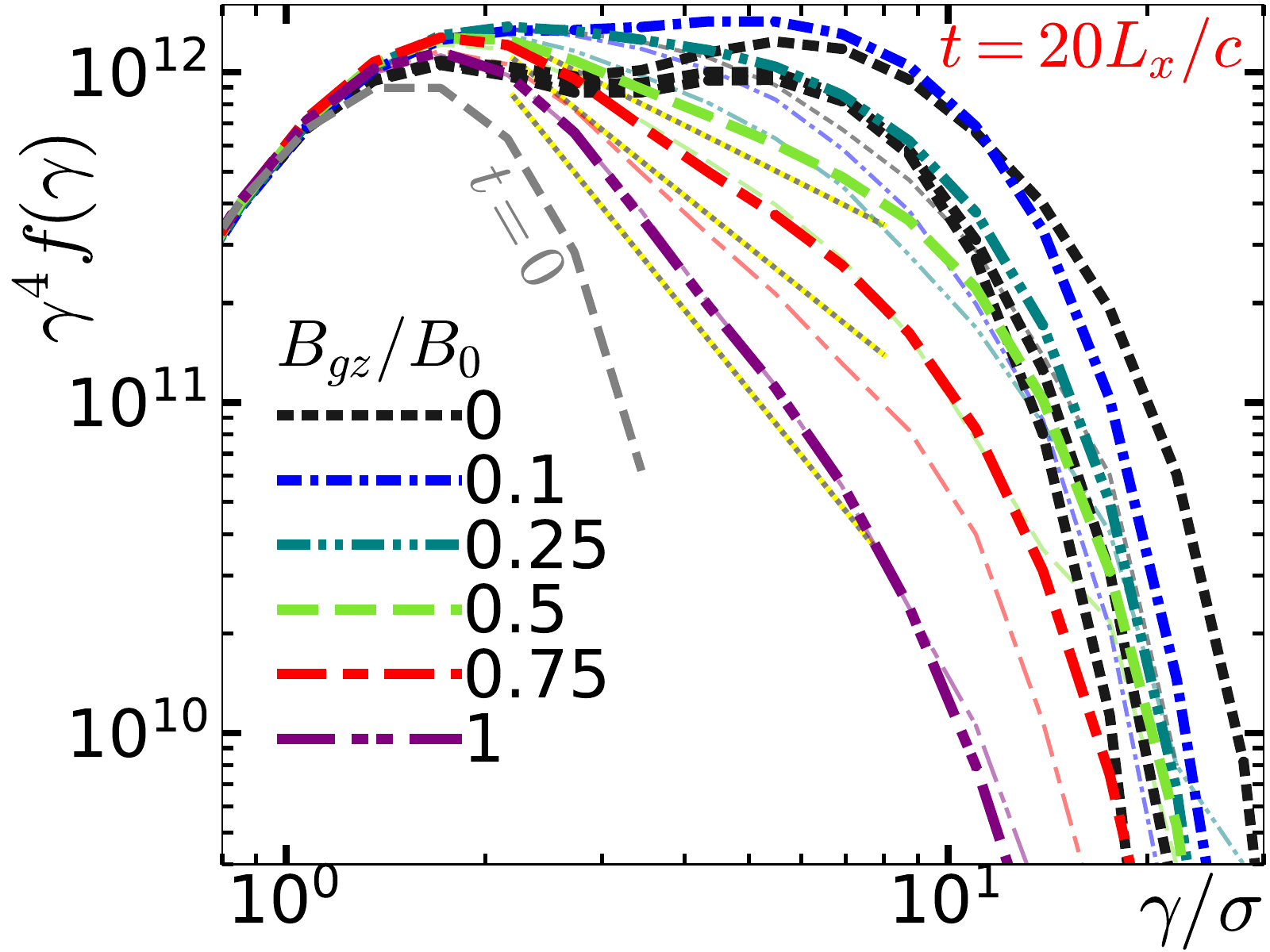}
}
\caption{ \label{fig:ntpaVsBgz3d}
Particle energy spectra compensated by $\gamma^4$ at a time (left) when $U_{Bt}=0.83U_{Bt0}$ (i.e., 17~per~cent of magnetic energy has been converted to plasma energy), and (right) at $t=20L_x/c$, for a range of guide fields $B_{gz}$, for 3D simulations with $L_z=L_x$ (thicker, darker lines) and, for comparison, for 2D (thinner, fainter).
Dotted grey-on-yellow lines show the slopes of $f(\gamma)\propto \gamma^{-p}$ for $p=5$, 5.6, and 6.5 ($p=4$ would be horizontal).
These simulations are the same as shown in Figs.~\ref{fig:energyVsBgz3d}, \ref{fig:fluxAndReconRateVsBz2d3d}.
The 2D simulation with $B_{gz}=B_0$ did not reach $U_{Bt}=0.83U_{Bt0}$, but remained around $U_{Bt}=0.85U_{Bt0}$ from $t=7L_x/c$ to $t=30L_x/c$.
}
\end{figure}

We have seen that both guide field and 3D effects can suppress reconnection, slowing the overall rates of reconnection and release of magnetic energy.
We find that an initial guide field suppresses NTPA---to the same extent in both 2D and 3D---whereas 3D effects do not suppress NTPA, despite slowing reconnection.
Figure~\ref{fig:ntpaVsBgz3d} shows the electron energy spectra $f(\gamma)$ (compensated by $\gamma^4$) for each $B_{gz}$, both at
the same amount of depleted magnetic energy, $U_{Bt}=0.83U_{Bt0}$ (left),
and at the same time, $t=20L_x/c$ (right), for the 3D simulations as well as for comparable 2D simulations.
Keeping in mind that graphing $\gamma^4 f(\gamma)$ enhances small differences in $f(\gamma)$, we consider the 2D and 3D spectra (for the same $B_{gz}$) to be fairly similar in magnitude [separated by less than a factor of 2 in either $\gamma$ or in $f(\gamma)$], especially at $U_{Bt}(t)=0.83U_{Bt0}$.
As the guide field increases, the spectra in 2D and 3D become even closer, until for $B_{gz}=B_0$ they are nearly identical.
We cannot make a general statement about whether NTPA is more efficient in 2D or 3D without going into much more detail regarding the exact times and energies at which spectra are compared.
However, it does appear that, because 3D simulations can convert more magnetic energy to plasma energy, over long times 3D simulations accelerate more particles.

\citet{Dahlin_etal-2017} observed enhancement of NTPA in 3D, relative to 2D, in subrelativistic electron-ion reconnection at $B_{gz}=0$, and the 3D enhancement became stronger with increasing guide field, until this effect saturated around $B_{gz}/B_0 \simeq 1$ [\citet{Li_etal-2019} also observed enhancement in 3D for $B_{gz}/B_0=0.2$].
As we discussed in~\S\ref{sec:Lz3d}, 
our results are not inconsistent with these, but the differences between 2D and 3D spectra do not seem very significant if compared at times with the same amount of magnetic energy depletion; however, because more magnetic energy can be converted to plasma energy in 3D, 3D simulation can accelerate more particles to high energies.
However, as the guide field becomes stronger, we expect 3D simulations to become essentially identical to 2D simulations, and then we expect NTPA to be identical as well.

\begin{table}
\centering
\begin{tabular}{lcccccccc}
$B_{gz}/B_0$ & 0 & 0 & 0 & 0.1 & 0.25 & 0.5 & 0.75 & 1 \\
\hline
loss in $U_{Bt}$ (2D) & 27\% &&& 27\% & 25\% & 21\% & 18\% & 15\% \\
loss in flux (2D) & 44\% &&& 43\% & 43\% & 40\% & 36\% & 33\% \\
\hline 
loss in $U_{Bt}$ ($L_z=L_x$) & 26\% & 23\% & 24\% & 41\% & 38\% & 36\% & 32\% & 18\% \\
loss in flux ($L_z=L_x$) & 20\% & 16\% & 17\% & 31\% & 33\% & 31\% & 29\% & 25\% \\
\end{tabular}
\caption{\label{tab:finalEnergyAndFluxVsBgz3d} 
The fractional loss in transverse magnetic field energy $U_{Bt}$ and
unreconnected magnetic flux $\psi$ over $30L_x/c$ of reconnection, for various guide fields,
for simulations with $L_x=256\sigma\rho_0$, both 2D and $L_z=L_x$ (for $L_z=L_x$, three simulations are shown
with $B_{gz}=0$ 
to give some idea of stochastic
variation).
}
\end{table}

We have not yet investigated the strong guide field regime, $B_{gz}/B_0\gg 1$, because of the high computational cost associated with larger $L_z/L_x$ and longer simulation times; 
it will be important in future work to explore this regime and in particular to confirm or deny conclusively that reconnection with very strong guide field rigorously approaches the ideal 2D limit with perfect uniformity in $z$.

\section{Comparison with large magnetization}
\label{sec:highSigmah}

In this paper we have found that, for $\sigma_h=1$, 3D reconnection is significantly slower than 2D reconnection, and lasts for much longer times over which it eventually converts more magnetic energy to plasma energy 
(but that NTPA is quite similar in 2D and 3D).
However, relativistic reconnection with large magnetization, $\sigma_h\gg 1$, has been found to convert similar amounts of magnetic energy to plasma energy at similar rates in 2D and 3D \citep{Guo_etal-2014,Guo_etal-2015,Werner_Uzdensky-2017,Guo_etal-2020arxiv,Zhang_etal-2021arxiv}.\footnote{
For reasons that are not clear, \citet{Sironi_Spitkovsky-2014} measured a slower reconnection rate in 3D, about $1/4$ of the rate in 2D, but in spite of that
observed otherwise substantially similar reconnection in 2D and 3D.
We note that recent work with $\sigma_h=10$ and $B_{gz}/B_0=0.1$ found 3D
reconnection rates $(c/v_A)\beta_{\rm rec}\sim 0.075$, comparable to but less than $(c/v_A)\beta_{\rm rec}\sim 0.12$ found in~2D
\citep{Zhang_etal-2021arxiv}.
}
It thus may appear that 3D effects play less of a role for large-$\sigma_h$ reconnection.
However, as we will discuss in the following, there is a possibility that the 2D-like behaviour for 3D reconnection might be a result of the initial simulation configuration, and not an inevitable, intrinsic consequence of larger~$\sigma_h$.

The recent work by \citet{Guo_etal-2020arxiv} should be considered when comparing large and small $\sigma_h$, since it studied 3D relativistic pair reconnection with $\sigma_h$ ranging from around~1 to~200 (though starting from a transrelativistic temperature $\theta_b \sim 1$).
There, high reconnection rates and magnetic energy conversion rates (similar to rates in 2D) were observed for the entire range of~$\sigma_h$.
However, that study focused on very early times; those rates were measured within the first~$1\,L_x/c$.
Moreover, those simulations began with an initial field perturbation, uniform in the third dimension, and used a force-free configuration---i.e., an initial magnetic field with a substantial $B_z$ component localized to the layer.
We have shown that differences in magnetic energy conversion rates between 2D and 3D are suppressed by an initial perturbation (\S\ref{sec:pertAndEta3d}) and by an initially globally-uniform guide field $B_{gz}$ (\S\ref{sec:Bz3d}), particularly at early times. 
Therefore, \citet{Guo_etal-2020arxiv} does not address our particular question about whether lower energy conversion rates might be observed in 3D if simulations ran for long times, especially starting with no initial perturbation and no guide field.

The very recent work by \citet{Zhang_etal-2021arxiv} studied reconnection in 3D pair plasma with $\sigma_h=10$ (and $B_{gz}/B_0=0.1$).
They observed slightly slower reconnection in 3D (compared with 2D), but not substantially slower.  This adds to the evidence suggesting that for larger~$\sigma_h$, 3D reconnection is more similar to 2D reconnection.
However, their use of an initial perturbation and outflow boundary conditions led to an X-line running down the centre of the box the entire length in~$z$, which could conceivably be the dominant driver of 2D-like reconnection (i.e., and not the larger~$\sigma_h$).  
The guide field, though weak, might also favour 2D-like reconnection, but the paper states that simulations with $B_{gz}=0$ did not differ significantly.
Thus we cannot confidently attribute the 2D-like speed of 3D reconnection purely to the difference in~$\sigma_h$. 
\cite{Zhang_etal-2021arxiv} also found NTPA to be very similar in 2D and 3D---except that, similar to the earlier finding in subrelativistic electron-ion reconnection \citep{Dahlin_etal-2017}, the highest-energy particles in 3D were accelerated to higher energies (than in 2D) because they escaped from flux ropes (with finite extent in~$z$) and thus experienced additional acceleration in the reconnection electric field.  Although a full comparison of this acceleration behaviour is beyond the scope of our work, preliminary analysis suggests that our $\sigma_h=1$ simulations show similar behaviour of high-energy particles experiencing (in 3D) additional direct electric field acceleration.

Our own simulations of large-$\sigma_h$ relativistic reconnection in 3D pair plasma from a previous work \citep{Werner_Uzdensky-2017}, with $\sigma_h=25$, also started with a magnetic field perturbation ($a=2.7$, $s/\delta=1.9$), but used a Harris sheet set-up (as in this paper) and ran for about~$5L_x/c$.
We also ran some simulations without perturbation, but mentioned them only in passing.  
One of those simulations---3D ($L_z=L_x$), without guide field ($B_{gz}=0$) and without initial perturbation ($a=0$)---did show exceptionally slow reconnection (perhaps like our $\sigma_h=1$ simulations in this paper) and much-suppressed NTPA (unlike our $\sigma_h=1$ simulations).
At the time we attributed the slow reconnection to the influence of the initial current sheet; i.e., we believed this effect would have disappeared if we could have run simulations with larger $L_x/\delta$.
In light of our $\sigma_h=1$ results presented above, we revisit these $\sigma_h=25$ simulations.

Although the larger $\sigma_h$ is the most fundamental physical difference between the $\sigma_h=1$ simulations in this paper and those
in \citet{Werner_Uzdensky-2017} with $\sigma_h=25$, 
practical considerations necessitated other differences.
Large $\sigma_h$ required higher grid resolution ($\Delta x = \sigma\rho_0/12$, to resolve the Debye length of the colder upstream plasma), so the $\sigma_h=25$ 
simulations were significantly smaller, with $L_x=80\sigma\rho_0 = 120\delta$;
whereas, our $\sigma_h=1$ simulations with $\Delta x=\sigma\rho_0/3$ are 4--6 times larger, with some parameter scans using $L_x=341 \sigma\rho_0=512\delta$ (e.g.,~\S\ref{sec:Lz3d}) and the largest 3D simulation reaching $L_x=512\sigma\rho_0=768\delta$ (cf.~\S\ref{sec:overview3d}). 
Because the same overdensity $\eta=5$ was used in the large-$\sigma_h$ regime,
the $\sigma_h=25$ simulations required a much hotter initial current sheet, $\theta_d=10.5\theta_b$, to balance the relatively strong upstream magnetic pressure;
whereas for $\sigma_h=1$, the initial current sheet is cooler than the upstream plasma, $\theta_d=0.4\theta_b$.
The initial current sheet for $\sigma_h=25$ thus had, at its peak density, about~50 times the energy and inertia density of the upstream plasma; for $\sigma_h=1$, this ratio is only 2.
It is therefore reasonable to expect the initial current sheet to exert greater influence for $\sigma_h=25$ than for $\sigma_h=1$, especially considering that $L_x/\delta$ was several times smaller for $\sigma_h=25$.

The relatively high enthalpy of the initial current sheet plasma for $\sigma_h=25$, relative to the background plasma, may play a critical role in slowing down reconnection in 3D (with $a=0$ and $B_{gz}=0$).
Here is what we believe may be happening in that case.
When reconnection starts, the initial current sheet plasma is not trapped and swept away from reconnecting X-points, which ultimately control the reconnection rate \citep{Uzdensky_etal-2010}, but rather is dispersed widely in the vicinity of the layer.
Some of that hot, dense plasma recirculates into X-point inflows, where its high inertia---50 times higher than the background plasma's inertia---significantly lowers the Alfv\'{e}n velocity and slows reconnection.
Either an initial perturbation, or a weak guide field, or 2D-ness will facilitate the trapping of the initial current sheet plasma in plasmoids, sweeping it away from X-points and preventing this scenario.
Specifically, the 2D/3D difference disappeared either with a perturbation  of $a=2.7$ or with a guide field of $B_{gz}\geq B_0/4$---where, importantly, $B_{gz}=B_0/4$ was strong enough to suppress 2D/3D differences but weak enough that it did not substantially affect reconnection rates or NTPA in either 2D or in 3D.

\begin{figure}
\centering
\fullplot{
\includegraphics*[width=0.49\textwidth]{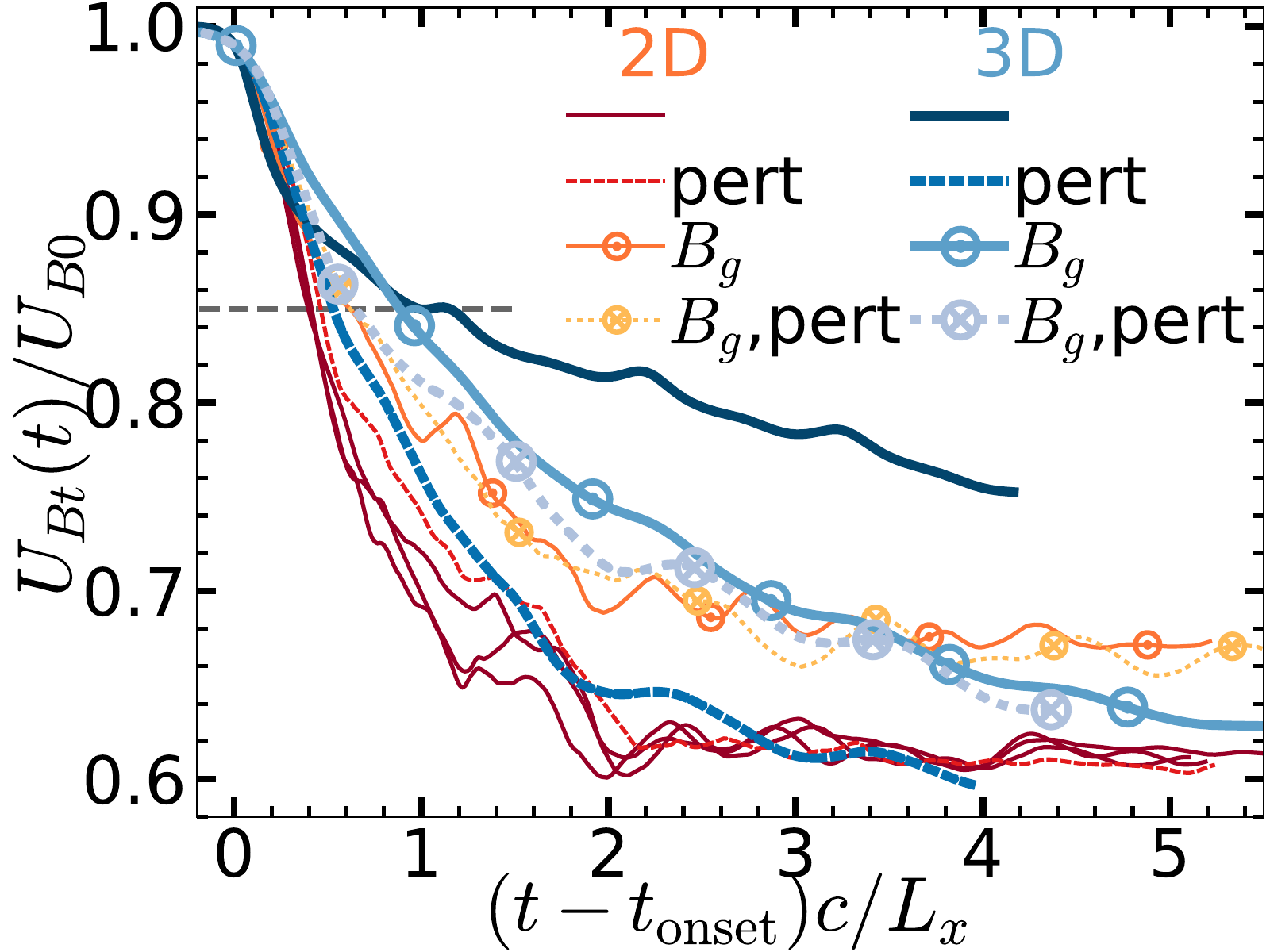}%
\hfill
\includegraphics*[width=0.49\textwidth]{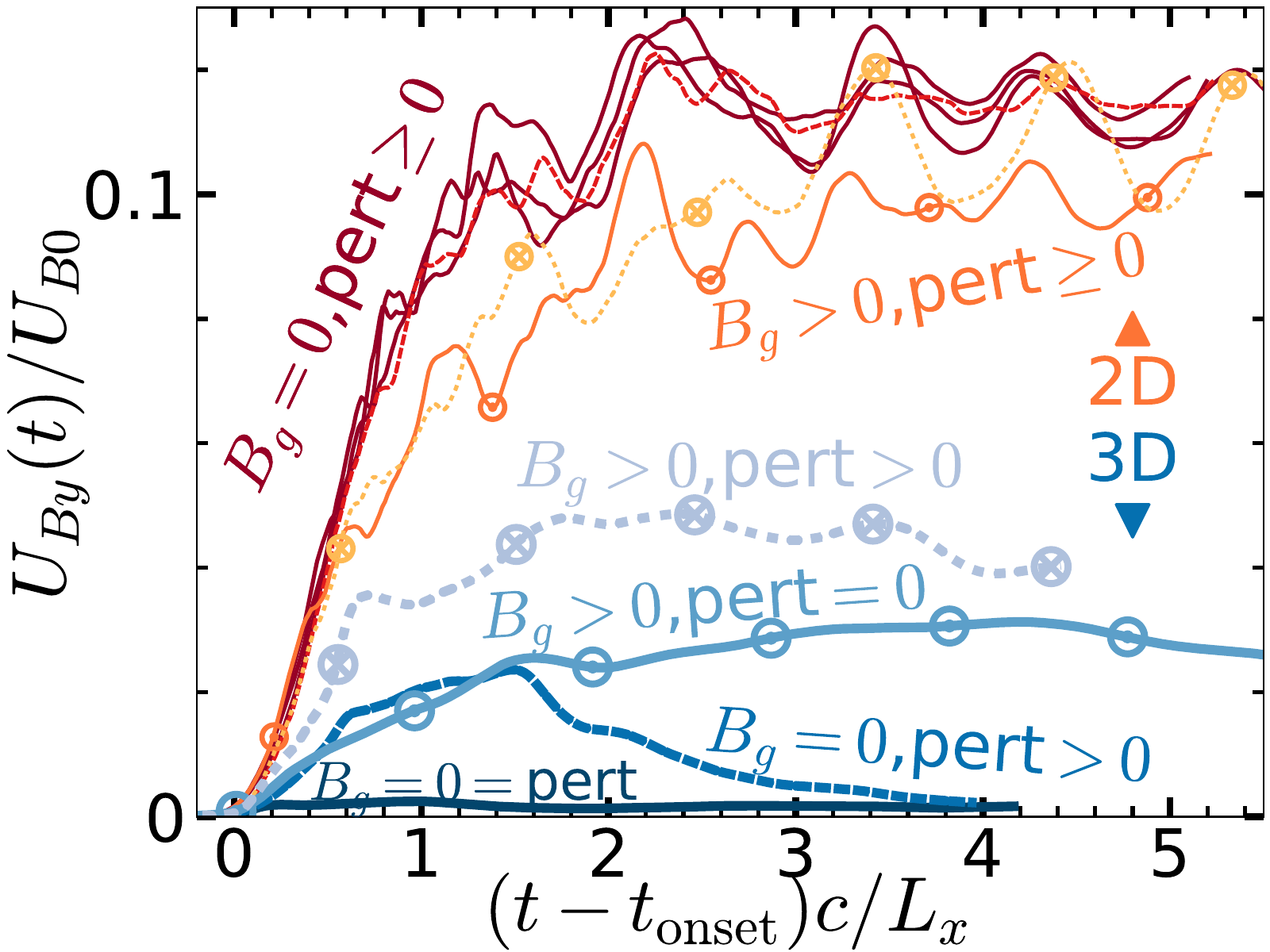}
}
\caption{ \label{fig:highSigmahEnergy2d3d}
Magnetic energy versus time (left) in $B_x$ and $B_y$ components, and (right) in $B_y$ components only (as a proxy for $U_{Bt,\rm layer}$), for eight different configurations of $\sigma_h=25$ simulations from
\citet{Werner_Uzdensky-2017}.
Shown are all combinations of 2D (thin, red/yellow lines) or 3D (thicker, blue-ish), $B_{gz}/B_0=0$ (no symbols) or~$1/4$ (symbols, `$B_g$'), without initial perturbation (solid lines) or with a perturbation such that the separatrix initially extends to $s=1.8\delta$ (dashed lines, `pert').
Results for three different 2D simulations with $B_{gz}=0$ and no perturbation are shown.  A dashed grey line marks the energy at which spectra are shown in Fig.~\ref{fig:highSigmahNTPA2d3d}.
}
\end{figure}

We now review the evidence that
the 3D simulation with $B_{gz}=0$ and $a=0$ was an outlier with slow reconnection and limited NTPA.
When we compare the transverse magnetic energy versus time, $U_{Bt}(t)$, for eight configurations with $\sigma_h=25$---all combinations of (1)~dimensionality: 2D and~3D ($L_z=L_x$), (2)~guide field: $B_{gz}=0$ and~$B_{gz}=B_0/4$, (3)~perturbation: $a=2.7$ and~$a=0$---we see three rough groups of similar $U_{Bt}(t)$ (Fig.~\ref{fig:highSigmahEnergy2d3d}, left).
First, there is a group of all four simulations with $B_{gz}=B_0/4$, which have similar $U_{Bt}(t)$, regardless of dimensionality or perturbation.
Second, there is a group of all simulations with $B_{gz}=0$---except for the outlier (i.e., 3D, $a=0$).  Comparing these two groups, we see, as expected, that the small guide field slows/suppresses reconnection by a modest amount.
Within each group, we see that the 3D simulations can ultimately convert a bit more magnetic energy to plasma energy than the 2D simulations (cf.~\S\ref{sec:Lz3d}), but this effect is not very significant, at least in the first~$5L_x/c$.
The third group contains just the single outlier---3D, $B_{gz}=0$, $a=0$---with much slower plasma energization.

\begin{figure}
\centering
\fullplot{
\includegraphics*[width=0.49\textwidth]{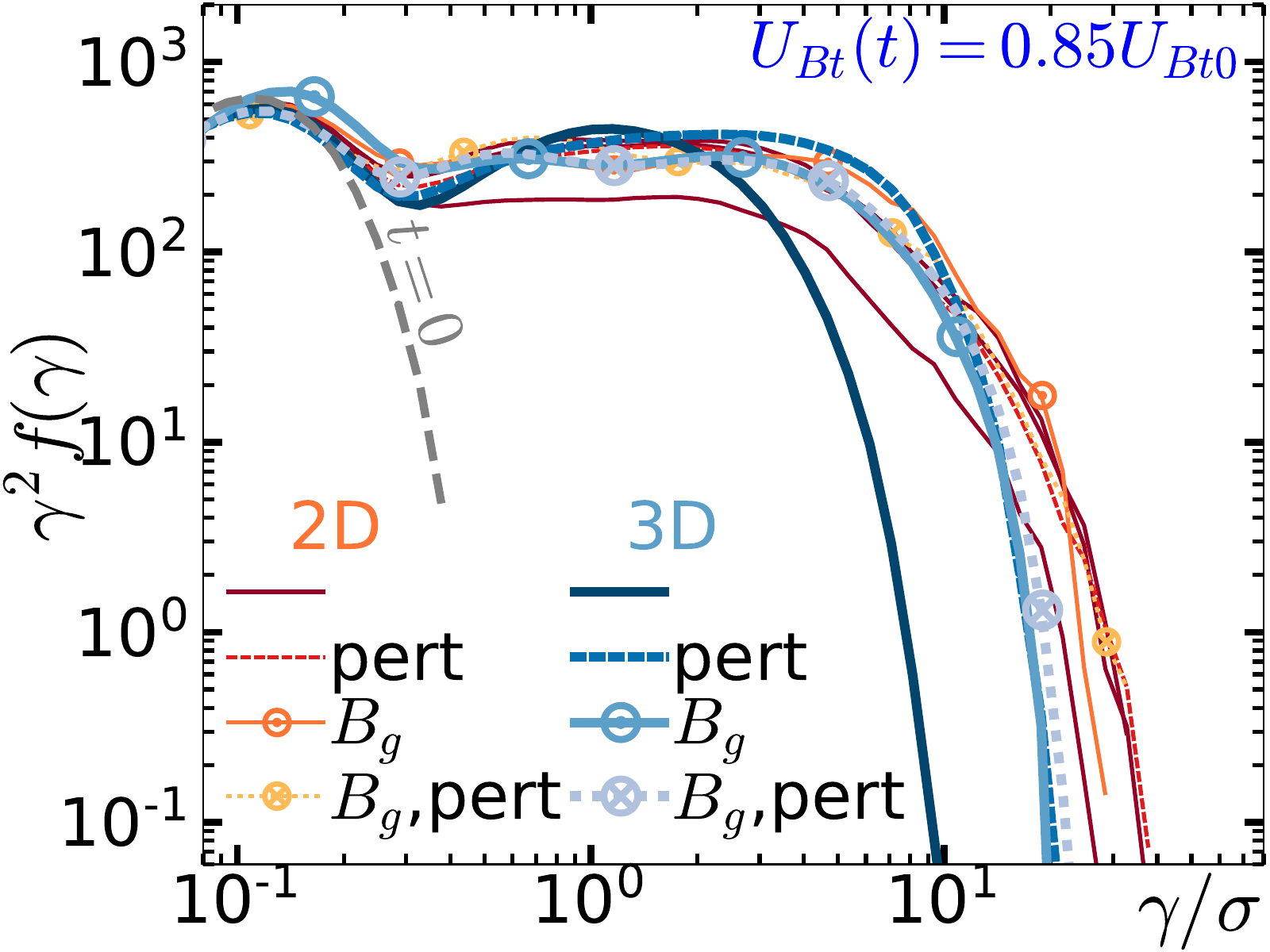}%
\hfill
\includegraphics*[width=0.49\textwidth]{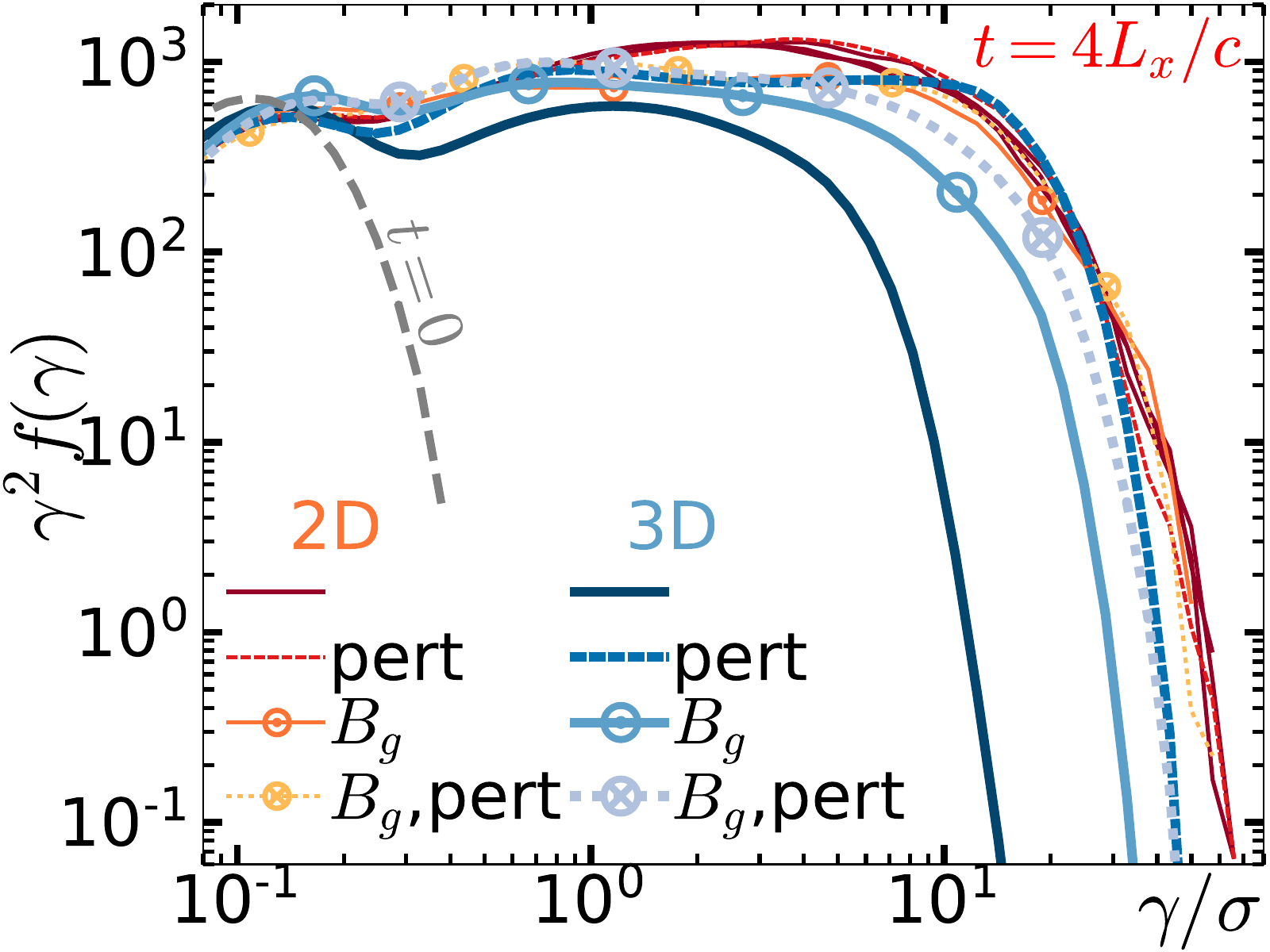}
}
\caption{ \label{fig:highSigmahNTPA2d3d}
Electron energy spectra for the simulations in Fig.~\ref{fig:highSigmahEnergy2d3d}: (left)
at a time when approximately 15~per~cent of the magnetic energy has been depleted (see dashed grey line in Fig.~\ref{fig:highSigmahEnergy2d3d}), and (right) at time $t=4L_x/c$.  A dashed grey line shows the spectra at $t=0$.
}
\end{figure}

\begin{figure}
\centering
\fullplot{
\includegraphics*[width=\textwidth]{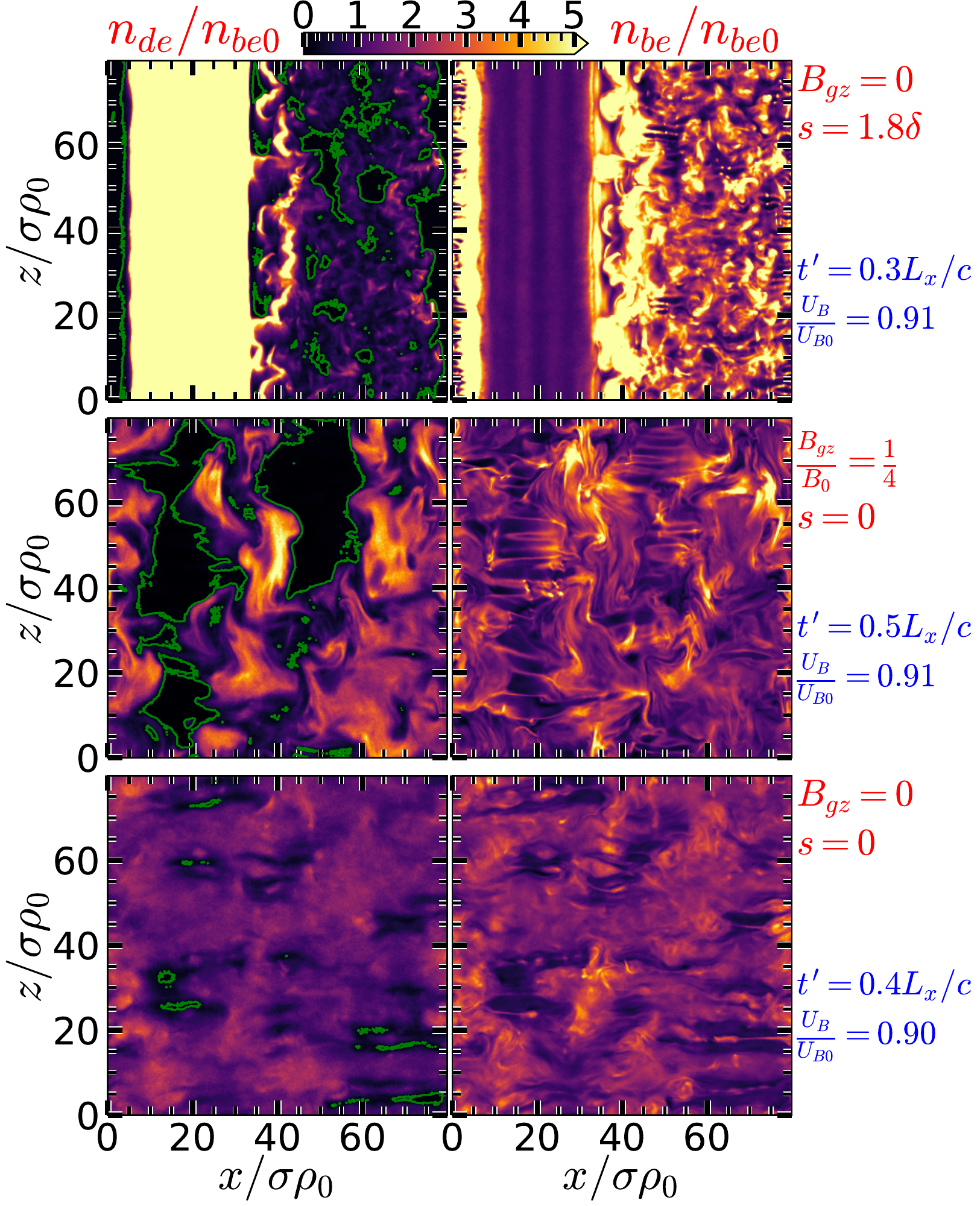}%
}
\caption{ \label{fig:highSigmahDensity3d}
In 3D simulations with $\sigma_h=25$, \emph{either} an initial perturbation \emph{or} a guide field can cause the drift particles to be trapped and cleared out of large areas, as shown by
the density $n_{de}$ of initially-drifting electrons in the original reconnection midplane (left column), roughly~$0.4L_x/c$ after reconnection onset when the magnetic energy has declined by about 10~per~cent, for:
(top row) 
$B_{gz}=0$ and perturbation, $a=2.7$ or $s/\delta=1.8$;
(middle) 
$B_{gz}=B_0/4$ and zero perturbation; 
(bottom) 
the outlier case with $B_{gz}=0$ and zero perturbation. 
The right column shows the density $n_{be}$ of background electrons for reference.
Green contours outline areas where $n_{de} < 0.2 n_{be0}$.
The time $t'$ since onset and the fraction of remaining magnetic energy are noted on the right side.
}
\end{figure}

Interestingly, this grouping falls apart when we look at $U_{By}(t)$, the energy in magnetic field components $B_y$ (Fig.~\ref{fig:highSigmahEnergy2d3d}, right).  Here, $U_{By}$ is a proxy for the magnetic energy $U_{Bt,\rm layer}$ in the layer, and we have seen (for $\sigma_h=1$, e.g.,~\S\ref{sec:Lz3d}) that 2D reconnection stores magnetic energy in plasmoids, whereas plasmoids tend to decay in 3D.
Here, there is a distinct difference between 2D and 3D, although the 3D case with $B_{gz}=0$ and $a=0$ is still an outlier with the lowest overall $U_{By}$.  Both guide field and initial perturbation make 3D simulations behave more---but not entirely---like 2D simulations, something that we have seen for $\sigma_h=1$ (cf.~\S\ref{sec:pertAndEta3d}, \S\ref{sec:Bz3d}).
This suggests, even for $\sigma_h=25$, 3D simulations may be more different from 2D simulations than previously suspected.
However, it is important to remember that 2D and 3D simulations with $\sigma_h=25$ have rather similar total magnetic energy evolution (unlike with $\sigma_h=1$), apart from the one outlier case.

Looking at NTPA for these same simulations, we again see that the 3D case with $B_{gz}=0$, $a=0$ is a clear outlier.
Particle energy spectra are shown in Fig.~\ref{fig:highSigmahNTPA2d3d}---at times when 15~per~cent of the initial magnetic energy has been depleted, i.e., $U_{Bt}(t)=0.85U_{B0}$ (left), and also at $t\approx 4L_x/c$ (right). 
All the simulations, except the outlier, have similar energy spectra, especially for $U_{Bt}(t)=0.85 U_{B0}$.  (When compared at the same, $t=4L_x/c$, one sees that, since even a weak guide field slows reconnection a little, NTPA for $B_{gz}=B_0/4$ has not quite caught up to NTPA for $B_{gz}=0$.)

We further observe that that this outlier---in reconnection rate and NTPA, as shown above---is also an outlier in the behaviour of the initially-drifting plasma (i.e., the initial current sheet plasma).
In all simulations but the outlier---e.g., 2D or $B_{gz} > 0$ or $a>0$---the initially-drifting particles are rapidly swept away from X-points and trapped within plasmoids \citep[see our discussion of plasmoid formation with and without perturbation in~\S\ref{sec:pertAndEta2d}, and also][]{Ball_etal-2018}.
This can be clearly seen in Fig.~\ref{fig:highSigmahDensity3d}, which shows $n_{de}(x,y_c,z)$, the density of initially-drifting electrons, as well as the density $n_{be}(x,y_c,z)$ of background electrons, in the original midplane, $y=y_c$.
With $a=2.7$ ($s=1.8\delta$, top row) or with $B_{gz}=B_0/4$ (middle row), it takes only a short time to evacuate most of the initially-drifting plasma from large areas (e.g., large areas with $n_{de}<0.2n_{be0}$ are outlined---$0.2n_{be0}$ is a very small density compared with the initial peak density $n_{de}=5n_{be0}$).
However, with no perturbation and no guide field (bottom row), the initially-drifting particles remain scattered roughly uniformly about the layer, while spreading in $y$, including in the vicinity of X-points. 
With 50 times the enthalpy density of the background plasma, the initially-drifting plasma can thus significantly load the inflows upstream of X-points, in this case lowering the effective~$\sigma_h$ and hence Alfv\'{e}n velocity, and thereby slowing reconnection.

Let us quickly review.  In ($\sigma_h=25$) simulations that have $B_{gz}>0$ or have $a>0$ or are 2D,
\begin{itemize}[leftmargin=3mm, itemindent=0mm, labelsep=1mm]
  \item the initial current sheet plasma is quickly swept away from the vicinity of X-points, 
  \item so that the plasma near X-points flows in from the upstream plasma with $\sigma_h \gg 1$,
  \item and fast reconnection and NTPA ensue as in 2D simulations.
\end{itemize}
In the (outlier) 3D simulation with $B_{gz}=0$ and $a=0$, 
\begin{itemize}[leftmargin=3mm, itemindent=0mm, labelsep=1mm]
  \item the initial current sheet plasma is dispersed about the layer and can flow back into X-points,
  \item reducing the~$\sigma_{h,X}$ of the plasma immediately upstream of X-points (since, for large~$\sigma_h$, the initial current sheet plasma has much higher enthalpy and inertia than the upstream plasma),
  \item and reconnection is slow and NTPA is substantially diminished.
\end{itemize}
When the initial current sheet plasma can flow back into X-points, we estimate the hot magnetization~$\sigma_{h,X}$, which controls reconnection rate and NTPA, as follows.
From pressure balance, the pressure of the initial current sheet plasma is $p_d\equiv n_{d0}\theta_d m_e c^2/\gamma_d = B_0^2/8\upi$; since the enthalpy density is $h\sim 4 p_d$, 
$\sigma_{h,d} \equiv B_0^2/(4\upi h) \sim O(1)$.
Even if the upstream has~$\sigma_h \gg 1$, $\sigma_{h,X}\sim [\sigma_{h}^{-1} + \sigma_{h,d}^{-1}]^{-1}$ may be~$O(1)$, roughly independent of the upstream~$\sigma_h$.

Lowering~$\sigma_h$ is known both to slow reconnection and to suppress NTPA \citep{Sironi_Spitkovsky-2014,Guo_etal-2015,Werner_etal-2016,Werner_Uzdensky-2017,Werner_etal-2018,Guo_etal-2020arxiv}.
In contrast, our 3D simulations with (asymptotic upstream) $\sigma_h=1$,
in which the initial current sheet plasma had only twice the enthalpy density of the background plasma, showed slow reconnection but \emph{no reduction} in NTPA (compared with 2D; cf.~\S\ref{sec:Lz3d}).
This suggests that the suppression of reconnection (and NTPA) 
in the outlier $\sigma_h=25$ simulation may not be caused directly by 3D effects (as manifested for~$\sigma_h=1$), but rather may be a consequence of the initial current sheet plasma behaviour, enabled indirectly by 3D effects (as long as $B_{gz}=0$ and~$a=0$).

This raises the important question: in a much larger simulation, or in a simulation with open outflow boundary conditions, would the initial current sheet plasma continue to load the upstream plasma at X-points for arbitrarily long times?
I.e., is this an artificial effect of the initial and boundary conditions of an idealized simulation that might not be important in realistic astrophysical systems, or is it an inevitable consequence of 3D reconnection in some regimes?
In a much larger simulation, or if particles could escape the simulation, the loading of the upstream plasma (by hot, dense plasma from the initial current sheet) might conceivably be a brief, transient phenomenon.
Eventually, the initially-drifting particles might escape or be trapped or become so diluted by background plasma that they cannot effectively continue to increase the plasma inertia at X-points.
Indeed, it was this reasoning that led us to present only results from simulations with $a>0$ in \citet{Werner_Uzdensky-2017}; the perturbation appeared to minimize the influence of the initial current sheet, allowing us to access the large-system regime, $L_x\gg \delta$, with computationally-feasible simulations.%
\footnote{
  Usually we refer to the large-system regime with respect to NTPA,
  with $L_x \gg 40\sigma\rho_0$ \citep{Werner_etal-2016}, 
  but here we use the
  term in a different sense, with ``large'' implying $L_x\gg \delta$, to 
  reduce the influence of the initial current sheet.
  A typical astrophysical system would almost
  certainly be very large in both senses.
}
On the other hand, the upstream plasma is heated and compressed as it flows through a reconnecting X-point, and---if that plasma can recirculate back into the X-point inflows---that could provide long-term suppression or self-regulation of reconnection.
Unfortunately we cannot conclusively answer this question with our previously-run $\sigma_h=25$ simulations and must leave it to future research.

Even if upstream loading by the initial current sheet plasma explains the suppressed reconnection and NTPA in 3D, and even if this effect vanishes for larger systems, this second look at $\sigma_h=25$ simulations suggests that we may still need to investigate whether and how 3D reconnection differs from 2D reconnection for much larger systems with~$\sigma_h\gg 1$.  Figure~\ref{fig:highSigmahEnergy2d3d} shows that, despite similar $U_{Bt}(t)$ and NTPA in 2D and 3D (for~$L_x=80\sigma\rho_0$), 3D reconnection with $\sigma_h=25$ nevertheless exhibits one very clear signature of fundamentally 3D effects---namely that magnetic energy stored in plasmoids (flux ropes) decays in 3D, but not in 2D.  
This could significantly affect the long-term evolution of a reconnection current sheet; for example, if plasmoids disintegrate in 3D (even for~$\sigma_h\gg 1$), that might potentially kill the acceleration mechanism (due to conservation of magnetic moment in a compressing plasmoid) for particles trapped in plasmoids \citep{Petropoulou_Sironi-2018,Hakobyan_etal-2021}.

In summary, a definitive answer to the question of whether 3D reconnection with $\sigma_h\gg 1$ and weak or zero guide field can be slow compared with 2D reconnection (as is the case for reconnection with $\sigma_h=1$) will require more investigation.
We can at least say that $\sigma_h\gg 1$ reconnection shows some distinctive 3D effects, most notably the decay of flux-rope structures, resulting in more magnetic energy release in 3D than in 2D.
However, in some regimes with $\sigma_h \gg 1$, simulations have nevertheless shown substantial similarities between 2D and 3D reconnection in terms of magnetic energy evolution and NTPA.  
It is possible that, in sufficiently large systems (larger than has been explored to date), $\sigma_h\gg 1$ reconnection is inevitably slow in 3D (compared with 2D), just as for $\sigma_h=1$.
Alternatively, it is possible that the slowing of reconnection and the less efficient NTPA for $\sigma_h=25$ (with zero guide field and no perturbation) is the result of a transient influence of the initial current sheet and that in more realistic astrophysical systems (at long times after transients effects have died away) 3D reconnection would eventually and inevitably resemble 2D reconnection.

\section{Summary}
\label{sec:summary}

Before a final discussion of the significance and some impacts of this study, we briefly summarize the most important results for magnetic reconnection in 2D and 3D ultrarelativistically-hot collisionless pair plasma with $\sigma_h=1$ (i.e., $\beta_{\rm plasma}=1/2$).
We remind the reader that we define $\sigma_h\equiv B_0^2/(16\upi n_b \theta_b m_e c^2)$ with respect to the upstream (ambient) plasma, excluding any guide magnetic field (here, $\theta_b m_e c^2$ is the temperature, and $\theta_b\gg 1$).
Thus the magnetic energy and thermal energy are comparable, and both greatly exceed the plasma rest-mass energy.
The results are organized in three subsections: (\ref{sec:sumQualitative}) major qualitative differences between 2D and 3D, (\ref{sec:sum2D}) 2D-specific results, and (\ref{sec:sum3D}) 3D-specific results.

\subsection{Qualitative similarities and differences in 2D and 3D}
\label{sec:sumQualitative}

Our observations of general, \emph{qualitative} aspects of current sheet evolution for (upstream) $\sigma_h=1$ are summarized below, emphasizing similarities and differences in 2D and 3D.
(We describe 2D behaviour here primarily to facilitate comparison with 3D; the qualitative 2D behaviour is familiar from previous works studying this and $\sigma_h\gg 1$ regimes.)
\begin{itemize}[leftmargin=3mm, itemindent=0mm, labelsep=1mm]
  \item In 2D and 3D, the initial current sheet breaks up into smaller current sheets because of the tearing instability and starts reconnecting.  Small plasmoids---structures of higher plasma density contained by magnetic field (possibly magnetic islands or flux ropes)---grow, fed by reconnection outflows from elementary reconnecting current sheets.
  (\S\ref{sec:overview2d},~\ref{sec:overview3d})
  \item In 3D, the current sheet ripples
    due to RDKI---but this can be suppressed by guide magnetic field.
  (\S\ref{sec:overview3d},~\ref{sec:Bz3d})
  \item In 2D and 3D, reconnection pumps plasma and magnetic energy from the upstream region through X-points (or X-lines) into plasmoids; in the process, some upstream magnetic energy is converted to plasma energy and some remains in magnetic form in plasmoids. 
  (\S\ref{sec:overview2d},~\ref{sec:overview3d})
  \item In 2D, plasmoids (magnetic islands) are distinct structural units; they have closed (eventually circular), reconnected magnetic field lines that permanently trap plasma and magnetic energy from reconnection outflows.  They are essentially stable in 2D and can move about along the layer.  Two colliding plasmoids will merge into one (while mostly conserving the trapped flux and energy).  As reconnection continues (in a closed system), a single large plasmoid eventually engulfs (i.e., merges with) all other plasmoids, growing (and never shrinking) as long as reconnection continues.
  (\S\ref{sec:overview2d},~\ref{sec:overview3d})
  \item In 3D, plasmoids (flux ropes) can decay, not only losing their individual integrity but also converting their magnetic energy to plasma energy.  Importantly, this means that upstream magnetic energy is more completely converted to plasma energy in 3D (because 2D reconnection hoards some magnetic energy in plasmoids).  Strong guide field inhibits flux rope disintegration.
  (\S\ref{sec:overview3d},~\ref{sec:Lz3d})
  \item An indication of plasmoid decay---and the clearest signature of novel 3D behaviour that we have observed---is the magnetic energy $U_{Bt,\rm layer}$ in the plasmoid-containing current layer (defined in~\S\ref{sec:terminology}).  In 2D, $U_{Bt,\rm layer}$ rises until reconnection ceases; in 3D, it rises at the onset of reconnection, but then declines as plasmoids decay faster than reconnection can inflate them.  $U_{Bt,\rm layer}$ is more sensitive to 3D effects than other measures, and it also exhibits less stochastic variation from simulation to simulation and more closely correlates with~$L_z/L_x$ (Fig.~\ref{fig:energyAndFluxVsLz}).
  As a rough indicator of 3D effects, the (much easier to calculate) magnetic energy in~$B_y$ components can be substituted for $U_{Bt,\rm layer}$. 
  (\S\ref{sec:overview3d},~\ref{sec:Lz3d})
  \item In 2D reconnection, flux is conserved in a precise sense: the total upstream flux plus the flux around the largest plasmoid is almost constant, equal to the initial upstream flux.  In 3D reconnection, flux may be outright annihilated.  The small amount of magnetic energy $U_{Bt,\rm layer}$ left in the layer in 3D shows that upstream flux is eventually, if not directly, annihilated. It is unclear how much upstream flux is annihilated directly without undergoing any sort of reconnection, and how much first undergoes reconnection to be annihilated later as plasmoids decay.  The resemblance between 3D simulations and 2D simulations in both the $x$-$z$ and $y$-$z$ planes suggests that both these processes occur in 3D, but the precise extents cannot be estimated without additional diagnostics. (\S\ref{sec:unreconnectedFlux},~\ref{sec:overview3d},~\ref{sec:RDKIamp},~\ref{sec:Lz3d})
  \item We speculate that, in 3D, for weak guide field, small plasmoids may decay faster than large, long plasmoids, and that the decay of plasmoids 
  may disrupt reconnection in nearby X-lines in thin current sheets.  Thus, if something (such as guide magnetic field) can cause reconnection to build up plasmoids faster than they decay, that may prevent interference with X-points, maintaining reconnection in a more 2D-like fashion.
  \item Whereas 2D reconnection yields a highly-structured plasmoid hierarchy, in which large (system-size-scale), highly-magnetized plasmoids do not obliterate nearby thin (kinetic-scale) elementary current sheets, 3D reconnection tends to develop a thick, turbulent layer, throughout which the magnetic field is greatly diminished.
  (\S\ref{sec:overview2d},~\ref{sec:overview3d})
  \item In 3D, with weak guide field, the depletion of upstream magnetic energy and flux may
    start rapidly (as in 2D, or perhaps a little more slowly), but then
    continue at an order-of-magnitude slower rate for an
    order-of-magnitude longer time than in 2D.
    This may be due to the thickened layer, and/or to reconnection 
    occurring only in sparse, small areas of the layer.
    Alternatively, true reconnection may operate to a lesser extent in 3D, 
    with other, slower mechanisms (perhaps turbulent diffusion) 
    depleting magnetic field.
  (\S\ref{sec:pertAndEta3d},~\ref{sec:Lz3d})
  \item In 2D, the current sheet evolution from initial to final state 
    (in a closed system) is in a broad sense
    inevitable, despite stochastic plasmoid behaviour.
    In 3D, however, stochastic variability in the
    early current sheet evolution can lead to substantially different
    states at much later times 
    (e.g., different layer thicknesses with different amounts
    of magnetic energy converted to plasma energy and different
    continuing rates of energy conversion).
    (\S\ref{sec:variability3d}, ~\ref{sec:RDKIamp})
  \item 2D reconnection (in a closed system) ceases when the major plasmoid become large (of order of the system size) and a stable magnetic configuration is realized.  In 3D, we have not observed the cessation of magnetic energy depletion, even after 30--50~light-crossing times;
  the final 3D state may depend strongly on the early evolution.
   (\S\ref{sec:overview2d},~\ref{sec:Lx2d},~\ref{sec:Lz3d})
  \item In 3D, increasing the initial guide magnetic field suppresses 
    3D effects, and
    3D reconnection becomes increasingly similar to 2D reconnection.
    (\S\ref{sec:Bz3d})
  \item In 2D, guide magnetic field suppresses both reconnection and
    NTPA.  This is also true in 3D; however, 
    because guide field also suppresses 3D effects (which suppress
    reconnection), increasing the guide field enhances reconnection
    up to a point where 3D effects are effectively suppressed; 
    beyond that point, stronger guide field has the same suppression 
    effect as in~2D.
    (\S\ref{sec:Bz2d},~\ref{sec:Bz3d})
  \item Despite differences in 2D and 3D current sheet evolution, {\bf NTPA is robust} and remarkably similar in 2D and 3D as well as in initially-similar 3D simulations with manifestly different current sheet evolution.  This is the case regardless of guide field.
    (\S\ref{sec:Lz3d}, \ref{sec:Bz3d})
  \item Moreover, NTPA is robust in a way that depends on the upstream
   $\sigma_h=1$ and $B_{gz}/B_0$; 
   particle energy spectra are steeper than
   those in simulations with larger~$\sigma_h$, and stronger guide field results in steeper power-law particle spectra.
   (\S\ref{sec:highSigmah})
\end{itemize}

\subsection{Current sheet evolution in 2D}
\label{sec:sum2D}

Besides providing a baseline against which to compare
the~3D simulations, our~2D simulations constitute the first systematic study across a broad swath of the multidimensional parameter space describing~2D, $\sigma_h=1$ (ultrarelativistic pair-plasma) reconnection, which is of interest 
in its own right. Here we summarize the most important 2D results.
\begin{itemize}[leftmargin=3mm, itemindent=0mm, labelsep=1mm]
  \item Current sheet evolution depends significantly on the ambient~$\sigma_h$ and guide field strength~$B_{gz}/B_0$.
   (\S\ref{sec:Bz2d}, \ref{sec:highSigmah})
  \item The evolution of magnetic and plasma energy, and resulting NTPA,
    is relatively (but not completely) 
    insensitive to the initial current sheet configuration,
    including the initial magnetic field perturbation, the 
    density, temperature, and thickness of the initial current sheet,
    and the drift speed of the current sheet plasma.
    Thus reconnection is governed mostly by the ambient background plasma as described below.  
    (\S\ref{sec:pertAndEta2d})
  \item For zero guide field (and, importantly, $\sigma_h=1$), reconnection rates (normalized to $B_0c$) are typically around $\beta_{\rm rec}\approx 0.02$--$0.03$, or (normalized to $B_0v_A$), $(c/v_A)\beta_{\rm rec}\approx 0.03$--$0.04$, the same order of magnitude but still significantly less than the often-assumed value of~0.1 \citep{Cassak_etal-2017}. [Values of $(c/v_A)\beta_{\rm rec}\approx 0.1$ are actually realized in 2D reconnection in relativistic pair plasma with $\sigma_h \gg 1$ \citep[e.g.,][]{Sironi_Spitkovsky-2014,Guo_etal-2015,Werner_etal-2018,Sironi_Beloborodov-2020}.]
    (\S\ref{sec:Lx2d})
  \item The magnetic energy evolution in systems with $L_x \gtrsim 160\sigma\rho_0$ already appears to
    depend only very weakly on system size $L_x$ (for $B_{gz}=0$, up to $L_x=2560\sigma\rho_0$), although larger simulations experience slightly
    slower reconnection rates than smaller simulations.
    (\S\ref{sec:Lx2d})
  \item For $B_{gz}=0$ (and, importantly, $\sigma_h=1$), 
    we observe NTPA (for $\sigma_h=1$) with a steep power-law slope 
    of roughly $f(\gamma)\sim \gamma^{-4}$.
    (\S\ref{sec:Lx2d})
  \item NTPA also has a fairly weak dependence on $L_x$ 
    for $L_x\gtrsim 160\sigma\rho_0$;
    although precise measurement of steep power laws is difficult, 
    our results (for $B_{gz}=0$) 
    suggest that the power-law index varies within only about
    10~per~cent for $80 \lesssim L_x/\sigma\rho_0 \lesssim 2560$.
    The maximum particle energy (or the high-energy cutoff of the 
    particle power-law energy spectrum) clearly increases 
    sublinearly with~$L_x$,
    from around~$7\sigma$ for $L_x=80\sigma\rho_0$ to 
    perhaps~$33\sigma$ for $L_x=2560\sigma\rho_0$.
    I.e., for a 32X increase in system size, the particle cutoff energy 
    increases by~$\sim$5X,
    consistent with a~$\sim\sqrt{L_x}$ scaling 
    \citep[cf. ][]{Petropoulou_Sironi-2018,Hakobyan_etal-2021}.
    (\S\ref{sec:Lx2d})
  \item Guide magnetic field slows reconnection.
    Increasing the guide field from zero to $B_{gz}=4B_0$ reduces the
    reconnection rate 
    (normalized to $B_0c$; cf.~\S\ref{sec:unreconnectedFlux}) from 
    $\beta_{\rm rec}\approx 0.03$ to~$0.007$.
    This is roughly (within 30~per~cent) consistent with 
    $\beta_{\rm rec} \approx 0.04 v_{A,x}/c$ where
    $v_{A,x}^2/c^2=\sigma_{h,\rm eff}/(1+\sigma_{h,\rm eff})$ and
    $\sigma_{h,\rm eff}\equiv (1/\sigma_h + B_{gz}^2/B_0^2)^{-1}$.
    (\S\ref{sec:Bz2d})
  \item Guide magnetic field inhibits NTPA,
    yielding steeper power-law spectra.
    The power-law index
    steepens from~$\sim \gamma^{-4}$ at~$B_{gz}=0$ to~$\sim \gamma^{-6}$
    at~$B_{gz}=B_0$, and it continues to be steeper for higher~$B_{gz}$.
    Keeping in mind the difficulty of measuring steep power laws,
    we find that the power-law index is roughly
    $p\approx 4.4 + 2B_{gz}/B_0$; this should be taken as a rough guide
    and not confirmation of a linear dependence on~$B_{gz}$.
    For $B_{gz}/B_0 \gtrsim 1.5$, we measure $p\gtrsim 8$
    and it is debatable whether the spectra are significantly nonthermal.
    (\S\ref{sec:Bz2d})
\end{itemize}

\subsection{Current sheet evolution in 3D}
\label{sec:sum3D}

Our most important results for 3D simulations with $\sigma_h=1$ 
(with system sizes in the range $256 \leq L_x/\sigma\rho_0\leq 512$):
\begin{itemize}[leftmargin=3mm, itemindent=0mm, labelsep=1mm]
  \item 3D effects can be introduced gradually by increasing the length 
    $L_z$ of the system in the third dimension.  We find that,
    for $B_{gz}=0$ and the range of system sizes considered, simulations
    with $L_z\gtrsim L_x/8$ can exhibit fairly 
    obvious departures from 2D reconnection, 
    while those with $L_z\lesssim L_x/16$ almost always closely resemble 2D.
    (\S\ref{sec:overview3d},~\ref{sec:Lz3d})
  \item As $L_z/L_x$ is increased, 3D effects first appear in
    the time evolution of the magnetic energy in the layer,
    $U_{Bt,\rm layer}(t)$.
    (The rest of this point applies to~$B_{gz}=0$ and~$a=0$.) 
    In 2D or very small $L_z/L_x \ll 1/32$,
    $U_{Bt,\rm layer}(t)$ increases as long as reconnection
    continues (up to around 25~per~cent of $U_{B0}$), 
    and does not decrease.  
    In 3D, $U_{Bt,\rm layer}(t)$
    increases at very early times, but then starts to decrease because
    plasmoids can disintegrate in~3D.
    Even when~$L_z/L_x$ is small enough to avoid
    other departures from 2D reconnection, $U_{Bt,\rm layer}$ may be
    significantly less than in~2D.
    The larger~$L_z/L_x$ is, the smaller~$U_{Bt,\rm layer}(t)$ is
    (at any given~$t$), up to around $L_z/L_x\lesssim 1$.
    For $L_z/L_x \gtrsim 1$, $U_{Bt,\rm layer}$ may rise to just a couple
    per~cent of~$U_{B0}$ before falling to lower values.
    (\S\ref{sec:overview3d},~\ref{sec:Lz3d})
  \item 
    Random variability is substantially greater in 3D than in 2D;
    identical simulations, aside from different random particle 
    initialization, can yield very different behaviour 
    (e.g., in terms of magnetic energy versus time) and very different
    states even after long times.  This contrasts strongly with the 
    variability in 2D, where similarly identical simulations will still
    yield the same magnetic energy release 
    versus time (when measured over long timescales) 
    and end up in nearly the same final state after the same amount
    of time.  Although more familiar, it is the 2D case that is more
    remarkable here---the stable plasmoids trap energetic plasma from reconnection outflows and prevent it from interfering with regions upstream of thin elementary current sheets even though the plasmoids are orders of magnitude larger.
    (\S\ref{sec:variability3d})
  \item 
     3D effects increase the sensitivity to the details of the initial current sheet configuration.
    (\S\ref{sec:pertAndEta3d})
  \item An initial magnetic field perturbation, uniform in $z$, can
    suppress 3D effects on reconnection (at least for $B_{gz}=0$), especially at earlier times.
    For this reason, our 3D simulations were 
    initialized with no perturbation (i.e., $a=0$) except when specifically studying the effect of~$a$.
    (\S\ref{sec:pertAndEta3d})
  \item For $B_{gz}=0$, increasing the initial current sheet overdensity $\eta$ 
    (while maintaining pressure balance) might weakly increase the early-time reconnection rate and make reconnection more 2D-like, but a larger
    ensemble of simulations will be needed to measure this effect beyond
    stochastic variation.
    (\S\ref{sec:pertAndEta3d})
  \item For weak guide field, RDKI modes cause the current sheet to ripple.  In the linear phase of the instability, a mode grows to a rippling amplitude comparable to its wavelength, without releasing much magnetic energy.  The mode may subsequently enter the nonlinear phase of the instability, during which the current sheet becomes highly distorted, folding over on itself and rapidly releasing magnetic energy within the rippling amplitude.
    This can occur even in a 2D~$y$-$z$-plane simulation (but not in the 2D~$x$-$y$ geometry used to study reconnection), and
    the magnetic energy depletion can occur as fast as in
    2D reconnection, even though driven by a manifestly
    different large-scale mechanism.
    (\S\ref{sec:RDKIamp}, especially Fig.~\ref{fig:kinkAmpVsBetad3d})
  \item For $B_{gz}=0$, varying the initial current sheet drift speed, $\beta_d c$ (while keeping~$\eta \beta_d$ hence~$\delta/\sigma\rho_0$ fixed), has
    a dramatic effect on current sheet evolution due to its influence
    on the RDKI.  Increasing $\beta_d$ increases 
    the growth rate of longer-wavelength RDKI modes, which can release more magnetic energy (than shorter-wavelength modes) if they grow fast enough to enter the nonlinear stage---simply because they can grow to larger amplitude.
    (\S\ref{sec:pertAndEta3d},~\ref{sec:RDKIamp})
  \item 
    The influence of~$\beta_d$ is only partly explained by its effect on
    the linear-phase growth rate. 
    A slower-growing, 
    longer-wavelength mode may overtake a faster-growing, 
    shorter-wavelength mode because the latter saturates nonlinearly at smaller
    amplitude.  
    The most influential mode may have much longer wavelength than the
    (linearly) most unstable RDKI mode.
    (\S\ref{sec:RDKIamp},~\ref{sec:Lz3d})
  \item The nonlinear development of RDKI may play a key role in the
    stochastic variability of 3D current sheet evolution (for weak guide field).
    Although in some cases an RDKI mode with long wavelength
    $\lambda_{z,\rm RDKI}\gg \delta$ 
    will grow to a large amplitude and enter nonlinear development, rapidly
    and dramatically changing the layer structure
    while converting magnetic energy to plasma energy,
    growth to the nonlinear stage is not inevitable.
    Nonlinear RDKI development may be triggered, e.g., in only a fraction of an ensemble of
    identically-initialized simulations 
    (up to random particle initialization), 
    thus putting them on very different evolutionary tracks.
    (\S\ref{sec:RDKIamp},~\ref{sec:Lz3d}, Fig.~\ref{fig:layerOffsetY})
  \item For $B_{gz}=0$, the reconnection rate (normalized to $B_0 v_A$; cf.~\S\ref{sec:unreconnectedFlux}) in the early stage
    of 3D development (typically $\lesssim 10L_x/c$) varies from
    at most $(c/v_A)\beta_{\rm rec}\sim 0.05$ (at small $L_z/L_x$, the same as 
    2D) to as little 
    as~$(c/v_A)\beta_{\rm rec}\sim 0.005$ for large $L_z/L_x$;
    however, stochastic evolution results in large variability, and 
    even for large~$L_z/L_x$, higher rates may sometimes be observed.
    (\S\ref{sec:Lz3d})
  \item (For $B_{gz}=0$) For $L_z/L_x\gtrsim 1/8$, most simulations exhibit an early stage
    of fast magnetic energy release 
    (but nonetheless slower than in 2D) lasting
    tens of~$L_x/c$, followed by a slower stage with reconnection
    rates (or more precisely, upstream flux depletion rates) 
    an order of magnitude lower, i.e., $(c/v_A)\beta_{\rm rec}\sim 0.0005$.
    The slower stage can last at least up to~$50L_x/c$, and perhaps
    much longer.
    (\S\ref{sec:Lz3d})
  \item Our results are consistent with the suggestion 
  of \citet{Yin_etal-2008} that local reconnection rates are as high as
  in 2D, but in 3D only a relatively small area of the layer undergoes
  active reconnection, resulting in a low global reconnection rate.
  (However, we did not measure local reconnection rates.)
    (\S\ref{sec:overview3d})
  \item NTPA is nearly the same
    in 3D as in 2D, if compared at times when the same amount of 
    magnetic energy has been converted to plasma energy.
    It is equally remarkable that NTPA is nearly the same in
    different 3D simulations that, despite being macroscopically identical at~$t=0$, undergo very different current sheet
    evolution (and 2D $y$-$z$ simulations that suffer 
    large-wavelength nonlinear RDKI also yield similar NTPA).
    (\S\ref{sec:RDKIamp}, \ref{sec:Lz3d}, \ref{sec:Bz3d})
  \item For $B_{gz}=0$ (and, importantly, $\sigma_h=1$), 
    particle energy spectra show power laws 
    $f(\gamma)\sim \gamma^{-p}$ with $p \in [3.4,4.6]$.
    NTPA actually appears to be slightly
    more efficient in 3D than in 2D, sometimes yielding slightly harder power 
    laws (e.g., $\gamma^{-3.4}$), and with the fraction of particles
    attaining $\gamma>10\sigma$ increasing by almost an order of magnitude
    from 2D (or $L_z/L_x \ll 1/8$) to 3D ($L_z/L_x \gtrsim 1/8$).
    Enhanced NTPA in~3D has been previously supposed to result from the weaker trapping of particles in plasmoids which, in 2D, ends particle acceleration \citep{Dahlin_etal-2015,Dahlin_etal-2017,Li_etal-2019,Zhang_etal-2021arxiv}.
    However, because of the steep power law, these differences in NTPA
    between~2D and~3D may be
    ultimately unimportant relative to other effects and uncertainties.
    (\S\ref{sec:Lz3d})
  \item Weak guide field can enhance 3D reconnection because it suppresses 3D effects (that slow reconnection); this is noticeable in reconnection rates and the evolution of magnetic energy even for weak guide field, $B_{gz}/B_0\gtrsim 0.1$.
    However, as the guide field becomes strong enough to suppress 3D effects substantially, reconnection behaves as in 2D, where increasing the guide field slows reconnection and inhibits NTPA.  We believe that, for $\sigma_h=1$, guide field causes 3D reconnection to resemble 2D reconnection strongly around $B_{gz}/B_0 \gtrsim 1$ (but see the parenthetical caveat in the next bullet).
    (\S\ref{sec:Bz3d})
  \item
    Stronger guide field therefore raises the 
    reconnection rate as long as 3D effects are more suppressive than
    guide field effects.  The reconnection rate rises from 
    $\beta_{\rm rec}\approx 0.002$ at~$B_{gz}=0$ to a maximum
    around $\beta_{\rm rec}\approx 0.02$ for 
    $0.25 \lesssim B_{gz}/B_0 \lesssim 0.75$.  
    For even stronger guide field
    the reconnection rate falls (however, because of system-size 
    limitations, we explored only up to $B_{gz}/B_0=1$ in~3D, with just a 
    single simulation for each value of~$B_{gz}/B_0$, and so this
    conclusion needs to be confirmed by future studies with better 
    statistics and higher~$B_{gz}$).
    (\S\ref{sec:Bz3d})
  \item For each value of $B_{gz}/B_0$ investigated ($0\leq B_{gz}/B_0 \leq 1$), NTPA is practically the same in 3D as in 2D.
    (\S\ref{sec:Bz3d})
\end{itemize}

\section{Discussion}
\label{sec:discussion}

\subsection{Differences between 2D and 3D: comparison with previous works}
\label{sec:discussionPrevWork}

Perhaps the most important general observation of this work is that, in the $\sigma_h=1$ regime, thin current sheets can evolve very differently in 2D and in 3D.  However, they do not always evolve differently.  In particular, a guide field---possibly even $B_{gz}/B_0\gtrsim 0.25$---can suppress 3D effects, and so can a small initial perturbation that is uniform in the third dimension.

Especially in cases with very weak guide field, 3D systems might not be well-modelled by 2D simulations.  Important consequences or signatures of 3D effects include slower rates of magnetic-to-plasma energy conversion and disintegration of plasmoid or flux rope structures (which releases additional magnetic energy).  Moreover, magnetic energy can be converted to plasma energy via a completely different mechanism---the nonlinear development of the RDKI (see~\S\ref{sec:discussionRDKI}).  Nevertheless, despite these 3D effects, NTPA remains essentially the same in 3D as in 2D.

As pointed out in~\S\ref{sec:highSigmah}, the dramatic differences in~3D might be a bit of a surprise because they were much less evident in reconnection in the~$\sigma_h\gg 1$ regime, despite clear presence of RDKI.
However, some of these effects have been previously observed in~$\sigma_h\sim 1$ pair-plasma PIC simulations.
Of particular interest for their close relevance and identification of similar 3D effects, are: \citet{Yin_etal-2008}---hereafter, Yin08; \citet{Liu_etal-2011}---Liu11; and \citet{Kagan_etal-2013}---K13.

Yin08, Liu11, and K13 all used PIC simulation to study 3D reconnection in pair plasma with~$\sigma_h\sim 1$; simulation parameters are compared in table~\ref{tab:simParams}.  Yin08 initialized subrelativistic plasma with $\theta_b=0.016$ and $\sigma_h\approx 0.2$ in simulations of size $L_x=240\sigma\rho_0$ and $L_z\leq L_x$, while Liu11 and K13 both used transrelativistic plasma with $\theta_b=1$ and $\sigma_h \in [0.5, 1.5]$; Liu11 had $L_x=60\sigma\rho_0$ and $L_z\leq L_x$, and K13 studied two configurations, one with $L_x=52\sigma\rho_0$ and $L_z=1.6L_x$ and another with $L_x=110\sigma\rho_0$ and $L_z=1.1L_x$.
None of these simulations used any initial perturbation.  K13 was the only one that did not use a standard Harris-sheet set-up; it was also the only one to study the effect of guide field.

\begin{table}
\caption{\label{tab:simParams}
Approximate (ranges of) parameters for the largest 3D
simulations of Yin08, Liu11, and K13 (with two series of simulations, S1 and S2), translated to the terminology of this paper; values for 3D simulations of this paper are also shown.  Parameters not defined in~\S\ref{sec:setup} or table~\ref{tab:etaEffect} are as follows: $v_A$ is the upstream Alfv\'{e}n velocity; 
$\beta_{\rm plasma} \equiv 8\upi n_b \theta_b m_e c^2 / B_0^2$ is the upstream plasma beta;
$d_{e}^{\rm NR}\equiv (m_e c^2/4\upi n_b e^2)^{1/2}$ and $d_{e,d}^{\rm NR}\equiv (m_e c^2/4\upi n_d e^2)^{1/2}$ are the nonrelativistic collisionless background and drifting plasma skin depths (these nonrelativistic scales are not applicable to this paper); 
$\ell_y$ is $L_y$ divided by the number of layers simulated;
$t_{\rm run}$ is the final simulation time.
}
\begin{tabular}{llllll}
& Yin08 & Liu11 & K13(S1) & K13(S2) & this work \\
\hline
$\sigma$   & 0.21 & 6.7 & 2 & 4 & $10^4$ \\
$\theta_b$ & 0.016 & 1   & 1       & 1       & $2.5\times 10^3$ \\
$\sigma_h$ & 0.20 & 1.5 & 0.46     & 0.92    & 1.0 \\
$v_A/c$    & 0.41 & 0.78 & 0.56    & 0.69    & 0.71 \\
$\beta_{\rm plasma}$ 
           & 0.15 & 0.30 & 1 & 0.5 & 0.5 \\
$B_{gz}/B_0$ & 0  & 0    & 0--1    & 0--1    & 0--1 \\
\hline
$\eta=n_d/n_b$     & 3.3  & 3.3  & $^* 1.5$ & $^* 2.3$ & 0.5--5 \\
$\theta_d/\theta_b $ & 1 & 1   & $^*$ ?     & $^*$ ?     & 0.4--5\\
$\beta_d$  & 0.36 & 0.82 & $^*\!\lesssim 0.36$ & $^*\!\lesssim 0.36$ & 0.075--0.6 \\
\hline
$\delta/\sigma\rho_0$  & 0.18  & 0.21 & 3.7 & 3.9 & 0.67--3.3 \\
$\rho_b/\sigma\rho_0$  & 0.96 & 0.48 & 1.6 & 0.79 & 0.75 \\
$d_{e}/\sigma\rho_0 \phantom{|}^\dagger$   
                       & 2.2 & 0.72 & 1.3 & 0.92 & 0.87  \\
$d_{e}^{\rm NR}/\sigma\rho_0$ 
                       & 2.2 & 0.39 & 0.71 & 0.50 & N/A 
\\
$d_{e,d}^{\rm NR}/\sigma\rho_0 \phantom{}^\ddag$ 
                       & 1.2 & 0.21  & $^*$0.58 & $^*$0.33 & N/A 
\\
\hline
$L_x/\sigma\rho_0$     & 240 & 60 & 52 & 110 & 256--512 \\
$\ell_y/L_x$ & 1 & 1 & 0.5 & 0.5 & 1 \\
$L_z/L_x$    & 0--1 & 0--1 & 1.6 & 1.1 & 0--1.5 \\
$t_{\rm run} v_A/L_x$ & 4.1 & 5.0 & $> 4.2$  & 2.6  & 7--35\\
\hline
\end{tabular} \\
$^*$K13 used a uniform density~$n_b$ with varying $\beta_d=\beta_d(y)$ to satisfy Ampere's law (without a separate drifting component), but pressure balance was not initially satisfied, resulting in an immediate compression of the current sheet by a factor~$\simeq\eta$ (and unknown temperature increase).
\\
$^\dagger$K13 referenced lengths to $\lambda_p \equiv d_{e}$.
\\
$^\ddag$Yin08 referenced lengths to $d_i\equiv d_{e,d}^{\rm NR}$.
\\
$^\ddag$Liu11 referenced lengths to $d_i\equiv \sqrt{2} d_{e,d}^{\rm NR}$.

\end{table}

All these previous $\sigma_h\sim 1$ studies observed tearing and reconnection as well as an initial linear growth and later nonlinear growth of RDKI modes.
Yin08 and Liu11 report similar reconnection rates 
with no clear/significant difference between~2D and~3D, measuring
$(c/v_A)\beta_{\rm rec}\approx 0.06$--0.08 (with respect to the magnetic field~$B$ averaged over the ``inflow surface'' rather than the asymptotic upstream~$B_0$); 
K13 reports 3D rates of $(c/v_A)\beta_{\rm rec}\approx 0.05$--0.08.
These reconnection rates for~$\sigma_h\sim 1$ are all roughly half of $(c/v_A)\beta_{\rm rec}$ measured for~$\sigma_h\gg 1$, 
and they are consistent with our results for~$L_z\ll L_x$.
However, for $L_z\sim L_x$ we measured significantly lower values with $(c/v_A)\beta_{\rm rec} \lesssim 0.02$ (see Fig.~\ref{fig:reconRateVsLz}).
This may be because the previous works measured the rates in smaller systems at earlier times, or because (for Yin08 and Liu11) they normalize to a local magnetic field smaller than~$B_0$.
In addition to calculating similar reconnection rates in~2D and~3D, Yin08 shows very similar magnetic energy evolution in~2D and~3D
while hinting that the energy depletion appeared to be ``somewhat slower'' in the simulation with largest~$L_z$ because of the limited size of the reconnecting patch.

Of these three, Liu11 is the only one that (like us) specifically reports an early stage of faster magnetic energy depletion (attributed to tearing/reconnection) \emph{followed by a stage of slower depletion} (attributed to nonlinear RDKI development).  In~Liu11, the slower stage results in significantly more release of magnetic energy in~3D than in~2D---a signature of flux rope disintegration (in contrast, Yin08 sees similar magnetic depletion in~2D and~3D; K13 shows $U_{B}(t)$ only for $B_{gz}=B_0/4$, and not for a very long time, but if anything, the rate of magnetic depletion speeds up at later times).

Both Yin08 and Liu11 (and possibly K13) appear to observe more growth and merging of flux ropes in~3D than we do; however, Liu11 and K13 run long enough to see the end of the ``period'' of flux rope merging, leaving a much less structured, more turbulent layer (which we have also observed).
Apparent qualitative differences among the various simulations may have more to do with simulation size than any fundamental difference; in fact, K13 reports (for $B_{gz}=B_0/4$) that in a smaller simulation flux ropes merge into a final, single flux rope (as in~2D), while in a larger simulation this merging is disrupted before reaching a single, large flux rope.
Although these simulations all ran in somewhat different regimes with different initial current sheet configurations (which, we have shown, can make a difference---see~\S\ref{sec:pertAndEta3d}--\ref{sec:RDKIamp}), the combined results appear consistent with the notion that as systems become increasingly larger, the stage of 2D-like reconnection becomes increasingly transient.
This may suggest that the later, slower stage---with significant~3D effects---is more relevant for astrophysically-large systems.

All these papers also observe nonlinear RDKI in some form, but always following a stage of tearing and reconnection (whereas we have shown that for some initial current sheet configurations, RDKI may drive the earliest stage of rapid magnetic energy release; see~\S\ref{sec:RDKIamp}).  Interestingly, in Yin08, a long (in~$z$) current sheet forms between growing flux ropes, undergoes extreme distortion and folding via nonlinear RDKI (which Yin08 calls ``secondary'' kinking), \emph{and subsequently reforms} as a thin, reconnecting current sheet, suggesting such behaviour may be cyclic.
Liu11 also observes ``self-organization'' of initially-patchy reconnecting regions into a ``highly-elongated'' current sheet between growing flux ropes, although this sheet appears to be permanently disrupted by nonlinear RDKI (i.e., it does not reform as in Yin08).

Both Liu11 and K13 investigated particle acceleration and report nonthermal particle energy spectra.
The spectra are consistent with power laws~$\gamma^{-p}$ with (very roughly), for Liu11, $p\approx 2.5$ up to a cutoff $\gamma_c\sim 4\sigma$, and for K13, $p\approx 3.5$ up to $\gamma_c\sim 7\sigma$ (for the  largest simulation, labelled S2K025L, with $\sigma=4$, $B_{gz}/B_0=0.25$, and $L_x=110\sigma\rho_0$; for a smaller simulation, $L_x=55\sigma\rho_0$, the spectra were nearly identical for $B_{gz}/B_0=0$ and~$0.25$).
Because the cutoffs of power laws can be misleading to compare for different power-law slopes, we also note a more straightforward measure of the (high) energy at which $f(\gamma)$ becomes small, namely the value~$\gamma_{-4}$ for which $f(\gamma_{-4})\equiv 10^{-4} \textrm{max}_\gamma f(\gamma)$; for Liu11, $\gamma_{-4}=7.1\sigma$ and for K13 (S2K025L), $\gamma_{-4}=7.8\sigma$; for our 3D simulations with $L_x=512\sigma\rho_0$, $\gamma_{-4}=10\sigma$.

Considering the somewhat different parameters, including system size and run time, these results seem reasonably consistent.
Liu11 and (to a lesser extent) K13 might measure less steep spectra than we do, but the cause of this could easily be the larger~$\sigma_h$ in Liu11 and/or the sub- or trans-relativistic upstream plasma, and/or the smaller system sizes and different initial current sheet configurations.
K13 investigated the effect of guide field, running simulations with $B_{gz}/B_0=0,$ 0.25, and~1.  Although K13 does not report on relative reconnection or magnetic energy depletion rates, there is a suppression of NTPA for $B_{gz}/B_0=1$ relative to the weaker guide fields, also consistent with our results.

Thus it appears that the most important 3D effects that we have observed in an evolving current sheet are consistent with previous observations in similar parameter regimes, although the 3D effects have generally become clearer with our larger system size and scan over a broader range of simulation parameters.
Importantly, we have studied 3D effects with respect to different current sheet configurations, which can dramatically alter the strength of these effects, especially at early times.
In particular, we have highlighted the importance of (sufficiently fast-growing) long-wavelength RDKI modes---namely that they can grow to large amplitudes before entering the nonlinear stage in which the current layer can be dramatically transformed while rapidly releasing magnetic energy (see~\S\ref{sec:RDKIamp} and~\S\ref{sec:discussionRDKI}).
In addition, we have shown that NTPA can be driven by 2D-like reconnection, or the slower stage of reconnection (and possibly RDKI), or by rapid nonlinear RDKI, yielding very similar particle energy spectra in all cases.

Having discussed the $\sigma_h\sim 1$ regime, we briefly comment on differences between 3D reconnection in the~$\sigma_h=1$ regime compared with the~$\sigma_h \gg 1$ regime, where 3D reconnection has been observed to be quite similar to 2D reconnection \citep[e.g.,][]{Sironi_Spitkovsky-2014,Guo_etal-2014,Guo_etal-2015,Werner_Uzdensky-2017,Guo_etal-2020arxiv}---more details were given in~\S\ref{sec:highSigmah}.
Based on these $\sigma_h\gg 1$ studies, it is likely that large~$\sigma_h$ tends to enhance reconnection and/or suppress 3D reconnection-disrupting effects.
However, most of the large-$\sigma_h$ studies have used an initial perturbation of some kind, which we have shown can suppress 3D effects, especially at early times.
The extent of the role of the initial perturbation requires further study; regardless of the results, however, what is really needed is a better understanding of system-size dependence in 3D.
In the~$\sigma_h\sim 1$ regime, a 2D-like reconnection stage may be very transient in larger simulations; and due to resolution requirements, feasible simulations with $\sigma_h\sim 1$ can be relatively larger than those with $\sigma_h \gg 1$.  It may be, therefore, that larger 3D simulations with $\sigma_h\gg 1$ begin to show distinctly 3D effects---such as the slowing of reconnection and disruption by RDKI, as well as the possibility of magnetic energy release through RDKI---at much larger sizes than anyone has yet simulated.

\subsection{Another magnetic energy release mechanism: nonlinear RDKI}
\label{sec:discussionRDKI}

We have found that, in at least some cases, a thin~3D current sheet can release magnetic energy by a mechanism that may have nothing to do with magnetic reconnection---namely the \emph{nonlinear} RDKI, which causes a highly-kinked (or rippled) current sheet to fold over on itself \citep[see~\S\ref{sec:RDKIamp};][]{Zenitani_Hoshino-2007}.
This mechanism can operate in 2D $y$-$z$ geometry (perpendicular to the magnetic field), where large-scale magnetic reconnection is impossible and upstream flux is directly annihilated, and it can release magnetic energy as fast as 2D reconnection while generating similar NTPA.
Importantly, the earlier \emph{linear} development of RDKI releases little magnetic energy and does not yield significant NTPA (and it might not even interfere with reconnection).
Because the nonlinear RDKI causes the rippling to saturate at an amplitude of the order of the instability wavelength, longer-wavelength RDKI modes can reach larger amplitudes and thus have greater impact, even if they grow more slowly; but of course they still have to grow enough to reach the nonlinear stage to trigger rapid magnetic energy release.
While it is possible that, in the nonlinear RDKI development, small-scale reconnection (or, alternatively, turbulence-enhanced magnetic diffusion) plays a role in releasing magnetic energy, it is clearly RDKI at large scales that drives the process of energy conversion.

As an alternate channel for releasing magnetic energy, RDKI may offer a neat solution to the ``triggering'' or ``onset'' problem faced by reconnection \citep[e.g.,][]{JiAstro2020Decadal-2019etal}.
This is the problem of reconciling the slow timescale of current sheet formation and magnetic energy build-up with the fast timescale of magnetic energy release; the current sheet must be relatively stable as it slowly forms, until at some point it suddenly becomes unstable---e.g., reconnection is triggered and releases energy rapidly \citep[e.g.,][]{Pucci_Velli-2014,Tenerani_etal-2016,Uzdensky_Loureiro-2016,Comisso_etal-2017,Huang_etal-2017,Huang_etal-2019}.
With reconnection, solving this problem likely requires understanding the mechanism behind current sheet formation.
With RDKI, there is a simpler explanation: the current sheet may form and RDKI (at sufficiently long wavelength) may grow slowly, because the linear stage of RDKI does not release much magnetic energy.
The trigger occurs when the RDKI amplitude becomes comparable to its wavelength, and nonlinear development begins.

However, to understand its potential astrophysical importance, it will be critical to determine the dependence of the nonlinear RDKI on system size.
For any given wavelength, the nonlinear RDKI cannot continue forever; it dissipates magnetic energy within a layer of thickness (at most) comparable to the wavelength.  It might not continue to release upstream magnetic energy in the way that 2D reconnection does (as discussed in~\S\ref{sec:RDKIamp}, one might suppose 2D reconnection would saturate after plasmoids grow to the size of the elementary current sheet; however, simulations show that plasmoids detach and move away without disrupting the elementary current sheets, so that reconnection does not saturate until plasmoids grow to the system-size scale).  However, RDKI could potentially continue to develop at different, longer wavelengths.

For instance, an RDKI mode with wavelength~$\lambda_{z,\rm RDKI}$ may grow in a current sheet of thickness~$\sim \delta$; if $\lambda_{z,\rm RDKI}\gg \delta$ and the mode reaches an amplitude of order~$\lambda_{z,\rm RDKI}$, nonlinear development may release the magnetic energy within a layer of thickness~$\delta' \sim \lambda_{z,\rm RDKI}$, resulting in a new current sheet of greater thickness,~$\delta' \gg \delta$.
Because shorter-wavelength modes are limited to smaller amplitudes, longer-wavelength modes tend to be more influential.
Indeed, in some of our simulations, current sheet evolution was eventually dominated by the RDKI mode with the longest-possible wavelength, $\lambda_{z,\rm RDKI}\sim L_z$; if this behaviour continues up to systems of arbitrary size, nonlinear RDKI could potentially release astrophysically-large amounts of energy.
This could occur in one fell swoop, or in a series of stages (bursts) cascading up to the system size: the thickened layer with $\delta'$ might subsequently be destabilized by a new mode with $\lambda_{z,\rm RDKI}' \gg \delta'$, which might grow enough to enter the nonlinear phase, suddenly releasing energy in a layer of thickness $\delta'' \sim \lambda_{z,\rm RDKI}'$, etc., until the entire system is substantially depleted of magnetic energy.
Alternatively, this process could slow to an effective halt well before reaching the system size, with further energy release requiring other processes to thin out the layer (e.g., the processes responsible for the current sheet formation in the first place). 
Whether and how fast nonlinear RDKI, operating in sufficiently large systems, can release astrophysically-large amounts of energy, are important questions that we leave to future research.

\subsection{Observable astrophysical consequences}

In this paper we have investigated the plasma dynamics of current sheet evolution as well as resulting NTPA, which are unlikely to be directly observable in any astrophysical source; however, synchrotron and inverse Compton emission from high-energy particles may be observable.
Although we must leave a detailed study of radiation to future work, we can infer some observable consequences under simplifying assumptions.
Synchrotron and inverse Compton emission from high-energy electrons and positrons will depend on the shape and normalization of the distribution of accelerated particles.
Importantly, we have shown that the magnetization~$\sigma_h$ and guide field strength~$B_{gz}/B_0$ strongly affect both energy conversion rates and the NTPA spectrum; on the other hand, the dimensionality (2D vs.~3D) and initial current sheet configuration can affect energy conversion rates but, at most, weakly affect the NTPA spectrum.
This is a double-edged sword.
On one hand, we may be able to infer~$\sigma_h$ and~$B_{gz}/B_0$ from observations conveniently without knowing other details.
On the other hand, we may not be able to determine less influential details of the current sheet or even the effective dimensionality of astrophysical current sheets from observations.  Here we say ``effective dimensionality'' because we have seen that, under some circumstances---e.g., strong guide field, small $L_z/L_x$, or perhaps some sort of perturbation that encourages uniformity in~$z$---a 3D current sheet can evolve as if it were in~2D.

The most basic observable signatures are simply the radiation intensity and (for a flaring event) duration.
For this reason the study of astrophysical reconnection has been justifiably obsessed with measuring and understanding the reconnection rate to determine whether reconnection can explain rapid magnetic energy releases.
We have observed that the reconnection rate in~2D simulations is about what would be expected for ``fast'' reconnection, i.e., $(c/v_A)\beta_{\rm rec}\sim 0.1$.
If more precision is desired, we have shown that the actual value is somewhat less than~0.1 for $\sigma_h=1$ and $B_{gz}=0$, and guide field can slow the rate even more, up to about a factor of~3 for strong guide field.
The reconnection rate in~3D can be substantially lower, leading to lower radiated power and longer duration (for a given source size); depending on precisely how 3D reconnection is triggered, it may yield two stages (as in our simulations), an initial flare somewhat less bright than one would expect from 2D simulations, following by a much longer, much dimmer afterglow.
Observed intensity could thus be an important diagnostic of effective dimensionality.
Unfortunately, it is complicated by the sensitivity to initial current sheet configurations, and also the difficulty of relating the total radiated power to the observed intensity at a particular viewing angle (discussed below).

The next most basic signature is probably the spectral energy distribution of observable radiation. 
The photon spectra are a consequence of the particle spectra and should therefore depend on~$\sigma_h$ and~$B_{gz}/B_0$, but may be fairly insensitive to dimensionality and initial current sheet details.
For instance, for~$\sigma_h=1$ and $B_{gz}/B_0= 0$, we have seen that the particle energy power-law index is around~$p\approx 4$, and we can infer (assuming a steady-state, uniform, isotropic, weakly-radiative system) that the emitted photon energy power-law index would be around~$\alpha = (p-1)/2 \approx 1.5$.
For stronger~$B_{gz}$, the photon spectrum would steepen accordingly.

However, more detailed computation of emitted radiation from self-consistent PIC simulations is needed to determine whether these conclusions are valid outside of a much-simplified radiation emission model.
In particular, we need to consider the spatial- and angular-dependence of accelerated particles.
In addition, if radiation is very strong/efficient, the radiation reaction force will alter the high-energy particle distributions and possibly even the current sheet evolution.

Because ultrarelativistic particles emit synchrotron and inverse Compton emission narrowly beamed around their directions of motion, the angular dependence of particle spectra may have important consequences for observation of radiation along a particular line of sight \citep[see, e.g.,][]{Jaroschek_Hoshino-2009,Cerutti_etal-2012a,Cerutti_etal-2012b,Cerutti_etal-2013,Cerutti_etal-2014a,Cerutti_etal-2014b,Kagan_etal-2013,Christie_etal-2018,Werner_etal-2019,Mehlhaff_etal-2020,Sironi_Beloborodov-2020}.
For example, kinetic beaming---i.e., a strong energy-dependent anisotropy of particle spectra and hence radiation---observed in high-$\sigma_h$, 2D reconnection \citep{Cerutti_etal-2012b,Mehlhaff_etal-2020} may be sensitive to 3D effects and/or the current sheet configuration, even if the isotropically-averaged spectra are not.
Indeed, kinetic beaming has previously been observed to be present but weaker in 3D \citep[][]{Kagan_etal-2013,Cerutti_etal-2014b}; however, this issue needs systematic study over a range of $\sigma_h$, current sheet configurations, guide fields, and radiation strengths.

Since we have seen that the spatial distribution of magnetic field can be quite different in 2D and 3D (cf.~Fig.~\ref{fig:BvsXy2D3D}),
the spatial dependence of particle distributions may also be important in determining synchrotron emission.
For example, in 3D the nonlinear RDKI can result in a thickened current layer---a relatively large volume with greatly-diminished magnetic field.  As a result, particles accelerated by the nonlinear RDKI may radiate very little within the layer, emitting and cooling only as they exit the layer, when they suddenly experience nearly the full ambient magnetic field.  This could affect both the average radiated power (because radiation is suppressed in the layer) and the spectrum (because the magnetic field is mostly either full strength or low strength).
In contrast, there is less reason to believe that overall inverse Compton radiation would be different in 2D and 3D, unless the spectrum of soft photons (to be upscattered by high-energy electrons and positrons) is different---e.g., for synchrotron self-Compton, where the ``soft'' photons may be a result of synchrotron radiation.

If radiative cooling of particles is strong (compared with accelerating forces), the radiation reaction force can alter the spectra of accelerated particles.
This is especially interesting in the case of synchrotron radiation, because particles accelerated by 2D reconnection may experience significant acceleration in regions of low magnetic field, allowing them to exceed the synchrotron burnoff limit \citep[i.e., to exceed the energy at which the radiation reaction force in the ambient/upstream magnetic field would cancel electric-field acceleration,][]{Uzdensky_etal-2011,Cerutti_etal-2012a,Cerutti_etal-2012b}.
The different geometry of 3D current sheet evolution, not to mention different large-scale driving mechanisms (e.g., nonlinear RDKI vs. 2D reconnection), might affect whether and how particles can exceed the burnoff limit.
In \citet{Cerutti_etal-2014b}, particles accelerated by RDKI in 2D$yz$ geometry did not exceed the burnoff limit, although particles accelerated by a 3D evolving current sheet did (although not nearly as much as in 2D$xy$ reconnection); however, that was for one particular current sheet configuration and~$\sigma_h$ (though multiple guide field strengths were investigated).
It is possible that this conclusion would change for other systems, especially considering that we have shown that nonlinear RDKI is dramatically affected by the initial current sheet configuration~(see~\S\ref{sec:pertAndEta3d},~\ref{sec:RDKIamp}).

By showing that 3D effects as well as the initial configuration can substantially affect current sheet evolution, we have demonstrated the need to simulate radiation from these systems to determine observable signatures.
Although the insensitivity of NTPA spectra to these details may result in photon spectra that depend primarily on~$\sigma_h$ and~$B_{gz}/B_0$, this cannot be determined until we also understand the spatial and angular dependence of NTPA spectra and the emitted photon spectra.

\section{Conclusion}
\label{sec:conclusion}

Magnetic reconnection is a plasma process that is important in large part because it converts magnetic energy to plasma (particle) energy.
It may play a key role in relativistically-hot plasmas (including electron-positron pair plasmas) in a variety of astrophysical sources, accelerating particles to very high energies where they can emit observable radiation.
In general, but especially for astrophysical applications, we seek to understand the rate and amount of magnetic energy conversion as well as the development of nonthermal particle energy distributions that can be correlated with emitted radiation.
The primary way in which we can infer the role of reconnection in astrophysical sources, which cannot be directly probed, is to connect observable consequences of reconnection (i.e., radiation) with the source plasma conditions (such as magnetic field and plasma density), which can be inferred therefrom.
Most of the systematic studies of the effects of varying reconnection parameters in astrophysically-relevant relativistically-hot plasma have focused on highly-magnetized regimes.
We add to the literature this study focusing on the moderately-magnetized regime, $\sigma_h=1$, in which the ambient magnetic energy is roughly in equipartition with the ambient plasma energy.
It is especially important to understand the effect of
different values of the ``hot'' magnetization~$\sigma_h$ because it
exerts a substantial influence on reconnection: not only does it place an upper bound on the relative energy gain of particles, but it also determines the Alfv\'{e}n velocity, which controls the rate of reconnection.

This paper presents the first study that systemically investigates energy conversion and NTPA in reconnection across a wide variety of parameters in the ultrarelativistic $\sigma_h=1$ pair plasma regime.
Specifically, we explore the effects of different 
initial current sheet configurations, system-sizes (in 2D), guide
magnetic fields, and (in 3D) aspect ratios $L_z/L_x$.
To compare simulations we look particularly at the rates at which
magnetic energy is converted to plasma energy, as well as the resulting 
NTPA.
An extensive summary of briefly-stated important results can be found in~\S\ref{sec:summary}.

Our 2D simulations show that reconnection with $\sigma_h=1$ \emph{qualitatively} resembles relativistic reconnection in the $\sigma_h\gg 1$ regime;
however, they also measure significant quantitative differences---differences that could ultimately allow one to constrain~$\sigma_h$ in an astrophysical source based on observed radiation.

In contrast, our 3D simulations reveal substantially different qualitative behaviours, as well as (in many cases) significantly slower conversion of magnetic to plasma energy.
We find that in 3D there are are other mechanisms besides large-scale (2D-like) reconnection that can drive current sheet evolution and plasma energization.
It would therefore be more precise to say that we are studying the evolution of a thin current sheet in 3D rather than 3D reconnection.
Fascinatingly, regardless of the large-scale driver, NTPA remains robust, very similar in 2D and 3D---when compared at equivalent fractions of converted magnetic energy.

Results characterizing reconnection and energy conversion rates and NTPA in~2D for $\sigma_h=1$ are described in~\S\ref{sec:2d} and compared with the~$\sigma_h\gg 1$ regime in~\S\ref{sec:highSigmah}. By comparing to other works studying $\sigma_h\gg 1$, we see that $\sigma_h$ strongly influences these key outcomes.  In 2D, for $\sigma_h=1$ and zero guide field, we find that the dimensionless reconnection rate (i.e., normalized to $B_0 v_A$) is around~0.03, which is somewhat less than the~$\sim 0.1$ observed for $\sigma_h \gg 1$.  
Reconnection with $\sigma_h=1$ clearly yields NTPA, but generates a high-energy power-law energy spectrum $f(\gamma)\sim \gamma^{-4}$, significantly steeper than for~$\sigma_h\gg 1$.
Increasing the guide field $B_{gz}$ slows reconnection and suppresses NTPA, yielding still steeper power-law spectra.
We believe the guide field slows down reconnection by reducing the effective Alfv\'{e}n speed in the outflow direction, while the reduced compressibility of the guide-field-threaded plasma causes reconnection to end with more unreconnected flux remaining (in a closed system).
We expect these guide field effects to become significant for $B_{gz}/B_0 \gtrsim 1/\sigma_h$ (thus they should become evident only for much stronger guide field in nonrelativistic reconnection, where $\sigma_h\ll 1$ because $\sigma_h$ includes rest-mass enthalpy).
Moreover, we find that these outcomes in 2D are relatively insensitive to the details of the initial current sheet, indicating that the primary results of 2D reconnection (energy conversion and NTPA) may be robust functions of the ambient plasma, as hoped (to connect observations with astrophysical source conditions).

In 3D, the story is much different.  Unlike relativistic reconnection with $\sigma_h\gg 1$, 3D effects substantially alter reconnection with $\sigma_h=1$.
In fact, true reconnection may have a much diminished role in the 3D evolution of thin current sheets, with other processes (likely driven by the RDKI) competing to convert magnetic energy to plasma energy.
As a result, the morphology of the current sheet does not feature the intricately hierarchical plasmoid structure familiar from 2D reconnection simulations.
In 3D, plasmoids can decay, and instead of storing magnetic energy as in 2D, they release their magnetic energy to the plasma.
Thus 3D ``reconnection'' can convert upstream magnetic energy more completely to plasma energy.
However, the rate of energy conversion may be significantly 
(an order of magnitude) slower in 3D, even at early stages, and still 
slower at later stages.

Reconnection simulations in both 2D and 3D evolve stochastically or chaotically, following different paths from the initially random particle distributions.
However, the competition among different magnetic-energy-releasing mechanisms leads to a variety of long-term behaviours in 3D, in contrast to the relative inevitability with which 2D reconnection evolves to a common final state
(despite unpredictable chaotic plasmoid behaviour along the way).
In~3D, current sheet evolution is significantly more sensitive to the initial current sheet configuration than in~2D.  For example, an initial magnetic field perturbation (relatively inconsequential in 2D simulations) tends to force 3D evolution to resemble 2D-like reconnection.
The main competition to reconnection appears to come from the RDKI; importantly, we find that it might not be the fastest-growing RDKI mode that presents the stiffest competition, but rather the longest-wavelength, ``sufficiently fast-growing'' mode.
While the linear stage of the RDKI does not deplete much magnetic energy,
its nonlinear development---especially in large-amplitude modes---competes with reconnection to convert significant amounts of magnetic energy to plasma energy.
In particular, we have observed the RDKI-induced rippling of the current sheet grow to an amplitude comparable to its wavelength (without depleting much magnetic energy) and subsequently fold over on itself, rapidly converting most of the magnetic energy within the rippling amplitude to plasma energy.  
Because longer-wavelength modes reach larger amplitudes, they can have a larger impact; whereas short-wavelength modes, though they may grow faster, saturate at small amplitudes without obstructing reconnection.
We find that a completely different mechanism can thus deplete magnetic energy as fast as reconnection.  However, whereas reconnection results in a highly-structured chain of plasmoids that---despite monstrously large plasmoids---preserves thin (kinetic-scale) inter-plasmoid current sheets, this RDKI-triggered process results in a turbulent, generally-thickened current layer, which seems to disrupt reconnection.

Despite sometimes very large differences between 2D and 3D current sheet evolution---and even between different 3D evolutions---the resulting NTPA is remarkably similar in all cases (with the same guide magnetic field).

The presence of a guide magnetic field tends to suppress 3D effects, causing current sheet evolution to behave more like 2D reconnection.
This leads to an interesting non-monotonic trend in 3D with increasing guide field, since both 3D effects and guide field suppress reconnection.
A weak guide field tends to enhance the 3D reconnection rate by diminishing 3D effects;
but once the guide field is strong enough to suppress 3D effects nearly completely, further increase suppresses reconnection in 3D as in 2D.
Again, even with guide field, NTPA is very similar in 2D and 3D.
Increased guide field (unlike 3D effects) suppresses NTPA, leading to steeper power-law energy distributions. 

This investigation shows that 3D current sheet evolution is not necessarily a perturbation or modification of 2D reconnection, but rather involves a complicated interaction of linear and nonlinear stages of multiple instabilities.  This leads to a range of possible behaviours and an accompanying sensitivity to initial conditions that motivates further exploration of initial configurations and construction of parameter-space phase diagrams; these possibilities and sensitivities must then be considered in astrophysical modelling.  It also increases the urgency of understanding the current sheet formation process in the first place.  On the other hand, the diversity of behaviour highlights universalities in magnetic energy dissipation and resulting NTPA that could find important use in astrophysical models.  While the rate of magnetic energy dissipation can vary significantly depending on the details of the current sheet, all configurations yield, though by different means, ``fast'' magnetic energy release in the sense that it is tremendously faster than naive magnetic diffusion; and the nonthermal particle spectra are even more universal, determined primarily by the ambient~$\sigma_h$ and~$B_{gz}/B_0$.  These universalities will be invaluable for astrophysical modelling and may be important clues to a deeper understanding of magnetic energy dissipation and particle acceleration in plasmas.

\vspace{\baselineskip}
We would like to thank NSF, DOE, and NASA for supporting this work---in particular, grants NSF AST-1806084, NSF AST-1903335,
NASA ATP NNX16AB28G,
NASA ATP NNX17AK57G, NASA ATP 80NSSC20K0545.
In addition, this work would not have been possible without substantial supercomputing time.
A award of computer time was provided by the Innovative and Novel Computational Impact on Theory and Experiment (INCITE) program; for the 3D simulations, this research used resources of the Argonne Leadership Computing Facility, which is a DOE Office of Science User Facility supported under contract DE-AC02-06CH11357.
This work also used the Extreme Science and Engineering Discovery Environment (XSEDE), which is supported by National Science Foundation grant number ACI-1548562 \citep{XSEDE2014}; in particular, the large 2D simulations were run at the Texas Advanced Computing Center (TACC) at The University of Texas at Austin.

\appendix

\section{Grid resolution}
\label{sec:res2d}

Throughout this paper, we use a grid resolution of 
$\Delta x = \sigma\rho_0/3$.
For $\sigma_h=1$, this $\Delta x$ resolves the fundamental plasma 
scales: the (upstream) Debye length
$\lambda_D=(1/2)\sigma_h^{-1} \sigma\rho_0$ is marginally resolved, while the 
(upstream) collisionless
skin depth $d_e = \sqrt{3}\lambda_D$ and the Larmor radii of typical
upstream particles, 
$\rho_b \approx 3\theta_b \rho_0 = (3/4)\sigma_h^{-1} \sigma\rho_0$,
are slightly better resolved.
Although we explore different values of the 
initial current sheet half-thickness $\delta$, most simulations presented 
here use $\delta = (2/3) \sigma \rho_0$ (corresponding to $\eta=5$), giving 
4 cells over the initial current sheet thickness $2\delta$.

The main motivation to study the $\sigma_h \sim 1$ pair plasma regime is its
astrophysical application; however, it is also of interest because 
the small scales ($\lambda_D$, $d_e$, $\rho_b$) are nearly the same,
allowing maximum dynamic range (between the kinetic scales and system size)
with minimum computation.

\begin{figure}
\centering
\fullplot{
\includegraphics*[width=0.49\textwidth]{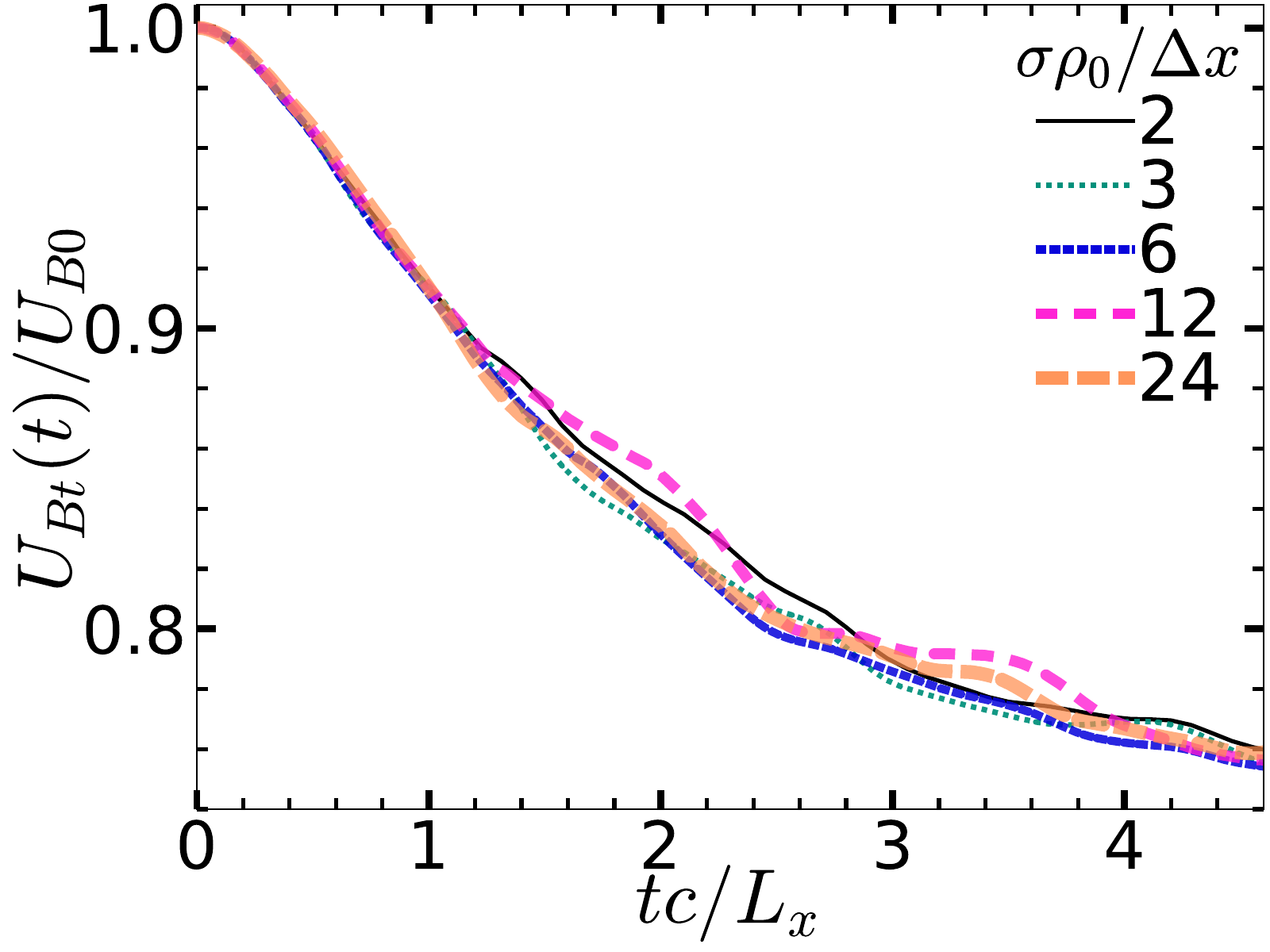}%
\hfill
\includegraphics*[width=0.49\textwidth]{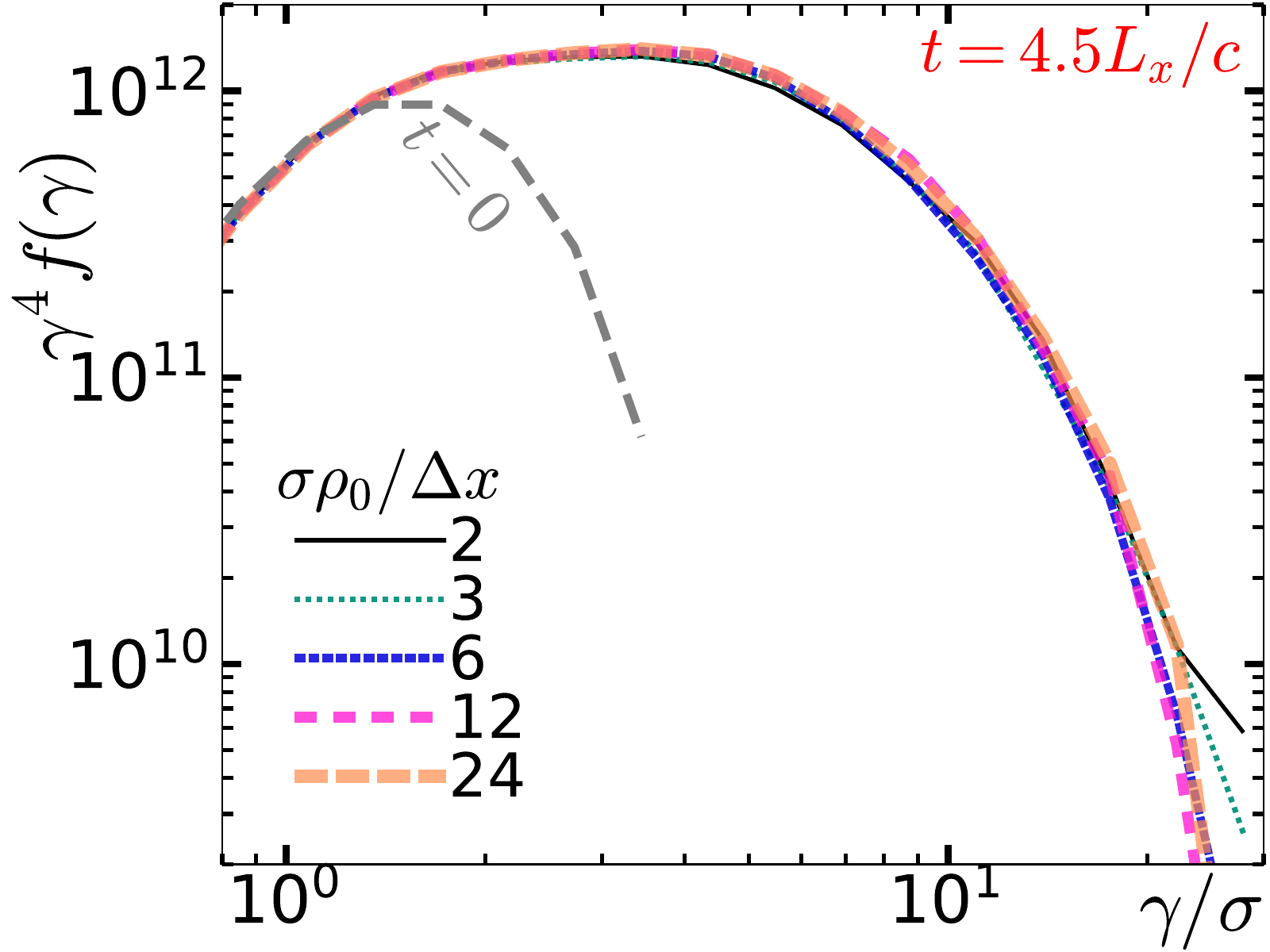}
}
\caption{ \label{fig:res2dLx320}
(Left) Transverse magnetic energy versus time, and (right) particle
energy spectra $f(\gamma)$, compensated by $\gamma^4$, at $t=4.5L_x/c$ (with the initial spectra at $t=0$ shown for comparison in dashed grey),
for the same 2D simulation with resolutions from 
$\Delta x=\sigma\rho_0/2$ to $\Delta x=\sigma \rho_0/24$.  These 
simulations have $L_x=320\sigma \rho_0$, $B_{gz}=0$, $\eta=5$ 
($\delta/\sigma\rho_0=2/3$), and an initial perturbation
$a=11$ ($s/\delta=5.4$), with 40 background particles per cell and 40 initially-drifting particles per cell.  
}
\end{figure}

We naturally 
want to use the coarsest resolution possible that captures the desired
plasma physics and yields stable simulation, so that we can reach the
largest possible system sizes with the least computation.
Figure~\ref{fig:res2dLx320} demonstrates the effect of changing 
resolution for 2D simulations of size $L_x=320\sigma \rho_0$, showing that the two
most telling diagnostics, magnetic energy evolution and NTPA spectra, are
the same (up to inherent stochasticity) for resolutions ranging from 
$\Delta x = \sigma \rho_0/2$ to $\Delta x = \sigma\rho_0/24$; simulations
were only run for $4.5L_x/c$ to save computation time.
For the rest of the paper, we use $\Delta x=\sigma\rho_0/3$.
This resolution becomes insufficient for very large systems
($L_x\gtrsim 2560 \sigma\rho_0$), which begin to exhibit poor 
energy conservation, where the total simulation energy, which
should be conserved, increases by more than a fraction of a per~cent.
However, we can show that even when the total energy grows by
$O$(1~per~cent), it does not significantly affect the results, 
aside from a small amount of extra heating (cf. Fig.~\ref{fig:cutoffVsLx2d}, right, where we compare $\Delta x=\sigma\rho_0/3$ and $\Delta x=\sigma\rho_0/4$ for a very large simulation).  

We note that another relevant length scale is the most unstable wavelength
of the RDKI (in the absence of guide magnetic field), 
$\lambda_{\rm RDKI} = 2\upi/k_{z,\rm RDKI} \simeq  2\upi (8\gamma_d \beta_d^2) \delta$ 
\citep{Zenitani_Hoshino-2007}.
For $\beta_d=0.3$, $\lambda_{\rm RDKI} \simeq 5 \delta$;
thus this most unstable mode is resolved somewhat better than the 
initial current sheet (of half-thickness $\delta$)---which is 
marginally resolved for $\Delta x = \sigma \rho_0/3$.  

In (2D) $y$-$z$ simulations, we do see that, when the initial current
sheet is marginally resolved, as for $\delta/\sigma \rho_0=2/3$ ($\eta=5$) 
and $\Delta x = \sigma \rho_0/3$, that RDKI is suppressed and, e.g., 
greater instability and magnetic depletion occurs for higher resolution,
e.g., $\Delta x = \sigma\rho_0/6$.  However, such a thin initial current
sheet requires initially-drifting particles with a temperature below that
of background particles ($\theta_d/\theta_b = 2\gamma_d/\eta \approx 0.4$).
In a 3D simulation, tearing and reconnection rapidly moves (hotter) 
background particles into the current sheet, thickening it.
Moreover, in~\S\ref{sec:pertAndEta3d}, where we specifically study the initial current sheet evolution in~3D, we include cases where the current sheet and linear RDKI modes are much better resolved.

\providecommand{\noopsort}[1]{}

\end{document}